\documentclass{aa}  
\usepackage{graphicx}
\usepackage{txfonts}
\usepackage{longtable}
\usepackage{natbib}
\bibpunct{(}{)}{;}{a}{}{,}
\begin{document}

\title{SNLS Spectroscopy: Testing for Evolution in Type Ia Supernovae}

\author{T.~J.~Bronder\inst{1}\fnmsep\thanks{The Figures in Appendix A are only available in electronic form via http://www.edpsciences/anada.org}\thanks{E-mail:jtb@astro.ox.ac.uk, current address: AFRL Directed Energy 3550 Aberdeen Ave SE, Kirtland AFB, Albuquerque, NM 87117}
 \and I.~M.~Hook\inst{1} \and P.~Astier\inst{2} \and D.~Balam\inst{3} \and C.~Balland\inst{2} \and
 S.~Basa\inst{4} \and R.~G.~Carlberg\inst{5} \and A.~Conley\inst{5} \and
 D.~Fouchez\inst{6} \and J.~Guy\inst{2} \and D.~A.~Howell\inst{5} \and
 J.~D.~Neill\inst{7} \and R.~Pain\inst{2} \and K.~Perrett\inst{5} \and
 C.~J.~Pritchet\inst{3} \and N.~Regnault\inst{2} \and M.~Sullivan\inst{5} \and
 S.~Baumont\inst{2} \and S.~Fabbro\inst{8} \and M.~Filliol\inst{9} \and
 S.~Perlmutter\inst{10} \and P.~Ripoche\inst{6}
}

\institute{
University of Oxford Astrophysics, Denys Wilkinson Building, Keble Road, Oxford OX1 3RH, UK 
\and                                        
LPNHE, CNRS-IN2P3 and University of Paris VI \& VII, 75005 Paris, France
\and
Department of Physics and Astronomy, University of Victoria, PO Box 3055, Victoria, BC V8W 3P6, Canada
\and
LAM, BP8, Traverse du Siphon, 13376 Marseille Cedex 12, France
\and
Department of Astronomy and Astrophysics, University of Toronto, 50 St. George Street, Toronto, ON M5S 3H4, Canada
\and
CPPM, CNRS-Luminy, Case 907, 13288 Marseille Cedex 9, France
\and
SRL California Institute of Technology, 1200 East California Blvd.~, Pasadena CA 91125 \and
CENTRA - Centro Multidisciplinar de Astrof\'{\i}sica, IST, Avenida Rovisco Pais, 1049 Lisbon, Portugal
\and
LAM CNRS, BP8, Traverse du Siphon, 13376 Marseille Cedex 12, France
\and
University of California, Berkeley and Lawrence Berkeley National Laboratory, Mail Stop 50-232, Lawrence Berkeley National Laboratory,
1 Cyclotron Road, Berkeley CA 94720 USA}

   \offprints{T. J. Bronder}

   \date{Received ; accepted }

 
  \abstract
{}
{We present a quantitative study of a new data set of high redshift
Type Ia supernovae spectra, observed at the Gemini telescopes
during the first 34 months of the Supernova Legacy Survey.  During this time 
123 supernovae candidates were observed, of which 87 have been identified
as SNe Ia at a median redshift of $z=0.720$. Spectra from the
entire second year of the survey and part of the third year (59 total
SNe candidates with 46 confirmed SNe Ia) are published here for the
first time.  The spectroscopic measurements made on this data set are used 
determine if these distant SNe comprise a population similar to those observed
locally.  }
{Rest-frame equivalent width and ejection velocity
measurements are made on four spectroscopic features. Corresponding
measurements are presented for a set of 167 spectra from 24 low-$z$
SNe Ia from the literature.}
{We show that there exists a sample at high redshift with properties
similar to nearby SNe. The high-$z$ measurements are consistent with
the range of measurements at low-$z$ and no significant 
difference was found between the distributions of measurements at low
and high redsift for three of the features. The fourth feature
displays a possible difference that should be investigated
further. Correlations between Type Ia SNe properties and host galaxy
morphology were also found to be similar at low and high $z$, and
 within each host galaxy class we see no evidence for
redshift-evolution in SN properties. A new correlation between SNe Ia
peak magnitude and the equivalent width of SiII absorption is presented. 
Tests on a sub-set of
the SNLS SNe demonstrates that this correlation 
reduces the scatter in SNe Ia luminosity distances in a manner
consistent with the lightcurve shape-luminosity corrections that are
used for Type Ia SNe cosmology.}
{We show that this new sample of SNLS SNe Ia has spectroscopic properties similar to nearby objects.}


   \maketitle

\section{Introduction}
\label{sec:intro}

Hubble diagrams making use of the `standardizeable' nature of Type Ia
SNe \citep{b47,b45,b32,b3} indicate the significant presence of a
`dark energy' that is driving the accelerating expansion of the
universe.  This result is additionally constrained by observations of
the power spectrum of the cosmic microwave backround \citep{b53,b52},
properties of massive clusters \citep{b1} and other large scale
structure observations \citep{b54}. The equation of state parameter
of this dark energy (defined as $w = p/\rho$, the ratio of dark energy
pressure $p$ to density $\rho$) has been investigated by recent
surveys \citep{b32,b4,b48}, but the results are still consistent with
a wide range of dark energy models. The CFHT Supernova Legacy Survey
[SNLS, \citet{b3}] aims to increase the measured precision of
this parameter by finding and following about 500 distant Type Ia SNe
over a five-year survey lifetime.

The fundamental assumptions of the methods used to estimate the
cosmological parameters with SNe are that the selected low-$z$ and high-$z$
objects have similar peak magnitudes (\emph{i.e.} within the locally observed
dispersion) and that the established lightcurve shape-luminosity
relations are applicable to these SNe at all redshifts. However, the current
understanding of SNe Ia explosions [see \citet{b23} for a review] and
their progenitors (PGs) illustrates that there may be some
differences, or evolution, in their inherent properties at high
redshifts (Hoflich et al 1998, Hatano et al 2000, Lentz et al 2000).
Such effects could distort the cosmological results of SNe Ia surveys. 
The differences in properties of SNe in hosts of
different morphology \citep{b61,b60,b21b,bhowell,b56,b17,b58} provides
additional evidence of SNe Ia heterogeneity. The luminosity-lightcurve
shape calibrations used for SNe Ia cosmology \citep{b121,b46,b20,b111}
make an empirical accomodation for the dispersion of SNe properties
observed locally, but these methods may be vulnerable to systematic
changes in the SNe Ia population at high redshifts \citep{brvw}.
Thus, it is imperative to undertake a quantitative investigation to
determine whether there are any changes in these objects that may affect
their reliability as distance indicators for cosmology.

The spectra of SNe Ia reveal more subtle characteristics than
photometric observations and are well suited for this investigation.
For example, measurements of photospheric ejection velocities
($v_{\rm{ej}}$) explore the kinetic energy and distribution of
elements in the expanding SN ejecta.  These measurements provided some
of the first evidence of intrinsic differences between Type Ia SNe
\citep{b9} and have also been shown to correlate with
luminosity \citep{b69,b70}. More recently, \citet{b6} used cluster
analysis of measurements of the decline rate of SNe Ia $v_{\rm{ej}}$
to yield a new way of grouping low-$z$ SNe that perhaps illuminates
the role of different explosion mechanisms in SNe Ia diversity.
Comparisons of SNe Ia ejection velocities from objects at different
redshifts have also been made [e.g. \citet{b27,b7,b123}]  
demonstrating that the range of velocities
measured in SNe Ia ejecta are comparable at all redshifts.  Folatelli
(2004, F04 hereafter) illustrated the utility of equivalent width
(EW) measurements on low-z SNe Ia spectra and \cite{bgabri} extended 
this work to compare 12 high-$z$ objects to the low-$z$ population.
\citet{b113} also illustrated correlations between these EW measurements 
and the velocity-gradient classification scheme from \citet{b6}.  \citet{b114} 
presents additional evidence for a temperature or explosion mechanism sequence
in SNe Ia with comparisons of different low-$z$ objects in EW-space.

These studies all illustrate how quantitative spectroscopic
measurements can be used to explore the factors that affect the
heterogeneity and homogeneity of SNe Ia.  The utility of these
parameters for exploring differences in high-$z$ SNe Ia is limited,
however, as the majority of these measurements cannot be reproduced in the 
high-$z$ data.  These limitations are due to the reduced signal-to-noise (S/N), redshift
effects (which shift the features to redder
regions where sky noise is dominant or moves features out of optical
detection limits entirely), host galaxy contamination and the smaller
number of observations which all characterise high-$z$ data.  

The analysis presented here includes rest frame equivalent width 
and ejection velocity measurements on a large set of SNe Ia spectra
observed for the SNLS with methods that are specifically tailored for
high-$z$ data. These measurements are similar to the recently published
work of \citet{bgabri} but are utilized on a much larger high-$z$ set and include
a different treatment of host galaxy contamination and estimates of the
low-$z$ trends.  The studied set of SNe were selected from the objects considered
`confirmed' SNe Ia by the SNLS [see Howell \emph{et al.} (2005), H05 hereafter]
and used in the survey's cosmological analysis.  
These high-$z$ results are compared to corresponding
measurements on a set of low-$z$ supernovae from the literature.  The
results of this investigation address the cosmological
implications from high-$z$ SNe surveys by determining whether this SNLS-set of 
distant objects comprises a population that is spectroscopically similar to their 
low-$z$ counterparts.  An overview of the data sets used in
this analysis is presented in Sect.~\ref{sec:data}.  The methods for
measureing EW and $v_{\rm{ej}}$ are presented in
Sect.~\ref{sec:analysis}. The results and a comparison of the two
SNe sets are presented in Sect.~\ref{sec:results}, in Sect.~\ref{sec:mgII_diff}
these results are discussed in more detail and the
conclusions are presented in Sect.~\ref{sec:conclusion}.

\section{Spectroscopic Data Sets --- Overview}
\label{sec:data}

A large number of the SNe Ia that were used to provide the latest
constraints on $w$ and other cosmological parameters \citep{b3,b52}
are studied in this paper. Specifically, this includes the SNLS
objects observed at the Gemini telescopes during the first year of the
survey. This set is augmented by Gemini spectra from the entire
2$^{\rm{nd}}$ year and part of the 3$^{\rm{rd}}$ year of SNLS
observations, which are published here for the first time.  This
section briefly describes the observation, data reduction, and
identification of these SNLS SNe spectra in
Sect.~\ref{sec:hiz_data}.  In Sect.~\ref{sec:lowz_data} a summary
of the set of low-$z$ SNe that were quantitatively compared to the
distant SNe Ia is provided.

\subsection{High Redshift SNLS Supernovae}
\label{sec:hiz_data}

\subsubsection{SNLS Spectra -- Observation, Data Reduction and Classification}
\label{sec:snls_spect_sum}

Full details of the methods used for SNLS SN candidate selection and
observation can be found in \citet{b58} and \citet{b30}, respectively.  In summary, the
Canada-France-Hawaii Telescope Legacy Survey (CFHTLS), using the
Megacam wide field imager, acquires real time lightcurves of possible
SNe by repeatedly imaging four, one square-degree fields approximately
every four days (observers' frame) in several filters (a combination
of g', r', i', and z') as part of the Deep component of the CFHTLS.
Spectroscopic observations are then made at 8m-class telescopes in
order to identify the SN-type and redshift of these candidates.

The objects studied here were observed at the Gemini North and South
telescopes with the Gemini Multi Object Spectrograph [GMOS; \citep{b25}]
in longslit mode.  Typically, three 20 minute to 30 minute
exposures are taken of each candidate.  The GMOS R400 grating (400
lines per mm) and 0.75'' slit are used, which gives a 6.9 \AA\ spectral 
resolution. A central wavelength of 680nm was chosen for objects at
an estimated $z \lesssim 0.4$ and objects at higher estimated $z$ are
observed with a central wavelength setting of 720nm.  This central
wavelength selection is important in order to ensure that the defining
SiII features (located at rest wavelengths of $\sim 4100$ \AA\ and
$\sim 6150$ \AA) of SNe Ia spectra are observed.  The 680 nm setting
was paired with the GG455\_G0305 order-sorting filter to provide
wavelength coverage from 4650 \AA\ to 8900 \AA.  The 720 nm setting
was used along with the OG515\_G0306 order-sorting filter to cover
wavelengths from 5100 \AA\ to 9300 \AA.  Queue observing was employed
to observe the SNLS candidates, which require better than 0.75'' image
quality (corrected to zenith), in either classical or nod-and-shuffle
(N\&S) observing modes (see H05).

The reduction of the spectra was completed using IRAF\footnote{IRAF is
written and supported by the IRAF programming group at the National
Optical Astronomy Observatories (NOAO). NOAO is operated by the
Association of Universities for Research in Astronomy (AURA),
Inc. under cooperative agreement with the National Science
Foundation}, independently from that of H05, but with very similar
methods. The independent reductions were made to check for any
systematic differences in the pipelines and the results from all
reductions were in very close agreement.

To classify the SNLS objects, the final spectra are quantitatively
compared to template supernovae from large library of low-$z$ objects
with a $\chi^2$ matching program [see \citet{b29} and H05 for
details].  The results from this matching program and the opinions of
SNLS members are used to rank the identifications according to one
`confidence index' (CI) from the six categories defined in H05, with a
5 being a `Certain Ia' and a 0 as `Not a Ia'.  Objects that are ranked
at a CI of 3 or greater are considered `confirmed' Type Ia SNe.  These
are the objects that were used in the cosmology fits from \citet{b3}
and also make up the 80 high-$z$ SNe Ia studied here.

The selection criteria of the SNLS sample means that Type Ia SNe at
high-$z$ with properties that are very different from those seen
nearby are likely to be rejected. Thus in the comparisons below the
existence of a new population of high-z SNe with unusual properties
cannot be ruled out. However as will be seen later, the selection
process does allow a range of SNIa properties through, and thus tests
for evolution within this range are possible.
 
\begin{figure*}
\begin{minipage}{2.0\textwidth} 
\begin{tabular}{cc}
\includegraphics[width=7.85 cm] {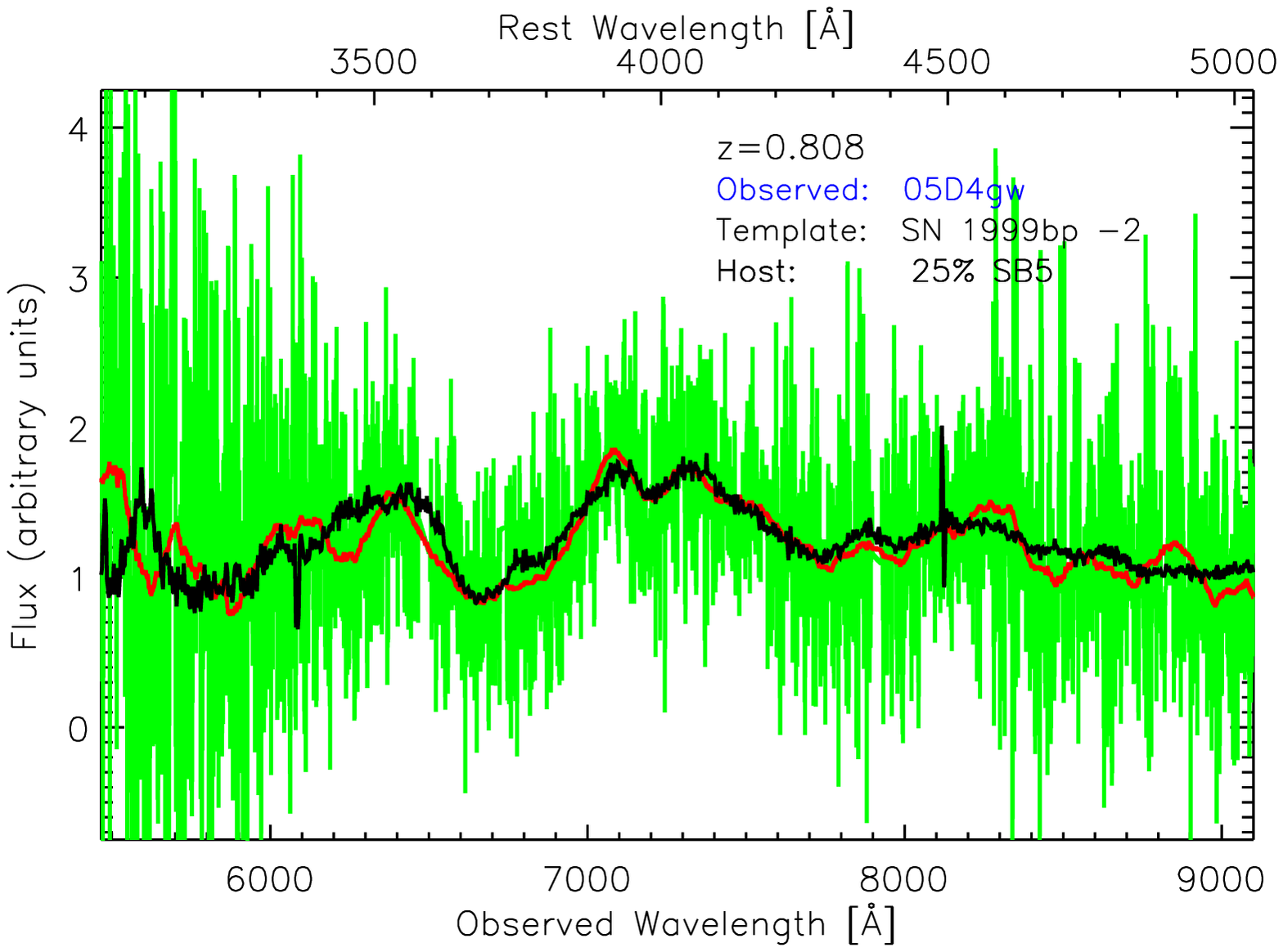} & \includegraphics[width=7.85 cm] {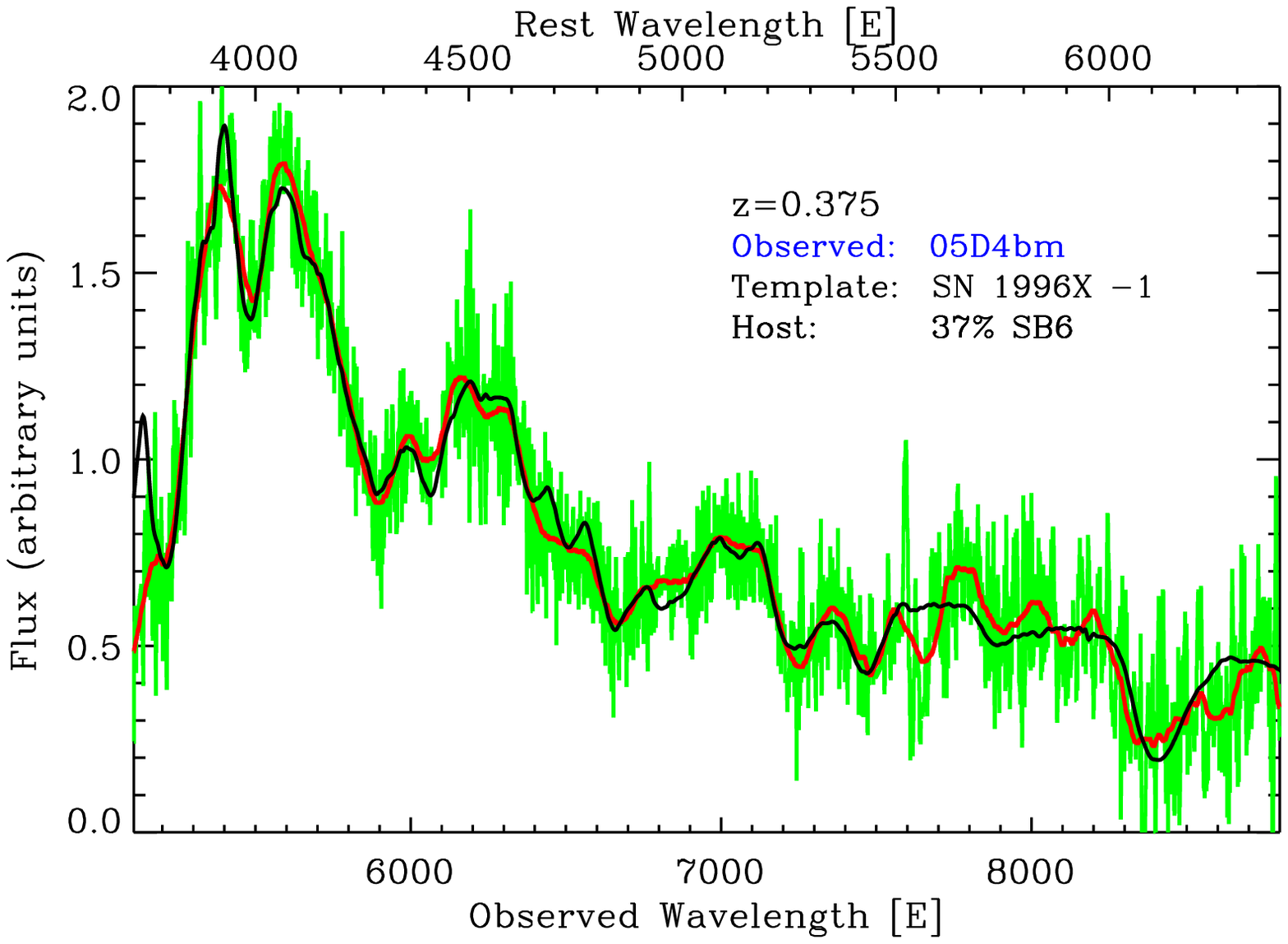} \\
\includegraphics[width=7.85 cm] {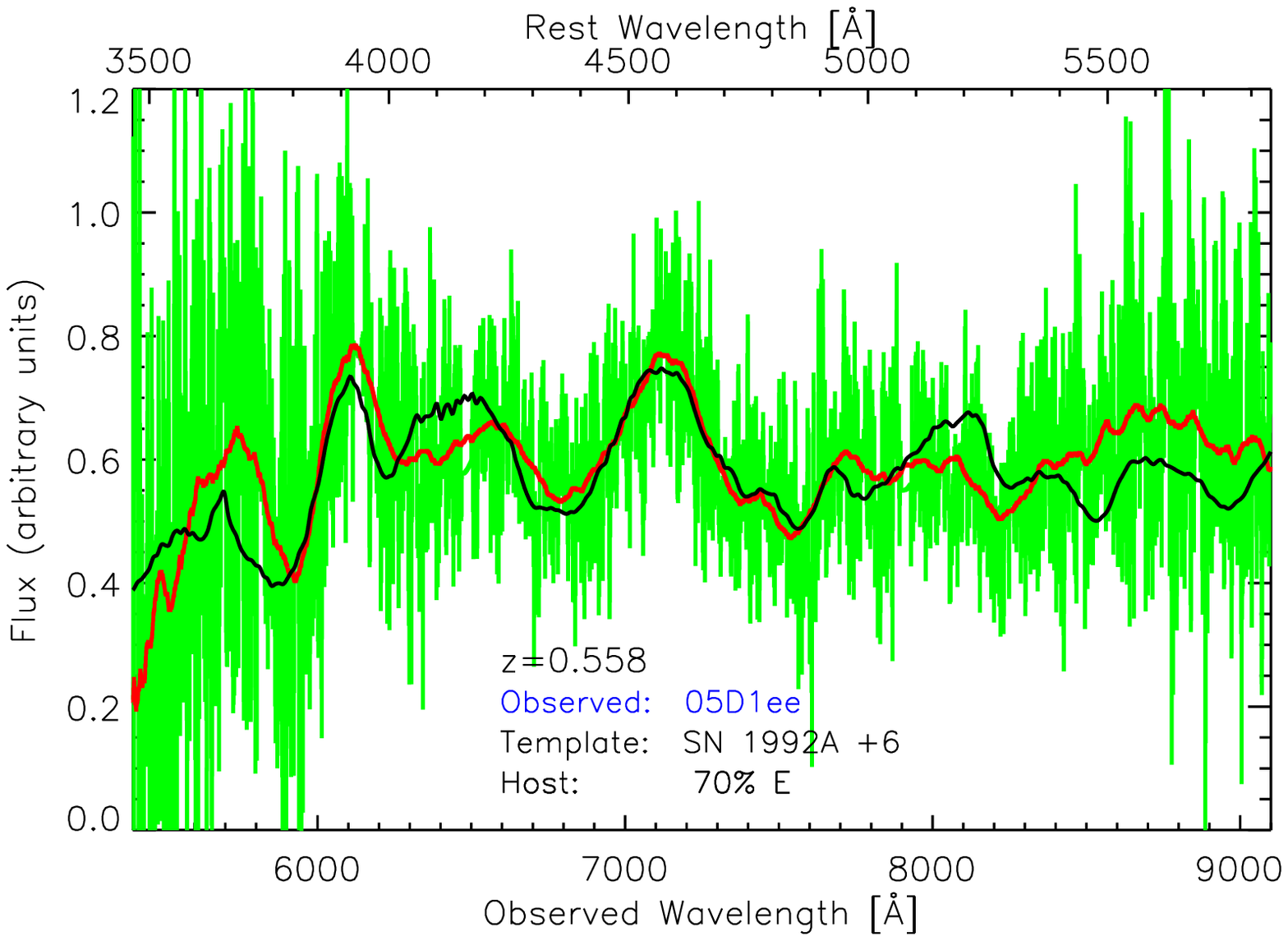} & \includegraphics[width=7.85 cm] {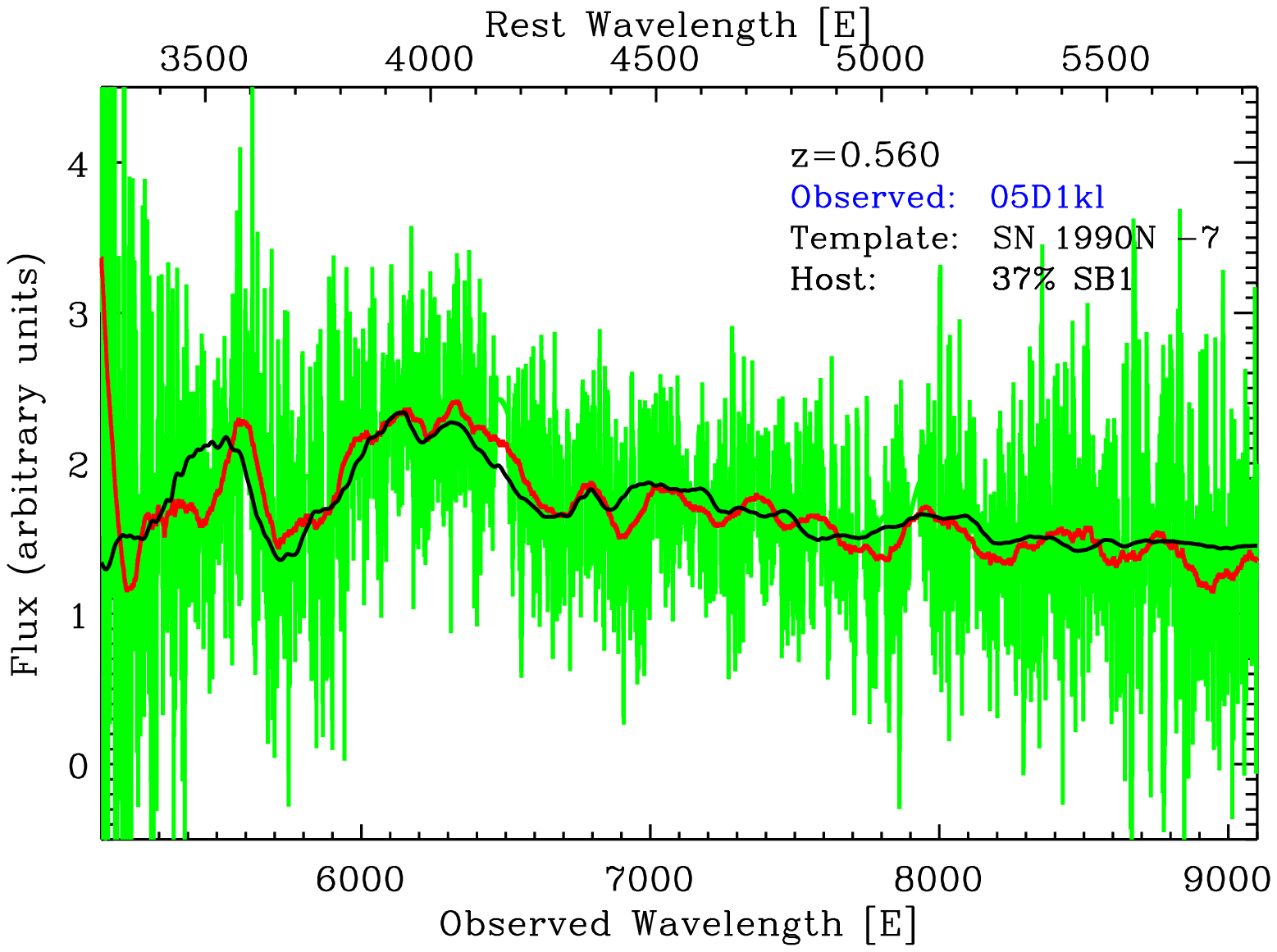} \\
\end{tabular}
\end{minipage}
\caption{\label{fig:hiz_IDs} Example spectra of SNLS candidates
observed by Gemini.  The observed spectra after subtraction of the
best-fitting host galaxy template are displayed in green.  The
smoothed spectra are shown in red and the best fitting template SNe
are overplot in black.  The top two objects are `certain type Ia' SNe
(CI = 5) as they have clear signs of SiII.  The bottom objects are
SNLS candidates with CI = 4 (\emph{left}) and CI = 3 (\emph{right}).
These spectra match SNe Ia templates but have ambiguous SiII
absorption.  All of the other spectra are available online.  }
\end{figure*}

\subsubsection{Summary of the SNLS Dataset}
\label{sec:snls_yr1}

The SNe candidates observed during the first year of the survey at the
Gemini telescopes were published in H05. Table 2 of H05 lists the
derived properties of these spectra.  This set includes 64 candidates
observed from August 2003 through October 2004.  Within this set, 41
objects were classified with as confirmed SNe with a CI $\geq 3$ and
are studied here.

Spectra from the entire second year and part of the third year of SNLS
observations at Gemini are presented here for the first time.  These
spectra were observed between November 2004 and May 2006.  During this
period 59 SNe candidates were observed and 46 of these objects were
confirmed as SNe Ia. The setup and conditions of the observations of
these objects are listed in Table \ref{tab:gemobs}.  The derived
properties and final classifications of these spectra are summarized
in Table \ref{tab:gemprop}. Some example spectra are presented in
Fig. \ref{fig:hiz_IDs} and all of the spectra are available in the
on-line version of this paper.

For the analysis in this paper we make use of not only spectral
properties but also photometric properties. The latter are derived
from lightcurve fits to the SNLS CFHT photometry, following the
methods described in \cite{b3}. The main lightcurve parameters used
are the date of B-band maximum light, and percentage increase relative
to the host galaxy (interpolated to the date of spectroscopy).

\subsection{Low Redshift Type Ia Supernovae}
\label{sec:lowz_data}

\subsubsection{Low-$z$ Spectra}
\label{sec:lowz_spect}

The spectra that make up the low redshift sample in this study were
selected from the literature. A majority of these spectra can be found
on the online Supernova Spectrum Archive (SUSPECT\footnote{{\small
\texttt{http://bruford.nhn.ou.edu/suspect/index1.html}}}).  The main
goal of this study is to quantify the similarities and differences
between high-$z$ SNe Ia and those discovered locally, so objects with
epoch and wavelength coverage that approximated that of the SNLS
sample were selected.  These spectra were corrected for peculiar
velocities and redshift where appropriate.  All of the low-$z$ objects
that met this criteria are listed in Table \ref{tab:lowzsrc}.

\begin{table*}
\begin{minipage}{150mm}
\caption{\label{tab:lowzsrc} The low redshift spectra analyzed in this
study. The dates correspond to the epoch relative to maximum
luminosity in the B-band.  Many of these spectra are available in the
SUSPECT archive as well as their listed publications.}
\begin{center}
 \begin{tabular}{cll}
 \hline \hline
Name & Date(s)    & Reference\\
     &            &  \\
 \hline
1981B & 0,20 & Branch \emph{et al.} 1983 \\ 
1986G & -5,-4,-3,0 & Phillips \emph{et al.} 1987 \\ 
1989B & -7,-5,-3,-2,-1,3,5,8,9,11,12,13,16,17,18,19 & Barbon \emph{et al.} 1990, Wells \emph{et al.} 1994 \\ 
1990N & -14,-7,2,3,17 & Mazzali \emph{et al.} 1993 \\ 
1991M & 3 & Gomez \& Lopez 1998 \\ 
1991T & -12,-11,-10,-8,-7,-6,0,10,15 & Mazzali, Danziger, \& Turatto 1995 \\ 
1991bg & 0,1,2 & Turatto \emph{et al.} 1996 \\ 
1992A & -5,-1,3,5,6,7,9,11,16,17 & Kirshner \emph{et al.} 1993 \\ 
1994D & -11,-10,-9,-8,-7,-5,-4,-2,2,4,10,11,12,13,20 & Patat \emph{et al.} 1996 \\ 
1994Q  & 10 & Gomez \& Lopez 1998 \\ 
1996X & -2,-1,0,1,2,7 & Salvo \emph{et al.} 2001 \\ 
1997br & -9,-8,-7,-6,-4,8 & Li \emph{et al.} 1999 \\ 
1997cn & 3 &  Turatto \emph{et al.} 1998 \\ 
1998aq & -8,0,1,2,3,4,5,6,7 & Branch \emph{et al.} 2003 \\ 
1998bu & -4,-3,-2,8,9,10,11,12,13 & Capellaro \emph{et al.} 2001 \\ 
1999aa & -11,-3,-1,5,6 & Garavini \emph{et al.} 2004 \\ 
1999ac & -15,-9,0,2,8,11 & Garavini \emph{et al.} 2004 \\ 
1999aw & 3,5,5,12 & Strolger \emph{et al.} 2002 \\  
1999by & -5,-4,-3,-2,3,4,5,6,7,8,11 & Garnavich \emph{et al.} 2004 \\ 
1999ee & -11,-9,-8,-6,-4,-2,-1,0,3,5,7,8,9,12,14 & Hamuy \emph{et al.} 2002 \\ 
2000E & -2,3,5,14 & Valentini \emph{et al.} 2003 \\ 
2000cx & -4,-3,-2,-1,0,1,5,6,7,9,11 & Li \emph{et al.} 2001 \\ 
2002bo & -14,-13,-11,-7,-6,-5,-4,-3,-2,-1,4 & Benetti \emph{et al.} 2004 \\ 
2003du & -11,-7,13 & Anupama, Sahu, \& Jose 2005 \\ 
\hline
\end{tabular}
\end{center}
\end{minipage}
\end{table*}

The low-$z$ sample includes 167 spectra from 24 different objects.
The set includes the so-called `core normal' \citep{b114} objects
[\emph{i.e.} SNe similar to SN 1981B \citep{b101}, SN 1989B \citep{b124,b69},
SN 1992A \citep{b125} or SN 1994D \citep{b126}] as well as
spectroscopically peculiar SNe Ia such as the overluminous
1991T-like objects and underluminous 1991bg-like SNe.

\subsubsection{Low-$z$ photometry and distance estimation}
\label{sec:lowz_phot}

The photometric properties of these nearby objects are listed in
Table \ref{tab:lowz_phot}. These properties have been presented in the literature
\citep{b45,b46,b109,b74} with slightly different results depending on
the color corrections used or the lightcurve fits to the published
photometry.  To remove systematic differences when comparing results, the photometry from these SNe was re-fit by SNLS members to derive the magnitudes and lightcurve shape parameters used later in this analysis.  This lightcurve fitting used a new template spectrum for K-corrections [see \citet{bhsiao} and Conley \emph{et al.} (\emph{in prep}) for full details].  The parameters derived from these fits are summarized in Table \ref{tab:lowz_phot}.

These re-derived photometric properties were also used to estimate the
absolute peak (B') magnitudes, $M_{\rm{B_{peak}}}$, of the low-$z$ SNe
Ia.  A new estimate of these magnitudes is of interest as the
published values have [in general, see \citet{b46} and \citet{b109} for exceptions] already been corrected for a
derived lightcurve shape-luminosity correction.  In order to fully
explore the homogeneity of SNe at different redshifts and to investige
any possible spectroscopic sequences in SNe Ia, an uncorrected, or
intrinsic, $M_{\rm{B_{peak}}}$ is desired.

To make this estimation, the observed magnitudes from Table \ref{tab:lowz_phot} 
were first corrected for host galaxy reddening using the relation between
$\Delta m_{15}(B)$ (estimated from the re-fit lightcurves) and $B-V$
colour described by Eq. 7 in \citet{b46}.  Combined with the relations
in Eq. 9 and Eq. 11 from the same publications, the observed and
estimated `true' $B-V$ colours were approximated to constrain the host
galaxy reddening for each SN. This is the only step of the estimate
that made use of a SNe Ia lightcurve shape-luminosity correction and
is unavoidable as some assumption of the intrinsic colour of these
objects is necessary to infer the host galaxy reddening.  This
reddening was then converted to a B'-extinction with $R_B = 3.0$.  This is
not the standard value used in this exercise and does not take into account
extremely red objects \citep{bnancy}, but it is approximately the median 
value of the different $R_B$ estimates in the literature.  The selection 
of this $R_B$ value does not have a large effect on the results in 
this paper.  A different $R_B$ would
systematically shift all the values in Table \ref{tab:lowz_phot} a slight
amount, but leave the relative relationship between the magnitude of the
different SNe the same, which is the true parameter of interest in this 
study as illustrated in Fig. \ref{fig:siII_mags}.

Next, the distance modulus, $\mu$, to the host galaxy was used to convert to an absolute magnitude
via $\mu = m - M$ (where $m$ and $M$ are the observed and absolute
magnitudes, respectively).  To yield the most correct distance
modulus, the authors used the most reliable Cepheid variable based
distances to the SNe host galaxies that were available in the
literature.  These distances were available for 9 of the SNe in Table
\ref{tab:lowzsrc} and another 5 low-$z$ SNe had reliable host distances from the
\emph{Nearby Galaxies Catalogue} \citep{b112}.  The remaining objects
did not have any independent distance constraints and were not
considered for these $M_B$ estimates. The errors in each of these
steps were propagated through this estimate to provide a final error
on $M_{\rm{B}}$.  The final column in Table \ref{tab:lowz_phot} lists these
intrinsic peak magnitude values for the 14 low-$z$ SNe that had
reliable, independent distance estimates and secure observed
photometric properties.

\begin{table*}
\centering
 \begin{minipage}{150mm}
\caption{\label{tab:lowz_phot} The photometric properties of the
low-$z$ spectra analyzed in this study. The published photometry from
these objects was re-fit to determine the magnitudes and lightcurve
shape parameters shown here.  SN 1991M did not have published
photometry in B-band, so the parameters for this object were not
determined.  The SNLS models did not fit the sub-luminous SN 1991bg
with any accuracy; the values given here for this unusual object are
from \citet{b109}.  The precision of the derived values for the
unusual SN 2000cx are also in doubt and were not used in any of the
analysis in this paper. Uncertainties in each quantity are given in
brackets.}
 \begin{tabular}{lcccccc}
 \hline \hline
Name  &  $z$   & MJD$_{\rm{max}}^{\rm{a}}$     &  $m_{\rm{B}}^{\rm{b}}$   & $(B - V)_{\rm{max}}^{\rm{c}}$  & $s^{\rm{d}}$ & M$_{\rm{B}}^{\rm{e}}$ \\
\hline
1981B & 0.0060 & 44671.84(0.09) & 11.83(0.01) & -0.06(0.01) & 0.95(0.01) & -19.58(0.22)$^{(1)}$ \\
1986G & 0.0027 & 46560.53(0.05) & 11.91(0.01) & 0.83(0.01) & 0.73(0.01) &  $\cdots$ \\
1989B & 0.0024 & 47564.39(0.27) & 12.12(0.02) & 0.30(0.02) & 0.94(0.01) &   -19.03(0.31)$^{(1)}$ \\
1990N & 0.0034 & 48081.92(0.03) & 12.53(0.01) & -0.09(0.01)& 1.07(0.01) &  -19.59(0.23)$^{(1)}$ \\
1991M & 0.0072 & $\cdots$       & $\cdots$    & $\cdots$    & $\cdots$  & $\cdots$ \\
1991T & 0.0060 & 48374.25(0.04) & 11.30(0.01) & -0.04(0.01) & 1.05(0.01) &  -20.07(0.27)$^{(2)}$ \\
1991bg& 0.0030 & 48604.50(1.0) & 14.75(0.04) & 0.00(0.08)  & .68(0.10) & $\cdots$ \\
1992A & 0.0063 & 48639.71(0.02) & 12.41(0.01) & -0.09(0.01) & 0.85(0.01) &  -18.73(0.22)$^{(7)}$ \\
1994D & 0.0015 & 49431.41(0.01) & 11.64(0.01) & -0.19(0.01) & 0.84(0.01) &   -19.44(0.22)$^{(7)}$ \\
1994Q & 0.0294 & 49494.35(0.90) & 16.16(0.09) & -0.14(0.04  & 1.11(0.04) & $\cdots$ \\
1996X & 0.0070 & 50190.65(0.10) & 12.86(0.01) & -0.12(0.01) & 0.90(0.01) &  -19.38(0.22)$^{(7)}$ \\
1997br & 0.0053 & 50558.57(0.03) & 13.23(0.02) & 0.12(0.01) & 0.89(0.01) & $\cdots$ \\
1997cn & 0.0162 & 50593.36(0.32) & 18.12(0.02) & 0.92(0.01) & 0.89(0.02) & $\cdots$ \\
1998aq & 0.0037 & 50930.16(0.02) & 12.18(0.01) & -0.25(0.01) & 0.98(0.01) &  -19.87(0.23)$^{(6)}$ \\
1998bu & 0.0030 & 50951.89(0.04) & 11.97(0.01) & 0.15(0.01)  & 0.99(0.01) &  -19.18(0.26)$^{(1)}$ \\
1999aa & 0.0144 & 51231.82(0.03) & 14.59(0.01) & -0.18(0.01) & 1.12(0.01) & $\cdots$ \\
1999ac & 0.0095 & 51250.07(0.08) & 13.99(0.01) & -0.08(0.01) & 1.02(0.01) & $\cdots$ \\
1999aw & 0.0380 & 51253.34(0.08) & 16.59(0.01) & -0.17(0.01) & 1.22(0.01) & $\cdots$ \\
1999ee & 0.0114 & 51468.89(0.01) & 14.69(0.01) & 0.11(0.01)  & 1.05(0.01) &  -19.70(0.30)$^{(3)}$ \\
2000e & 0.0047 & 51576.44(0.03)  & 12.68 (0.01) & 0.01(0.01) & 1.03(0.01) &   -20.08(0.25)$^{(7)}$ \\
2000cx & 0.0079 & 51752.03(0.03) & 12.92(0.01) & -0.09(0.01) & 0.91(0.01) &  $\cdots$ \\
2002bo & 0.0042 & 52356.43(0.04) & 13.82(0.01) & 0.29(0.01)  & 0.95(0.01) &  -19.04(0.46)$^{(7)}$ \\
2003du & 0.0064 & 52766.22(0.19) & 13.36(0.01) & -0.25(0.01) & 1.02(0.01) &  -19.53(0.45)$^{(5)}$ \\
\hline
\end{tabular}

\medskip

\small

a -- date of maximum luminosity (rest-frame B') 

b -- maximum apparent magnitude in rest-frame B' magnitudes,
corrected for Milky Way extinction with E(B-V) values from Schlegel et
al. (1998)

c -- $B-V$ color at (B-band) maximum luminosity

d -- stretch values, based on fitting a fiducial $s = 1.0$ Type Ia
template

e --- absolute peak luminosities (B'), \textbf{not} adjusted for
lightcurve shape-luminosity corrections. These made use of the
following distance modulus references - 1: \citet{blowz1} 2:
\citet{blowz2} 3: \citet{blowz3} 4: \citet{blowz4} 5: \citet{blowz5}
6: \citet{blowz6} 7: \citet{b112}
\end{minipage}
\end{table*}

\section{Spectroscopic Analysis Techniques}
\label{sec:analysis}

\subsection{Equivalent Widths}
\label{sec:ew_defn}  

\begin{figure}\label{fig:f04}
\begin{minipage}{.40\textwidth}
\centerline{\includegraphics[width=7.85 cm]{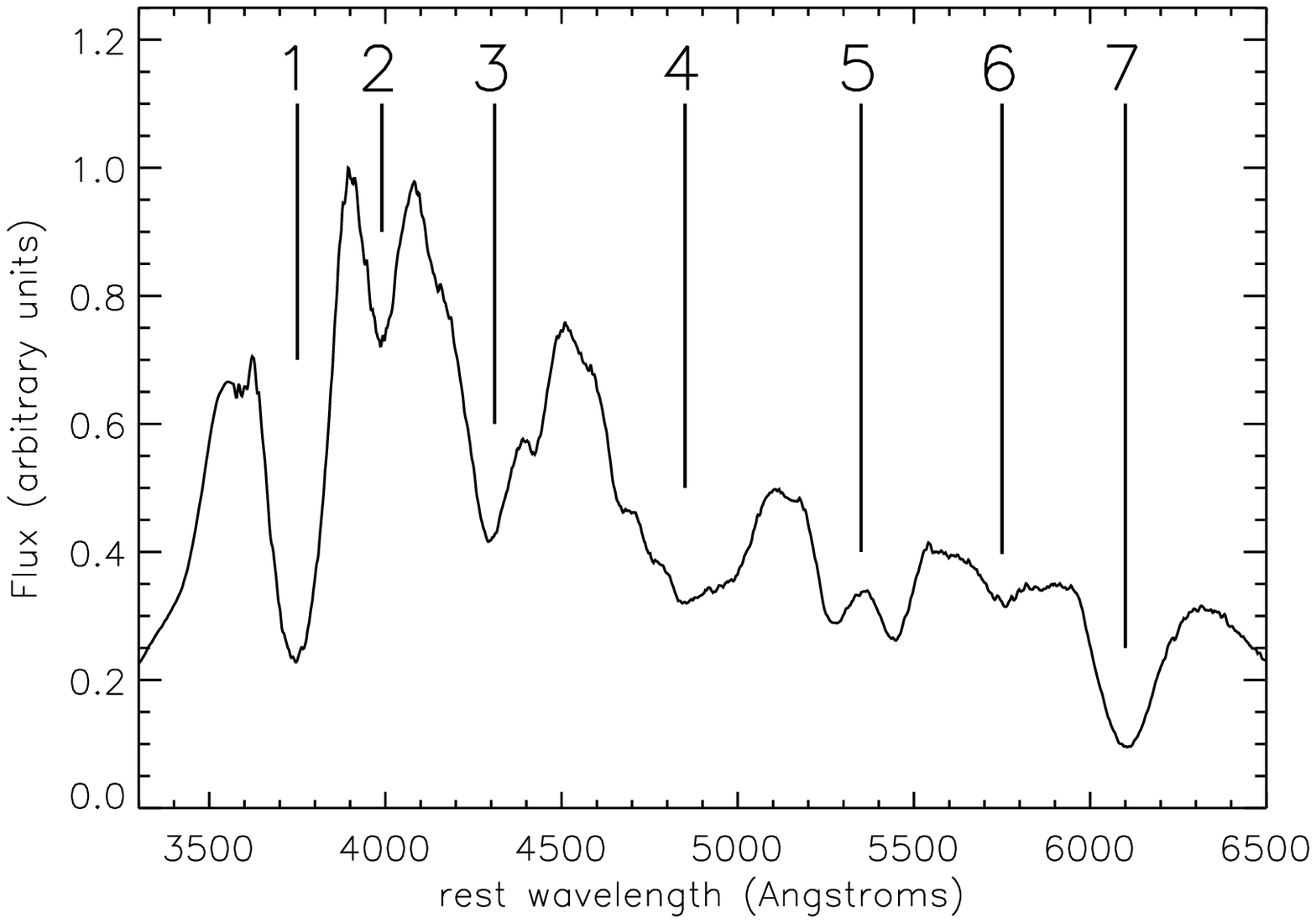}}
\end{minipage}

\caption{\label{fig:f04} The spectroscopic regions of specific
interest to SNe Ia EW measurements, as defined by F04, are shown here
on a core normal SN Ia near maximum light. The wavelength ranges and
nomenclature for these features is summarized in Table~\ref{tab:f04}.}

\end{figure}

\begin{table}
\caption{Wavlength regions for the SNeIa features defined by F04.}
\begin{tabular}{clcc}
\hline \hline
{\small Feature}&{\small Label}&{\small Blue-ward}     &{\small Red-ward}\\
{\small Number }&               &{\small Limit (\AA)} &{\small Limit (\AA)}\\
\hline
1 & `CaII H\&K'  &  3500 - 3800 & 3900 - 4100 \\
2 & `SiII 4000'  &  3900 - 4000 & 4000 - 4150 \\
3 & `MgII'       &  3900 - 4150 & 4450 - 4700 \\
4 & `FeII'       &  4500 - 4700 & 5050 - 5550 \\
5 & `SII W'      &  5150 - 5300 & 5500 - 5700 \\
6 & `SiII 5800'  &  5550 - 5700 & 5800 - 6000 \\
7 & `SiII 6150'  &  5800 - 6000 & 6200 - 6600 \\
\hline
\end{tabular}
\label{tab:f04}
\end{table}

Rest frame equivalent width measurements are a shape-independent
method of quantifying the strength of spectroscopic features. 
The method used in this EW analysis is similar to F04 and \citet{bgabri}, which defined eight regions
that contain the strongest absorption features observed in core normal
SNe Ia spectra near maximum luminosity.  The first seven regions are
illustrated in Fig. \ref{fig:f04}, and the nomenclature and
wavelength ranges used to define them are given in
Table~\ref{tab:f04}. For supernovae, the continuum needed to make the
EW measurement is not well defined, so it is estimated by making a
straight line fit between the local maxima that bound each feature.
The following rules govern how the continuum is set:

\begin{itemize}
 \item \noindent the pseudo-continuum is always set within the feature regions listed in Table~\ref{tab:f04}
 \item \noindent the feature boundaries are selected to maximise the wavelength span of the measurement without intersecting the boundaries of any neighboring features
 \item \noindent the EW is always calculated in the rest frame
\end{itemize}

The continuum is set automatically but a human judgement is made on
its placement to ensure the guidelines above are followed. The EW is
then calculated by numerically integrating underneath this
`pseudo-continuum' at every wavelength pixel $\lambda_i$ (for all $N$
points within the set continuum bounds):

\begin{equation}
EW = \Sigma^{N}_{i=1} \left( 1-\frac{f_\lambda(\lambda_i)}{f_c(\lambda_i)} \right) \Delta \lambda_i
\label{eq:ew_calc}
\end{equation}

\noindent where $f_\lambda$ is the flux level of the spectrum, and
$f_c$ is the flux of the pseudo-continuum.  The statistical
uncertainty in the EW measurement is calculated by propagating the
estimated uncertainties in the flux and pseudo-continuum:

\begin{equation}
\sigma_{EW}^2 = \Sigma^{N}_{i=1} \left( \{ \frac{\sigma_f(\lambda_i)}{f_c(\lambda_i)}\}^2 + \{\frac{f_\lambda(\lambda_i)}{f_c(\lambda_i)^2}\}^2 \cdot \sigma_{c_i}^2 \right)  \Delta \lambda_i^2
\label{eq:ew_sigma}
\end{equation}

\noindent here $\sigma_f$ is the measurement uncertainty in the flux
and $\sigma_{c_i}$ is the uncertainty in the pseudo-continuum flux.
The $\sigma_{c_{i}}$ is calculated via the propagation of the
correlated errors in the selection and calculation of this continuum.

In addition to the error calculated in Eq. \ref{eq:ew_sigma},
uncertainties can be caused by variations in the extent or the slope
of the pseudo-continuum.  The range and effects of these possible
pseudo-continuum variations were estimated by randomly shifting the
continuum boundaries within $\frac{R}{4}$ \AA\ of the selected maxima
(where $R$ is the range of the feature boundaries, see
Fig. \ref{fig:f04}) for the feature and spectrum in question.  The EW
was then re-measured, and the standard deviation of these measurements
(which were repeated 50 times) was accepted as an approximation of
this error in each EW measurement.  For the high S/N spectra, the
possible variations in the continuum were found to be the dominant
source of error.  The high-$z$ data was subject to some additional
systematic errors, so this general procedure had to be amended for
these objects; this is discussed in the following section.

In calculating the low-$z$ EW, any possible effects from noise
fluctuations were mitigated by setting the continuum end-points to an
average flux value that was computed over a 25 \AA\ window around the
selected maxima.  Observed error spectra were not available for most
of the objects in this nearby sample, so the statistical error could
not be calculated.  However, the dominant error from the possible
pseudo-continuum variations was estimated as described above.
In total, these uncertainties were on the order of $5-10 \%$ of the measured low-$z$ EW.
This measurement techniqe differs slightly from previously published
EW studies \citep{b16,b113,b114}, but the results here are in very
good agreement with these other references.  Based on the wavelength coverage available in the high-$z$
spectra, this EW analysis is focused on EW features 1, 2, and 3 - CaII
H\&K, SiII 4000, and MgII in the F04 nomenclature (EW\{CaII\},
EW\{SiII\}, \& EW\{MgII\} hereafter).  

\subsection{High-$z$ EW Measurements}
\label{sec:hiz_EWmeas}

\begin{figure}
\begin{minipage}{.40\textwidth}
\centerline{\includegraphics[width=7.85 cm]{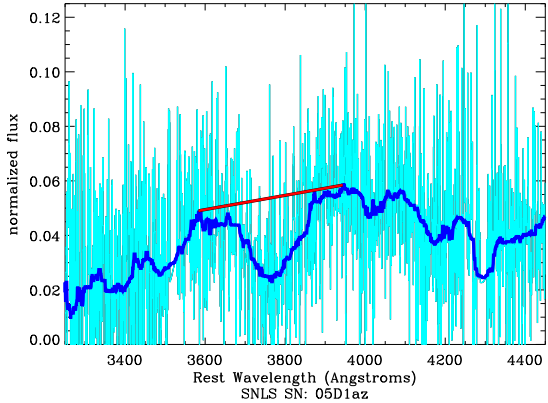}}
\end{minipage}

\caption{\label{fig:hiz_EWex} An example of an EW measurement (for
EW\{CaII\} in this case) on a high-$z$ SNLS object.  The median
filtered spectrum is plotted on top of the observed spectrum.  The
local maxima of the feature have been found within this median filter.
The EW is then computed by numerically integrating the weighted,
un-filtered flux underneath the pseudo-continuum defined by the two
maxima.}
\end{figure}

As explained in Sect.~\ref{sec:syserr1}, the high-z spectra were
median filtered in order to set the feature bounds. The EW measurement
was then made with the initial, un-filtered points as Fig.
\ref{fig:hiz_EWex} shows.  For the SNLS data set, the available
statistical uncertainty was used to weight each point so that the
equivalent width was calculated as

\begin{equation} 
EW_{\rm{SNLS}} = \left( \Sigma^N_{i=1} \left( 1 - \frac{f_\lambda(\lambda_i)}{f_c(\lambda_i)} \right) \Delta \lambda_i w_i \right) \frac{N} {\Sigma(w_i)}
\label{eq:ew_calc2}
\end{equation}

\noindent where $w_i$ is the weighting function at each point: 

\begin{equation}
w_i = 1/\sigma_f(\lambda_i)^2
\label{eq:ew_wt}
\end{equation}

\noindent The statistical errors were calculated as defined in
Equation \ref{eq:ew_sigma}.  The error induced by any variance in the
pseudo-continuum was computed with the method described in
Sect.~\ref{sec:ew_defn}.  A description of the errors caused by sky
noise and host galaxy contamination is presented in the next sections.

Note that if either the lower or upper range of an EW
measurement (defined in Table~\ref{tab:f04}) was outside of the
observed wavelength range then the measurement had to be removed from further analysis. This constraint affected results on a
measurement-by-measurement basis and caused the removal of 61 EW
measurements in the SNLS data set.  This left a total of 217 measurements from 73
different high-$z$ SNe in the final set of EW results. A number of
these measurements were later removed for host contamination
considerations, however (see Sect.~\ref{sec:syserr2}).

\subsubsection{High-$z$ EW Systematic Errors I -- Signal to Noise}
\label{sec:syserr1}

The sky noise affects the high-$z$ measurements by augmenting the
local maxima selected to set the pseudo-continuum.  This can result in
an overestimate of the equivalent width. A median filter was used to
correct this systematic effect. The size of the filter window was
selected via simulations in which spectra from the low-$z$ sample were
degraded with noise to match that in the high-$z$ SNe. The EW was then
re-calculated on each of the artificial spectra, using a median filter
set to various window sizes. Using a median filter over a 60 \AA\
window proved to be the most robust for all measurements at the wide
range of S/N levels covered by the high-$z$ spectra.

These simulations were also used to check that the uncertainties in EW
measurements behaved as expected for a given signal-to-noise input
spectrum.  Both the simulations and the final measurement results 
demonstrated that the statistical errors in the high-$z$ sample were
approximately $7\% - 20 \%$ of the measured EW value.  This is nearly
twice the uncertainty of the low-$z$ objects due to their reduced S/N.

\subsubsection{High-$z$ EW Systematic Errors II -- Host Galaxy Contamination}
\label{sec:syserr2}

In nearby objects, the angular separation between the host and SN
candidate is enough that most observations are not affected by host
galaxy light.  This is not the case with distant SNe.  In previous
high redshift spectroscopic studies, this error has been addressed by
subtracting host galaxy templates during the $\chi^{2}$-fitting
process \citep{b29,b27,bgabri}, during the extraction of the SN spectrum
\citep{b7}, or not at all. In reality, the host contamination tends
to `wash out' the absorption troughs in SNe spectra, which causes the
measured strength of these features to be lower than what would be
seen in a spectrum free of host galaxy flux.  The actual effects of
the contamination are dependent on the redshift and host galaxy type
of each object in addition to the amount of contamination.
The rolling search method of the SNLS provides extensive photometric
data that was used to estimate the amount of host galaxy
contamination at the epoch of spectroscopy for each SN.

Each object in the SNLS sample has epochs of non-detection before and
after the explosion. These are used as a baseline from which to
calculate the precentage increase of the SN light over that of the
host galaxy on each date when photometry was taken. A 1.10''
(diameter) circular aperture centred on the SN is used as
it is the basis for the percent contamination estimates that the 
SNLS photometric pipeline provides for each SN candidate.
By interpolating these measurements to the date of spectroscopy, the
amount of host contamination in the supernova spectrum can be
estimated.

In order to do this, the fact that the spectroscopic aperture
is not the same size and shape as the photometric aperture must be
taken into account. The spectroscopic aperture is rectangular, with a
width of 0.75'' and a length set by the extraction window used during
the spectroscopic data reduction. The amount of host light entering the slit
was simply scaled from that in the 1.10'' photometric aperture by assuming a
spatially flat profile. The amount of SN light entering the slit was
estimated by modelling the SN signal as a two-dimensional PSF with a
FWHM equal to the seeing in the appropriate GMOS acquisition image. The
photometrically determined percent increase can thus be converted to a
`percent contamination' of host flux in the observed spectrum.

The effects of host contamination on the EW results was determined by
artificially contaminating low-$z$ spectra from several different SNe
Ia with template galaxy spectra from E, S0, Sa, Sc, and starburst
galaxies. In each run of the simulation, the range of the observed
CFHT i'-band was shifted, based the redshift of each SNLS object, to
where it would fall in the rest frame. The i'-band was chosen
because this is the wavelength region where SNe Ia are brightest for the
redshift range of this survey; most of the SNLS observations are thus made 
in this band.  The flux of a chosen host
galaxy template was then scaled to match the level host contamination
(estimated as described above) and added to the low-$z$ supernova
spectrum.  The EW was re-measured to give the percent error caused by
the specific host type and contamination amount for every EW
measurement on each SNLS object.  

The results of these simulations were then used to correct for this
systematic error in the high-z spectra. The median value of the
percent error estimates among the different low-$z$ spectra was used
as the best estimate for the contamination effects on each SNLS
supernova, EW measurement, and galaxy type in question.  The
uncertainty of this percent error estimate was quantified
with the standard deviation of the contamination effects from
different host galaxy and SNe templates.  The effects within each
general host galaxy type were similar, so this uncertainty was only
$5\% - 7 \%$ of the measured EW.  Including the excess statistical 
uncertainty from reduced S/N, this error means that the
uncertainties in the high-$z$ EW results are roughly 2-3 times
those in the corresponding low-$z$ set.  
The galaxy type for the SNe was selected based on the observed host features in each affected
SNLS spectrum. The results of these corrections on the measured EW values
are displayed in Fig. \ref{fig:ew_gem_hiz3}.  

The results of these simulations helped determine that objects with
greater than 65\% host contamination were too biased to be included in
the final analysis of this data set.  This is because at this
contamination level, a majority of the high-$z$ EW results require
adjustments greater than 100\% of the measured EW and this amount of
host flux severely alters the shape of the SNe spectra so that the
local maxima for each feature cannot be identified in a manner
consistent with the rest of the sample. This removed measurements from 19
objects (in addition to the measurements removed for observed-wavelength restrictions). 
The final set of high-$z$ results thus includes a total 162 EW
measurements from 54 SNLS SNe Ia.

\subsection{Ejection Velocities}
\label{sec:v_defn}

The distribution of the ions in the homologously expanding SN
photosphere and the kinetic energy of the thermonuclear explosion are
expressed in the ejection velocities measured in the SN spectrum.
Measurements of the ejection velocities of the features attributed to
CaII H\&K, SII (near $\lambda$5640) and SiII (near $\lambda$6150) ---
our EW features 1, 5, and 7, respectively --- have previously been
studied in both nearby \citep{b9,b69,b70,b37,b6,b36} and high redshift
\citep{b27,b7} SNe Ia populations.  The redshifts of the distant
objects that make up the bulk of the SNLS sample are high enough that
the SII `W' ($\lambda5640$) and SiII $\lambda6150$ features are
generally not observed, so only the CaII H\& K feature was studied.

The ejection velocity, $v_{\rm{ej}}$, was calculated as the blueshift
of the minimum of the feature away from an estimated rest wavelength
of 3945.0 \AA\ (a weighted average of the two components of the CaII
H\&K feature).  The contribution of other ions to the absorption in
the wavelength range of CaII H\&K (most notably SiII), as well the
intrinsic distribution of CaII across velocity space, can give this
feature a `kinked' shape with two possible minima.  This shape makes
it difficult to determine where the mininum of the feature is located.
This difficulty is compounded by the low S/N of the high-$z$ objects.
To make a consistent measurement between the two sets, the entire CaII
H\&K feature was fit with a Gaussian function.  The best fit was
estimated with robust, non-linear least-squares MPFIT\footnote{see
\texttt{http://cow.physics.wisc.edu/$\sim$craigm/idl/}} routine in
IDL. The fit was weighted to the most dominant minima of the
absorption region by making small, initial Gaussian fits to the
deepest, bluest, part of the feature.  The wavelength range from this
initial fit was then expanded iteratively to span the entire
absorption trough, and the minimum of the final fit was used to
calculate $v_{\rm{ej}}$ as

\begin{equation}
v_{\rm{ej}} = -c \cdot \left( \frac{(\frac{\lambda_m - \lambda_0}{\lambda_0} + 1)^2 - 1}{(\frac{\lambda_m - \lambda_0}{\lambda_0} + 1)^2 + 1} \right)
\label{eq:v_calc}
\end{equation}

where $\lambda_m$ is the measured featured minimum, $\lambda_0$ is
3945.0 \AA\ and $c$ is the speed of light.  The statistical error for
this fit was computed by propagating the uncertanties in the final
Gaussian parameters into Eq. \ref{eq:v_calc}. An additional source of 
error, caused by uncertainty in the selection of points used to fit
the Gaussian curves, was estimated by taking the difference between the calculated velocities
from the initial and final fits. 

Without having measurement uncertainties available for the low-$z$
spectra, the statistical errors were not calculated for these objects.
For the high-$z$ data, the lower S/N meant these spectra had to be
smoothed before the Gaussian fits to the CaII H\&K feature could be
made. Again a 60 \AA\ median filter was used.  The filtered flux was
fit with a Gaussian model that was weighted by the error spectra.  The
statistical error in the final $v_{\rm{ej}}$ values was estimated by
propagating the estimated redshift uncertainties through the velocity
calculation.  

No corrections were necessary to account for the systematic
error from host galaxy contamination as this error was minimal for
$v_{\rm{ej}}$ measurements made on the SNe within the 65\%
contamination limit discussed in Sect.~\ref{sec:syserr2}.

\section{Spectroscopic Analysis Results}
\label{sec:results}

\subsection{Low-$z$ Results}
\label{sec:lowz_EW_results}

\begin{figure*}
\begin{minipage}{2.0\textwidth} 
\begin{tabular}{cc}
\includegraphics[width=7.85 cm] {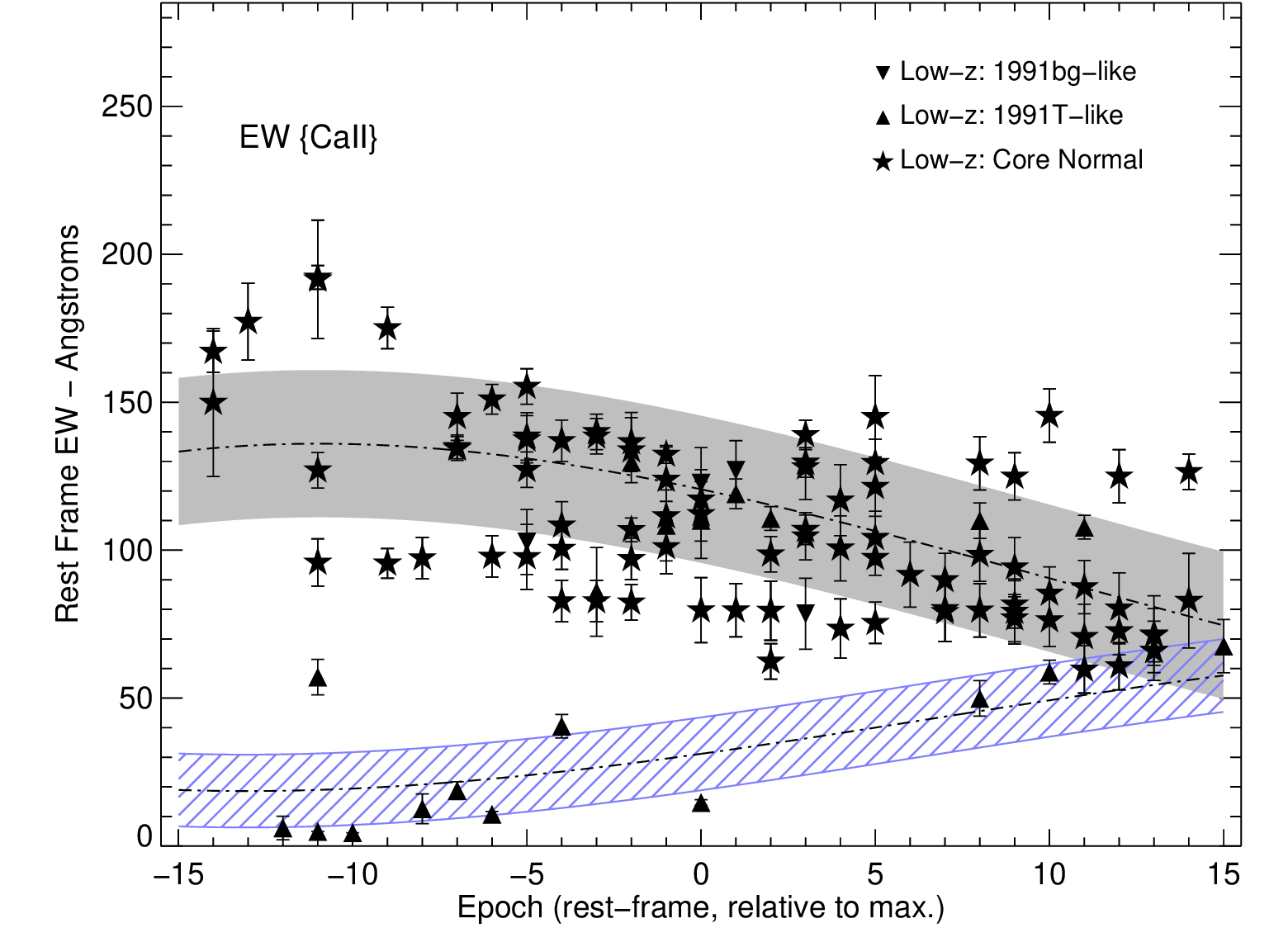} & \includegraphics[width=7.85 cm] {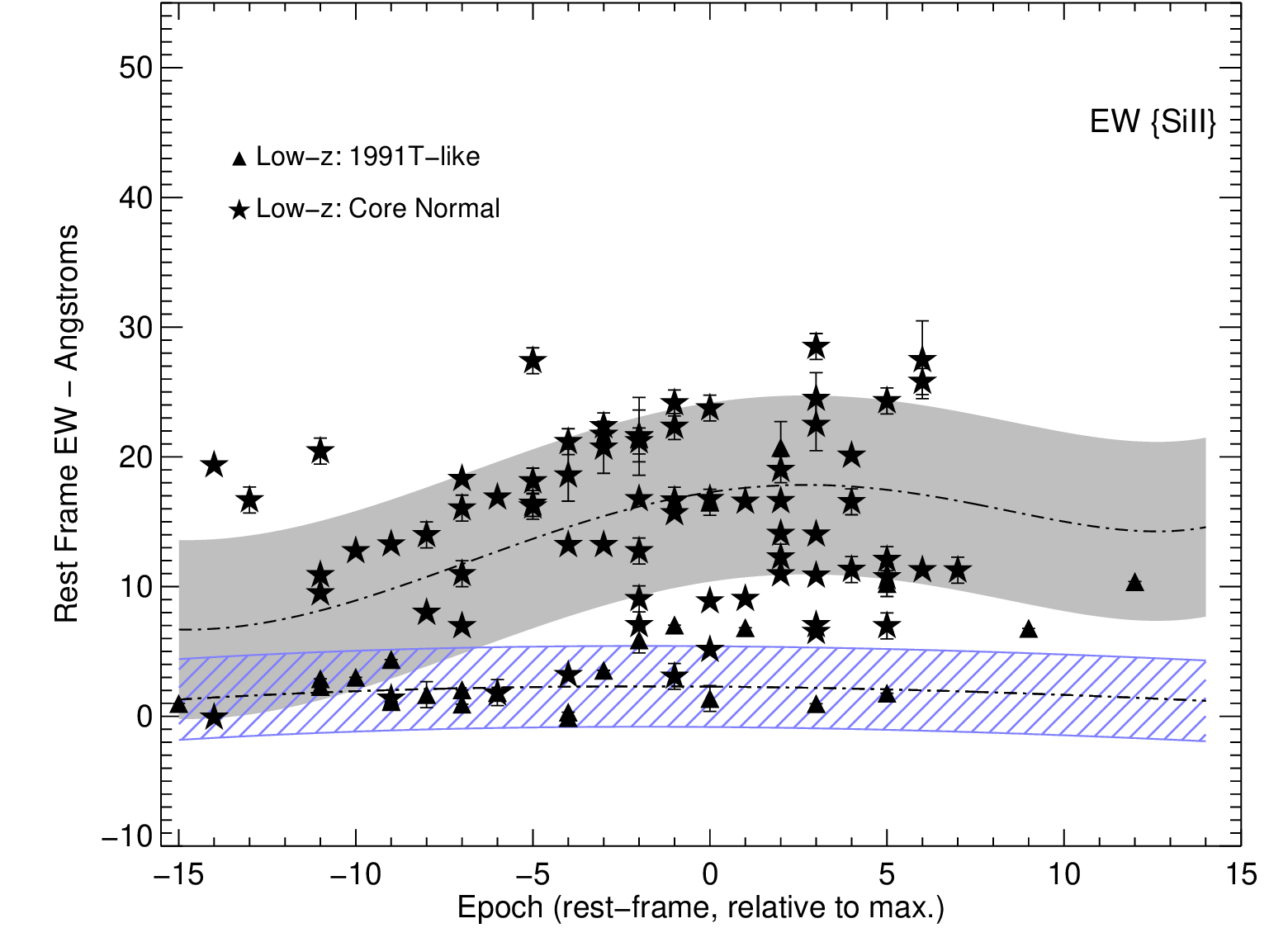} \\
\includegraphics[width=7.85 cm] {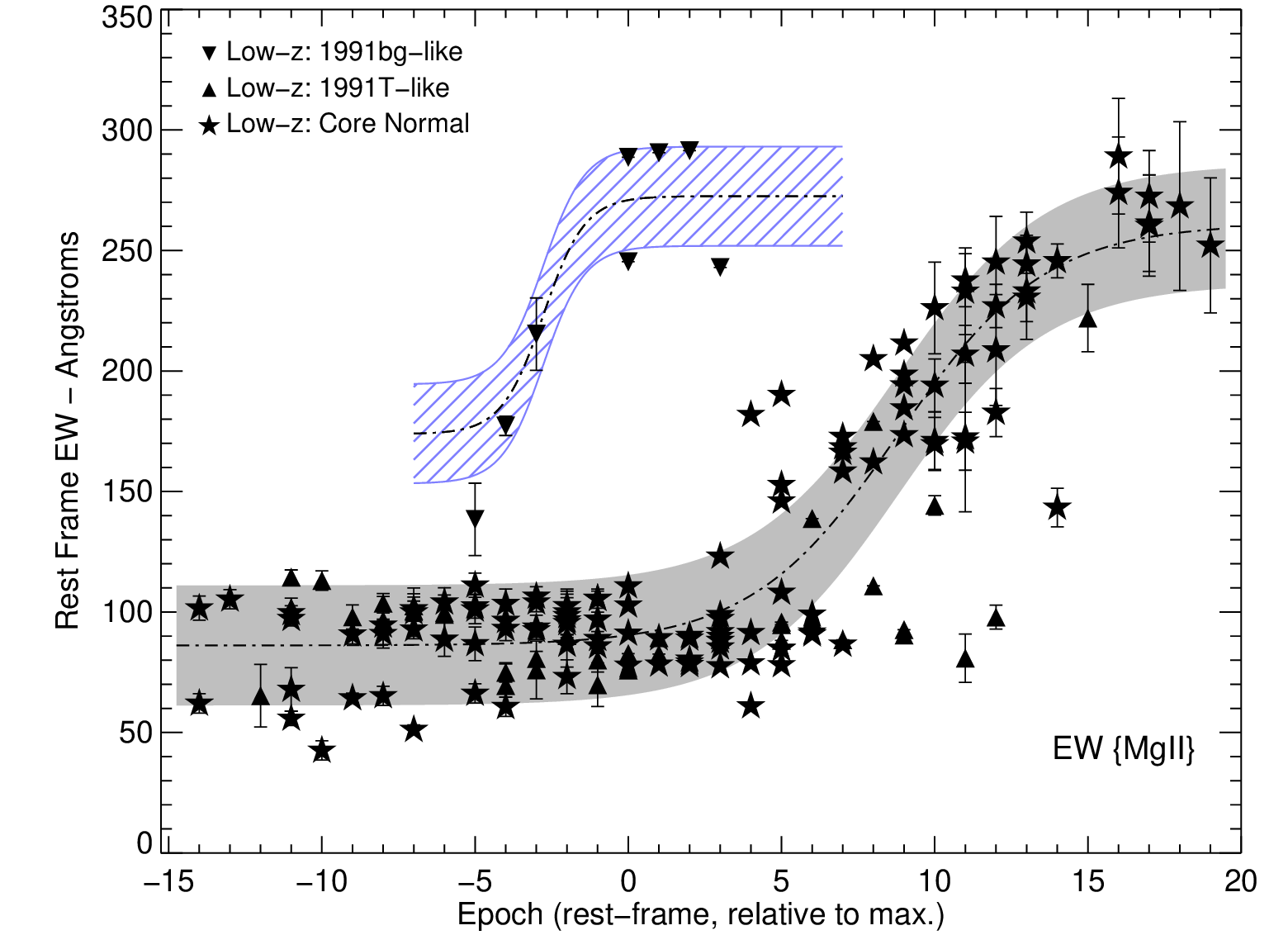} & \includegraphics[width=7.85 cm] {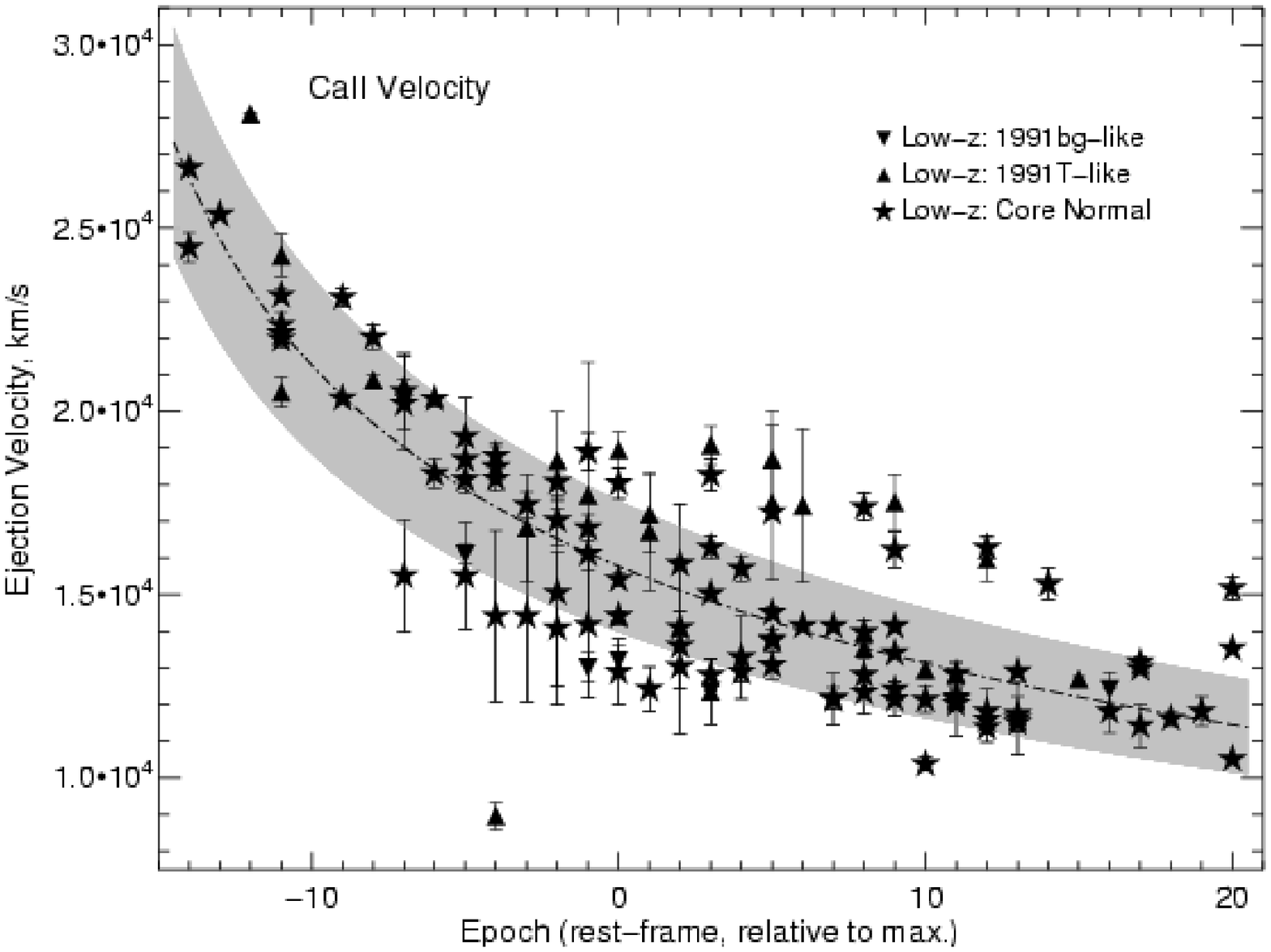} \\
\end{tabular}
\end{minipage}
\caption{\label{fig:lowz_4} EW Measurements for the `CaII H\&K' (top,
left), `SiII' (top, right), and `MgII' (bottom, left) features and
CaII H\&K $v_{\rm{ej}}$ (bottom, right) plotted versus rest-frame
epoch, for the low-$z$ sample.  The shaded, grey region marks the mean
trend and $1\sigma$ distribution of the epoch evolution of these
measurements in core normal objects.  The striped region marks the
same trend for 1991T-like objects in the EW\{CaII\} and EW\{SiII\}
plots and the trend for 1991bg-like objects in the EW\{MgII\} plot.
These trends were estimated with EW measurements on template SNe Ia
spectra from the literature.}
\end{figure*}

The results of the low-$z$ EW measurements, plotted against rest-frame
epoch (relative to B' maximum light) are shown in the first 3 panels
in Fig. \ref{fig:lowz_4}.  The mean trend (and 1$\sigma$ deviation)
exhibited by normal SNe Ia for each feature is shown in the grey
contour in each panel.  This trend was derived by making EW
measurements on a template core normal Type Ia SN spectrum made from a
combination of core normal spectra by \citet{b115}.  The mean trends
for the 1991T-like and 1991bg-like objects were derived from similarly
constructed template Type Ia spectra (\cite{b115,b116,b118}, Nugent
private communication) and are shown where appropriate.  The
$1\sigma$ deviation shown on these trends was estimated by expanding
the error bars on each trend until a fit to the appropriate low-$z$ SNe 
measurements produced a reduced $\chi^2$ of $1.0$.  This was done
as other methods of estimating the $1\sigma$ range would be more 
dependent on the varying epoch and object sampling of the low-$z$ data 
set.  A constant scatter as a function of epoch was thus assumed, hence
the lack of variance over epoch in Figs. \ref{fig:lowz_4} - \ref{fig:hiz_4}. 

The core normal objects exhibit fairly homogenous behavior in these EW
plots.  Not surprisingly, this parameter space also clearly
differentiates between the SNe Ia sub-types.  The lack of absorption
from IME (intermediate mass elements) in 1991T-like objects (which is due to inferred higher
temperatures during thermonuclear runaway in these objects) is clearly
seen in the EW\{CaII H\&K\} and EW\{SiII\} results.  Similarly, the
additional absorption from (primarily) TiII in cooler, 1991bg-like
objects results in a significant difference with regards to EW\{MgII\}
measurements.  These results are in agreement with the qualitative and quantitative
differences that have already been established for these different SNe
sub-types \citep{b13,b14,b6,b114}.

The changes of CaII H\&K $v_{\rm{ej}}$ with rest-frame epoch is shown
in the fourth panel of Fig. \ref{fig:lowz_4}.  The mean trend was
estimated by fitting a power law to all of the data points from core
normal objects.  The core normal objects display a broad trend about
this best fit and the SN sub-types are generally interspersed around
it.  Previous estimations of SN Ia ejection velocities implied that
over-luminous objects have significantly higher velocities
\citep{b31}, but the statistical analysis of SiII velocities by
\citet{b6} suggests that these objects perhaps differ in their
velocity gradients, which were not studied here as corresponding
measurements are not available on the high-$z$ set.

\subsection{High-$z$ Results and Comparison to Low-$z$ SNe}
\label{sec:hiz_stats}

\begin{figure}
\begin{minipage}{1.0\textwidth} 
\begin{tabular}{cc}
\includegraphics[width=7.85 cm] {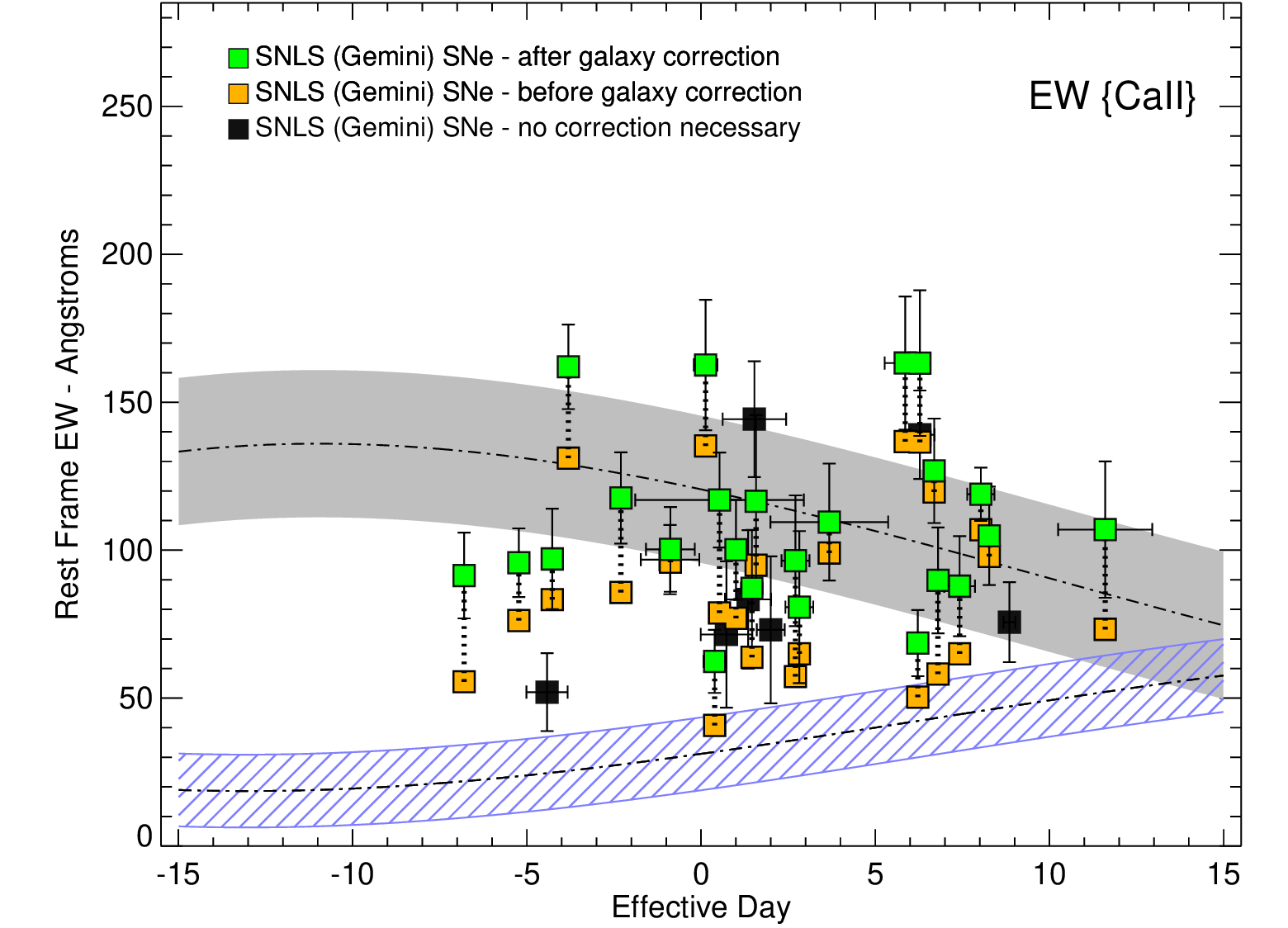} \\
\includegraphics[width=7.85 cm] {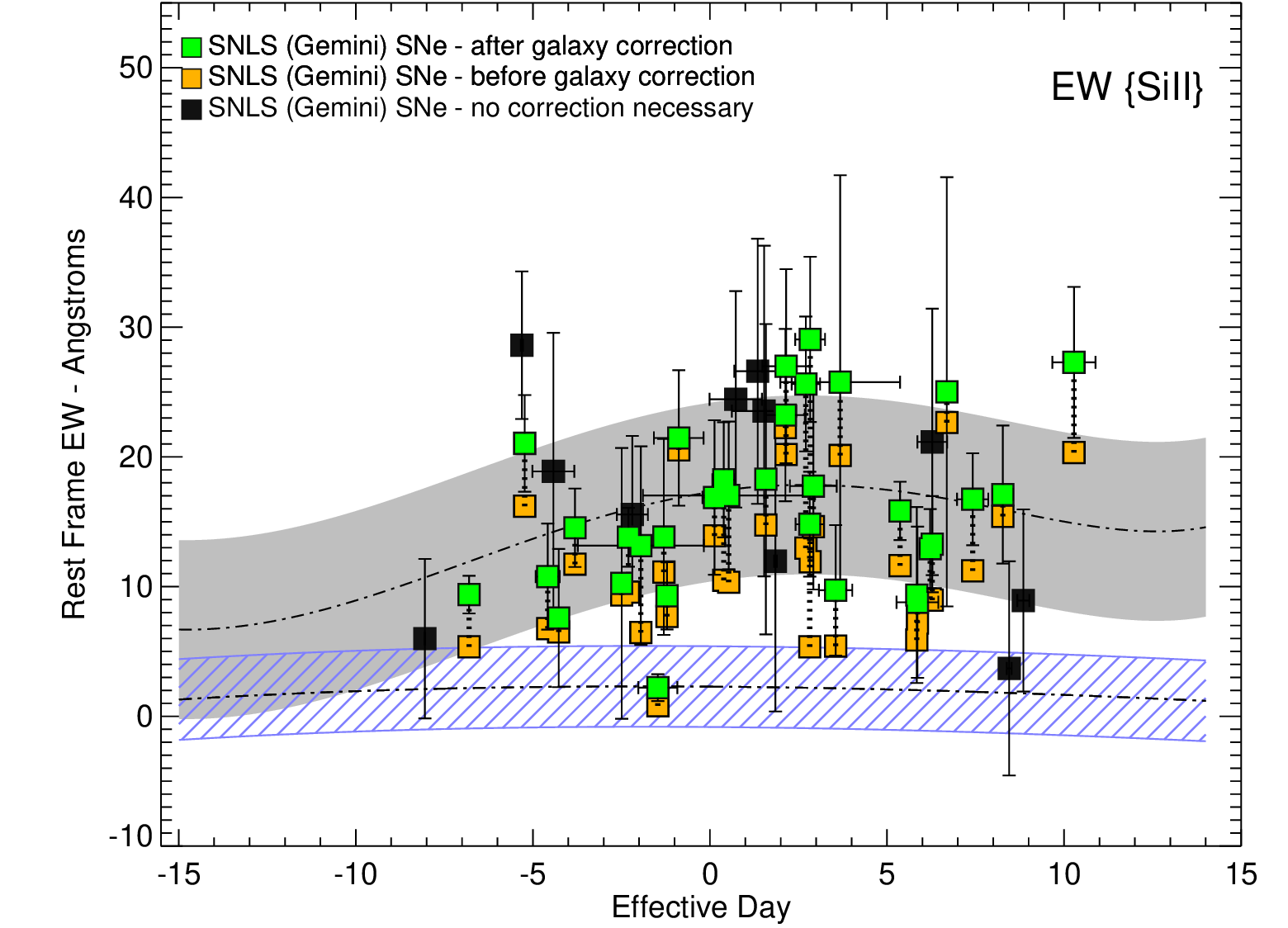} \\
\includegraphics[width=7.85 cm] {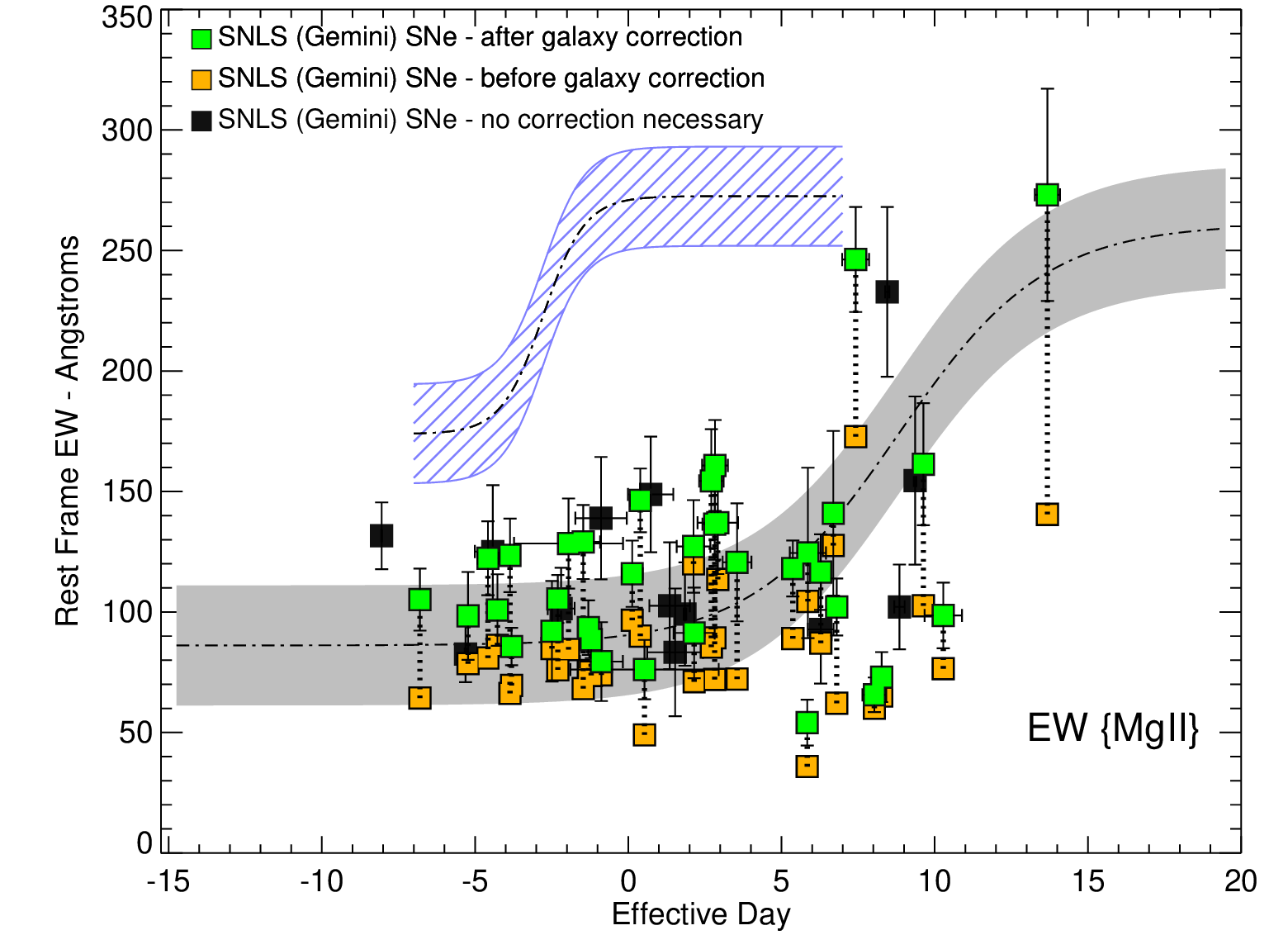} \\
\end{tabular}
\end{minipage}
\caption{\label{fig:ew_gem_hiz3} Measurements of the EW of the `CaII
H\&K', `SiII', and `MgII' features (from top to bottom), versus
rest-frame epoch, in the SNLS high-$z$ sample.  The shaded and striped
regions are the same as in Fig. \ref{fig:lowz_4}.  The EW
measurements that did not need any correction for the systematic
effects of host galaxy contamination are shown with the dark, filled
squares.  All of the other points illustrate the measurements that
were corrected for host contamination.  The adjusted EW values are
connected to the corresponding initial measurements with dotted
lines. }
\end{figure}

The EW measurements on the high-z data before and after correction for
host galaxy contamination are shown in Fig.~\ref{fig:ew_gem_hiz3}.
The corrected results are compared to the corresponding low-$z$
results in Fig. \ref{fig:hiz_4} and all of these high-$z$ results are listed in Table \ref{tab:gem_results} in the Appendix. 

Figure \ref{fig:hiz_4} shows that the general trends exhibited by the
low-$z$ sample are indeed followed by the high redshift objects and
all of the SNLS results are within the range of the low-$z$ SNe
measurements.  A few spectra (marked with open circles in Fig. \ref{fig:hiz_4})
had EW results and other characteristics 
that are more similar to the SN 1991T-like outliers; 
these objects are discussed in more detail in
Sect.~\ref{sec:hiz_subtypes}.

\begin{figure*}
\begin{minipage}{1.0\textwidth} 
\begin{tabular}{cc}
\includegraphics[width=7.85 cm] {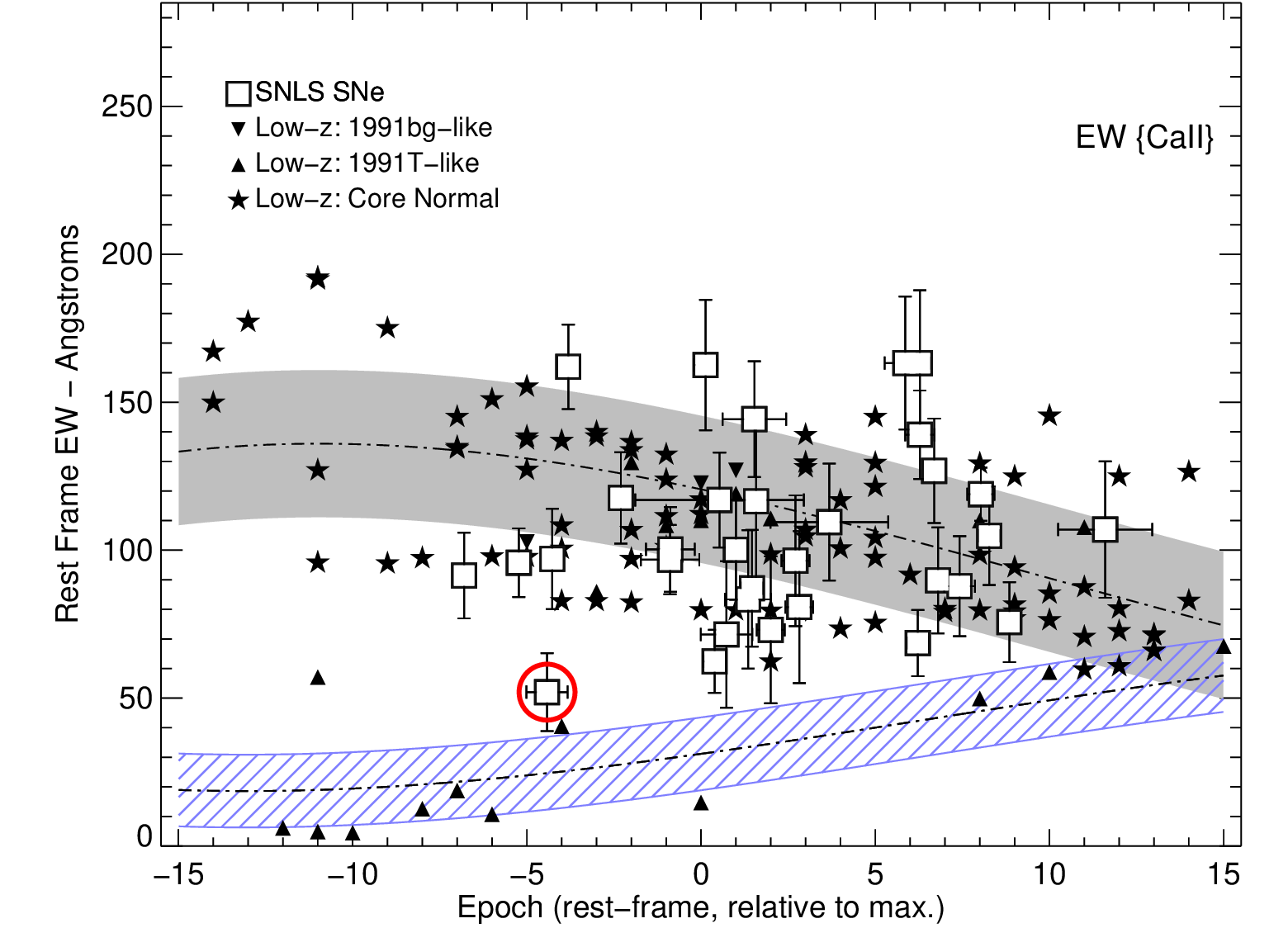} & \includegraphics[width=7.85 cm] {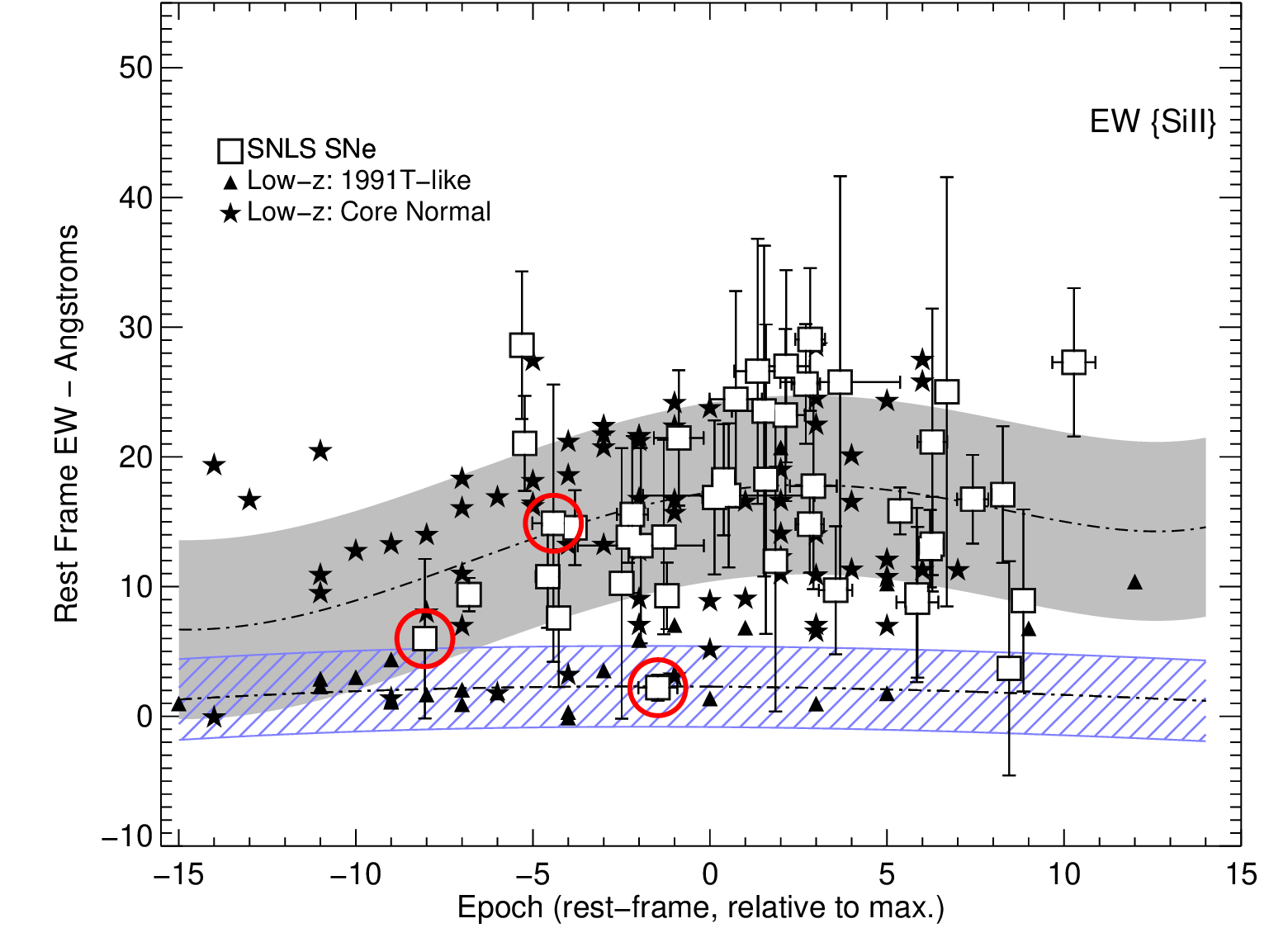} \\
\includegraphics[width=7.85 cm] {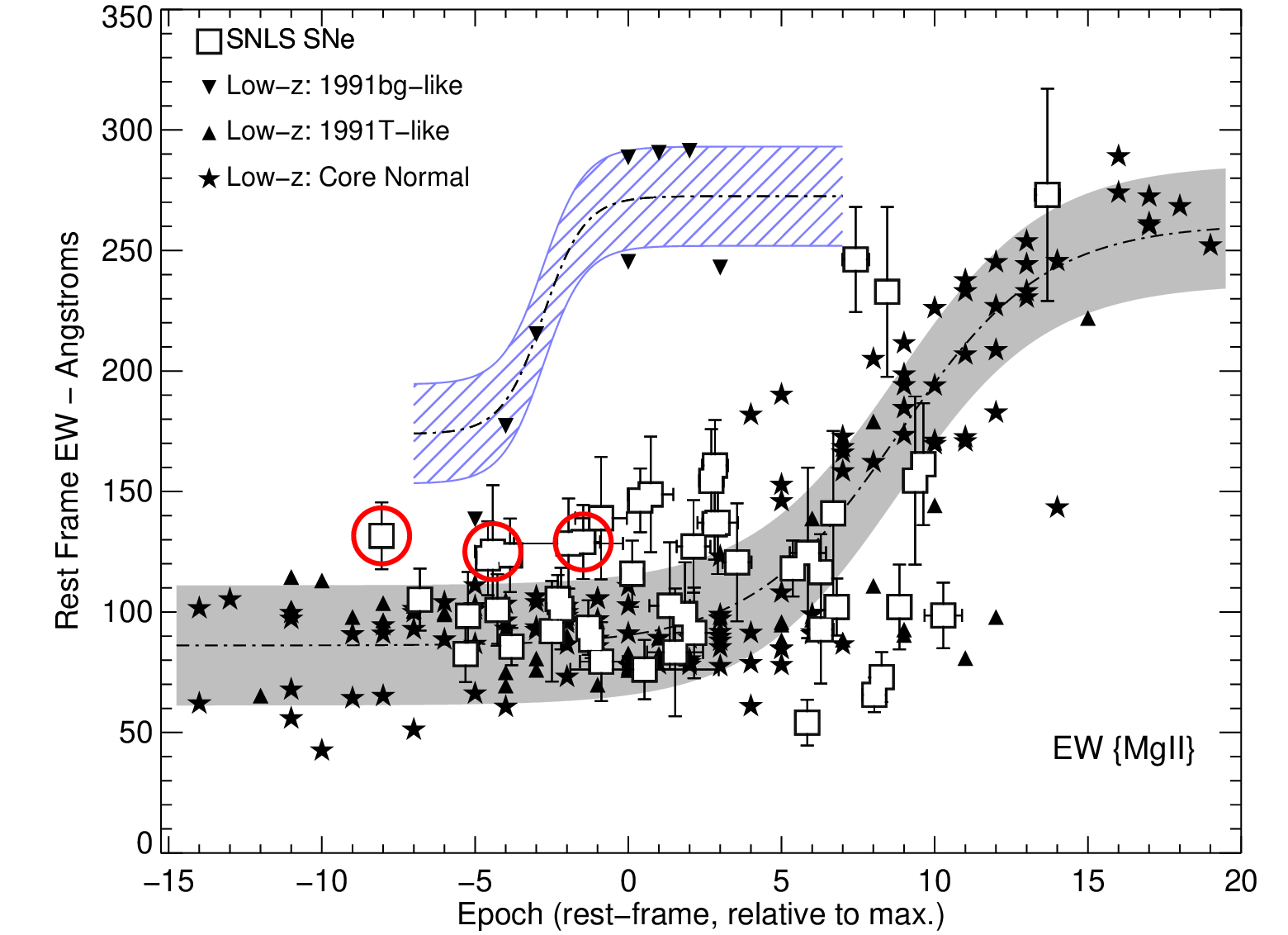} & \includegraphics[width=7.85 cm] {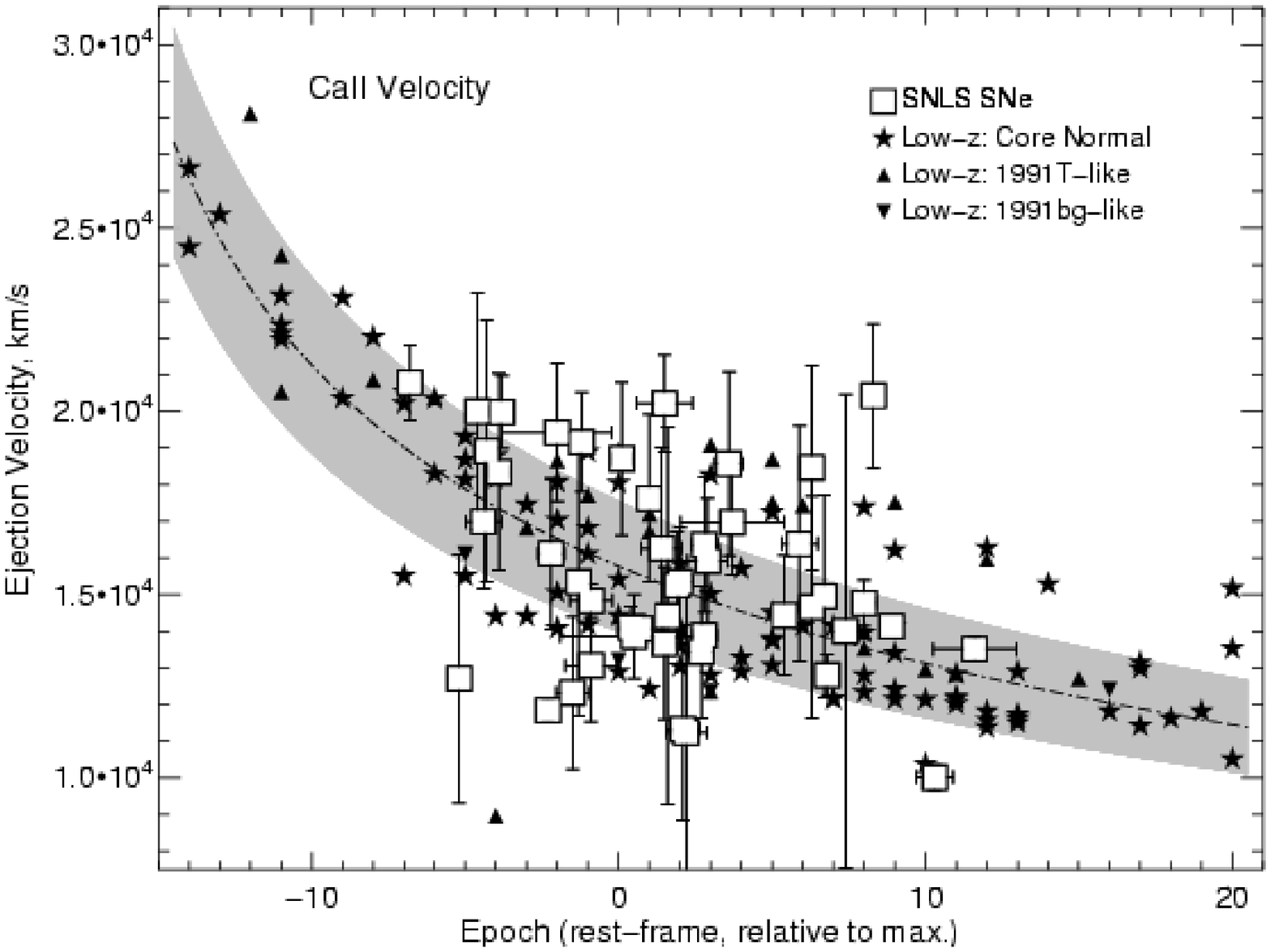} \\
\end{tabular}
\end{minipage}
\caption{\label{fig:hiz_4} The results from the SNLS high-$z$ SNe
(open squares) compared to the EW results from the low-$z$ sample
(filled symbols).  The contours in each plot are the same as
Fig. \ref{fig:lowz_4}.  The objects that were identified as 1991T-like
SNe are circled in red.  }
\end{figure*}

The fraternity between the high redshift SNe measurements and the
trends exhibited by the local, core normal objects was further
investigated with $\chi^2$ tests.  This test compared the high-$z$
measurements to the core normal low redshift `model' illustrated for
each feature in Figs. \ref{fig:lowz_4} - \ref{fig:hiz_4}.  When
making these $\chi^2$ calculations, the full uncertainties for each
measurement were used. The 1$\sigma$ dispersion in the low-$z$ trends
was also considered by adding this error (in quadrature) to the 
uncertainty of each high-$z$ measurement.  The uncertainty in the
estimated epoch for every high-$z$ result was also added (again, in
quadrature) according to the low-$z$ trend for each measurement.  The
results, shown in Table \ref{tab:chi2_results}, indicate that there
are no statistically significant differences between the spectroscopic
features of these SNe Ia for three of the parameters studied here.

\begin{table}
 \caption{\label{tab:chi2_results} A quantitative comparison of the
 spectroscopic features in high-$z$ supernovae (as measured with EW
 and CaII H\&K ejection velocities) and the mean trends exhibited for
 these quantities by local SNe Ia.  These results show that there are
 no significant differences between distant SNLS objects and nearby SNe, 
with the exception of EW\{MgII\}.}
\begin{tabular}{l|cccc}
\hline \hline Feature & Reduced$^{\rm{a}}$ & DOF$^{\rm{b}}$ & Prob$^{\rm{c}}$
\\ & $\chi^2$ & & \\ \hline EW\{CaII\} & 1.10 & 30 & 0.67 \\
EW\{SiII\} & 0.47 & 40 & $<$ 0.002 \\ EW\{MgII\} & 1.86 & 44 &
$>$0.999 \\ $v_{\rm{ej}}$ & 1.03 & 45 & 0.64 \\ \hline
\end{tabular}

\smallskip

{\small
a --- $\chi^2 / dof$ calculated for the measurement/data set in question and the corresponding low-$z$ mean trend

b --- degrees of freedom used when calculating the reduced $\chi^2$, $N_{\rm{data points}}-1$

c --- probability (0.00 - 1.00) that the null hypothesis, that the high-$z$ data are consistent with the trend displayed by low-$z$ SNe, can be rejected.  

}
\normalsize
\end{table}

The exception to this result is the comparison of EW\{MgII\}
measurements; however, the reality of this difference is questionable.
It is possible that this difference is a statistical artifact caused
by the differences in the epoch sampling and total number of objects
between the high-$z$ and low-$z$ data sets.  Figure \ref{fig:lowz_4}
shows that there are an appreciable number of low-$z$ core normal
points between 0 and +10 days past maximum (where the bulk of the
outliers in the SNLS set are located) which define the 1$\sigma$ range
on the empirically derived mean trend.  Yet these points are from 15
different SNe only; half of the number of SNLS objects in the same
range.  This sampling difference, combined with the fact that the
low-z sample is not well defined (i.e. it is not a complete sample)
could mean that the true amount of variance in this feature is
underestimated in the low-$z$ mean trends, which would artificially
increase the significance of the measured difference in the high-$z$
results.  If this epoch selection effect is not responsible for the difference 
then it is interesting to consider possible physical interpretations, 
which we discuss in Sect. \ref{sec:mgII_diff}.

\subsection{High-$z$ Results: Identification of SNe Ia Sub-Types}
\label{sec:hiz_subtypes}

The comparisons in Fig. \ref{fig:hiz_4} show that a handful of the
high-$z$ SNLS SNe Ia exhibit spectroscopic features that are more
consistent with the 1991T or 1991bg-like SNe from the low-$z$
surveys.

\begin{figure}
\begin{minipage}{1.0\textwidth} 
\begin{tabular}{cc}
\includegraphics[width=7.85 cm] {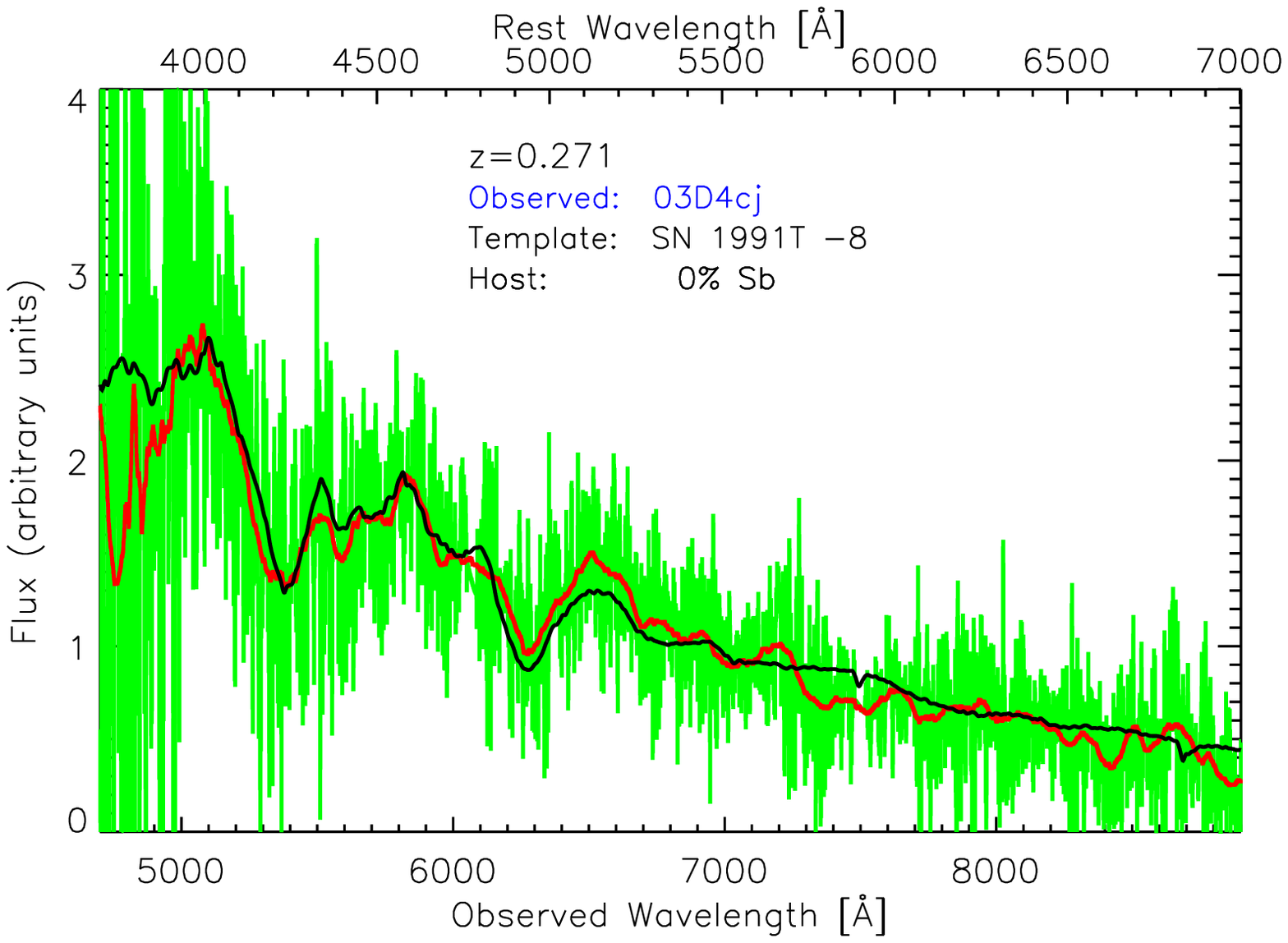} \\
\includegraphics[width=7.85 cm] {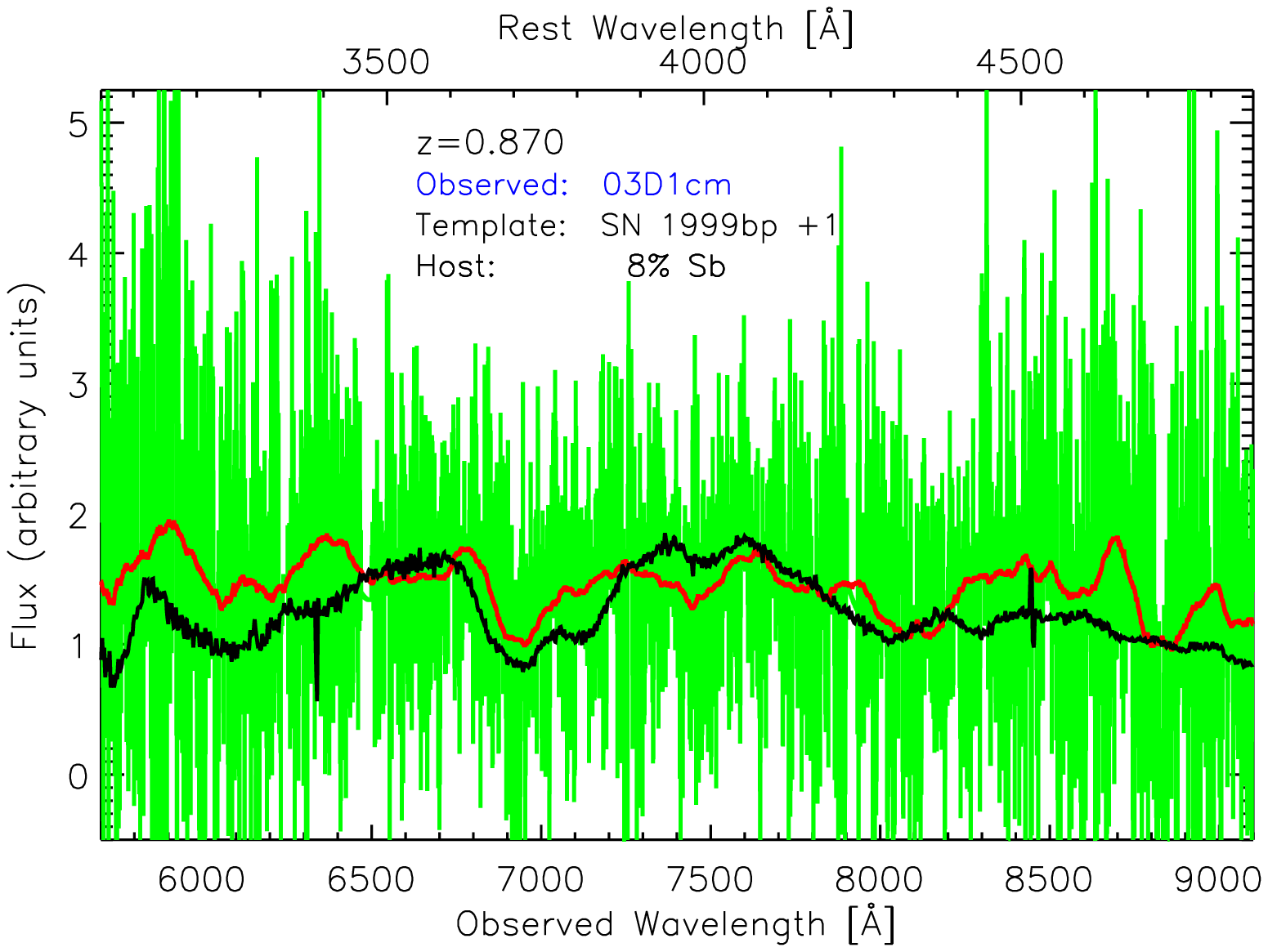} \\
\includegraphics[width=7.85 cm] {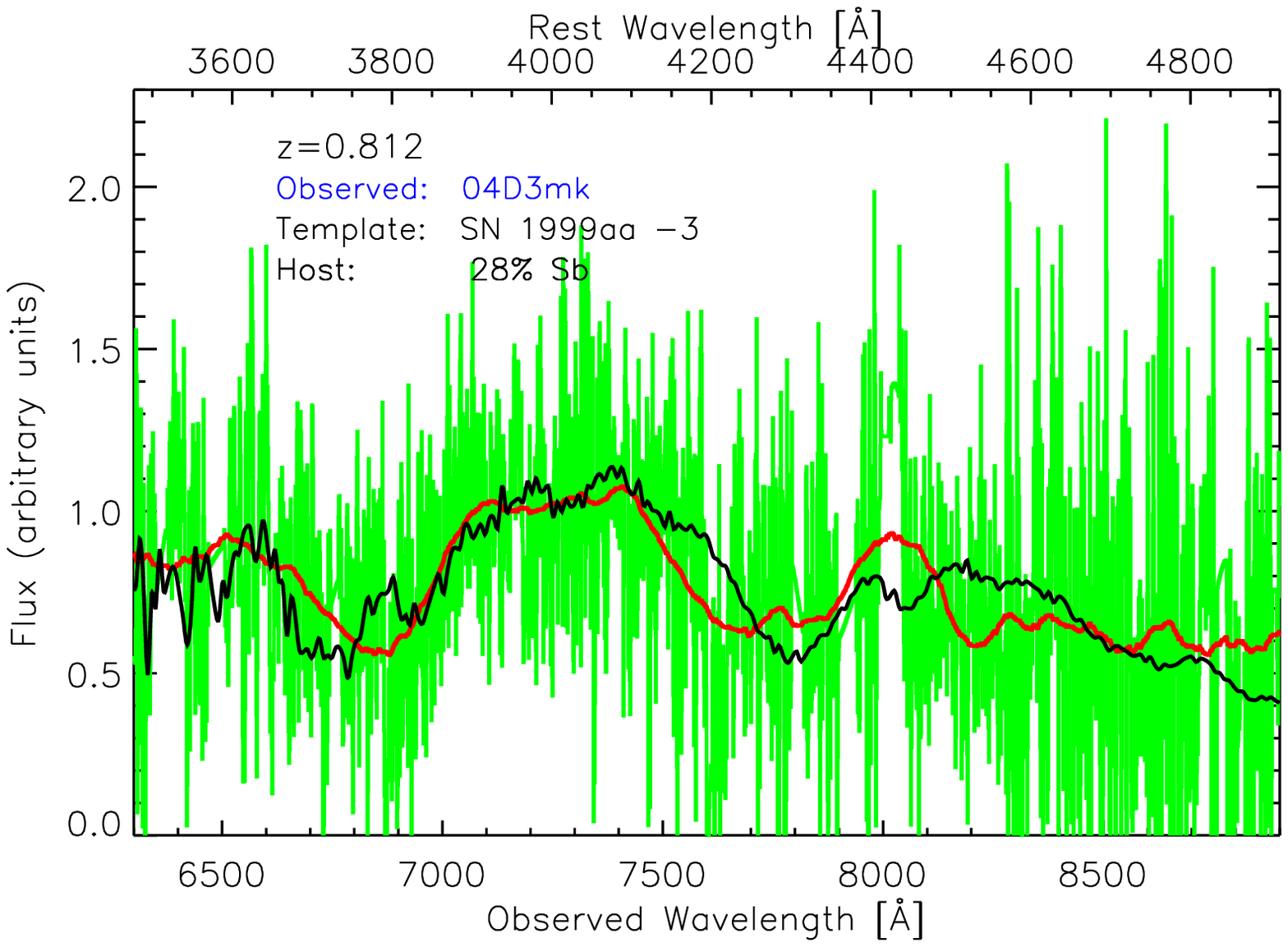} \\
\end{tabular}
\end{minipage}
\caption{\label{fig:hiz_91T} Three high-$z$ SNe Ia - 03D4cj, 03D1cm
and 04D3mk - that demonstrate a quantitative similarity to the
1991T-like SNe Ia sub-type. The green spectrum is the observed
spectrum without any binning. The black spectrum is the best-fitting
SN from the low-$z$ library and the red spectrum a smoothed version of
the observed spectrum.  The primary evidence for the similarity of
these objects to 1991T-like SNe is the EW\{CaII\} and EW\{SiII\}
measurements, although in each case the $\chi^2$ fits to SN 1991T or
SN 1999aa were quantitatively better than those of core normal SNe
Ia. Photometric information also supports this sub-type identification
for these three SNe.}
\end{figure}

Overluminous 1991T-like SNe Ia have less absorption from IME at
certain epochs. Three outliers in the EW\{CaII\} and EW\{SiII\}
results have evidence of being similar to the known class of
1991T-like objects. Their spectra are shown in
Fig.~\ref{fig:hiz_91T}. 03D4cj displayed a smaller amount of SiII
than a typical core normal object.  The error bars on this measurement
are relatively large (EW\{SiII\}$ = 5.9 \pm 6.1$ \AA), but the
photometric data for this object supports the conclusion that it is a
1991T-like SN; \citet{b3} stated that the lightcurve from this SN was
incompatible with the model used to fit all of the other SNLS SNe.
The $\chi^2$ template fits to 03D4cj also indicate that it is similar
to SN 1991T.

The 1991T-like classification also appears valid for SNLS supernova
03D1cm. This SN has a very low EW\{CaII\} measurement and an
EW\{SiII\} result that (within large error bars, EW\{SiII\} $= 14.8
\pm 10.6$ \AA) is also consistent with the low values measured in
1991T-like objects.  Photometrically, 03D1cm is also more like a
1991T-like SN as it had one of the highest stretch values from the
entire set of Gemini SNLS data at $s=1.173 \pm 0.061$ \citep{b3}.
The best matching template spectrum also corresponds to an
overluminous low-$z$ SNe.  The spectral and photometric dates do 
differ (+1 day and -4.42 days, respectively) in this case, but the 
low S/N on this object and inherent uncertainty on spectral dating 
with this template matching method (which is on the order of a few days) means that
this is not necessarily enough evidence to outweigh the 1991T-like
EW and $s$ values of 03D1cm.

The SNLS SN 04D3mk also had a very low, well constrained EW\{SiII\}
value.  At the epoch this object was observed, the EW\{MgII\} result
cannot differentiate between normal and overluminous SNe and the
EW\{CaII\} measurement was unusable as it fell on the gap between GMOS
detectors (e.g., Sect.~\ref{sec:hiz_EWmeas}).  The final photometry
of this object did not reveal any photometric peculiarities, but the
$\chi^2$ template matches to 04D3mk singled out overluminous SNe Ia
spectra only as the most similar low-$z$ objects.  This supports the
quantitative similarities of 04D3mk to this particular SN Ia sub-type.
With this evidence, SNLS objects 03D1cm and 04D3mk are the most
distant supernova (at $z=0.87$ and $z=0.812$, respectively) to be
shown to be spectroscopically similar to the 1991T-like SNe Ia
sub-types.  To reiterate, the primary source of this identification
for these objects is the EW measurements which are quantitatively more 
similar to the overluminous sub-class of SNe Ia than other sub-types. 

The SNLS SNe 03D1co and 05D3mq are 1991bg-like candidates as their large
EW\{MgII\} values are more consistent with the excess absorption
measured in the subluminous low-$z$ SNe.  However, these two high-$z$
SNe had measurable amounts of SiII, a feature that has yet to be
observed in any 1991bg-like SNe because it is obscured by additional
absorption from TiII and FeII.  This evidence, combined with the
normal photometric parameters \citep{b3} derived for these objects and
the fact that template matches to 1991bg-like objects resulted in poor
$\chi^2$ results, is enough to justify that these SNe are more like
core normal SNe than any of the underluminous SNe observed to date.

\subsection{Correlations With Host Galaxy Type}

\subsubsection{Low-$z$ spectra}
\label{sec:lowz_host}

The general trend of brighter SNe and a higher dispersion of SNe Ia
peak magnitudes in late-type galaxies is well documented
\citep{b21a,b21b,b56}. Yet only a few quantitative spectroscopic
corollaries \citep{b133,b134} have been made that agree or disagree
with this trend. Kolmogorov-Smirnov (KS) tests were thus used to
explore any differences between the spectroscopic parameters measured
in SNe from different host types in our low-$z$ sample. This test was
carried out on the residuals of these measurements, where the residual
for each EW and $v_{\rm{ej}}$ value is defined as the difference
between the measurement and the corresponding low-$z$ mean trend.

The host type for each low-$z$ SN was grouped into the broad
categories of `E/S0' and `Spiral' (Sa, Sb, Sc and irregular) galaxies.
The published morphology (from the Nearby Galaxy Catalogue and other
appropriate databases) of each SN host galaxy (based on the hosts
listed in the SNe reference papers, see Table \ref{tab:lowzsrc}) was
used to determine which group was appropriate for each SN host.

\begin{figure*}
\begin{minipage}{1.0\textwidth} 
\begin{tabular}{c}
\centerline{\includegraphics[width=15 cm] {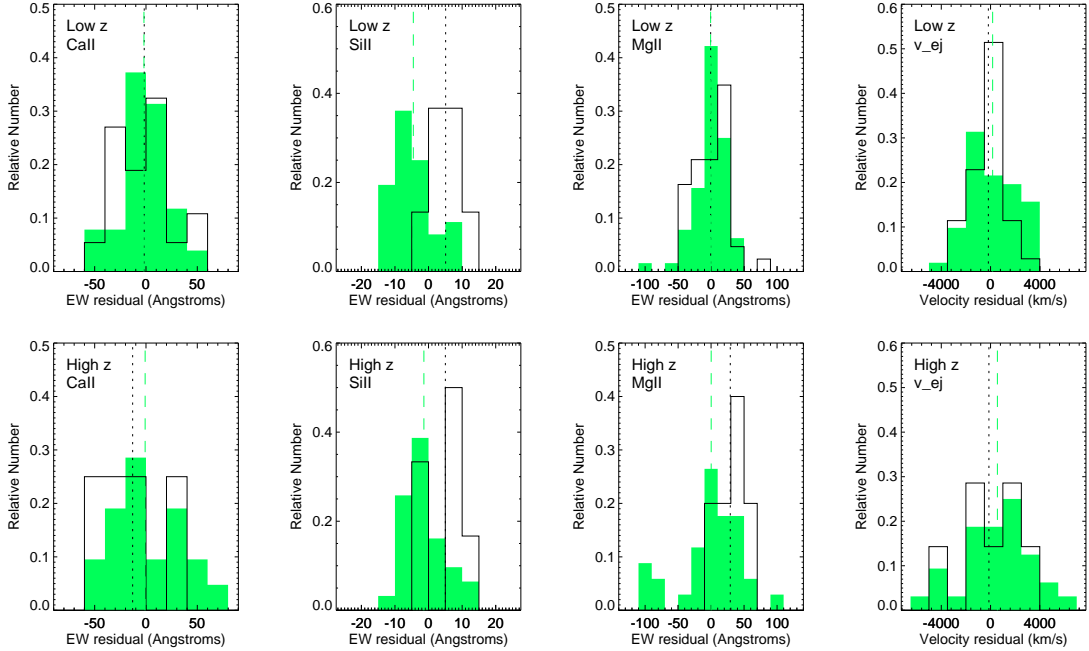}}
\end{tabular}
\end{minipage}
\caption{\label{fig:allz_ks_tests} Histograms comparing the residuals of EW and CaII $v_{\rm{ej}}$ measurements
in SNe from E/S0 (open histograms) and spiral (filled histograms) host galaxies.  The
residuals from low-$z$ SNe are displayed in the top row and the SNLS high-$z$ results
are on the bottom.  The vertical dashed line in each panel is the mean value for each distribution.
The histograms have been normalized to have the same area.  
The distribution of low-$z$ SNe results show a significant difference
for the objects from early and late hosts (as determined with KS tests).  The high-$z$ 
EW\{SiII\} results show a similar trend while EW\{MgII\} distributions display a difference
that is not seen at low-$z$.}
\end{figure*}

A comparison of the residuals for the four measurements discussed in this
paper, grouped according to these two broad host galaxy groups, are displayed
in Fig. \ref{fig:allz_ks_tests}.  At low-$z$ the only feature that shows a 
significant trend with host type is SiII.  



\subsubsection{High-z spectra}
\label{sec:hiz_hosts}

Another test for possible evolution is to determine whether the
correlations between low-$z$ SNe Ia characteristics and host
morphology are observed within the high-$z$ set. The host types
were estimated for the SNLS objects that had clear host galaxy
features in their spectra and placed into one of three broad
categories; `E/S0' for early (elliptical and S0 hosts), `Early Spirals'
for objects with spectroscopic features similar to Sa and Sb galaxies, 
and `Late Spirals' for Sc and irregular hosts.  The sub-set of SNLS 
SNe Ia with indentifyable host galaxies includes 45 of the 54 SNLS
objects that had EW or $v_{\rm{ej}}$ results.  

The distribution of these 
SNLS measurements, displayed in the histograms in
the bottom of Fig. \ref{fig:allz_ks_tests}, do show clear similarities
to the low-$z$ distributions (top, Fig. \ref{fig:allz_ks_tests}).  The
distant late type galaxies host SNe that display a smaller amount of
SiII than objects from early hosts and the ejection velocities
measured in spiral galaxies peak at a higher value.  A comparison of
the mean values for each measurement (again after subtracting the low-z
trend to remove dependence on SN phase) divided by host type supports
this conclusion, as can be seen from Table~\ref{tab:host_hiz}.

\begin{table}
\caption{\label{tab:host_hiz}
Comparison of SNe Ia spectral features in different host types at
low and high-$z$.  The values given are the mean and error in the mean of the 
residuals from the low-$z$ trend.}
\begin{tabular}{lcc}
\hline \hline
Feature   & \multicolumn{2}{c}{SNe Measurement (low-z trend subtracted)} \cr
          &     E/S0        &   Spiral        \cr
\hline
Low-$z$ \cr 
\hline
EW\{CaII\} -  \AA &  $-1.7\pm 4.7$           & $-2.0\pm  3.2$ \cr
EW\{SiII\} - \AA &  $ 5.1\pm 0.7$           & $-4.5\pm  0.9$ \cr
EW\{MgII\} - \AA &  $-0.6\pm 3.8$           & $ -0.7\pm  2.9$ \cr
$v_{\rm{ej}}$ - 1000 km/s   &  $ -0.18\pm 0.23$   & $ 0.17\pm  0.27$  \cr
\hline
High-$z$ \cr
\hline
EW\{CaII\} - \AA &  $-13.0\pm 16.0 $        &  $-0.8\pm  7.2$ \cr
EW\{SiII\} - \AA &  $ 5.0\pm 2.4 $          &  $-1.4\pm  1.1$ \cr
EW\{MgII\} - \AA &  $29.1\pm 9.5$           &  $ 0.1\pm  7.7$ \cr
$v_{\rm{ej}}$ - 1000 km/s  &  $ -0.13\pm 0.90$   & $ 0.55\pm  0.50$ \cr
\hline
\end{tabular}
\end{table}

Similar to the low-$z$ results, the high-$z$ data
show a trend with regards to host type and EW\{SiII\}, though
at a slightly lower significance because of the larger 
high-$z$ uncertainties.  We also see a possible difference
in MgII that is not seen at low-$z$.  Thus, the KS tests of the distribution of
high-$z$ EW\{SiII\} and $v_{\rm{ej}}$ results do not show as
significant of a statistical difference between SNe from different host
types.  This is probably due to the smaller number of measurements in
the high-$z$ data in addition to the larger errors.  For example, the measured maximum deviation
between the EW\{SiII\} distribution in the high-$z$ galaxies (KS
statistic $D$) is $D = 0.50$.  This value is comparable to the low-$z$
$D = 0.64$ and would indicate a significant difference if the high-$z$
sample was larger by only 10 measurements.

The trend for lower EWs and higher ejection velocities in late type
galaxies seen at low and now also high-z is a corrolary to the
broad-band photometric study of SNLS SNe host galaxies completed by
\citet{b57}, which showed that the lightcurve shapes of high-$z$ SNe
demonstrate a dependence on host galaxy star formation rates in a
manner consistent with low-$z$ objects.

\subsubsection{Dependence on Redshift}

\begin{figure*}
\begin{minipage}{1.0\textwidth} 
\begin{tabular}{cc}
\includegraphics[width=7.85 cm] {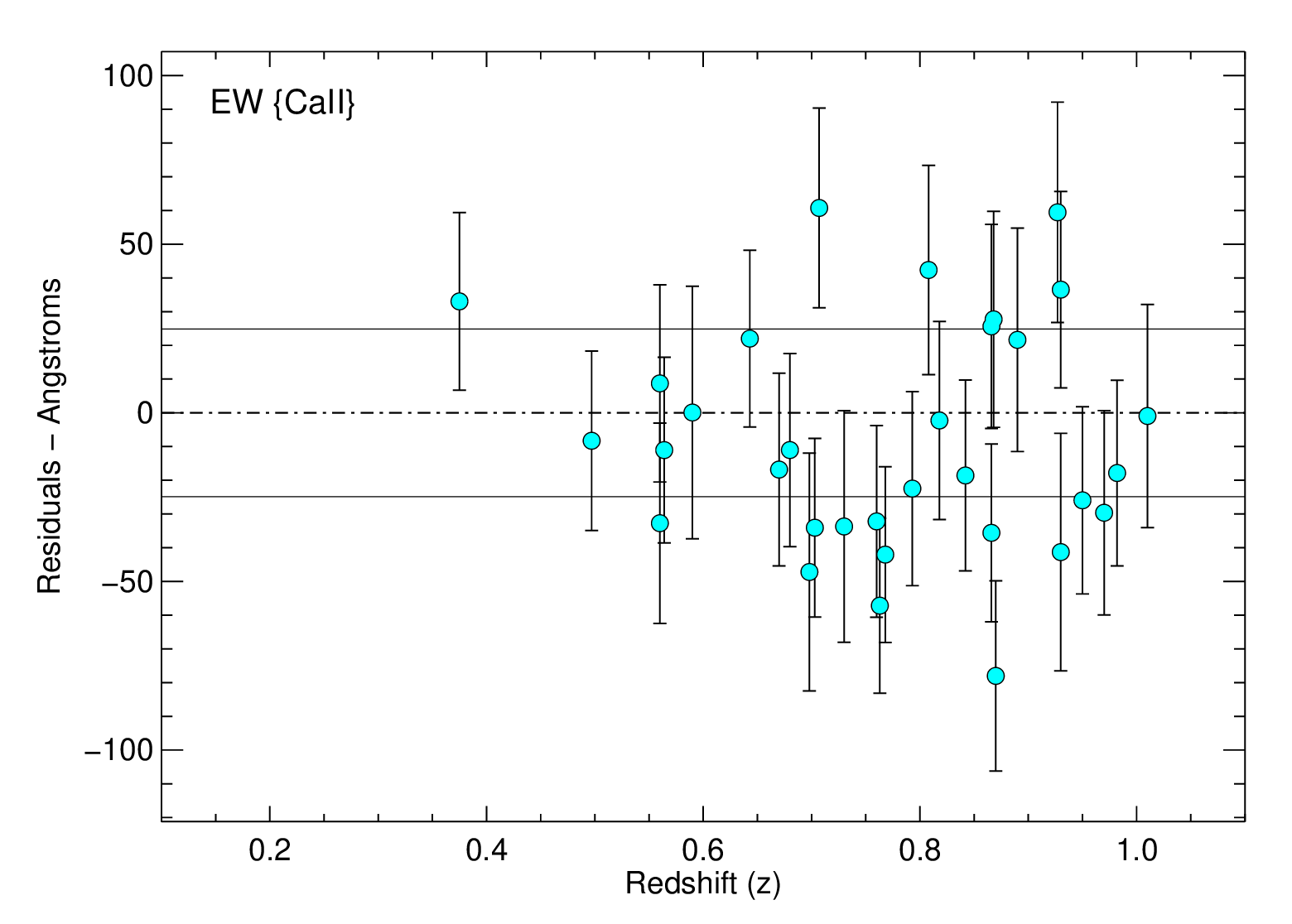}  & \includegraphics[width=7.85 cm] {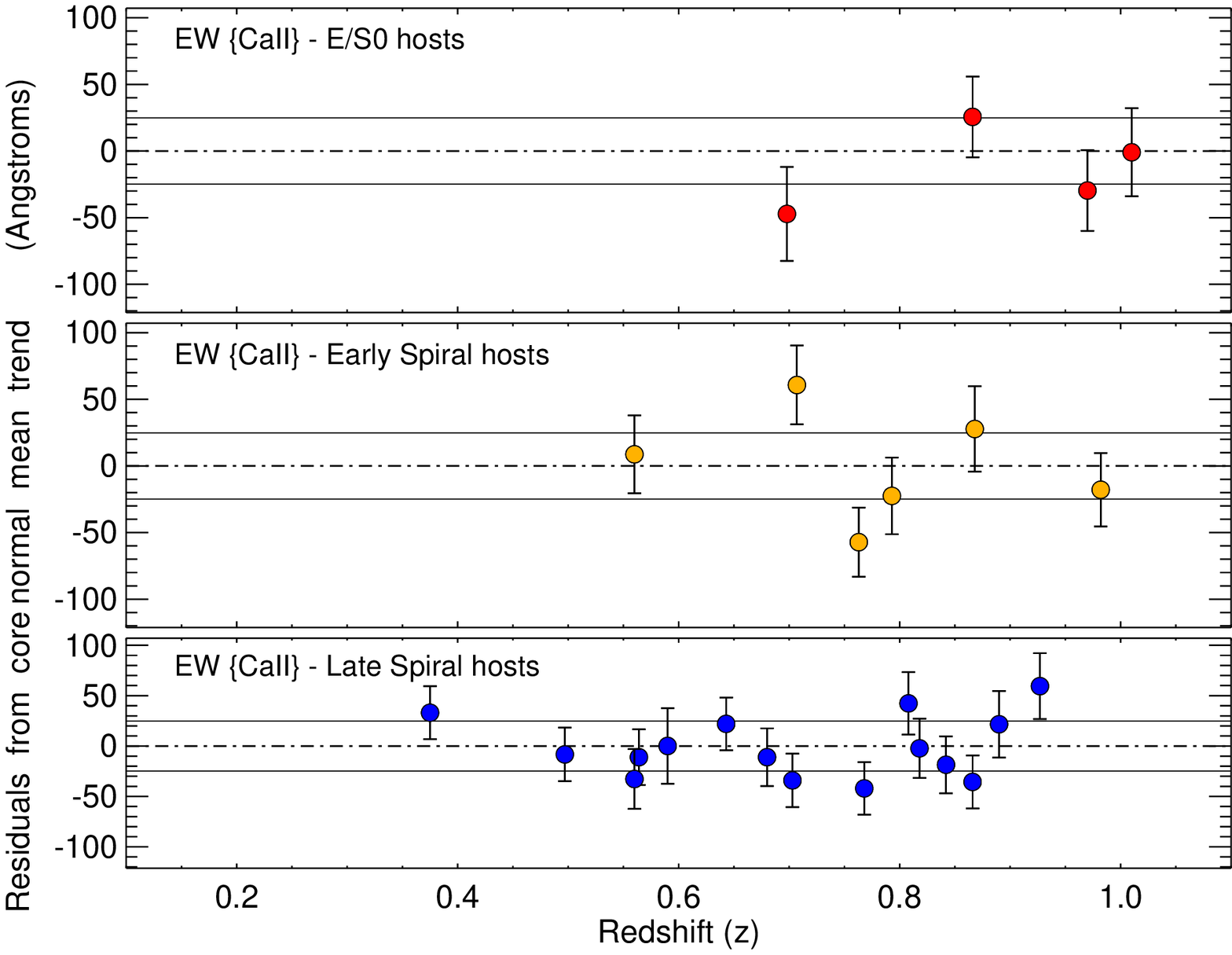} \\
\includegraphics[width=7.85 cm] {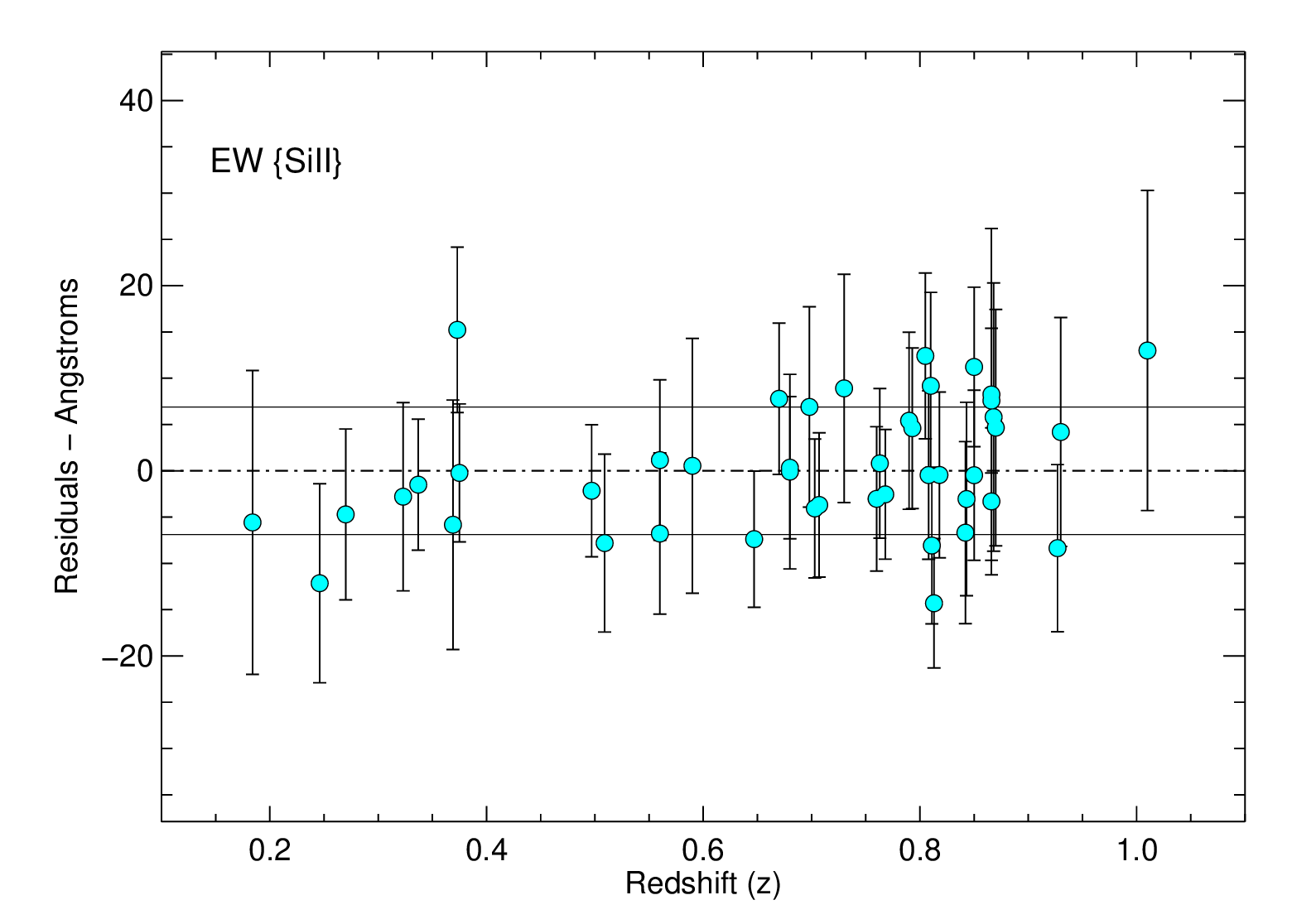}  & \includegraphics[width=7.85 cm] {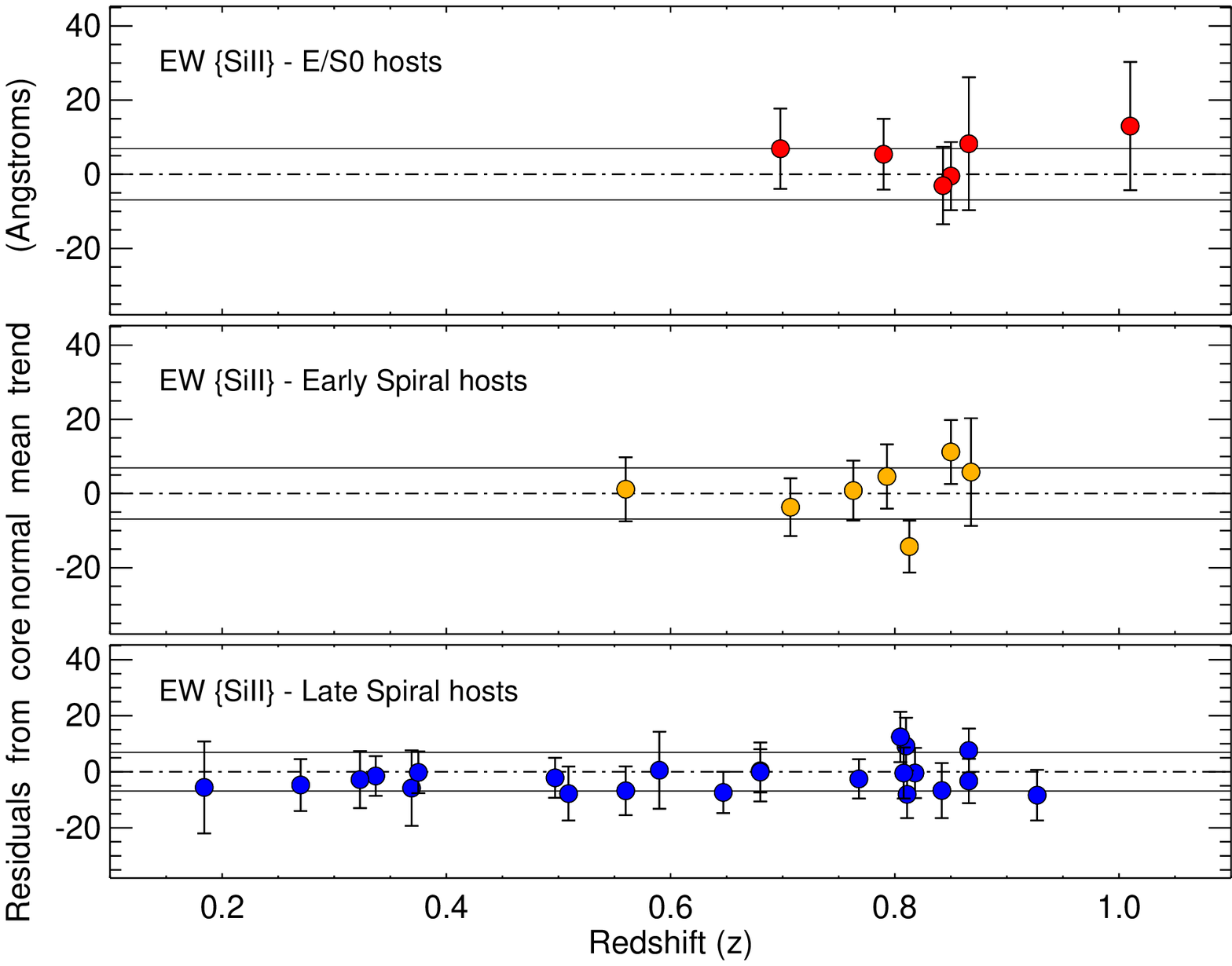} \\
\includegraphics[width=7.85 cm] {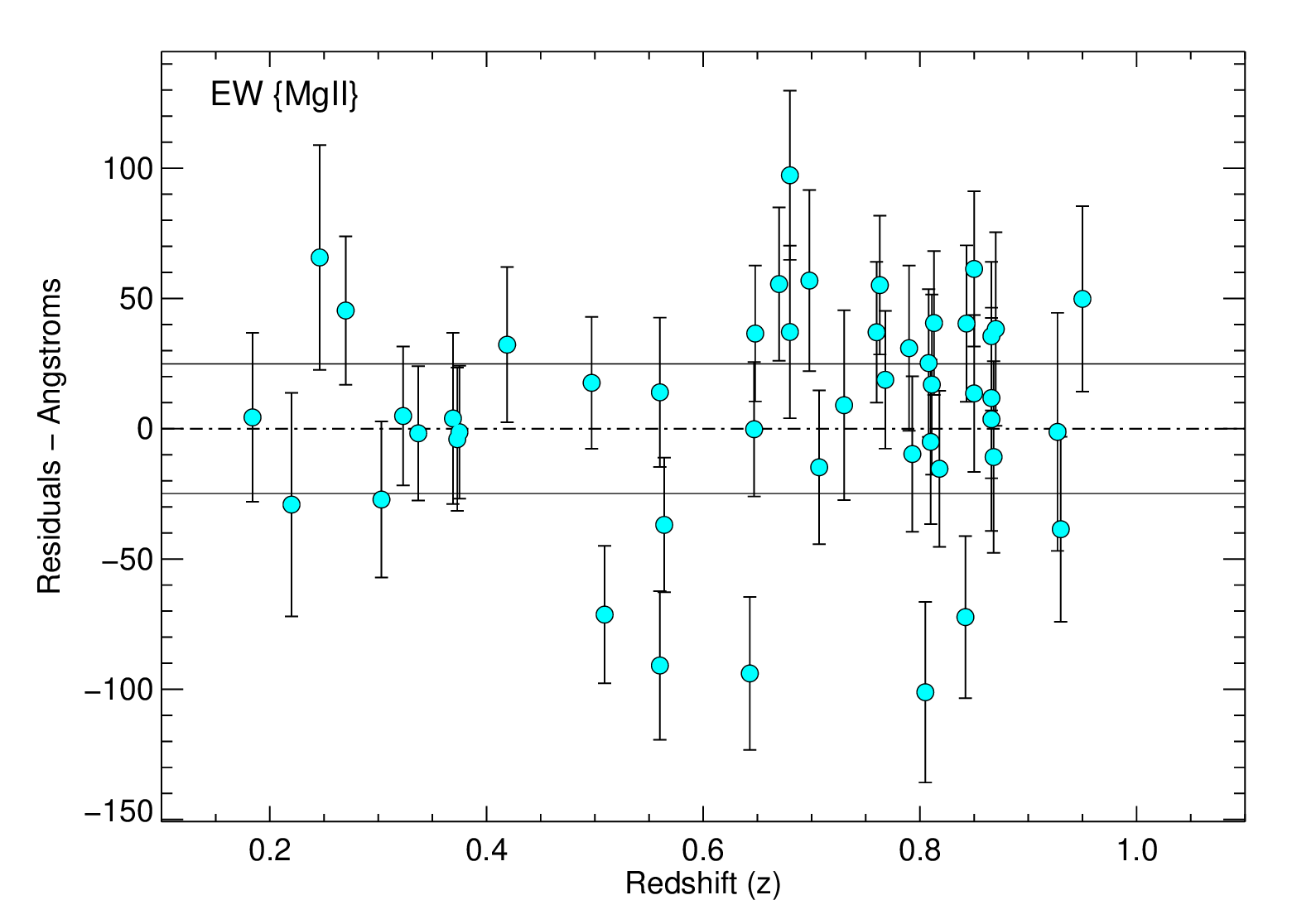}  & \includegraphics[width=7.85 cm] {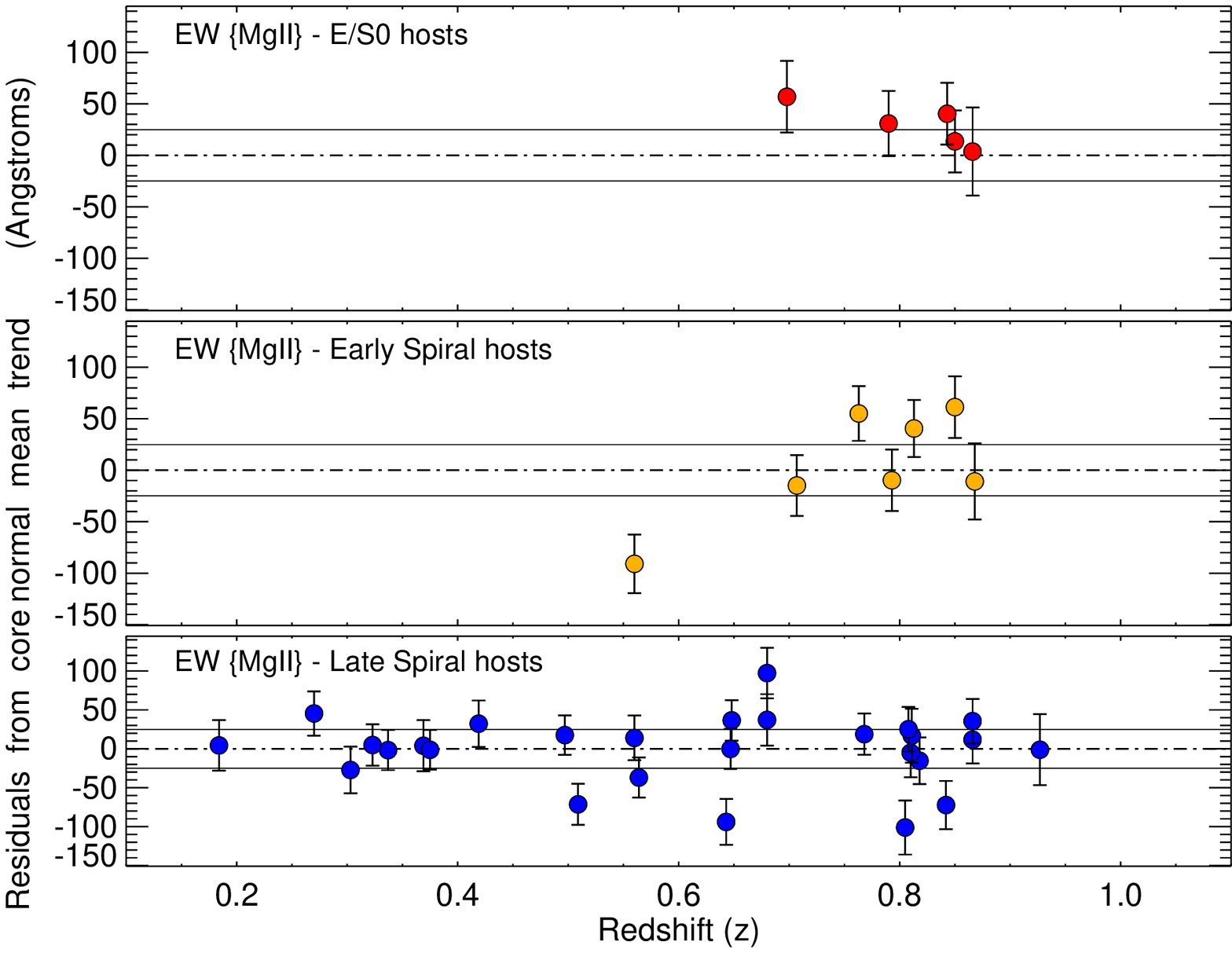} \\
\includegraphics[width=7.85 cm] {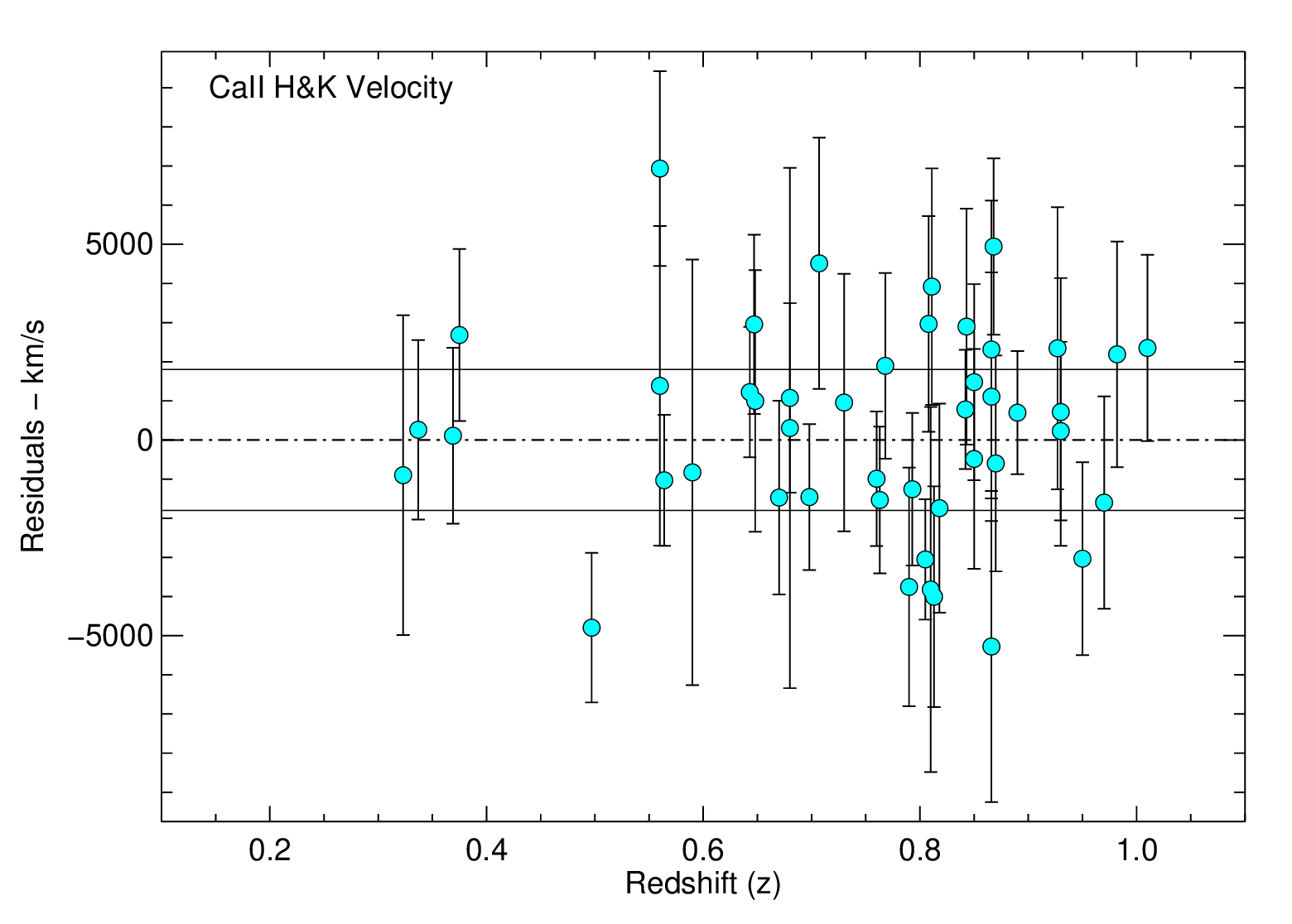}  & \includegraphics[width=7.85 cm] {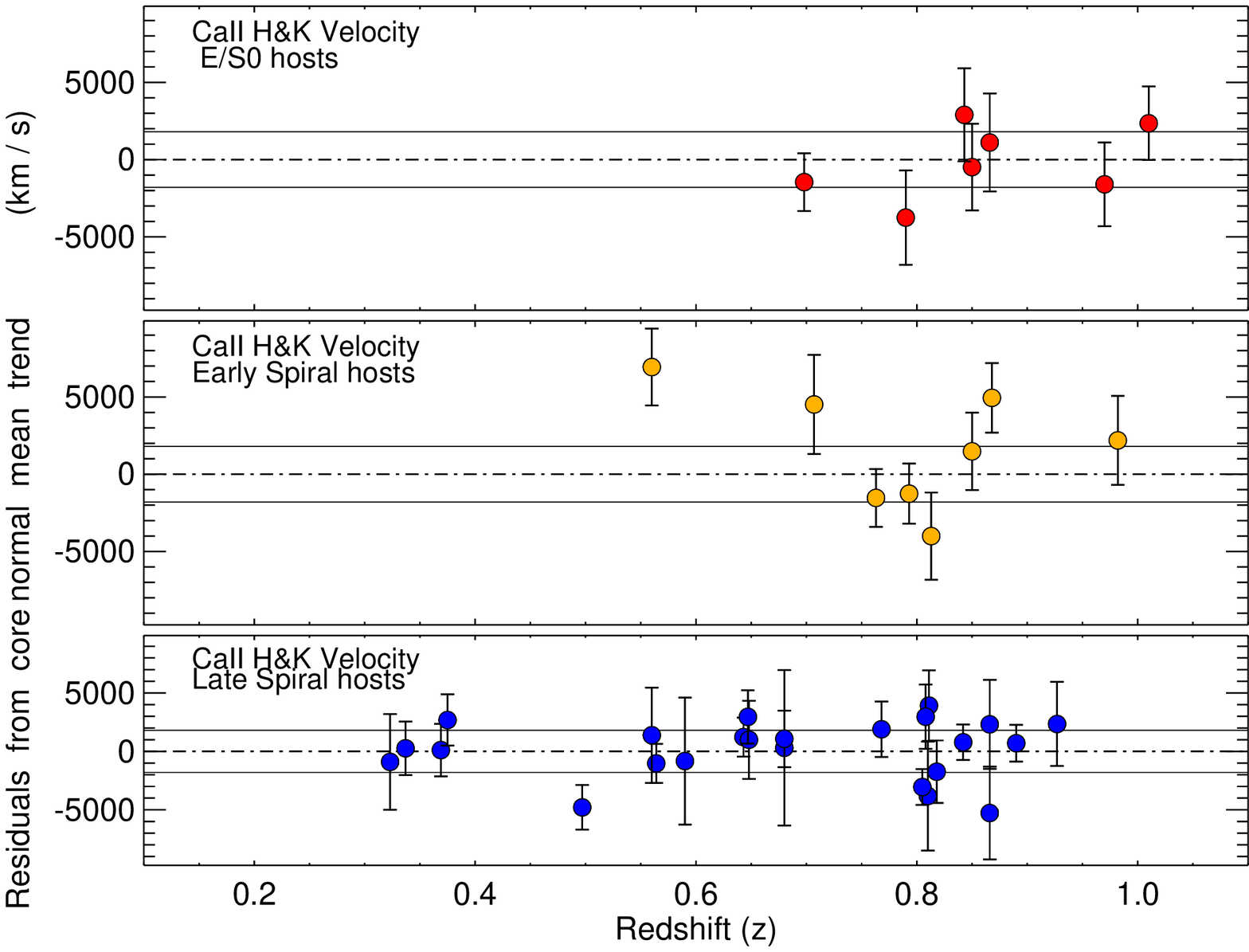} \\
\end{tabular}
\end{minipage}
\caption{\label{fig:all_resids} Left: The high-$z$ residuals (\emph{i.e.} the
difference from low-$z$ SNe Ia core normal mean trend) from the EW and
$v_{\rm{ej}}$ measurements are plotted against redshift. Right: the
same data divided by host galaxy type. }
\end{figure*}

The high-z EW and $v_{\rm{ej}}$ measurements (after subtracting the low-z
trend to remove the dependence on SN phase) are plotted against
redshift in the left-hand column in Fig. \ref{fig:all_resids}.  No
dependence on redshift is seen. Calculations of Spearman's $\rho$ rank
correlation and Kolmogorov-Smirnov tests attest that there are no
trends in the data with redshift that are significantly different from
random scatter.  Thus we see no evidence in our sample for systematic 
evolution of SNe Ia properties in the large redshift range covered 
by the SNLS.

On the right side of Fig. \ref{fig:all_resids}, the measurements are
divided by host galaxy type. The evolutionary histories of host galaxies within a given
morphological category have more in common than galaxies of different
classfications.  The comparison of SNe Ia spectroscopic features at
different redshifts but within similar host types should thus be more
sensitive to any evolutionary effects than tests using the bulk
properties of the high-$z$ sample as a whole.  The panels to the right
in Fig. \ref{fig:all_resids} illustrate that there are no
redshift-dependent trends within the SNe results from each host type.

\subsection{EW\{SiII\} - A New SNe Ia Spectroscopic Sequence}
\label{sec:siII_mags}

\begin{figure}
\begin{minipage}{.40\textwidth}
\centerline{\includegraphics[width=7.85 cm]{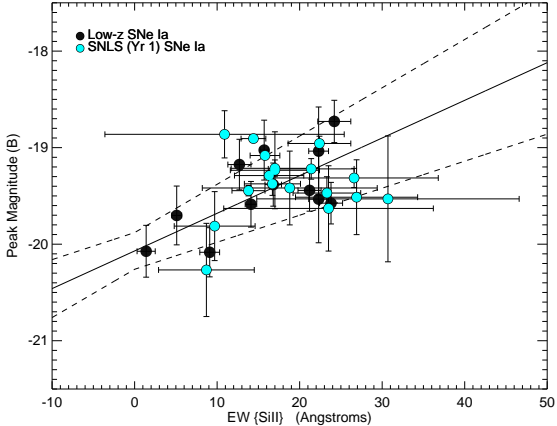}}
\end{minipage}
\caption{ \label{fig:siII_mags} A comparison of the EW measurement of
the `SiII 4000' feature to the absolute B' peak magnitudes of low-$z$
SNe Ia (black circles) and distant SNe (lighter circles). The
significant correlation in the low-z data reveals a new spectroscopic
sequence in SNe Ia that can be used to estimate SNe Ia peak
magnitudes.  The high-$z$ points have larger error bars with regards
to both EW and magnitude, but are still consistent with the low-$z$
trend (illustrated with the solid line). }
\end{figure}

During the course of this study, an interesting correlation was found
between EW\{SiII\} measurements and $M_{\rm{B}}$ for the low-z
SNe. Figure~\ref{fig:siII_mags} shows the low-$z$ SNe EW\{SiII\}
measurements, taken as close as possible to maximum light (within $\pm
7.0$ days), compared to the absolute peak (B') magnitudes (uncorrected
for any lightcurve shape-luminosity relationship) that were estimated
for 13 of the low-$z$ SNe (see Sect.~\ref{sec:lowz_phot}). A Spearman
rank correlation test found a correlation with $3\sigma$
significance between the measured EW\{SiII\} and calculated peak
magnitude, $M_{\rm{B}}$.  The best fit 
linear correlation of these quantities is described by the following:

\begin{equation}
M_{\rm{B}} = (-20.07 \pm 0.19) + (0.039 \pm 0.011) \times (\rm{EW\{SiII\}})  .
\label{eq:siII_mfit}
\end{equation}

The small residuals from this fit place this spectral indicator on par
with others in the literature \citep{b69,b8,b39,b113}. This spectroscopic
sequence is also relatively free of the epoch constraints on similar
sequences \citep{b39,b6,b16} because EW\{SiII\} evolves little near
maximum light and only one measurement is needed to constrain
$M_{\rm{B}}$ with this correlation. Any single EW measurement
near this date (the authors found that $\pm 7.0$ days was best as the
EW\{SiII\} values evolved little over this period) yielded the similar
results. The wavelength of this feature also makes it ideal as it
falls within the optical spectrum out to redshifts of about $1.2$ [unlike
measurements in the redder end of the spectrum such as the $\lambda5972$ feature studied in \citet{b113}].

One major physical parameter that
underlies this sequence is temperature \citep{b39,b114,b122}.  For an
assumed Chandrasekhar mass ($M_{\rm{CH}}$) WD explosion
scenario, the distribution of peak magnitudes is based primarily on
the different amounts of $^{56}$Ni synthesised.  Higher production of
this isotope, which is made at the expense of IME such as SiII,  results 
in a brighter SNe and causes the ejecta to be hotter.  
This tends to overionise the already smaller amount of IME
observed at maximum light thus results in the appearance of a smaller
amount of SiII in the spectra of hotter and brighter SNe at these epochs.  

Evidence that this spectroscopic sequence is
similar at high-$z$ is illustrated in the SNLS points in Fig.
\ref{fig:siII_mags}. In this Fig. the absolute magnitudes of the
SNLS high-$z$ SNe, uncorrected for lightcurve stretch or colour
adjustments, were computed assuming the best fit cosmology and
published $m_{\rm{B}}$ values from \citet{b3}.

\begin{figure*}
\centering
\begin{tabular}{cc}
\includegraphics[width=7.85 cm] {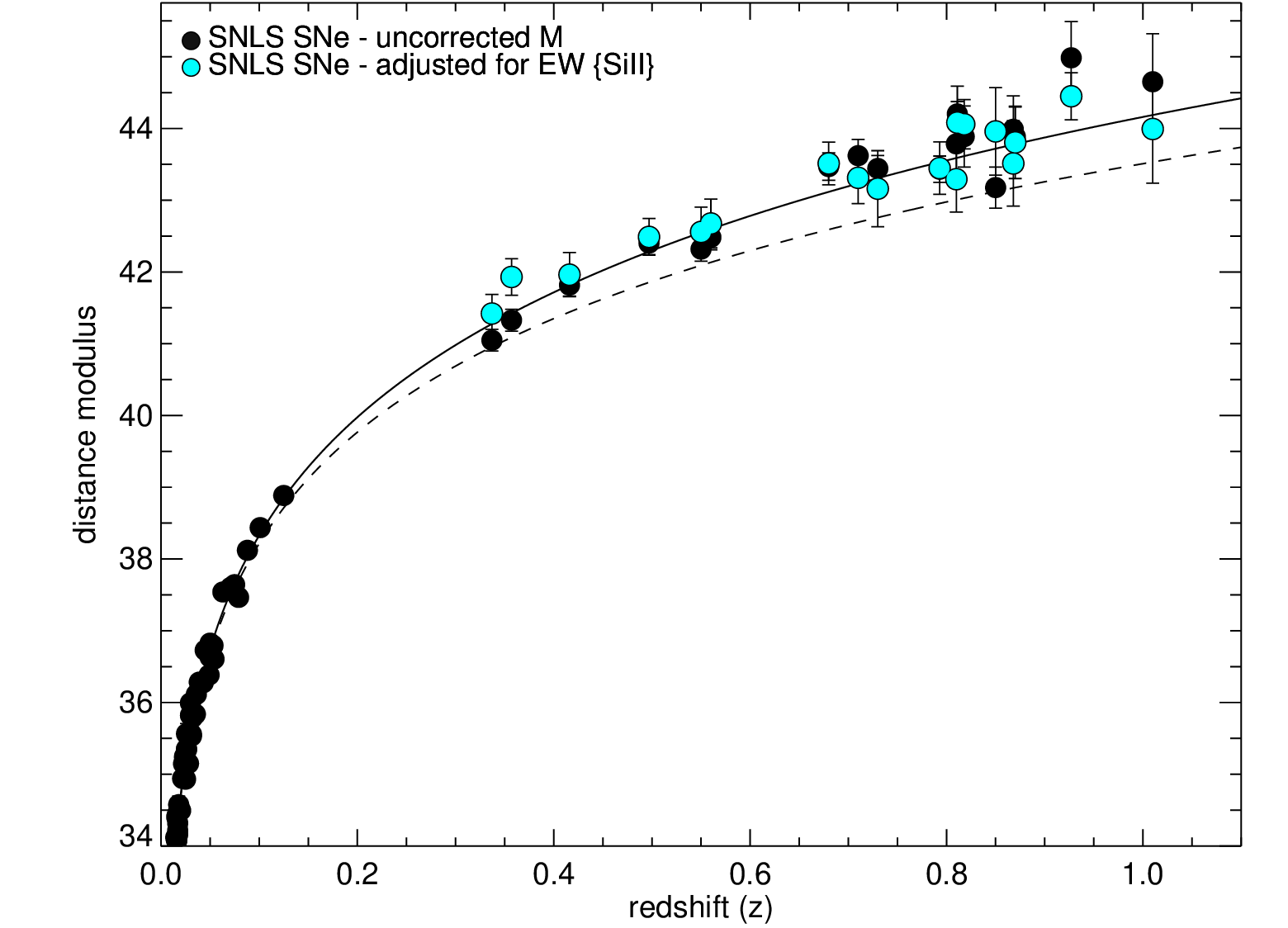} & \includegraphics[width=7.85 cm] {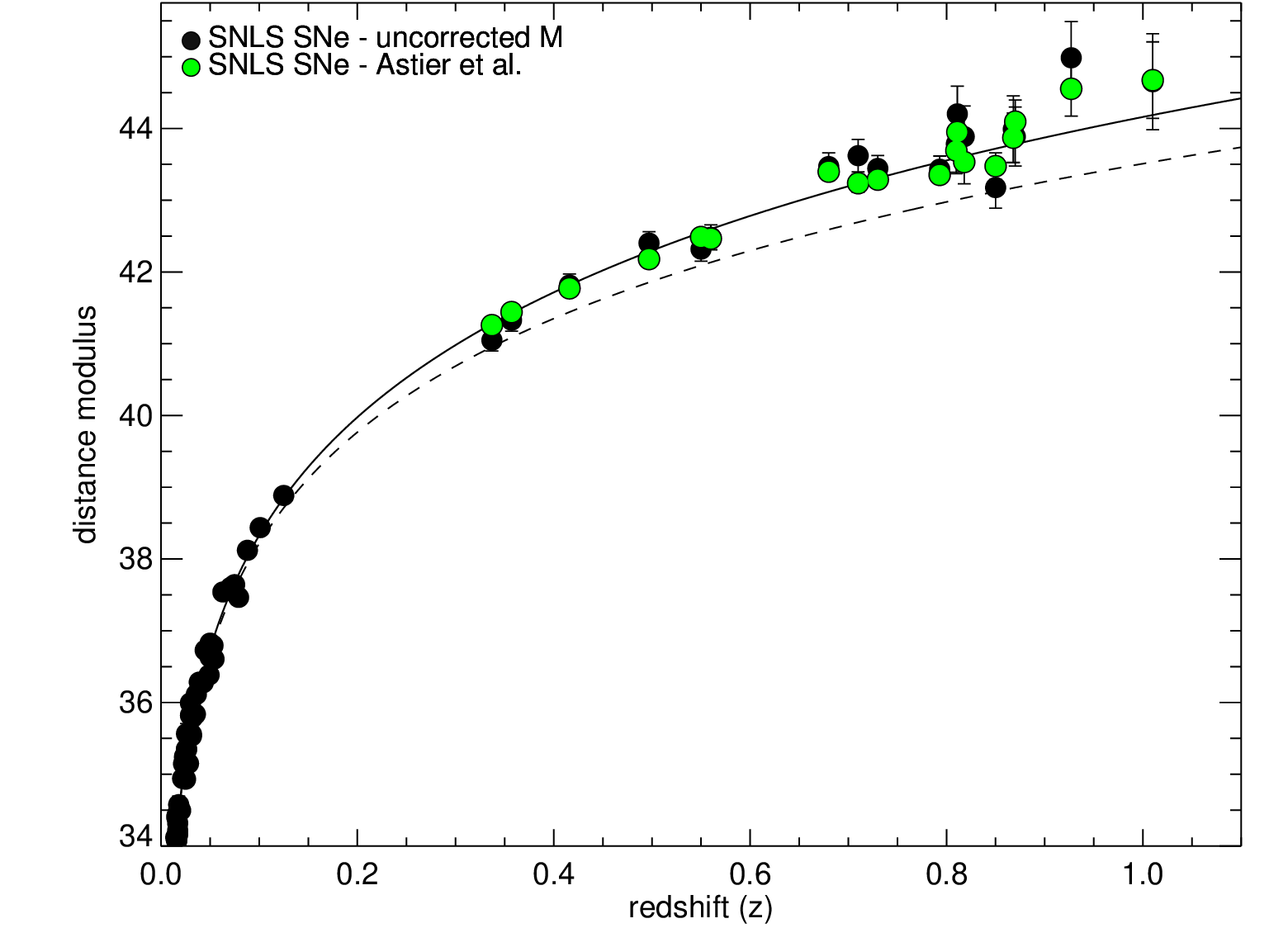} \\
\includegraphics[width=7.85 cm] {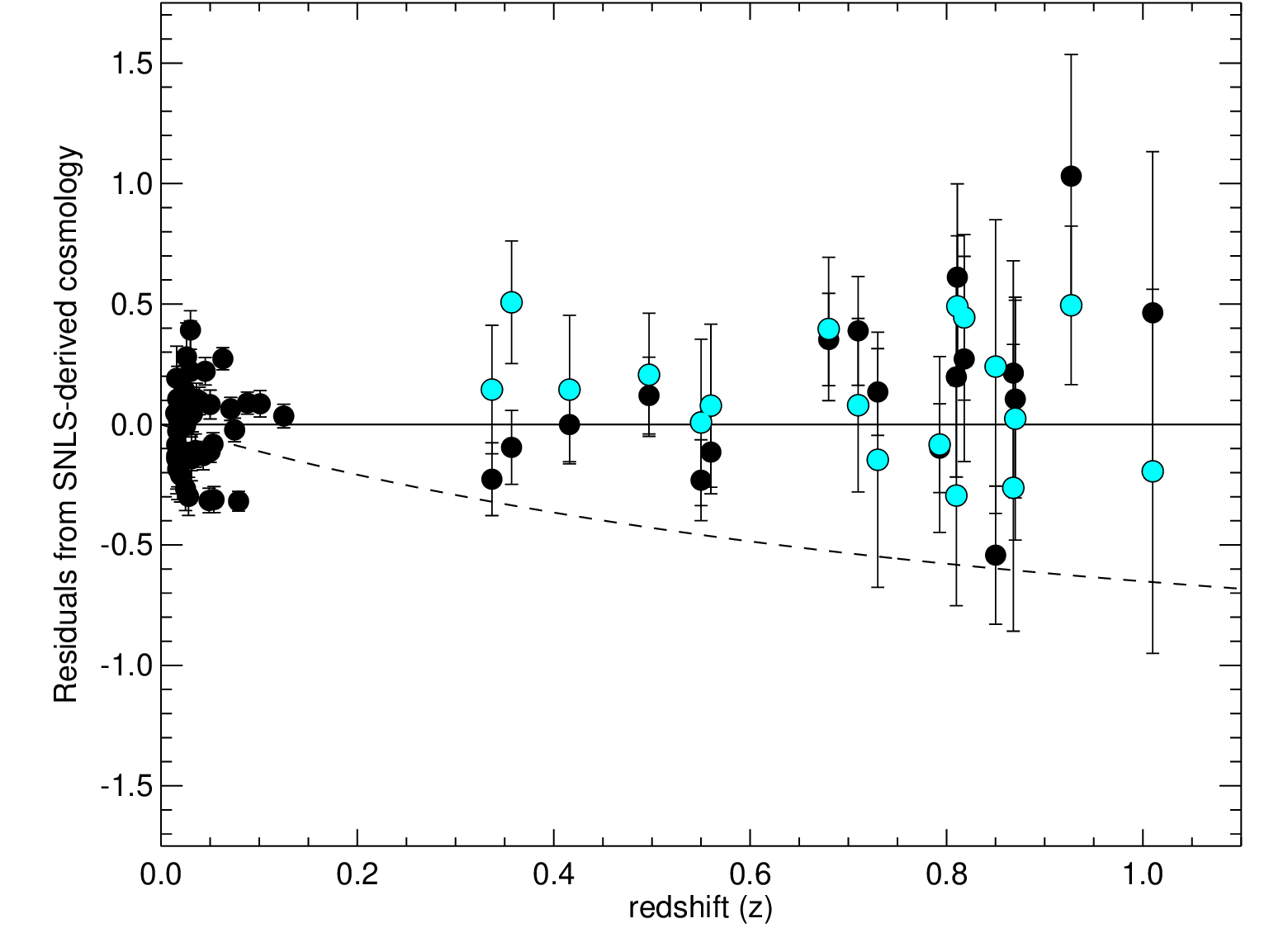} & \includegraphics[width=7.85 cm] {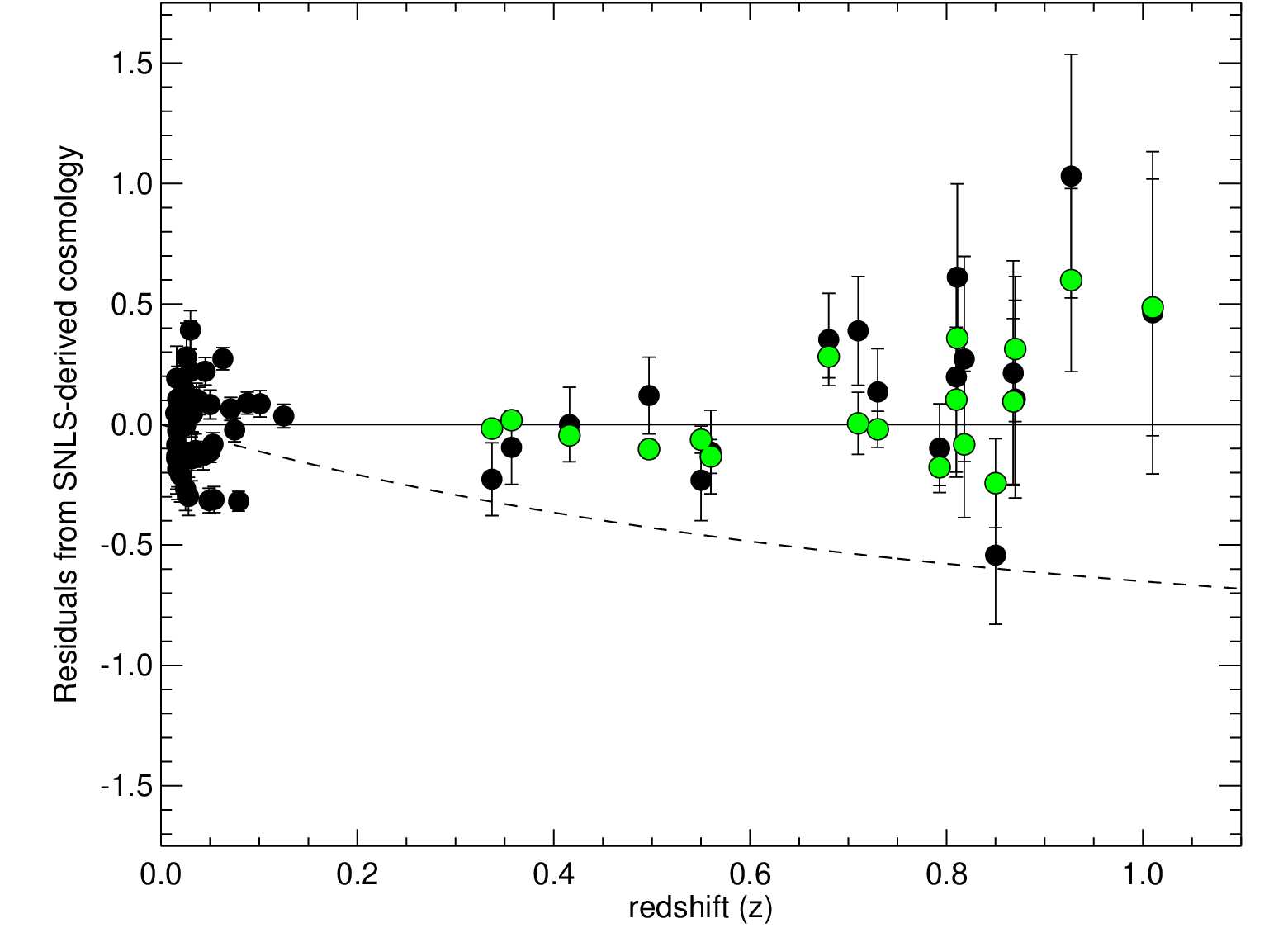} \\
\end{tabular}
\caption
{Hubble diagrams for a subset of SNe Ia from the SNLS first year
dataset. Solid points show the distance moduli uncorrected for stretch
or colour effects. Left: The lighter (or coloured) points show
distance moduli that were calculated with peak magnitudes estimated by
the EW\{SiII\} sequence. Right: The lighter (or coloured) points show
distance moduli that were corrected with $s$ (stretch) and $c$
(colour) corrections \citep{b3}.  The solid line indicates the best
fit cosmology from the first year SNLS results - a spatially flat
Universe with $\Omega_M = 0.263$ and $\Omega_{\Lambda} = 0.737$.  The
dashed line is the model from a flat Universe with $\Omega_M = 1.00$.
The residuals about this cosmology are shown in the bottom two panels;
the uncorrected high-$z$ SNe points have a dispersion about this
cosmology of $0.37$ magnitudes, while the EW\{SiII\} points have a
dispersion of $0.26$ magnitudes; this is comparable to the $0.23$
magnitude dispersion in the \citet{b3} results.  }
\label{fig:hubble_diagrams} 
\end{figure*}

Within the relatively large EW error bars, the SNLS supernovae are
still consistent with the low-$z$ trends in this parameter space.  The
next logical step is to determine whether the application of this
relation to high-$z$ SNe, just like the use of $s$ or
$\Delta_{m_{15}}$ parameters, decreases the scatter of luminosity
distances in a Hubble diagram.  This test is displayed in Fig.
\ref{fig:hubble_diagrams}.

For this test, the set of SNLS SNe that were published in the first
year SNLS cosmology results and had measured EW\{SiII\} values were
used.  A total of 18 SNe Ia from \citet{b3} met these criteria.  The
high-$z$ EW\{SiII\} measurements were used with the relation estimated
from the low-$z$ sample (Eq. \ref{eq:siII_mfit}) to estimate the
absolute peak B-band magnitudes (now termed $M_{\rm{SiII}}$).

The distance modulus ($\mu$) for these distant objects was then
calculated as

\begin{equation}
\label{eq:mu1}
\mu = m_{\rm{B}} - M_{\rm{SiII}} , 
\end{equation}

\noindent where $m_{\rm{B}}$ is the observed peak B' magnitude
previously published in Table 9 of \citet{b3}.  The uncertainties in
the linear $M_{\rm{B}}$-EW\{SiII\} fit and $m_{\rm{B}}$ were
propagated along with the EW errors during this calculation.  

Figure \ref{fig:hubble_diagrams} indicates that the EW\{SiII\}
sequence does provide a plausible method to use for estimating peak
SNe Ia distances for cosmology.  The dispersion in the
EW\{SiII\}-estimated distances about the best-fit cosmological model
from \citet{b3} is $0.26$ mag., which is near the computed residuals
of $0.23$ mag. from the \citet{b3} stretch and colour-corrected SNLS
SNe.  Both of these values are much improved over the $0.37$ magnitude
dispersion from the un-corrected SNe magnitudes in this sub-set of 18
high-$z$ SNLS SNe.  The application of this correlation to SNe Ia
cosmology is further discussed in \citet{b71}.

\section{Discussion}
\label{sec:mgII_diff}

As discussed in Sec. \ref{sec:results}, the spectroscopic measurements
and comparisons completed on this new set of distant SNe have shown that these
SNLS-observed objects display an overall similarity with the SNe Ia observed at 
low redshifts.  We find no significant difference between the SNe Ia samples for 
three of the four features studied and a possible difference in the fourth feature (EW\{MgII\}).  
We also find similar trends with host type at low and high redshift.  
Below we discuss these results in the context of the population of progenitors in addition
to the rates of various Type Ia sub-types in the SNLS high-$z$ sample.  

\subsection{Low-$z$ and High-$z$ Discrepancy - EW\{MgII\}} 

In Sect.~\ref{sec:hiz_stats} a possible difference in EW\{MgII\}
between low and high-z samples was presented. For the reason given in
that section (namely the differences in the time-sampling of observations 
in the two samples), the significance of this difference may not be high.
This was also sugested in a similar study by \citet{bgabri}, who 
also noted the need for an analysis of a larger sample of normal SNe Ia to more accurately
determine the true mean trend for this measurement.

However, it is interesting to consider possible physical explanations,
for example it is possible that it is a sign of metallicity
differences.  Explosion and nucleosynthesis calculations from
\citet{b143} and \citet{b136}, which were computed by modelling
thermonuclear burning by deflagration (as opposed to delayed
detonation or detonation only) of a C+O WD progenitor, showed that
Type Ia SNe from higher metallicity (Z) environments produce a
significantly lower (by a factor of 10) amount of $^{24}$Mg.  In their
results, the abundances of Si and Ca were relatively unaffected by
metallicity changes.  Thus, it may be that the group of high-$z$ SNe
that are below the mean trend for this feature are all SNe from higher
metallicity progenitors.  The normal EW\{SiII\} and EW\{CaII\}
measurements in these SNe support this explanation within the context
of these models.

Conversely, model SNe Ia spectra from progenitors of varying Z calculated
by \citet{b33} showed the opposite effect in the EW\{MgII\}
measurements.  The \citet{b33} spectra were calculated with a similar
deflagration model SN [W7 from \citet{b153}], with the metallicity
changes approximated by scaling the number abundances of elements
heavier than O in the unburned C+O layer.  The final IME abundances
were not published in \citet{b33}, so they could not be compared
directly to the \citet{b143} results.  Instead, the EW measurements
defined for this study were made directly on the final \citet{b33}
model spectra.  These measurements showed that the SNe with higher Z
had larger EW\{MgII\} measurements at pre-maximum epochs but then
evolved such that the different Z models were nearly indistinguishable
after maximum light.  

Another factor that may be contributing to this difference is the
trend for the SNLS to find more slightly bluer, brighter objects
\citep{b3}.  F04 noted that the evolution of different SNe along the
mean trend for EW\{MgII\} correlated with luminosity, with brighter
objects making the `jump' to higher EW at later times.  The small
group of outliers below the EW\{MgII\} mean trend between $+5$ and
$+10$ days may just be reflecting a slightly brighter group of SNe in
the high-$z$ sample. The average $s$ value within these SNe is just
above 1.0 [this corresponds to brighter SNe, see Fig.~9 in
\citet{b3}], which supports this hypothesis.  This feature comparison
may be more sensitive to slight differences in the strecth
distribution of the samples than the other EW results.  Further 
investigation into this parameter, both
observationally and theoretically, is certainly warranted.

\subsection{Rate of Peculiar SNe}

The EW results of this large high-$z$ set note 2 to 3 objects
(out of 54 distant supernovae with EW results, $\sim 3-6\%$) that are
spectroscopically similar to the class of 1991T-like SNe sub-types
that have been observed locally.  Previous studies of distant SNe Ia
have noted a similarly small percentage of these sub-types in the
high-$z$ samples; only spectra from SNe 2002iv ($z = 0.231$), 2003jn
($z = 0.33$), 2002fd ($z=0.27$) \citep{bgabri}, and 2003lh 
($z = 0.56$) \citep{b34,b129} have been shown
to be quantitatively similar to 1991T or 1999aa.  This is in contrast
to the apparent contribution of these sub-types to the low-$z$ SNe
sample, which is roughly 10\% to 20\% \citep{b1020,b1019a}.  This has
been interpreted as a possible sign of evolution
\citep{b1023,b1020,b1021} or of a significant difference between the
nearby and distant SNe Ia populations.  The identification of the 3
possible 1991T-like objects in this high-$z$ sample nearly doubles the number
of these sub-types observed at higher redshifts, but the percentage of
identified 1991T-like SNe is still much lower than at low-$z$.  It is
difficult to determine the significance of this apparent discrepancy
due to the unknown influence of selection effects.  For example, the
low-$z$ peculiar objects are often singled out for spectroscopic
follow up and thus may be over-represented in the nearby sample.  Another
complication is the uncertainty with regards to whether the overluminous 1991T-like
and 1999aa-like SNe are distinct sub-classes of Type Ia SNe or if they represent a
possible continuum of thermonuclear SNe \citep{b114}.  Note that the EW and $v_{\rm{ej}}$ 
measurements cannot readily address this latter issue; the overluminous high-$z$ objects are
referred to as `1991T-like' in the interest of brevity only and does not represent an attempt
to clarify the debate on the 1991T-1999aa relationship.

It may be that the difference between the fraction of these sub-types
at different redshifts is primarily an effect of the epoch coverage
differences between these data sets.  SNe 1991T and 1999aa displayed
unusual spectroscopic features primarily before maximum light and then
evolved to look little different than core normal SNe.  Within the
low-$z$ set considered here, 77\% of the SNe were observed at least
once prior to maximum light, compared to only 33\% (18 of 54 SNe) of
the high-$z$ sample.  With the simple assumption that this factor of
$\sim 2$ difference has reduced the number of overluminous SNe
possibly identified in the high-$z$ sample by a similar amount, the
disparity in the percentage of overluminous sub-type identifications between the
two data sets is explained.  

Even if this explanation is true, there is still a lack of
of the underluminous 1991bg-like SNe at high redshifts.  This is perhaps
expected for this data set as many factors work against SNLS spectroscopic
identification of these objects.  Mostly, it is because they are intrinsically much fainter and thus 
harder to observe photometrically and spectroscopically.  
This implies that even if the SNLS is observing these 
objects spectroscopically, they would be much harder to conclusively type.  Thus the 
lack of 1991bg-like SNe Ia in this set is not unexpected. 

The end of Sec. \ref{sec:snls_spect_sum} summarizes that, due to
the SNLS survey setup, high-$z$ SNe with vastly different
properties than the nearby objects would most likely not be
selected for spectroscopic observation.  Thus, the conclusions of
the tests made here cannot rule out that such a population of SNe
Ia exists at high-$z$, although none of the results show any
indication of new sub-classes of distant SNe. However, we note that
occasionally very rare and unsual objects are found [see SNLS
object 03D3bb in \citet{bhowellschandra}, for example]. 

\subsection{Implications for Progenitors}

Comparisons of SNe Ia across different host galaxy types can also reveal keys to the progenitor scenarios for these objects.
One model of SNe Ia formation \citep{bsb} has used the implications of observed SNe Ia environmental dependencies
\citep{b61,b60,b21b,bhowell,bmannucci} to create a model that describes SNe Ia as two distinct populations; a `prompt' component dependent
on the galaxy star formation rate and a `delayed' component whose rate is proportional to galaxy mass.  This model has been supported with SNLS
data \citep{b57} and may be useful in predicting any evolutionary shifts in Sne Ia properties (Howell \emph{et al.~in prep}).  The measurements made in
this study can also be put into the context of this `two-component' model.  Table \ref{tab:host_hiz} and Fig \ref{fig:allz_ks_tests} show a
trend of SiII with host type (at both low and high-$z$) in the direction expected from this two-component model.  At all redshifts, SNe in spiral
hosts show weaker SiII.  Assuming that these early type galaxies are star-forming, and hence host the `prompt' SNe Ia (which also tend to be
the brighter, higher stretch SNe Ia), and that the EW\{SiII\}-$M_{\rm{B}}$
correlation is valid, then the trend of lower EW\{SiII\} measurements in SNe from late type galaxies agrees with the main aspects of the two-component model. 

Another implication of this model is that as observations are extended to higher redshifts, 
a larger sampling of SNe Ia from the `prompt' star formation-based population would be expected.  
As these `prompt' SNe are brighter than their `delayed' counterparts, this should correspond to 
additional observations of SNe with high $s$ values and correspondingly low EW\{SiII\} measurements.  
This effect can be tested in the SNLS sample considered here by combining the high-$z$ results from all 
galaxy types.  $\chi^2$ tests and measures 
of correlation show no statistical difference in EW\{SiII\} between low and high-$z$ samples (see also 
Fig. \ref{fig:all_resids}).  However, this could be due to a lack of sensitivity to detect this subtle 
drift in the population.  A possible shift of $\approx 10\%$ in the brightness (or stretch) of the high-$z$ 
population (Howell \emph{et al.~in prep}) would correspond to a change of the order of a few ($\approx 4$) $\AA$ in EW\{SiII\}; 
Table \ref{tab:host_hiz} shows that the error in the means of the distribution of high-$z$ EW\{SiII\} is on the 
order of this shift and is much higher than the corresponding error in the low-$z$ results.  An SNLS sample 
with 2-3 times more SNe (which will be available before the survey is complete) should be able to detect this 
shift toward the `prompt' SNe. 

\section{Conclusions}
\label{sec:conclusion}

New results from a quantitative study of a large set of spectra from high-$z$
SNe Ia used for cosmological fitting  
have been compared to
corresponding measurements on a sample of SNe Ia observed locally.
The measurements from all of the studied high-$z$ spectra are within the
range of the results at low-$z$.  A statistical comparison of these data sets reveals
that there is no significant difference (less than 2$\sigma$) in the
strength of 2 SNe Ia-specific absorption features and the ejection
velocity of CaII H\&K. The 4$^{\rm{th}}$ feature -- which quantifies absorption 
from MgII and other ions near 4300 \AA\ -- shows a possible
difference that should be investigated further.

These methods also reveal a new spectroscopic sequence in SNe Ia based
on the strength of the SiII feature near 4000 \AA\ at epochs close to
maximum luminosity. An initial test of this sequence was made on a
sub-set of 18 SNLS SNe that had the appropriate EW\{SiII\} measurement
and were published with previous SNLS cosmology results \citep{b3}.
This demonstrated that the EW\{SiII\}-$M_{\rm{B_{peak}}}$ relation
reduces the scatter in high-$z$ luminosity distance estimates in a
manner consistent with the lightcurve shape-luminosity corrections
that are currently used to standardise SNe Ia peak magnitudes for
cosmology.

At present this type of spectroscopic analysis is one of the most
promising avenues of investigating the dispersion, homogeneity and
possible evolution of distant Type Ia supernovae.  This paper has used
empirical techniques on the largest set of distant SNe Ia data studied
to date.  Although the selection method used to generate the SNLS
high-$z$ sample implies that the existence of a new population of Type
Ia SNe at high-z with unusual spectral properties cannot be ruled out,
a large set of high-$z$ SNe have been shown to be spectroscopically
similar to nearby objects. At high-z the same trends of SN properties
with host type are seen as at low-z, and furthermore, within each host
galaxy class no evidence for evolution in SN properties with redshift
is seen. These results represent an important step in addressing one
of the most pressing issues for Type Ia cosmology.

\begin{acknowledgements}

The SNLS collaboration gratefully acknowledges the assistance of
Pierre Martin and the CFHT Queued Service Observations team.  We also
thank the Gemini queue observers and support staff for both taking the
data presented in this paper and making observations available
quickly. TJB is grateful to Exeter College and the Alberta-Bart
Holaday Scholarship.  TJB would also like to thank the EU Type Ia
SNe RTN `young researchers' -- particularly Gabriele Garavini and
Gaston Folatelli -- for all of their assistance.  
Canadian collaboration members acknowledge
support from NSERC and CIAR; French collaboration members from
CNRS/IN2P3, CNRS/INSU and CEA; Portuguese Collaboration members
acknowledge support from FCT-Fundacao para a Ciencia e Tecnologia;
UK collaborators acknowledge support from the Science and Technolgy
Facilities Council (STFC, formerly PPARC).

SNLS relies on observations with MegaCam, a joint project of CFHT and
CEA/DAPNIA, at the Canada-France-Hawaii Telescope which is operated by
the National Research Council of Candada, the Institut National des
Science de l'Univers of the Centre National de la Recherche
Scientifique of France, and the University of Hawaii.  This work is
based in part on data products produced at the Canadian Astronomy Data
Centre as part of the CFHT Legacy Survey, a collaborative project of
the National Research Council of Canada and the French Centre national
de la recherche scientifique.  The Work is also based on observations
obtained at the Gemini Observatory, which is operated by the
Association of Universities for Research in Astronomy, Inc., under a
cooperative agreement with the NSF on behalf of the Gemini
partnership: the National Science Foundation (United States), the
STFC (United Kingdom), the
National Research Council (Canada), CONICYT (Chile), the Australian
Research Council (Australia), CNPq (Brazil) and CONICET (Argentina).
This research used observations from Gemini program numbers:
GN-2005A-Q-11, GN-2005B-Q-7, GN-2006A-Q-7, GS-2005A-Q-11 and
GS-2005B-Q-6.

The views expressed in this article are those of the authors and do
not reflect the official policy or position of the United States Air
Force, Department of Defense, or the U.S. Government.

\end{acknowledgements}

\bibliographystyle{aa}


\longtab{6}{
\begin{longtable}{llllrclrrr}
\caption{\label{tab:gemobs} Observed properties and instrument settings for SNLS SNe
candidates observed from November 2004 to May 2006. The magnitudes and
``percentage increase'' are estimated from the lightcurves and correspond to the date of spectroscopy.}\\
\hline\hline
SN Name & RA (2000) & Dec (2000) & UT Date & Exposure (s) & Mode & $\lambda_{\rm{cent}}$ & seeing (``) & Mag (i') & $\%_{\rm{inc}}$ \\
\hline
\endfirsthead
\caption{continued.}\\
\hline\hline
SN Name & RA (2000) & Dec (2000) & UT Date & Exposure (s) & Mode & $\lambda_{\rm{cent}}$ & seeing (``) & Mag (i') & $\%_{\rm{inc}}$ \\
\hline
\endhead
\hline
\endfoot
04D1pu & 02:27:28.437 & -04:44:41.71 & 2004-12-12 & 7200 & N\&S & 720 & 0.49 & 24.08 & 39 \\
04D2mh & 09:59:45.872 & +02:08:27.94 & 2005-01-09 & 2400 & C & 720 & 1.00 & 22.90 & 730 \\
04D2mj & 10:00:36.535 & +02:34:37.44 & 2005-01-09 & 3600 & N\&S & 720 & 0.62 & 23.19 & 54 \\
05D1az & 02:25:12.492 & -04:36:08.17 & 2005-09-04 & 5400 & N\&S & 720 & 0.63 & 23.78 & 1246 \\
05D1by & 02:24:35.448 & -04:12:04.16 & 2005-09-03 & 5400 & N\&S & 720 & 0.62 & 23.56 & 9 \\
05D1bz & 02:25:00.685 & -04:12:00.74 & 2005-09-02 & 7200 & N\&S & 720 & 0.88 & 23.96 & 4310 \\
05D1cc & 02:26:31.173 & -04:09:53.40 & 2005-09-05 & 3600 & N\&S & 720 & 0.46 & 23.18 & 114 \\
       &              &              & 2005-09-06 & 1800 &      &     & 0.83 & 23.18 & 114 \\
05D1ee & 02:24:46.766 & -04:00:01.86 & 2005-10-31 & 3600 & N\&S & 720 & 0.59 & 23.24 & 25 \\
05D1ej & 02:26:06.317 & -04:43:45.79 & 2005-10-27 & 3600 & N\&S & 720 & 0.47 & 22.83 & 15 \\
05D1em & 02:24:05.503 & -04:56:23.35 & 2005-11-05 & 3600 & N\&S & 720 & 0.62 & 23.71 & 476 \\
       &              &              & 2005-11-06 & 1800 &      &     & 0.57 & 23.71 & 9524 \\
05D1eo & 02:25:23.116 & -04:38:33.36 & 2005-10-27 & 5400 & N\&S & 720 & 0.39 & 23.75 & 21 \\
05D1er & 02:25:41.466 & -04:00:16.32 & 2005-10-28 & 7200 & N\&S & 720 & 0.55 & 23.99 & 94 \\ 
05D1hn & 02:24:36.254 & -04:10:54.94 & 2005-12-04 & 2700 & C    & 680 & 0.63 & 20.86 & 249 \\
05D1ju & 02:26:57.303 & -04:08:02.78 & 2005-12-27 & 7200 & N\&S & 720 & 0.50 & 24.04 & 91 \\
05D1ka & 02:24:58.945 & -04:34:28.56 & 2005-12-28 & 5400 & N\&S & 720 & 0.86 & 23.21 & 1059 \\
05D1kl & 02:24:33.544 & -04:19:08.33 & 2005-12-31 & 5400 & N\&S & 720 & 0.59 & 23.61 & 247 \\
05D2ab & 10:01:50.833 & +02:06:23.02 & 2005-01-09 & 1800 & C & 680 & 0.67 & 21.83 & 263 \\
05D2ah & 10:01:28.704 & +01:51:46.18 & 2005-01-09 & 1800 & C & 680 & 0.96 & 20.83 & 1017 \\
05D2ck & 10:00:45.203 & +02:34:22.13 & 2005-02-16 & 5400 & N\&S & 720 & 0.65 & 23.88 & 7592 \\
05D2ja & 10:00:03.809 & +02:17:36.12 & 2005-05-07 & 5400 & C    & 680 & 0.94 & 22.15 & 115 \\
05D2nt & 10:00:58.234 & +02:22:21.62 & 2005-12-26 & 5400 & N\&S & 720 & 0.69 & 23.46 & 340 \\
05D2ob & 09:59:00.705 & +01:50:56.62 & 2005-12-28 & 5400 & N\&S & 720 & 0.73 & 24.22 & 242 \\
05D3ax & 14:19:17.595 & +52:41:15.07 & 2005-02-11 & 5400 & N\&S & 720 & 0.45 & 23.40 & 342 \\
05D3bt & 14:18:35.980 & +52:54:55.95 & 2005-03-06 & 3600 & C & 720 & 0.84 & 23.52 & 71 \\
05D3cf & 14:16:53.369 & +52:20:42.47 & 2005-03-08 & 3600 & C & 720 & 0.69 & 22.96 & 94 \\
05D3ci & 14:21:48.085 & +52:26:43.33 & 2005-03-09 & 5400 & N\&S & 720 & 0.76 & 23.03 & 70 \\
05D3cq & 14:18:46.173 & +53:07:55.55 & 2005-04-11 & 5400 & N\&S & 720 & 0.64 & 24.10 & 156 \\
05D3cx & 14:21:06.560 & +52:45:01.70 & 2005-04-09 & 5400 & N\&S & 720 & 0.50 & 23.52 & 65 \\
05D3jq & 14:21:45.462 & +53:01:47.53 & 2005-06-04 & 5400 & N\&S & 720 & 0.60 & 23.36 & 30 \\
05D3ki & 14:21:16.341 & +52:35:42.47 & 2005-05-14 & 5400 & N\&S & 720 & 0.67 & 23.68 & 49 \\
05D3km & 14:22:38.298 & +53:04:01.14 & 2005-06-04 & 5400 & N\&S & 720 & 1.0 & 23.89 & 536 \\
05D3kp & 14:20:02.952 & +52:16:15.28 & 2005-06-03 & 5400 & N\&S & 720 & 0.81 & 23.36 & 7848 \\
05D3kt & 14:19:53.730 & +52:44:34.64 & 2005-06-03 & 5400 & N\&S & 720 & 0.82 & 23.53 & 141 \\
05D3lb & 14:17:31.775 & +53:10:04.38 & 2005-06-05 & 5400 & N\&S & 720 & 0.51 & 23.38 & 192 \\
05D3lc & 14:22:22.902 & +52:28:44.11 & 2005-06-06 & 3600 & C & 720 & 0.63 & 22.67 & 17 \\
05D3lw & 14:17:44.021 & +53:06:24.98 & 2005-07-01 & 7200 & N\&S & 720 & 0.45 & 24.41 & 170 \\
05D3mn & 14:18:45.206 & +52:19:23.56 & 2005-07-12 & 7200 & N\&S & 720 & 0.77 & 23.31 & 77 \\
05D3mh & 14:18:59.805 & +52:40:03.50 & 2005-07-10 & 7200 & N\&S & 720 & 0.54 & 23.50 & 30 \\
05D3mq & 14:19:00.398 & +52:23:06.81 & 2005-08-03 & 1800 & C & 680 & 0.71 & 21.48 & 4748 \\
05D3mx & 14:22:09.078 & +52:13:09.35 & 2005-07-30 & 2700 & C & 680 & 0.83 & 23.10 & $>$10000 \\
05D3ne & 14:21:02.946 & +52:29:43.92 & 2005-07-14 & 2700 & C & 680 & 0.85 & 20.48 & 1074 \\
05D4av & 22:14:10.515 & -17:54:42.67 & 2005-07-10 & 3600 & N\&S & 720 & 0.53 & 23.2 & 77 \\
05D4bm & 22:17:04.621 & -17:40:39.45 & 2005-07-11 & 3600 & C & 680 & 0.55 & 22.04 & 144 \\

05D4ca & 22:14:11.350 & -17:48:15.47 & 2005-08-02 & 2700 & C & 720 & 0.63 & 22.24 & 156 \\
05D4cn & 22:13:31.454 & -17:17:19.92 & 2005-08-02 & 5400 & N\&S & 720 & 0.70 & 23.30 & 82 \\
05D4dt & 22:14:25.851 & -17:40:16.03 & 2005-09-02 & 3600 & N\&S & 720 & 0.55 & 22.62 & 20 \\
05D4dx & 22:13:39.387 & -18:03:20.99 & 2005-09-06 & 5400 & N\&S & 720 & 0.47 & 23.14 & 39 \\
05D4dy & 22:15:30.102 & -18:12:55.54 & 2005-09-06 & 5400 & N\&S & 720 & 0.53 & 23.87 & 684 \\
05D4fn & 22:17:01.076 & -17:49:25.16 & 2005-10-08 & 7200 & N\&S & 720 & 0.65 & 23.86 & 4474 \\
05D4fo & 22:15:20.925 & -17:16:05.25 & 2005-10-07 & 3600 & N\&S & 720 & 0.77 & 22.47 & 1845 \\
05D4gw & 22:14:47.365 & -17:31:54.99 & 2005-11-05 & 5400 & N\&S & 720 & 0.46 & 23.69 & 283 \\
05D4hn & 22:17:13.545 & -17:54:45.40 & 2005-12-01 & 5400 & N\&S & 720 & 0.50 & 23.50 & 474 \\
06D3bz & 14:17:10.044 & +53:01:29.31 & 2006-02-08 & 3600 & N\&S & 720 & 0.71 & 23.20 & 115 \\
06D3cb & 14:20:43.536 & +52:11:28.11 & 2006-02-03 & 5400 & N\&S & 720 & 0.88 & 23.61 & 213 \\
06D3cc & 14:17:31.604 & +52:54:44.79 & 2006-02-02 & 5400 & N\&S & 720 & 0.58 & 23.55 & 99 \\
06D3cl & 14:22:15.106 & +52:16:41.53 & 2006-03-06 & 5400 & N\&S & 720 & 0.61 & 23.75 & 264 \\
06D3cn & 14:19:25.826 & +52:38:27.78 & 2006-03-06 & 1800 & C & 680 & 0.40 & 21.26 & 372 \\
06D3du & 14:21:18.563 & +53:03:27.10 & 2006-05-24 & 5400 & N\&S & 720 & 0.45 & 24.27 & 54 \\
06D3dz & 14:18:14.731 & +52:56:23.11 & 2006-05-02 & 2400 & C & 680 & 0.98 & 22.19 & 66 \\
\end{longtable}
}


\longtab{7}{
\begin{longtable}{llllll}
\caption{\label{tab:gemprop} SNLS identifications and derived properties for SNe candidates observed from Nov.~2004 to May 2006.} \\
\hline \hline
SN     & $z$   & $\sigma_{z}^{\rm{a}} $ & CI \& Type$^{\rm{b}}$  & Template$^{\rm{c}}$ & $z$ from$^{\rm{d}}$ \\
\hline
\endfirsthead
\caption{continued.} \\
\hline \hline
SN     & $z$   & $\sigma_{z}^{\rm{a}} $ & CI \& Type$^{\rm{b}}$  & Template$^{\rm{c}}$ & $z$ from$^{\rm{d}}$ \\
\hline
\endhead
\hline
\multicolumn{3}{l}{\emph{continued on next page}} \\
\endfoot
\hline
\multicolumn{6}{l}{{\small a --- redshifts identified with host galaxy features are accurate to $\approx \pm 0.001$; redshifts identified}} \\ 
\multicolumn{4}{l}{{\small with template matches are accurate to $\approx \pm 0.01$}} \\
\multicolumn{4}{l}{{\small b --- final object identification and confidence index}} \\
\multicolumn{4}{l}{{\small c --- best low-$z$ spectral fit (name and days past max.)}} \\
\multicolumn{4}{l}{{\small d --- the host features used to estimate $z$}}  \\
\endlastfoot
04D1pu & 0.639 & 0.001 & 3 - SN Ia*& ...      & OII \\
04D2mh & 0.59  & 0.01  & 3 - SN Ia*& 1996X +2  & SN \\
04D2mj & 0.513 & 0.001 & 5 - SN Ia & 1992A -1  & OII, H$\beta$, OIII \\
05D1az & 0.842 & 0.001 & 5 - SN Ia & 1996X +2  & possible OII  \\
05D1by & 0.298 & 0.001 & 5 - SN Ia & 2002bo -1 & H$\alpha$, H$\beta$, OIII, SII \\
05D1bz & ...   & ...   & 2 - SN   & ...      & ... \\
05D1cc & 0.564 & 0.001 & 5 - SN Ia & 1996X +7  & H$\beta$, OII, OIII \\
05D1ee & 0.558 & 0.001 & 4 - SN Ia & 1992A +6  & H\&K \\
05D1ej & 0.312 & 0.001 & 4 - SN Ia & 1989B -5  & NII \\
05D1em & 0.866 & 0.001 & 5 - SN Ia & 1999ee +5 & nearby galaxy\\
05D1eo & 0.737 & 0.001 & 2 - SN & ...      & H\&K, OII \\
05D1er & 0.85  & 0.01  & 3 - SN Ia*& 1999aa -1& SN \\
05D1hn & 0.149 & 0.001 & 5 - SN Ia & 2003du -7 & H$\alpha$, SII, OII \\
05D1ju & 0.707 & 0.001 & 3 - SN Ia*& 1996X +7 & OII, OIII \\
05D1ka & 0.857 & 0.001 & 2 - SN & ...      & H\&K \\
05D1kl & 0.560 & 0.001 & 3 - SN Ia*& 1990N -7 & OII, OII \\
05D2ab & 0.323 & 0.001 & 5 - SN Ia & 1998bu -4 & H$\alpha$ \\
05D2ah & 0.184 & 0.001 & 5 - SN Ia & 1994D -5  & H$\alpha$, SII \\
05D2ck & 0.698 & 0.001 & 5 - SN Ia & 1992A +6  & H\&K \\
05D2ja & 0.303 & 0.001 & 5 - SN Ia & 1999aw +16 & H$\alpha$, SII, OIII \\
05D2nt & 0.757 & 0.001 & 4 - SN Ia & 1999ee +3 & OII,H$\beta$, OIII \\
05D2ob & 0.924 & 0.001 & 4 - SN Ia & 1989B -1 & OII \\
05D3ax & 0.643 & 0.001 & 5 - SN Ia & 1999aw +12& OII, OIII \\
05D3bt & 0.462 & 0.001 & 2 - SN  & ... &   H$\beta$, OII, OIII \\
05D3cf & 0.419 & 0.001 & 4 - SN Ia & 1999be +19& H$\beta$, OII, OIII \\
05D3ci & 0.515 & 0.001 & 5 - SN Ia & 1992A -5  & OII, H\&K \\
05D3cq & 0.890 & 0.001 & 3 - SN Ia*& 1999ao +5 & H$\beta$, OII \\
05D3cx & 0.805 & 0.001 & 4 - SN Ia & 1999bk +7 & possible OII \\
05D3jq & 0.579 & 0.001 & 4 - SN Ia & 2000E +5& H$\beta$, H$\gamma$, OII \\
05D3ki & 0.965 & 0.001 & 2 - SN & ...      &  OII \\
05D3km & 0.97  & 0.01 & 3 - SN Ia* & 1999bo +1& SN \\
05D3kp & ...   & ...  &  2 - SN    & ...      & ... \\
05D3kt & 0.648 & 0.001 & 5 - SN Ia & 1999ee -2 & H$\beta$, OII, OIII \\
05D3lb & 0.647 & 0.001 & 5 - SN Ia & 1999ee 0  & H$\beta$, OII, OIII \\
05D3lc & 0.461 & 0.001 & 3 - SN & ...       & H\&K \\
05D3lw & 0.601 & 0.001 & 2 - SN & ...      & OII \\
05D3mn & 0.760 & 0.001 & 3 - SN Ia*& 1999av +2   & possible OII \\
05D3mh & 0.67  & 0.01 &  4 - SN Ia & 1998aq -8 & SN \\
05D3mq & 0.246 & 0.001 & 5 - SN Ia & 1992A +9  & H$\alpha$ \\
05D3mx & 0.46  & 0.01 &  4 - SN Ia & 1998bu +8 & SN \\
05D3ne & 0.169 & 0.001 & 5 - SN Ia & 1994D -4  & SDSS host spectrum \\
05D4av & 0.509 & 0.001 & 3 - SN Ia*& 2002bo -4 & OII \\
05D4bm & 0.375 & 0.001 & 5 - SN Ia & 1996X -1  & H$\beta$, OII, OIII \\

05D4ca & 0.607 & 0.001 & 2 - SN & 1988A +3& H$\beta$, OII, OIII \\
05D4cn & 0.763 & 0.001 & 3 - SN Ia*& 2002bo -1& OII, H\&K \\
05D4dt & 0.407 & 0.001 & 4 - SN Ia & 1994D -2 & H$\alpha$,OII, H\&K, H$\beta$, OIII \\
05D4dx & 0.793 & 0.001 & 3 - SN Ia*& 1002A +5 & H\&K, H$\delta$ \\
05D4dy & 0.79  & 0.01  & 4 - SN Ia & 1994D +2 & SN \\
05D4fn & 0.4   & 0.1   & 2 - SN & ...      & SN \\
05D4fo & 0.373 & 0.001 & 5 - SN & 1994D -2 & H$\alpha$, H\&K \\
05D4gw & 0.808 & 0.001 & 5 - SN Ia & 1999bp -2 & OII \\
05D4hn & 0.842 & 0.001 & 4 - SN Ia & 1981B 0   & OII, OIII \\
06D3bz & 0.72  & 0.01  & 4 - SN Ia & 1989B -1 & SN \\
06D3cb & ...   & ...   & 2 - SN & ...      & ... \\
06D3cc & 0.90  & 0.01  & 3 - SN Ia*  & 1990N -7 & ... \\
06D3cl & 0.555 & 0.001 & 2 - SN & ...      & OII,OIII,H$\beta$ \\
06D3cn & 0.818 & 0.001 & 5 - SN Ia & 1996X -1 & OII,OIII,H$\beta$ \\
06D3du & 1.0   & 0.1   & 2 - SN & ...      & SN \\
06D3dz & 0.294 & 0.001 & 2 - SN & ... & H$\alpha$,SII \\
\end{longtable}
}

\appendix
\section{Results - Spectroscopic analysis of SNLS SNe Ia}

\longtab{3}{
\label{tab:gem_results}
\begin{longtable}{lccccc}
\caption{Equivalent width and ejection velocity results for the
published high redshift SNe Ia observed by the SNLS at the Gemini
telescopes.  This set includes all of the SNLS objects (reduced by the
author) that were confirmed as Type Ia SNe (a confidence index of 3 or
higher) and were subject to less than 65\% contamination from their
host galaxies.  The blank spaces (`$\cdots$') indicate where a
measurement had to be removed due to inadequate wavelength
coverage. The uncertainties in the EW results include measurement
errors, errors from any possible variance in the pseudo-continuum, and
the uncertainty from the host galaxy contamination correction.  } \\
\hline\hline 
Name  &  Day$^{\rm{a*}}$  & EW\{CaII\}$^{\rm{b*}}$ & EW\{SiII\}$^{\rm{b*}}$ & EW\{MgII\}$^{\rm{b*}}$ & $v_{\rm{ej}}^{\rm{c*}}/1000.0$ \\ 
\hline
\endfirsthead
\caption{continued.} \\
\hline\hline 
 Name  &  Day$^{\rm{a*}}$  & EW\{CaII\}$^{\rm{b*}}$ & EW\{SiII\}$^{\rm{b*}}$ & EW\{MgII\}$^{\rm{b*}}$ & $v_{\rm{ej}}^{\rm{c*}}/1000.0$ \\ 
\hline
\endhead
\hline 
\multicolumn{3}{l}{\emph{continued on next page}} \\
\endfoot
\hline 
\multicolumn{3}{l}{{\small a --- Rest-frame day, relative to B' max}} \\ 
\multicolumn{2}{l}{{\small b --- Equivalent Width, \AA\ }} \\
\multicolumn{2}{l}{{\small c --- Ejection velocity, km/s}} \\
\multicolumn{3}{l}{{\small{\emph{\textbf{*} $1\sigma$ errors listed in parentheses}}}} \\
\endlastfoot

03D1ax  &  -2.30 (0.11)  & 117.6(15.4) & 13.8(2.0) & 105.3(10.0) & 11.84(0.29)\\
03D1bk  &  -5.23 (0.25)  & 95.7(11.6) & 21.0(3.7) & 98.3(18.2) & 12.71(3.40)\\
03D1cm  &  -4.42 (0.59)  & 52.0(13.1) & 14.8(10.6) & 125.0(27.5) & 16.98(1.81)\\
03D1co  &   7.42 (0.44)  & 87.8(16.9) & 16.7(3.4) & 246.2(21.8) & 13.99(6.45)\\
03D1ew  &   1.53 (0.91)  & 144.2(19.5) & 23.5(12.7) & 83.2(26.4) & 20.21(1.32)\\
03D4cj  &  -8.05 (0.01)  & $\cdots$ & 5.9(6.1) & 131.6(13.8) & $\cdots$\\
03D4cn  &  0.53 (2.42)  & 116.9(16.0) & 17.0(5.4) & 76.0(12.3) & 13.85(1.14)\\
03D4cy  &   5.86 (0.59)  & 163.2(22.5) & 8.7(5.8) & 124.5(35.3) & 16.38(3.22)\\
03D4fd  & -0.88 (0.70)  & 100.2(14.3) & 21.4(5.2) & 79.4(16.4) & 14.83(0.45)\\
03D4gl  &   8.27 (0.28)  & 104.8(16.6) & 17.0(5.3) & 72.9(10.4) & 20.42(1.96)\\
04D1de  &  -6.80 (0.10)  & 91.4(14.5) & 9.4(1.3) & 105.1(12.9) & 20.78(1.01)\\
04D1hd  &   1.85 (0.04)  & $\cdots$ & 11.9(11.5) & 99.1(21.4) & 15.25(1.44)\\
04D1hy  &  -2.19 (0.44)  & $\cdots$ & 15.5(6.0) & 101.4(16.9) & 16.11(2.05)\\
04D1ow  &   6.28 (0.42)  & 139.0(14.9) & 21.1(10.2) & 92.7(22.4) & 14.66(3.02)\\
04D2ae  &  -1.95 (1.78)  & $\cdots$ & 13.1(7.5) & 128.4(18.7) & 19.41(1.87)\\
04D2mh  &   1.58 (0.24)  & 116.5(29.1) & 18.2(11.9) & $\cdots$ & 14.41(5.15)\\
04D3dd  &   3.68 (1.69)  & 109.5(19.8) & 30.7(15.9) & $\cdots$ & 16.97(1.42)\\
04D3fq  &   1.35 (0.66)  & 83.3(23.4) & 26.6(10.2) & 102.5(26.2) & 16.26(2.76)\\
04D3kr  &   5.37 (0.09)  & $\cdots$ & 15.8(1.8) & 118.0(11.6) & 14.43(1.63)\\
04D3lp  &   1.00 (0.20)  & 100.1(16.9) & $\cdots$ & $\cdots$ & 17.62(2.28)\\
04D3mk  &  -1.47 (0.55)  & $\cdots$ & 2.2(1.0) & 129.0(15.4) & 12.31(2.08)\\
04D3ml  & -0.89 (0.84)  & 96.8(11.6) & $\cdots$ & 138.9(25.3) & 13.06(1.53)\\
04D3nq  &   9.36 (0.04)  & $\cdots$ & $\cdots$ & 154.5(34.8) & $\cdots$\\
04D3ny  &   2.15 (0.67)  & $\cdots$ & 26.9(7.4) & 91.3(18.8) & 11.23(4.32)\\
04D4dm  &   3.55 (0.47)  & $\cdots$ & 9.7(4.9) & 120.6(24.5) & 18.56(2.50)\\
04D4hu  &   6.22 (0.29)  & 68.6(11.2) & 12.9(3.0) & $\cdots$ & $\cdots$\\
04D4ic  &   2.92 (0.66)  & $\cdots$ & 17.7(7.9) & 137.0(21.3) & 15.92(1.69)\\
04D4ii  &  -4.58 (0.34)  & $\cdots$ & 10.7(3.9) & 122.2(15.3) & 19.99(3.21)\\
04D4jy  &   2.00 (0.40)  & 73.9(24.8) & $\cdots$ & $\cdots$ & 15.34(1.48)\\
05D1az  &   8.85 (0.17)  & 75.6(13.5) & 8.9(7.0) & 102.1(17.5) & 14.13(0.15)\\
05D1cc  &   6.80 (0.15)  & 89.7(17.9) & $\cdots$ & 102.0(11.8) & 12.79(0.58)\\
05D1em  &   6.69 (0.23)  & 126.8(17.7) & 25.0(16.5) & 140.9(34.2) & 14.95(2.76)\\
05D1er  &   2.83 (0.42)  & $\cdots$ & 29.0(5.5) & 160.7(18.9) & 16.35(1.83)\\
05D1ju  &   6.28 (0.27)  & 163.2(24.6) & 13.2(3.6) & 116.50(15.7) & 18.46(2.79)\\
05D1kl  &  -4.27 (0.21)  & 97.0(17.0) & 7.6(5.3) & 100.8(14.9) & 18.92(3.56) \\
05D2ab  &  -1.30 (0.08)  & $\cdots$ & 13.8(7.5) & 93.5(11.3) & 15.34(3.64)\\
05D2ah  &  -2.49 (0.03)  & $\cdots$ & 10.2(10.4) & 92.0(21.0) & $\cdots$\\
05D2ck  &  0.730 (0.74)  & 71.5(24.7) & 24.4(8.3) & 148.7(24.0) & 14.07(0.36)\\
05D2ja  &   9.63 (0.12)  & $\cdots$ & $\cdots$ & 161.3(25.3) & $\cdots$\\
05D3ax  &   8.03 (0.39)  & 118.9(9.0) & $\cdots$ & 65.6(7.2) & 14.77(0.62)\\
05D3cf  &   13.68 (0.41)  & $\cdots$ & $\cdots$ & 273.1(44.1) & $\cdots$\\
05D3cq  &   11.6 (1.35)  & 106.9(23.1) & $\cdots$ & $\cdots$ & 13.51(0.32)\\
05D3cx  &   10.28 (0.61)  & $\cdots$ & 27.2(5.7) & 98.5(13.6) & 10.01(0.33)\\
05D3km  &   1.46 (0.31)  & 87.1(19.7) & $\cdots$ & $\cdots$ & 13.67(2.07)\\
05D3kt  &  -3.86 (0.30)  & $\cdots$ & $\cdots$ & 123.5(15.2) & 18.34(2.68)\\
05D3lb  &  -1.21 (0.24)  & $\cdots$ & 9.2(2.6) & 88.5(7.7) & 19.16(1.35)\\
05D3mh  &   2.71 (0.40)  & 96.4(22.1) & 25.6(4.6) & 154.3(21.5) & 13.43(1.79)\\
05D3mn  &   2.82 (0.40)  & 80.7(25.7) & 14.8(3.7) & 136.4(14.7) & 13.88(0.24)\\
05D3mq  &   8.45 (0.07)  & $\cdots$ & 3.6(15.1) & 232.8(35.2) & $\cdots$\\
05D4av  &   5.84 (0.15)  & $\cdots$ & 9.3(6.7)  & 54.1(9.5) & $\cdots$ \\
05D4bm  &  -3.81 (0.05)  & 162.0(14.3) & 14.5(2.9) & 85.7(7.8) & 19.98(0.99) \\
05D4cn  &  0.39 (0.32)  & 62.3(10.7) & 18.2(4.3) & 146.2(13.2) & 14.10(0.57)\\
05D4dy  &   2.13 (0.55)  & $\cdots$ & 23.2(6.6) & 127.1(19.2) & 11.33(2.50)\\
05D4gw  &  0.13 (0.34)  & 162.6(22.1) & 16.8(5.9) & 115.8(13.8) & 18.70(2.08)\\
05D4fo  &  -5.31 (0.03) & $\cdots$  & 28.6 (5.1)  & 82.6 (11.7) & $\cdots$ \\
\end{longtable}
}

\Online

\section{SNLS Observed Spectra}

Sprectra from each Supernova Legacy Survey SN candidate observed at the Gemini telescopes from 
December 2004 through May 2006.  The observed properties and observation setup for each object 
is summarized in Table \ref{tab:gemobs} and the final object identifications and redshift estimates 
are displayed in Table \ref{tab:gemprop}.  These spectra have been corrected for sky absorption, the 
narrow host galaxy features have been removed, and the GMOS chip gaps have been interpolated over.  
These images show the observed spectrum, re-binned to 7.5 \AA\ resolution, in dark blue with a smoothed 
version of the spectrum overplot in light blue.  The smoothed spectrum is for illustrative purposes 
only and was generated witha  Savitsky-Golay filter computed with a $3^{\rm{rd}}$ degree polynomial fit 
over 40 pixels of the observed spectrum.  Note that for some of these spectra a large amount of host 
galaxy contamination is clear.    


\begin{figure*}
\begin{minipage}{2.0\textwidth}
\begin{tabular}{ccc}
\includegraphics[width=5.25 cm] {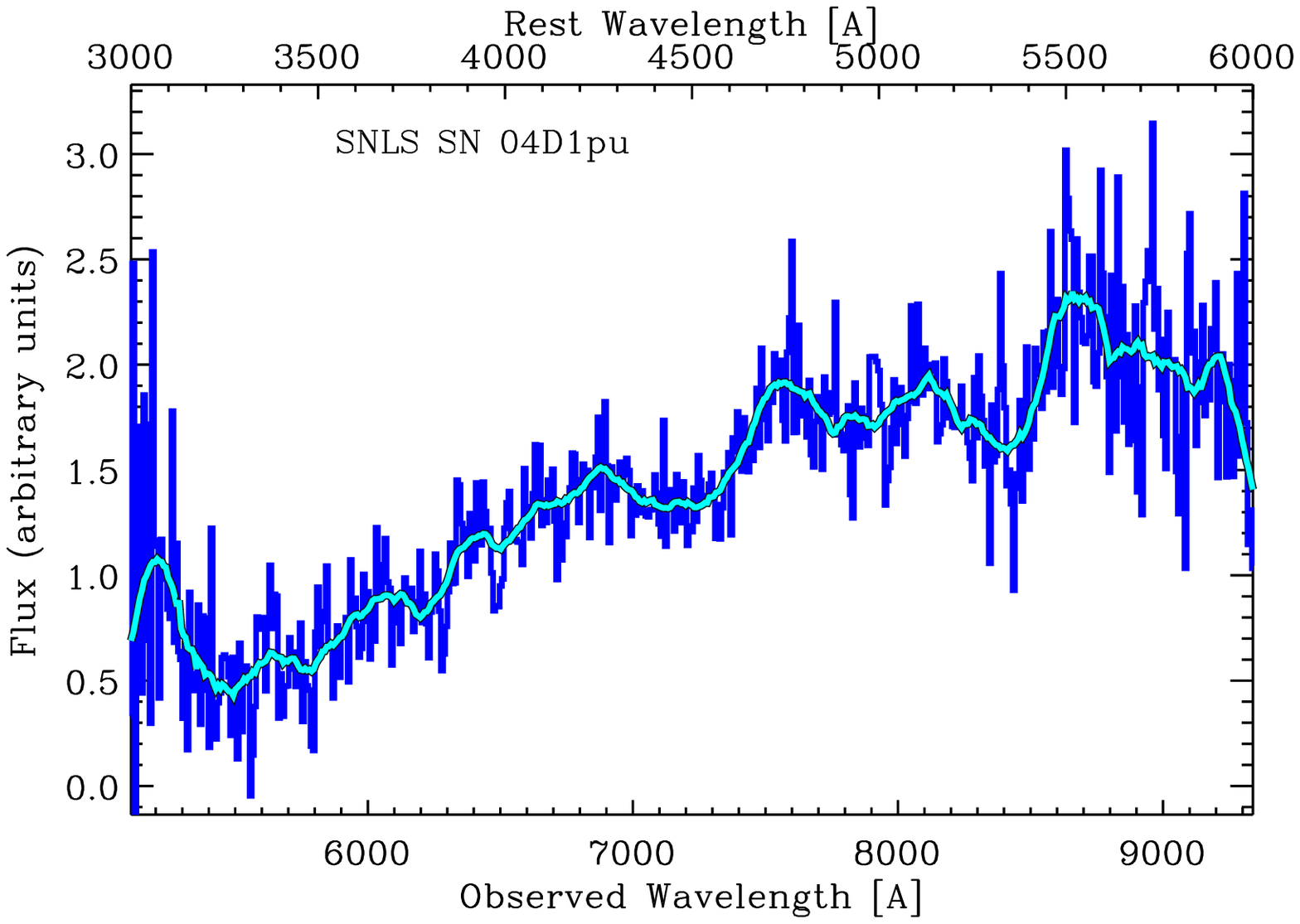} & \includegraphics[width=5.25 cm] {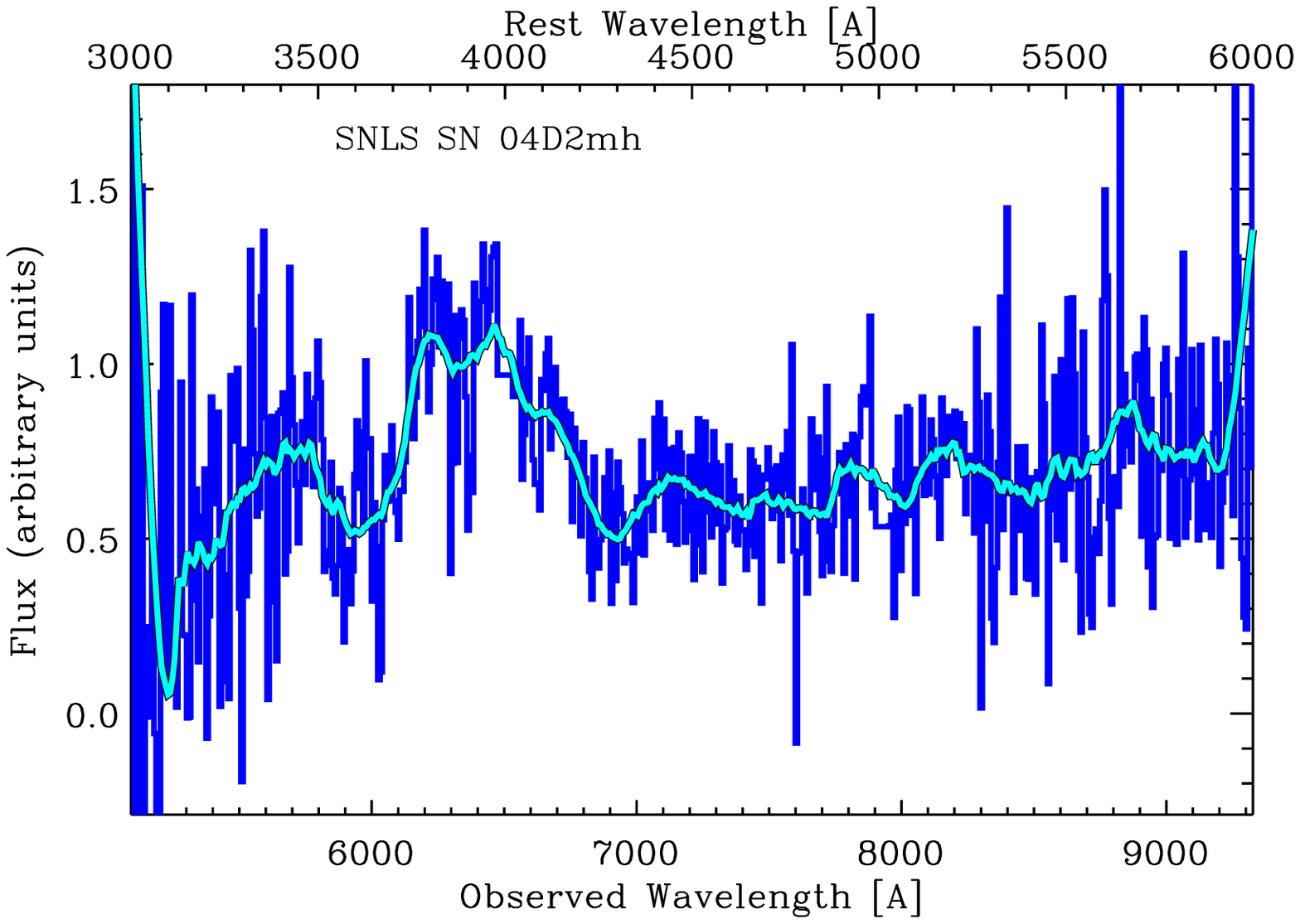} & \includegraphics[width=5.25 cm] {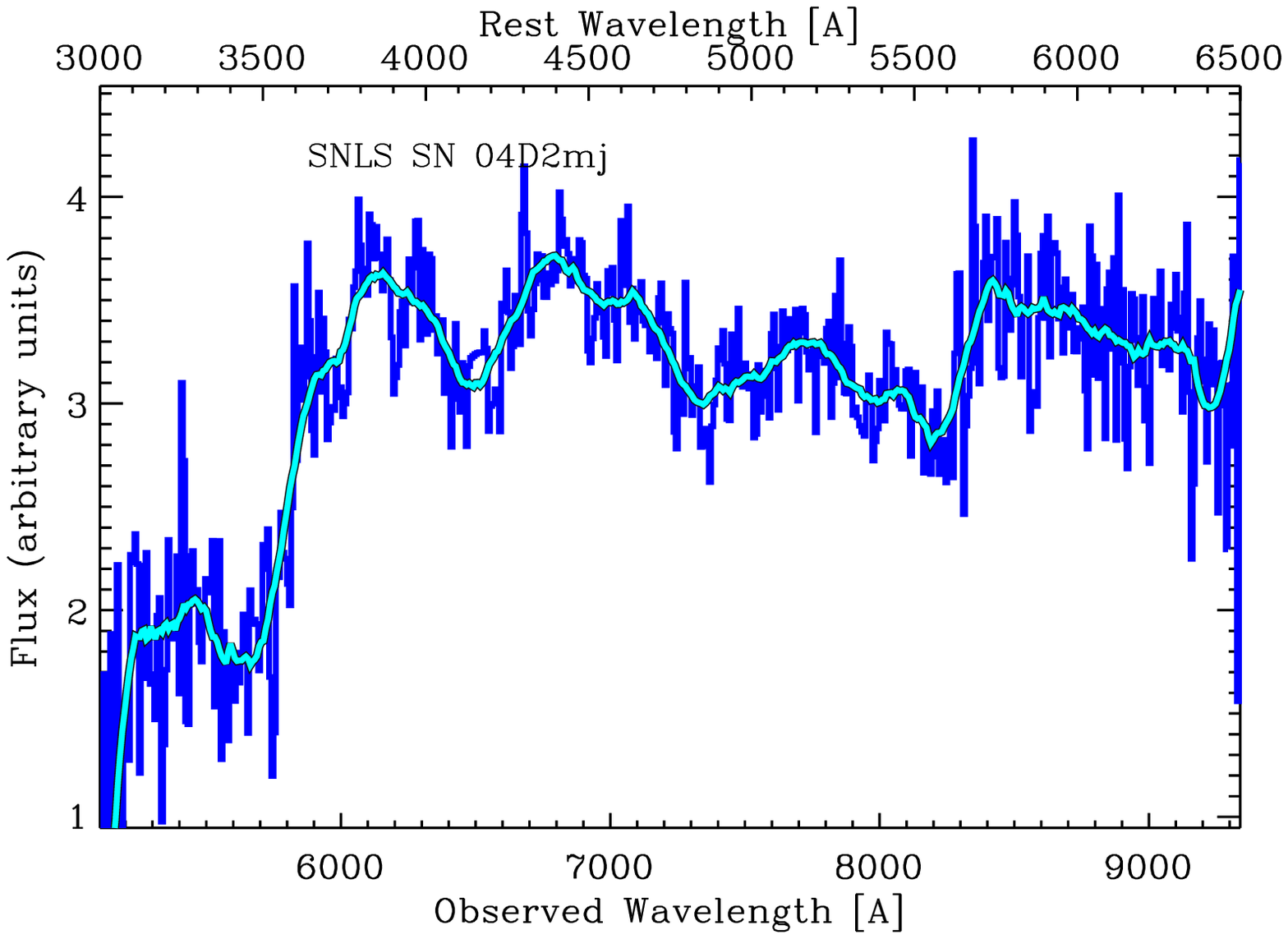} \\
\includegraphics[width=5.25 cm] {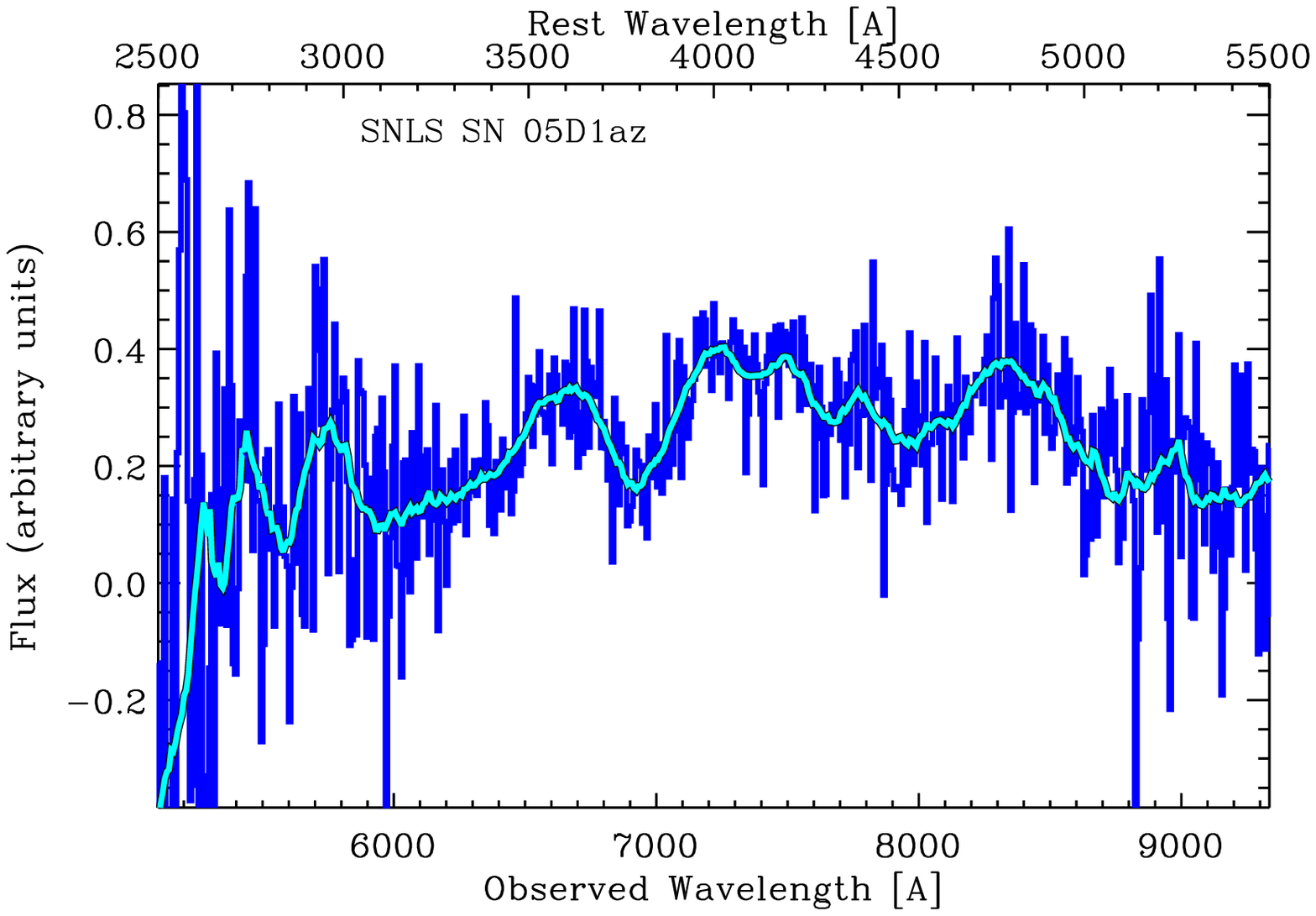} & \includegraphics[width=5.25 cm] {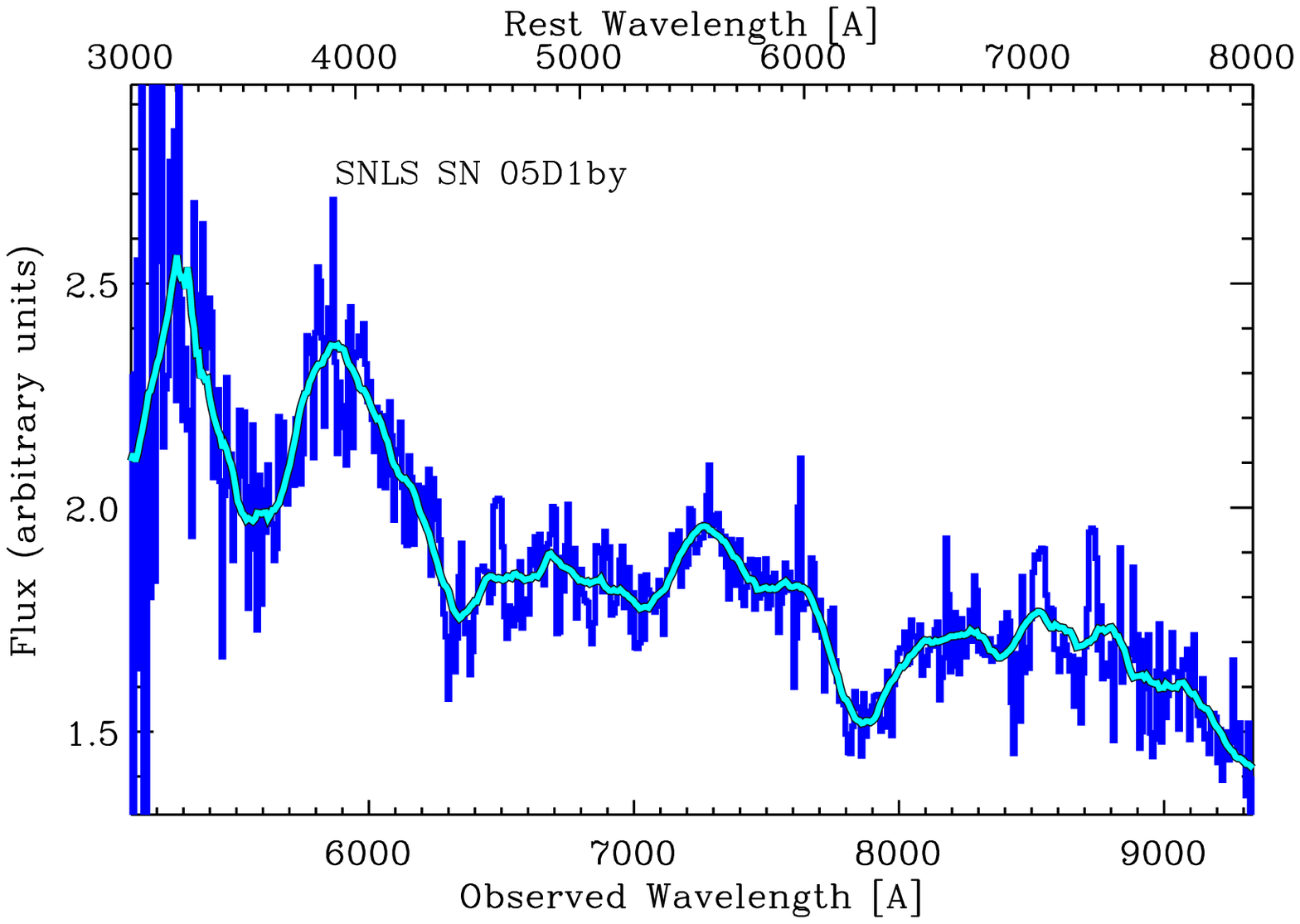} & \includegraphics[width=5.25 cm] {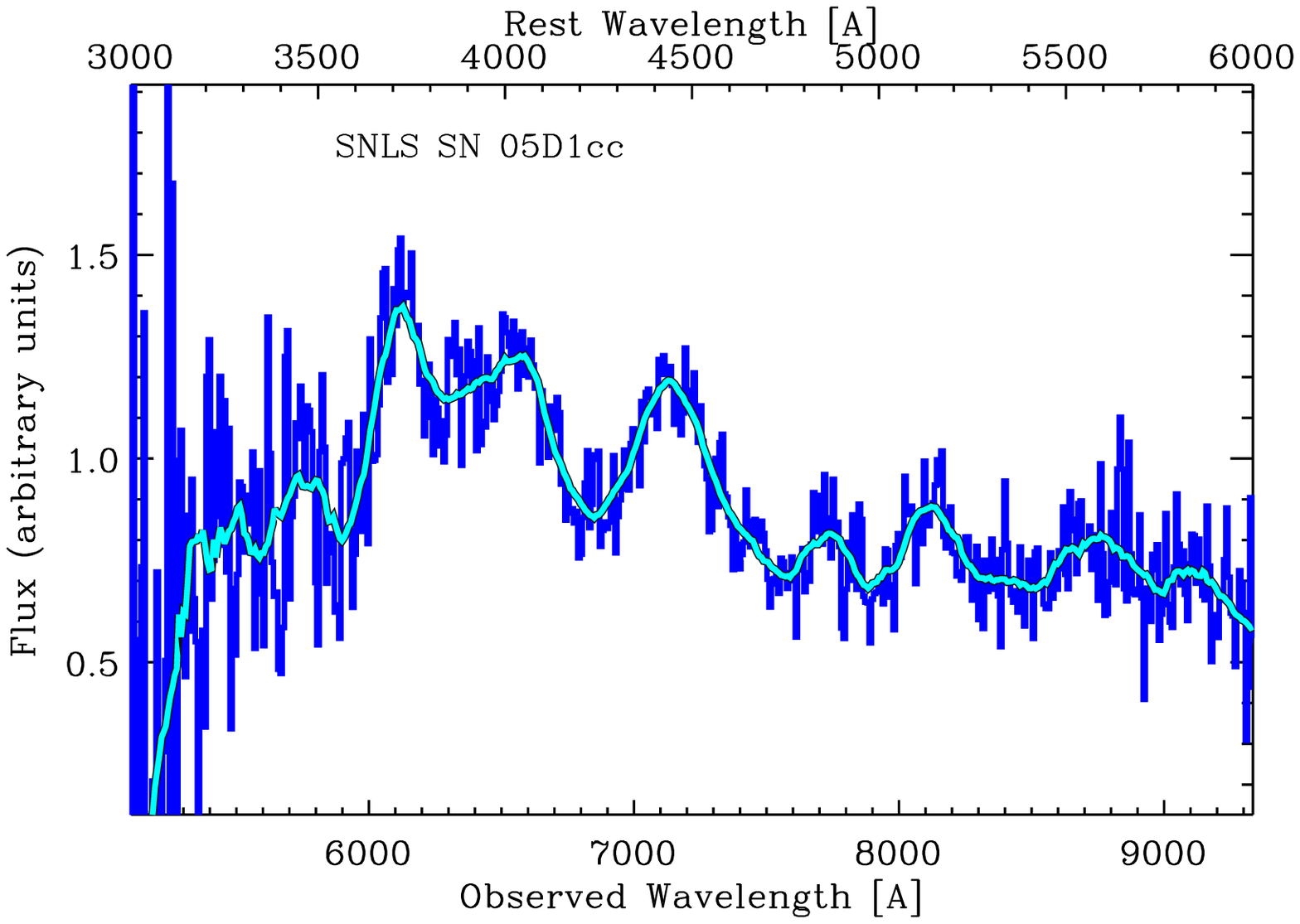} \\
\includegraphics[width=5.25 cm] {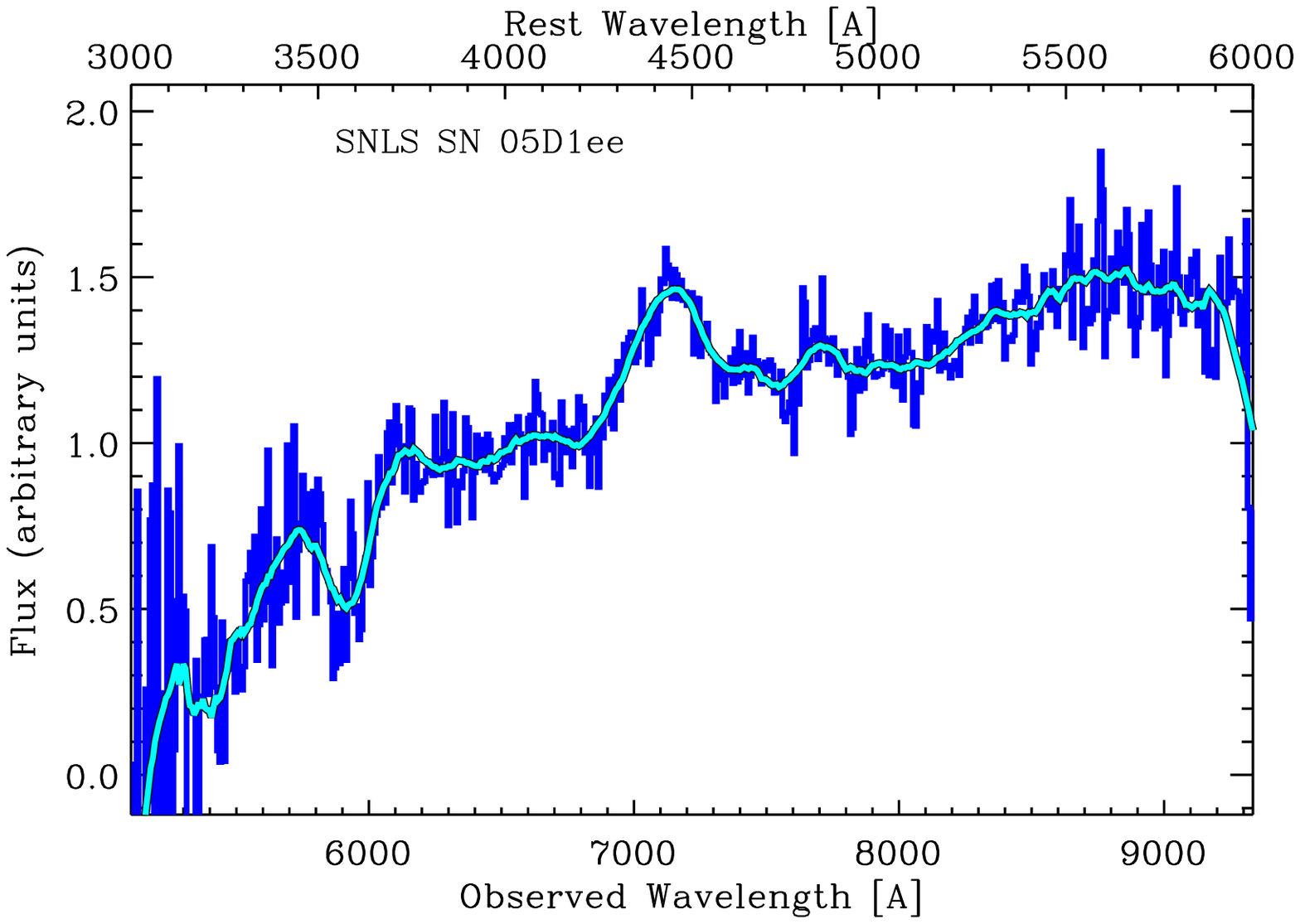} & \includegraphics[width=5.25 cm] {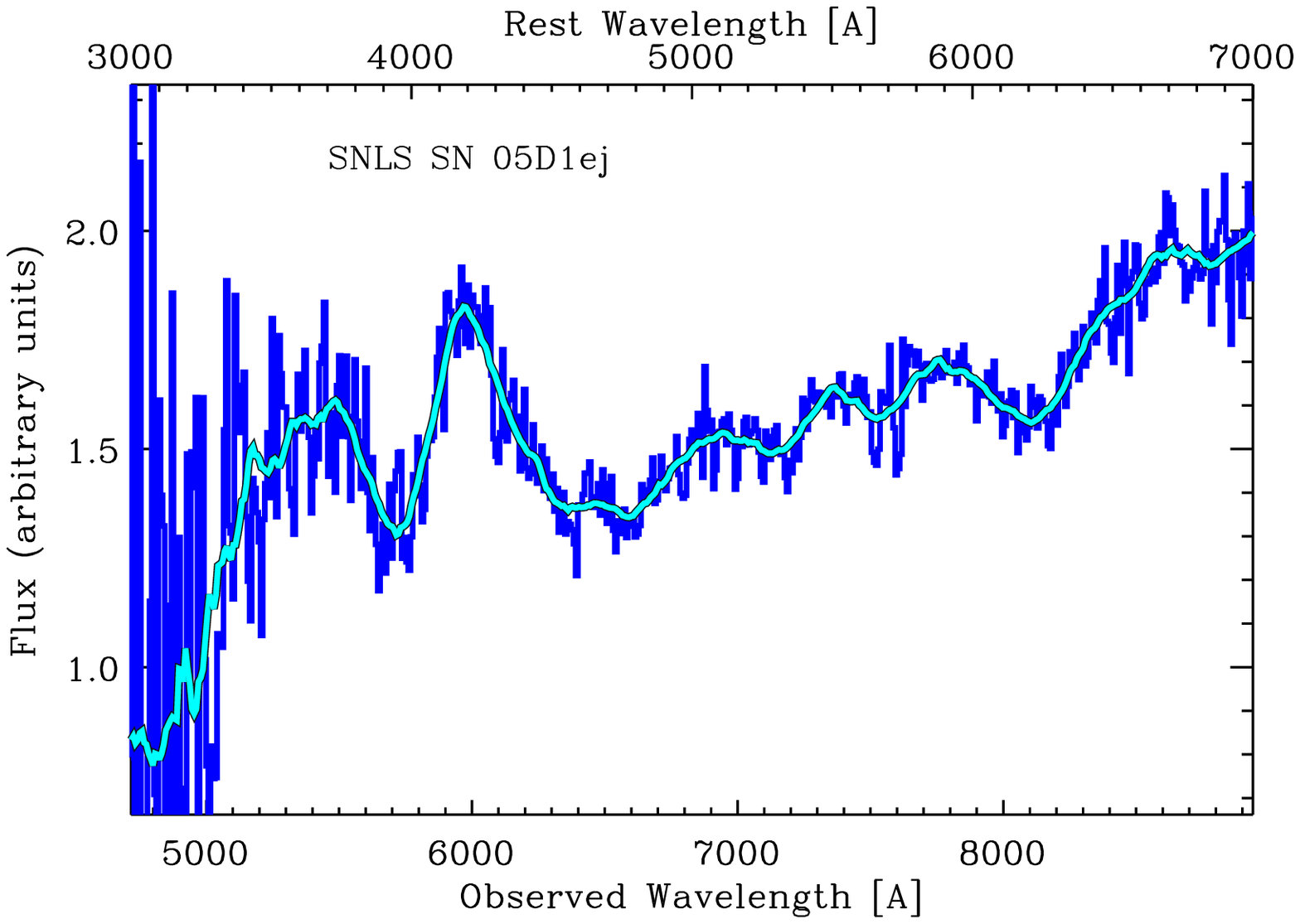} & \includegraphics[width=5.25 cm] {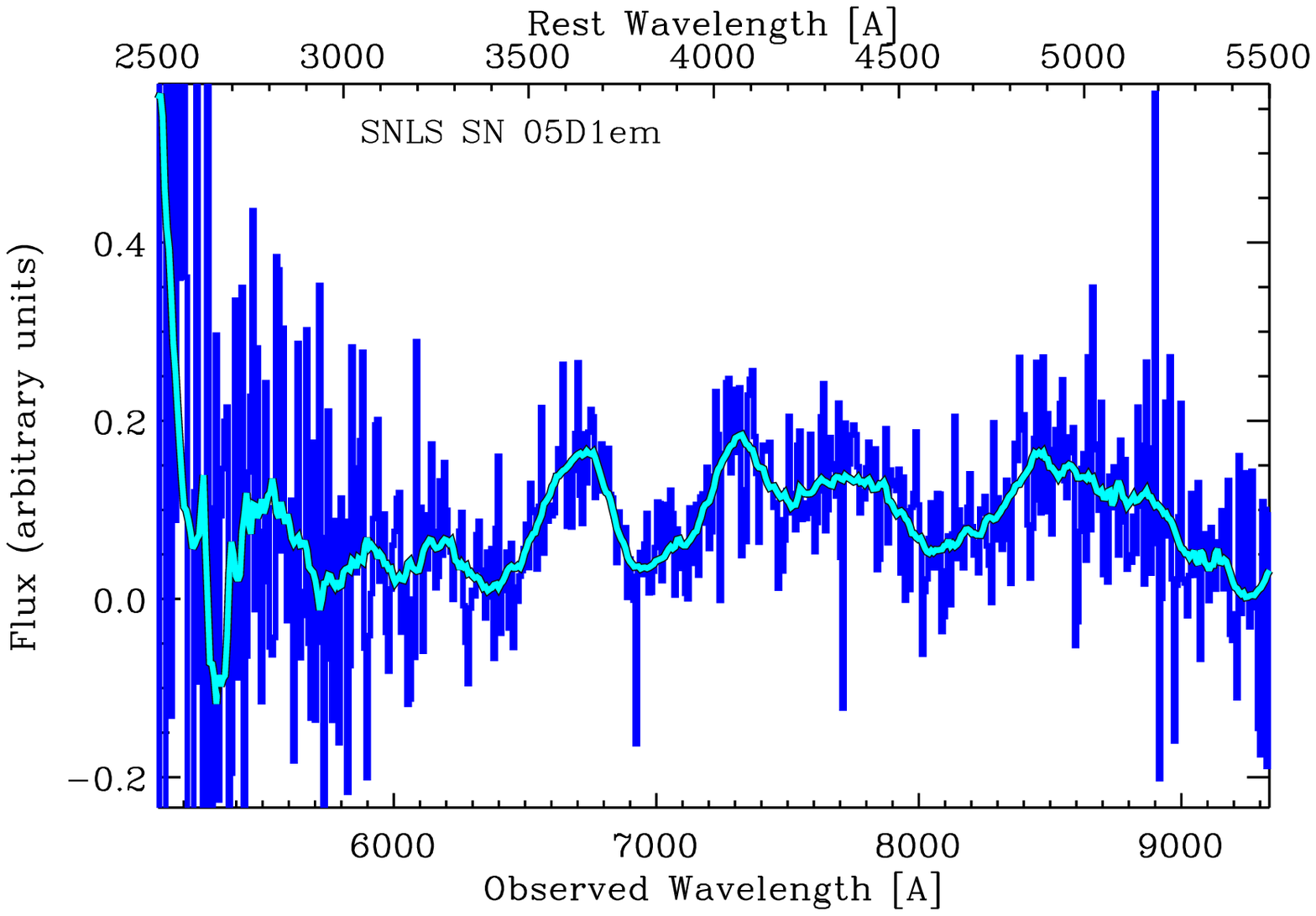} \\
\includegraphics[width=5.25 cm] {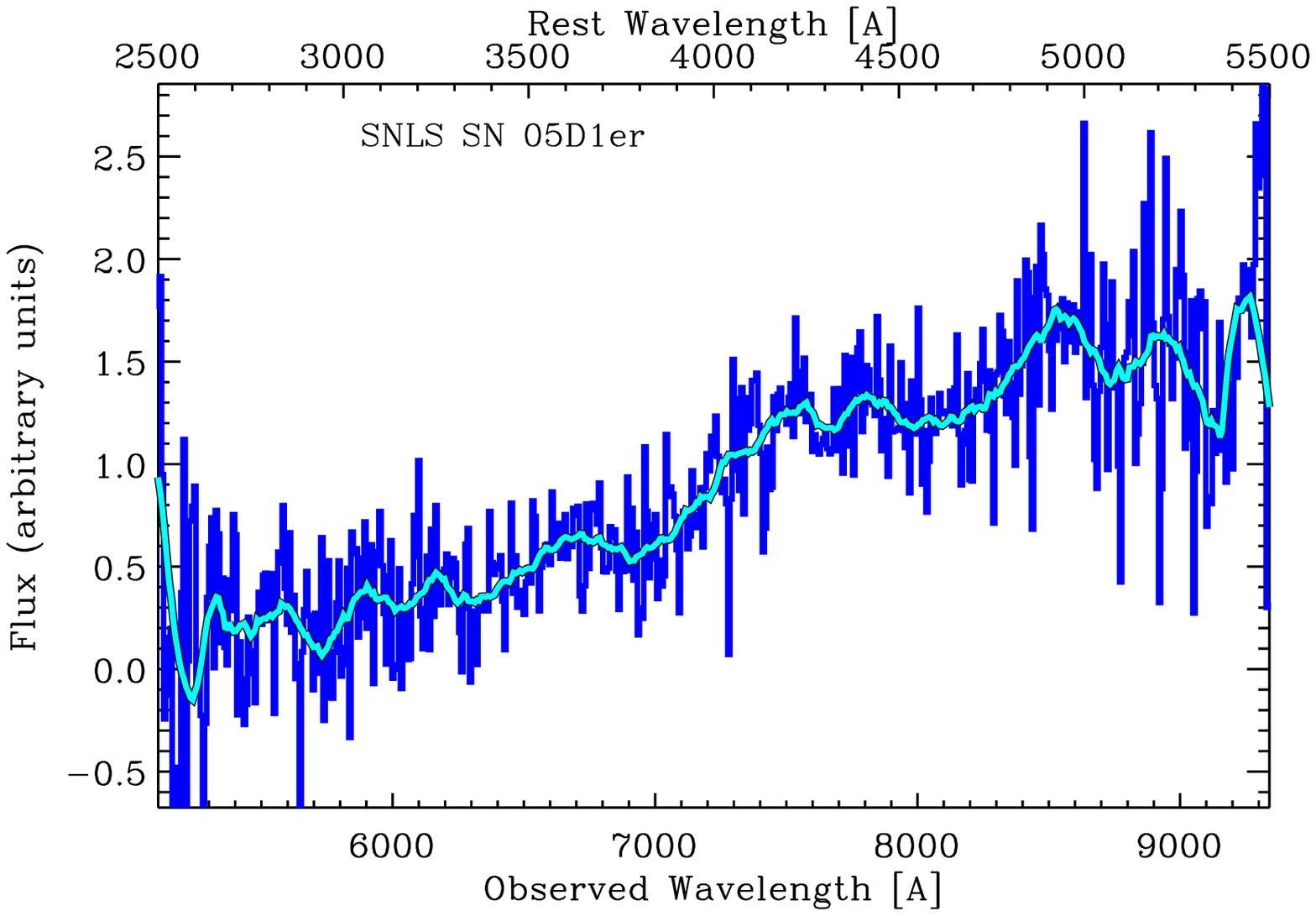} & \includegraphics[width=5.25 cm] {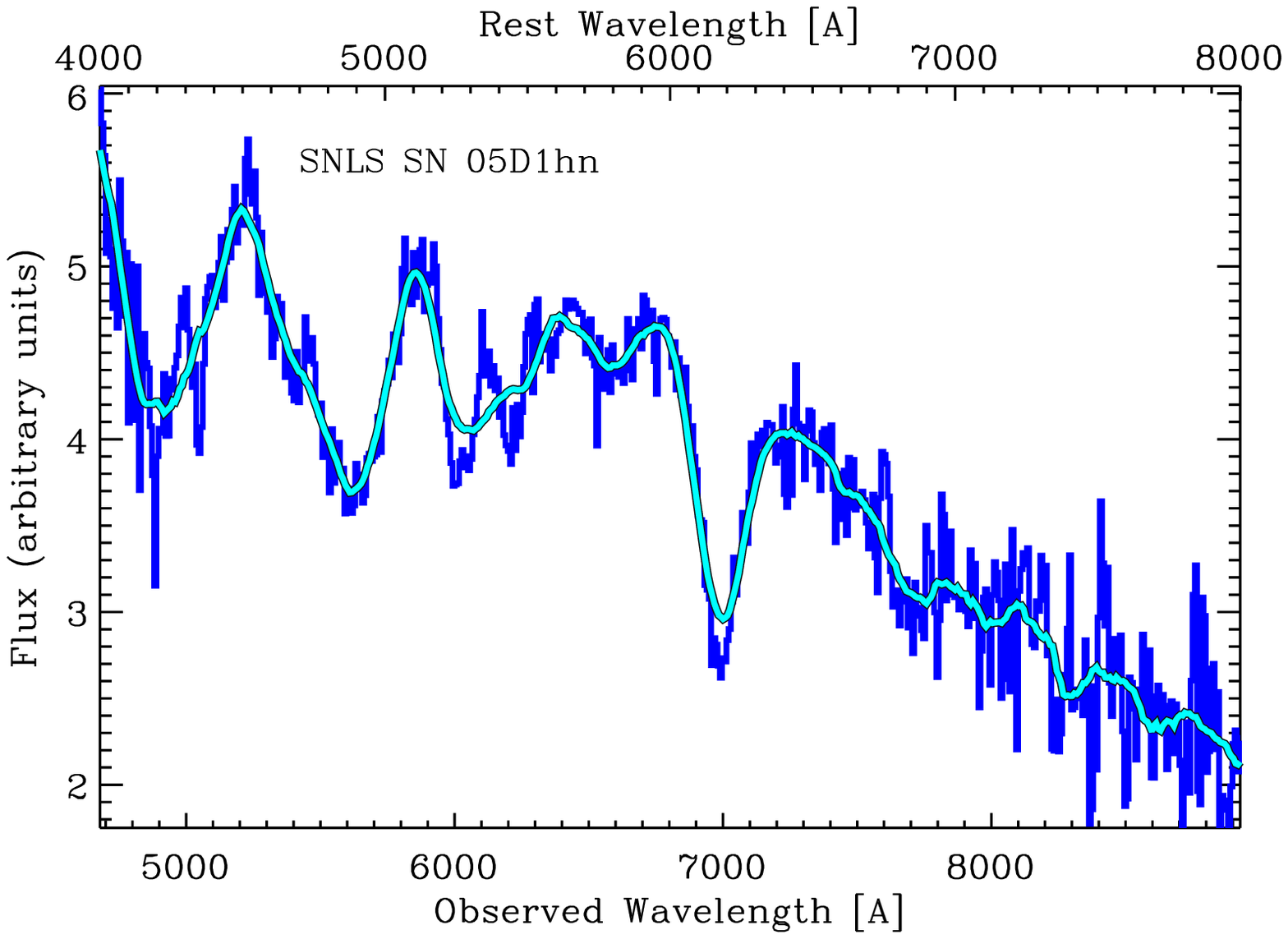} & \includegraphics[width=5.25 cm] {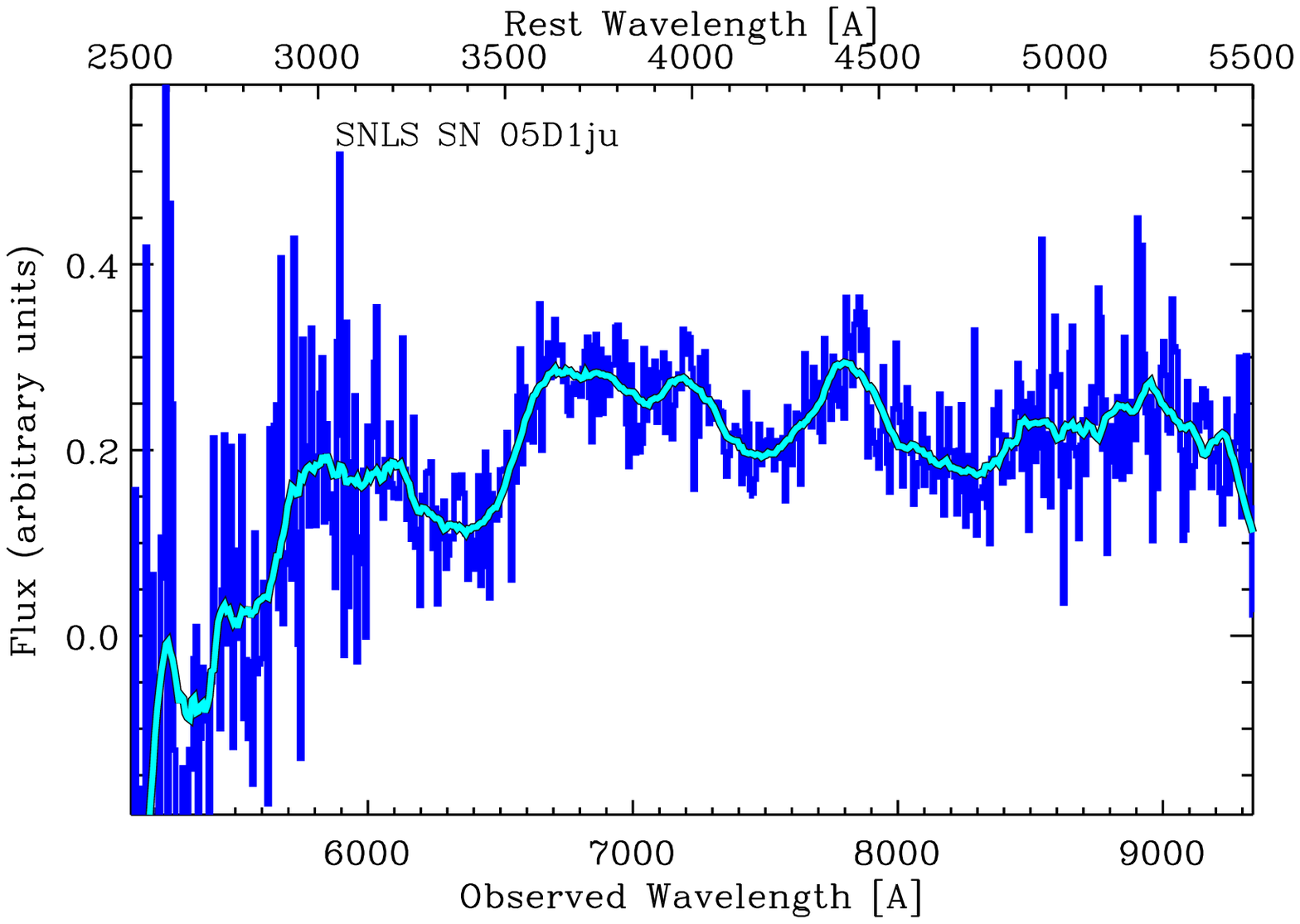} \\
\includegraphics[width=5.25 cm] {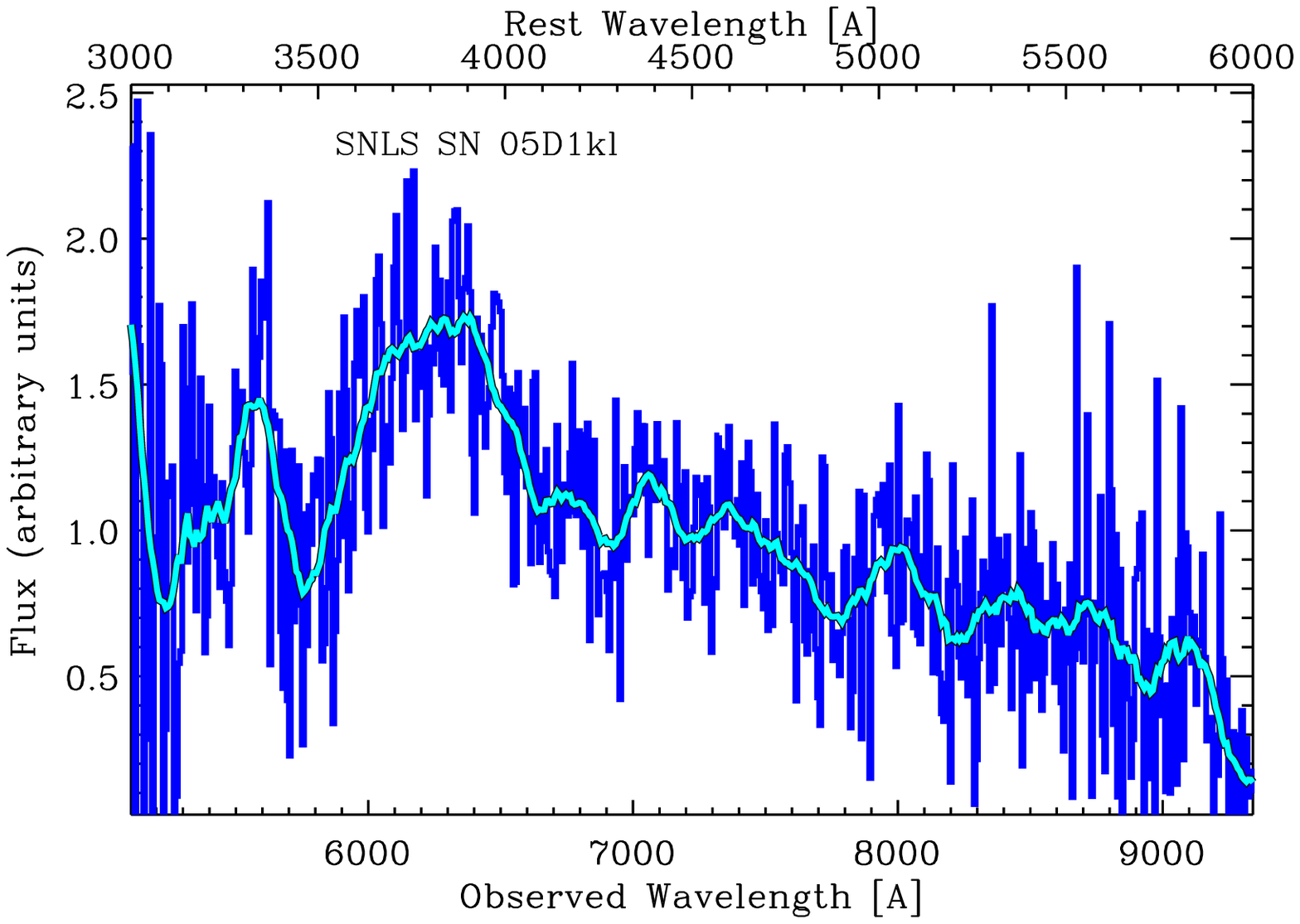} & \includegraphics[width=5.25 cm] {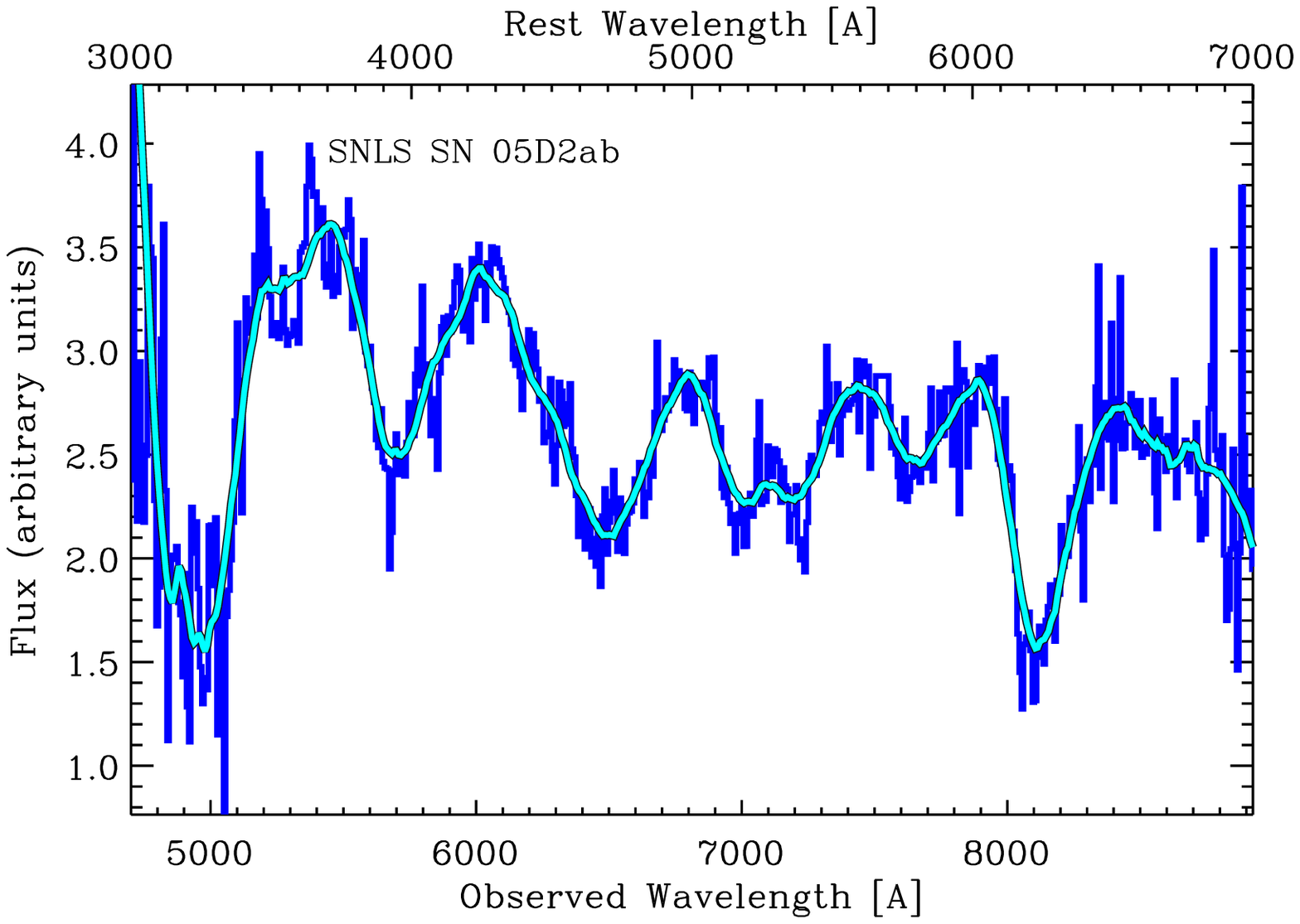} & \includegraphics[width=5.25 cm] {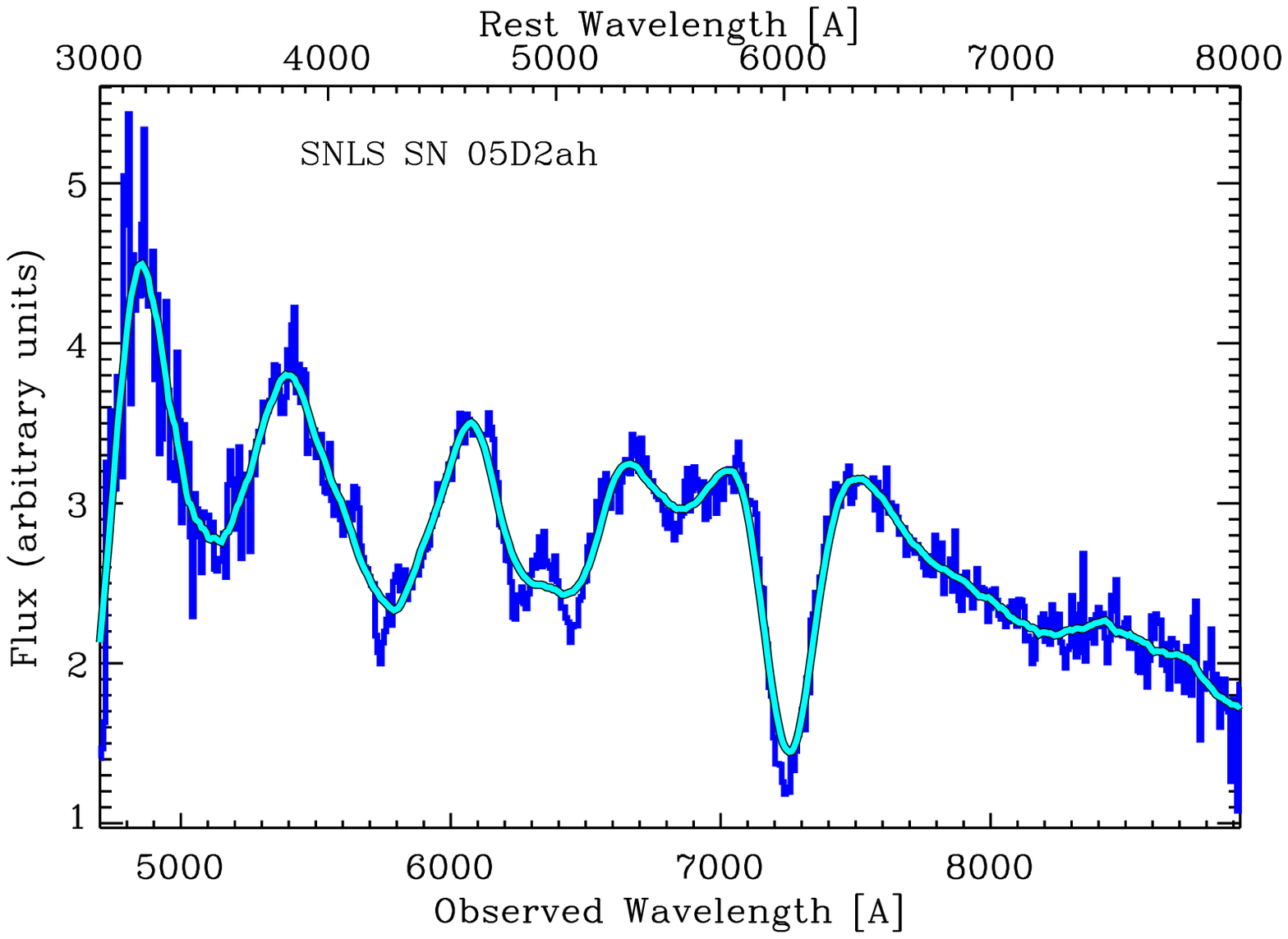} \\
\includegraphics[width=5.25 cm] {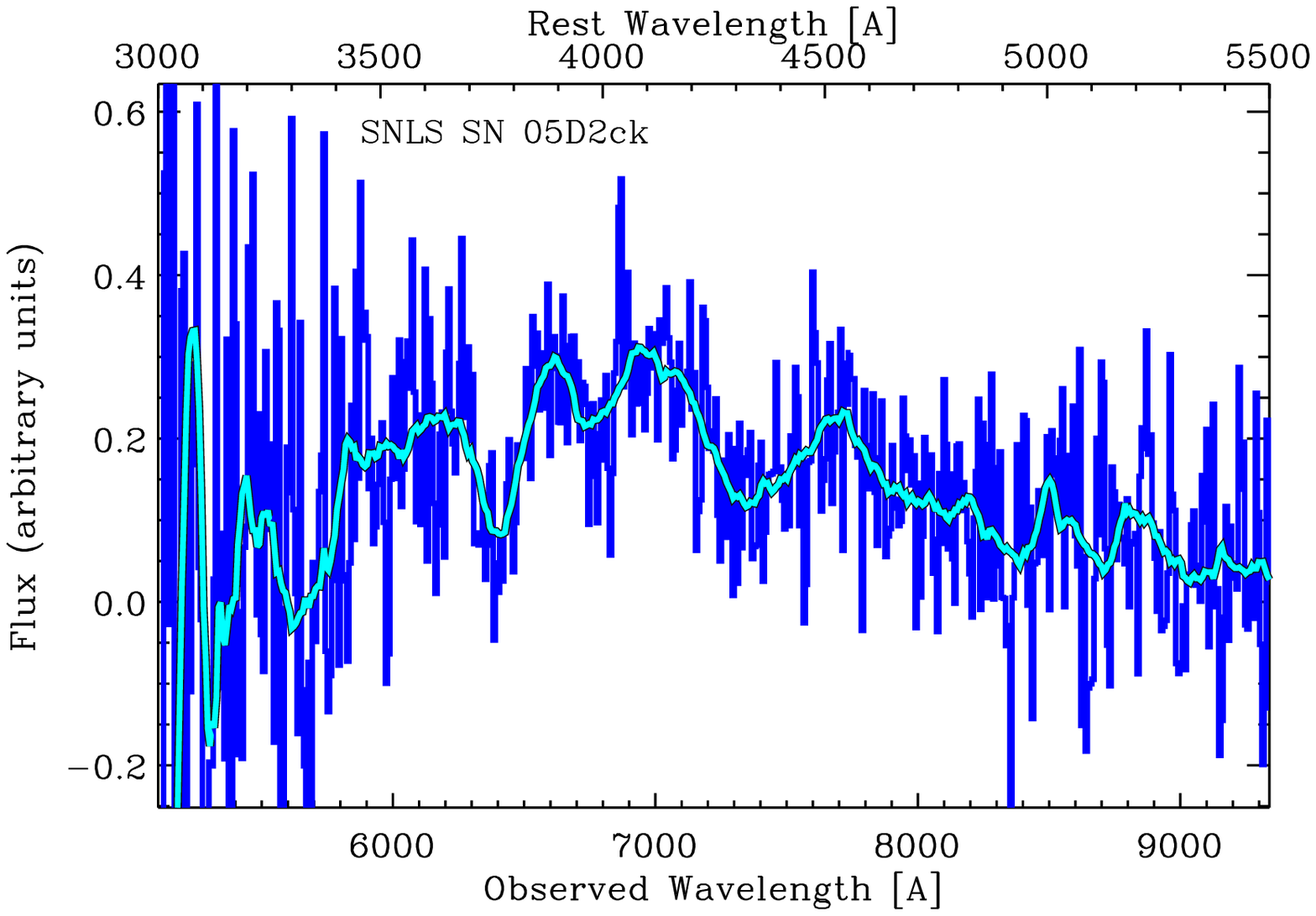} & \includegraphics[width=5.25 cm] {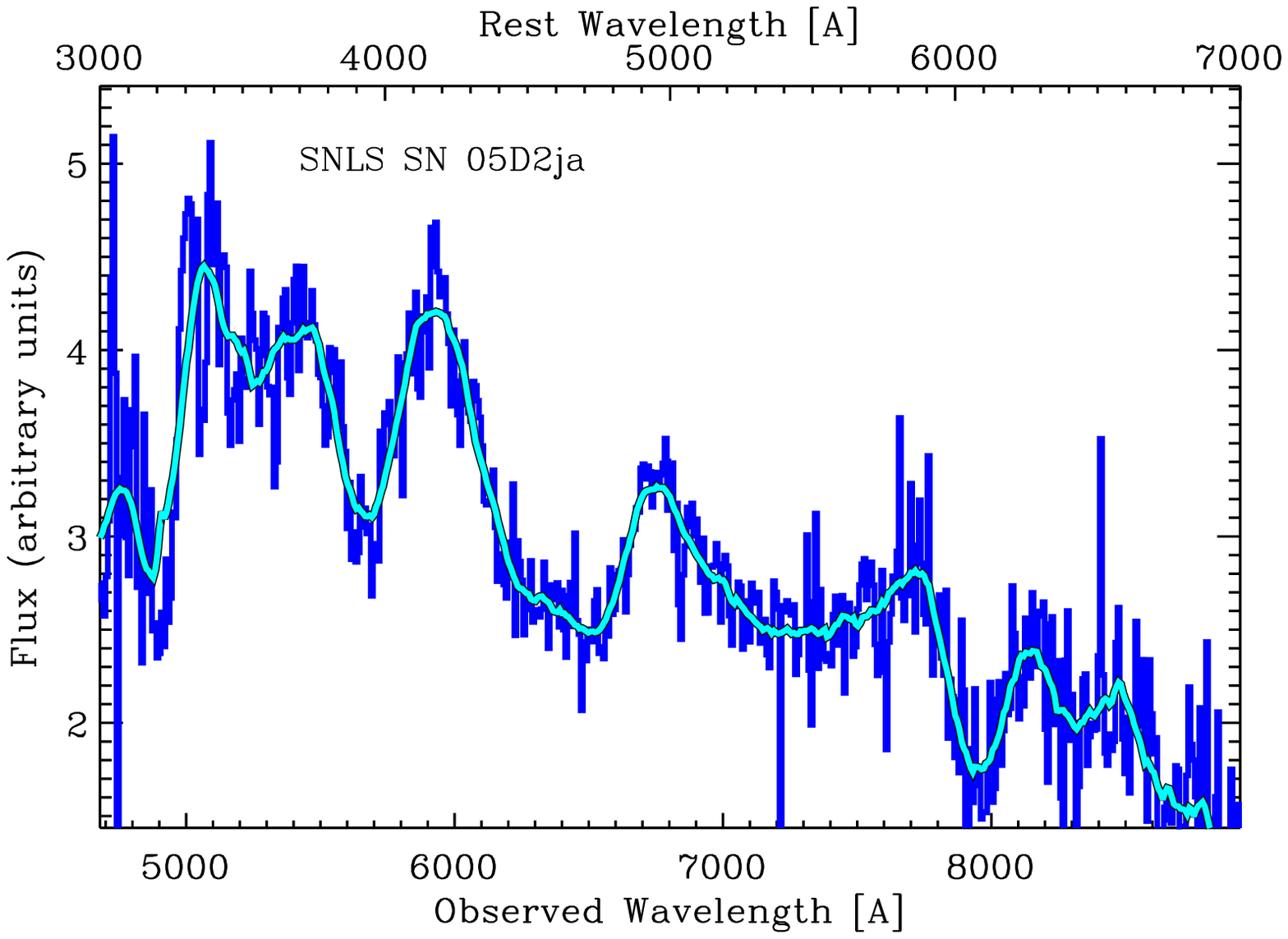} & \includegraphics[width=5.25 cm] {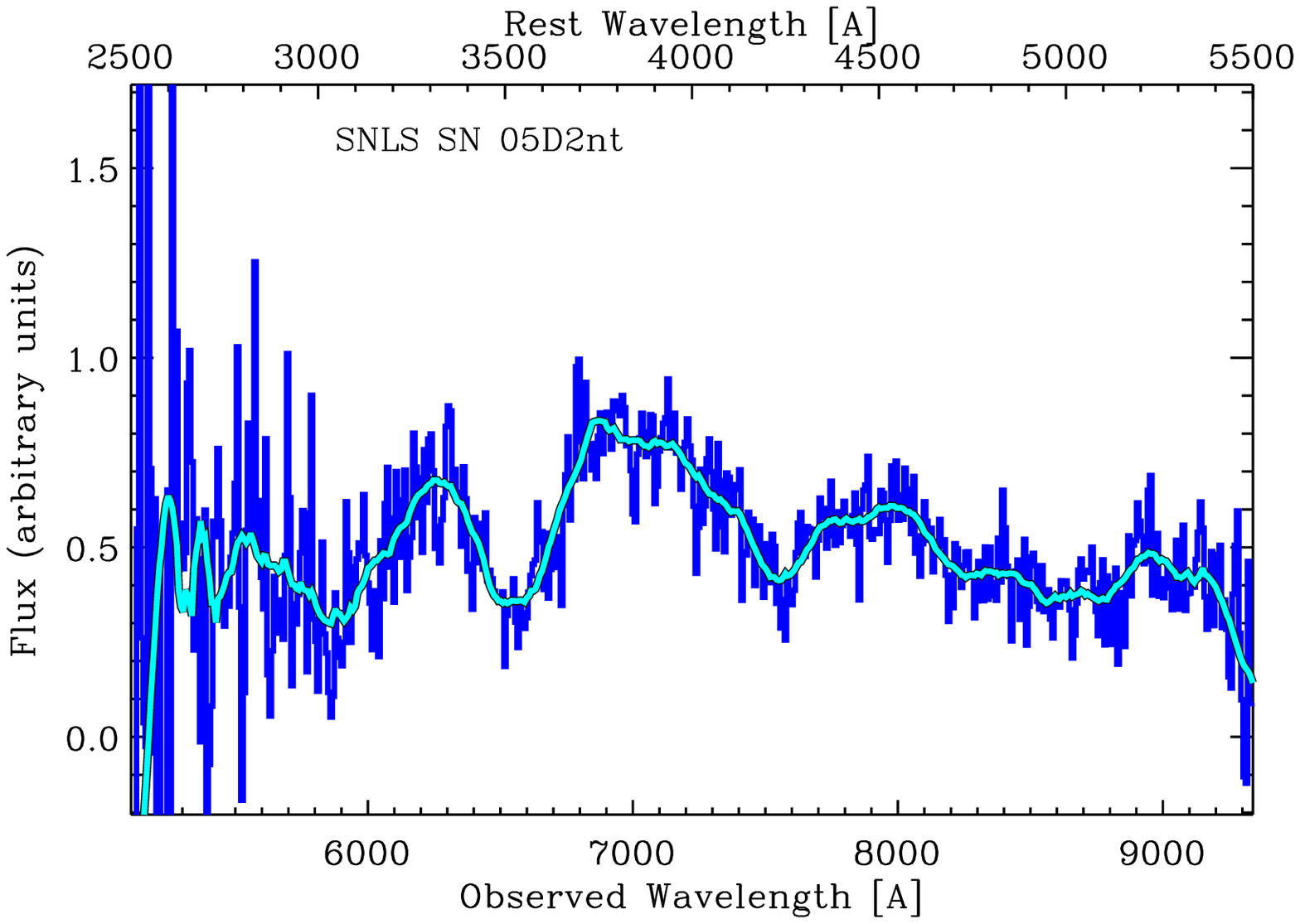} \\
\end{tabular}
\end{minipage}
\caption{Confirmed SNe Ia.  The observed, re-binned spectrum is shown in dark blue and the smoothed spectrum is overplot in light blue for illustrative purposes.}
\end{figure*}

\begin{figure*}
\begin{minipage}{2.0\textwidth}
\begin{tabular}{ccc}
\includegraphics[width=5.25 cm] {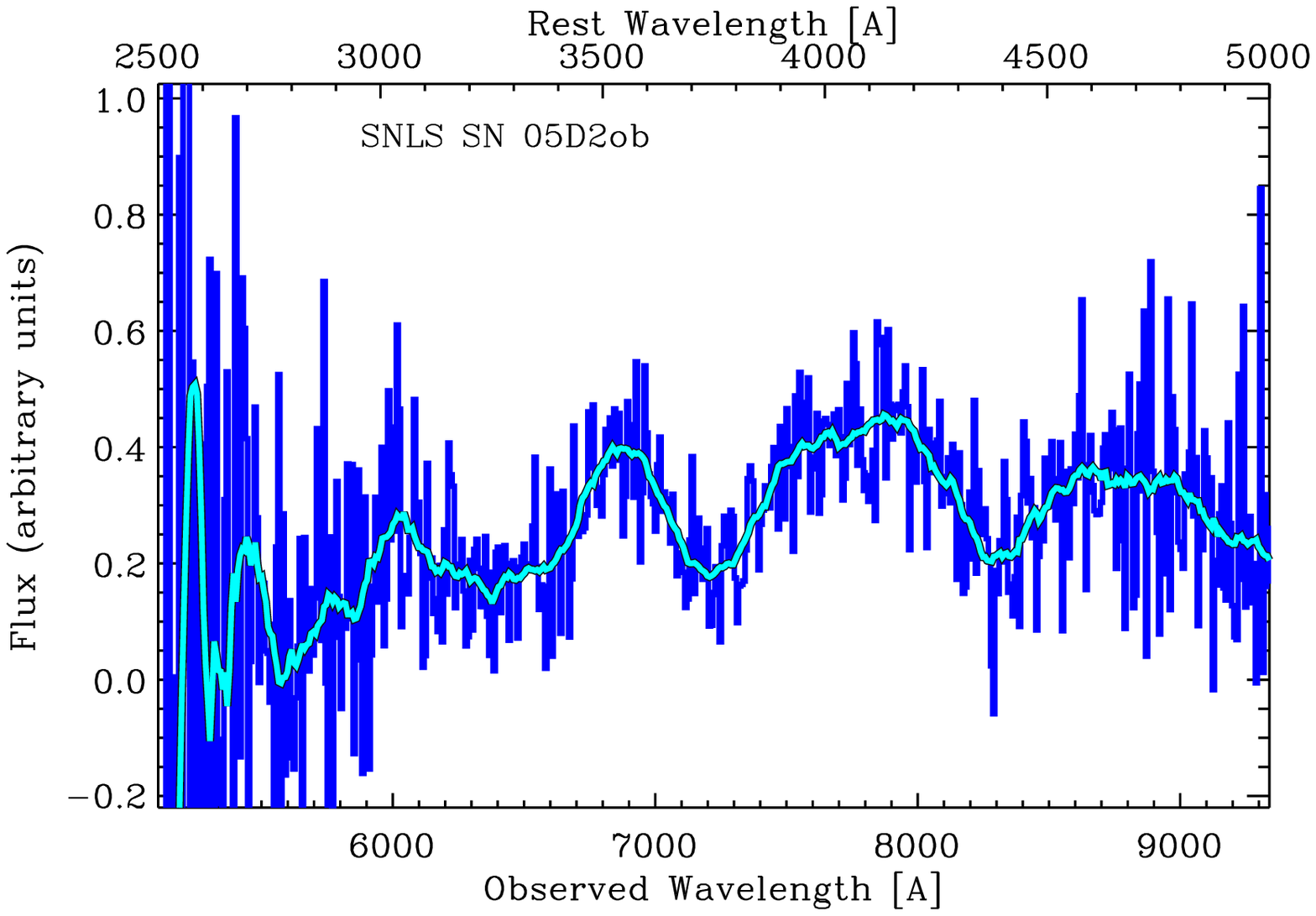} & \includegraphics[width=5.25 cm] {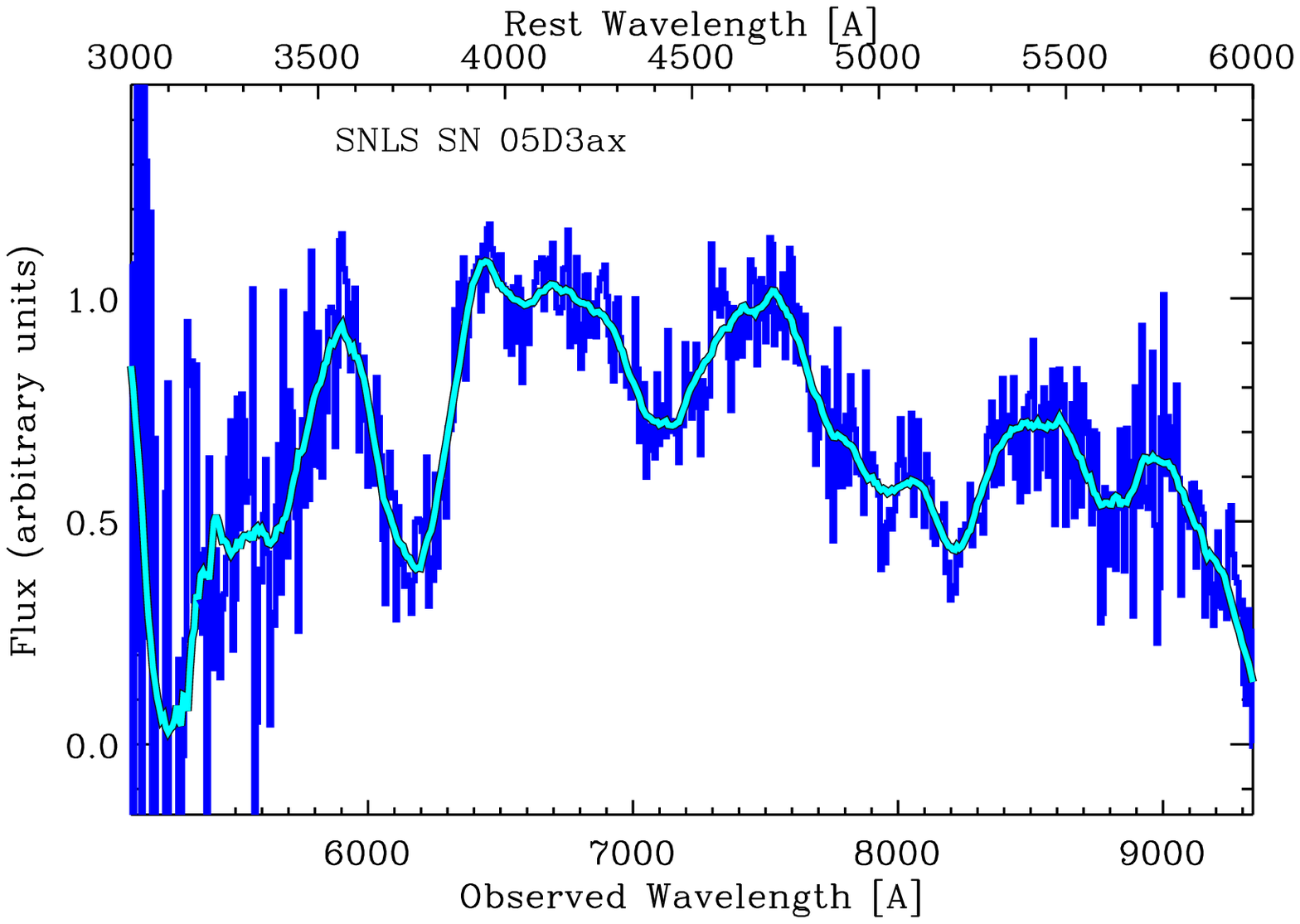} & \includegraphics[width=5.25 cm] {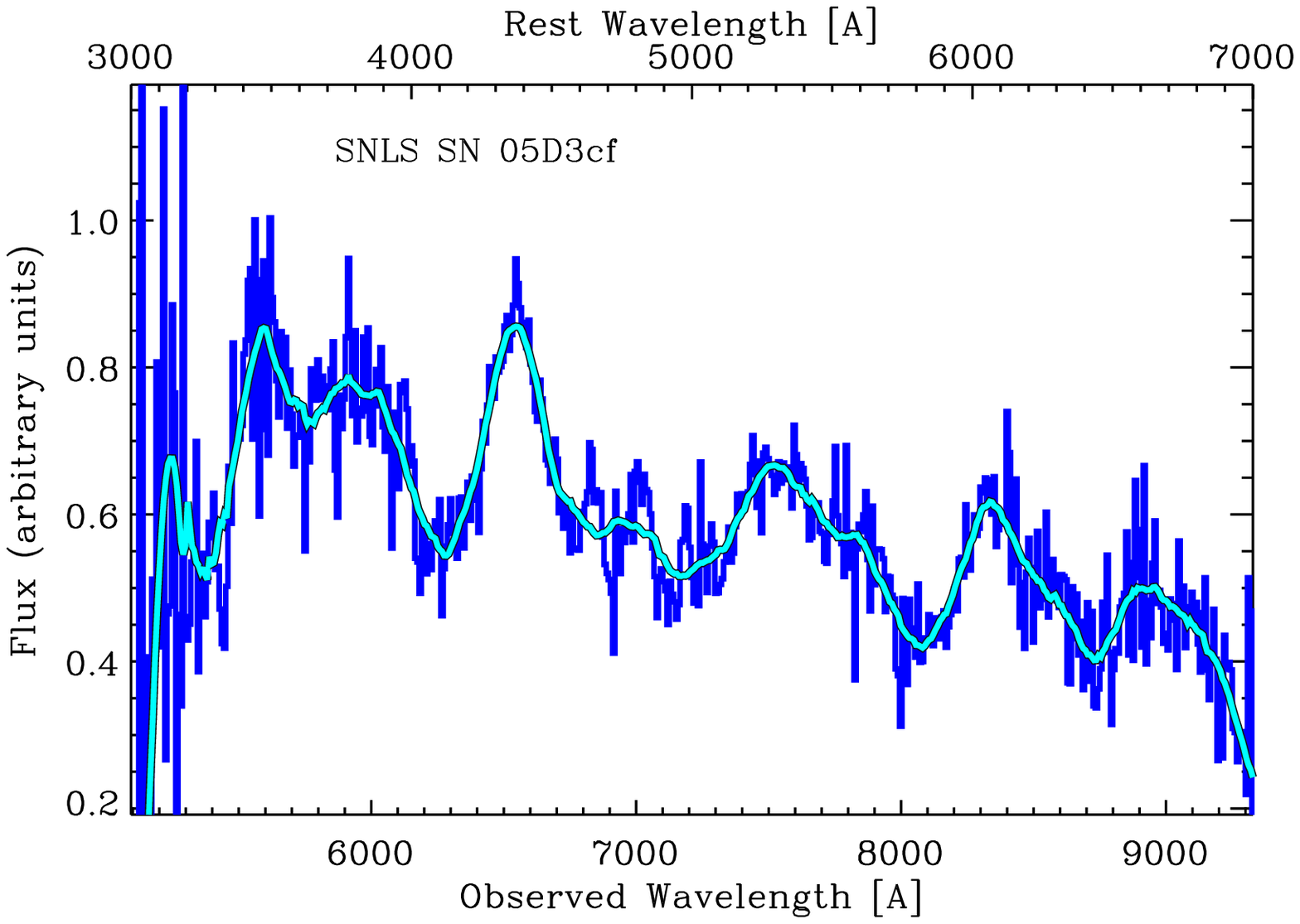} \\
\includegraphics[width=5.25 cm] {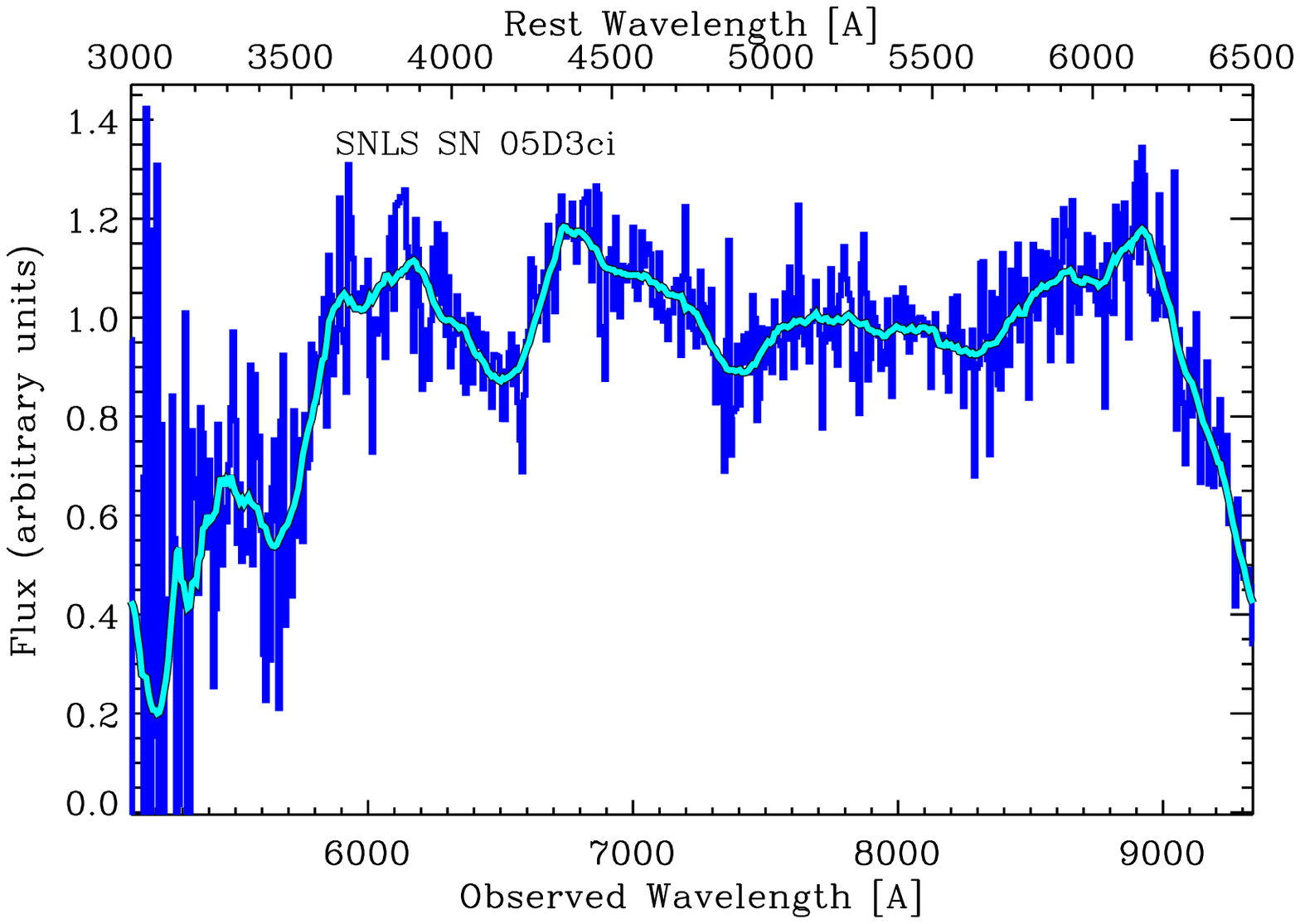} & \includegraphics[width=5.25 cm] {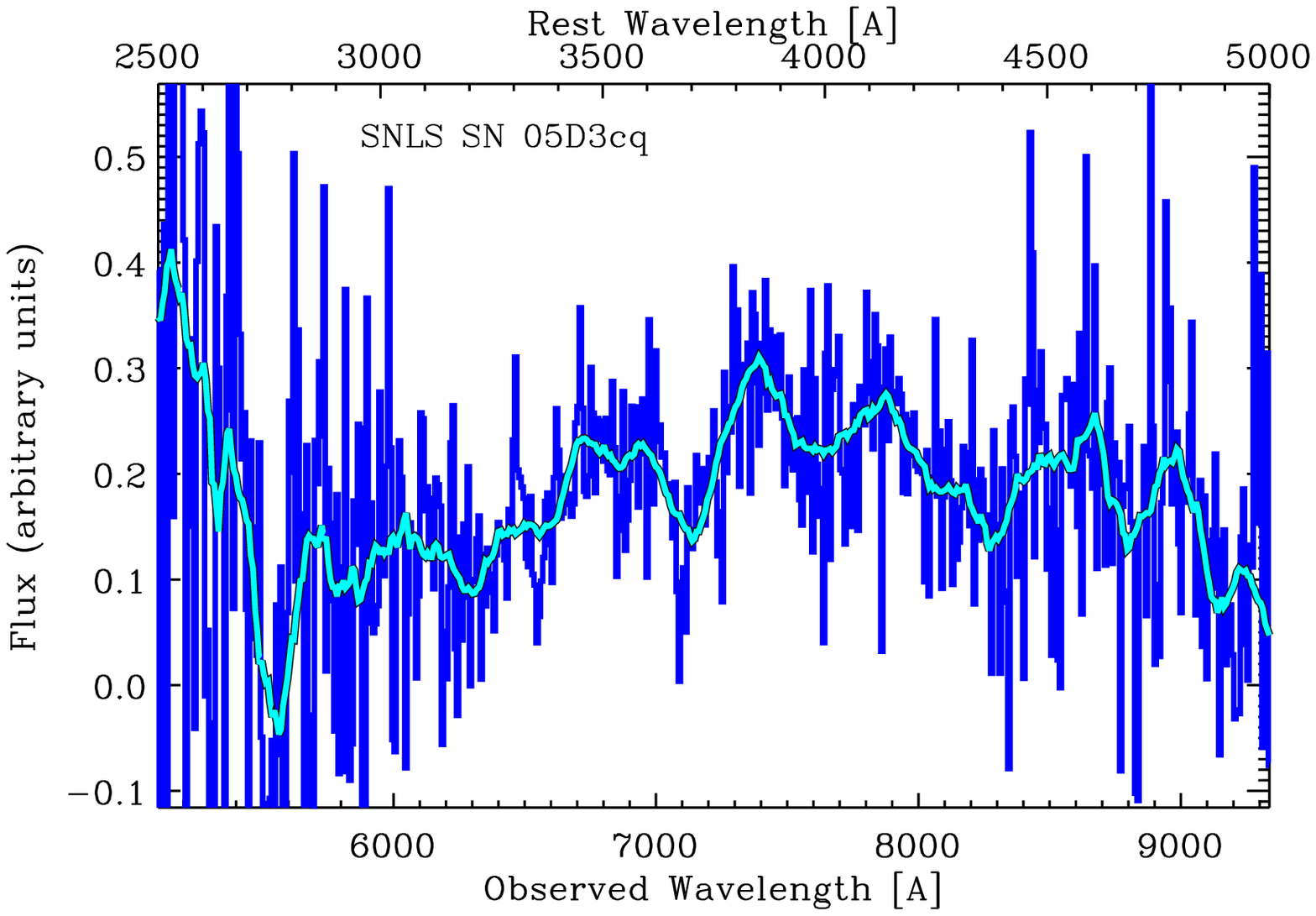} & \includegraphics[width=5.25 cm] {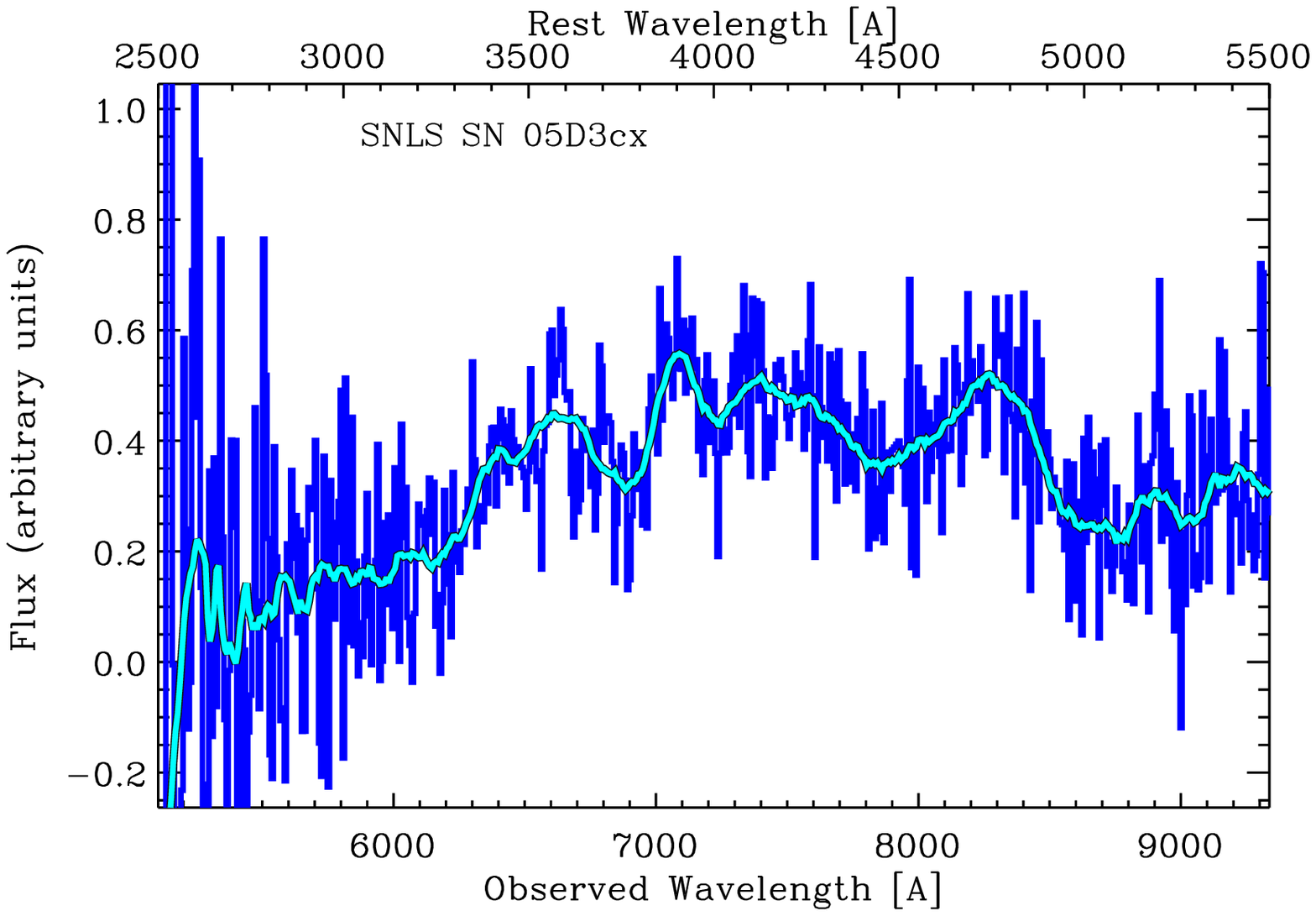} \\
\includegraphics[width=5.25 cm] {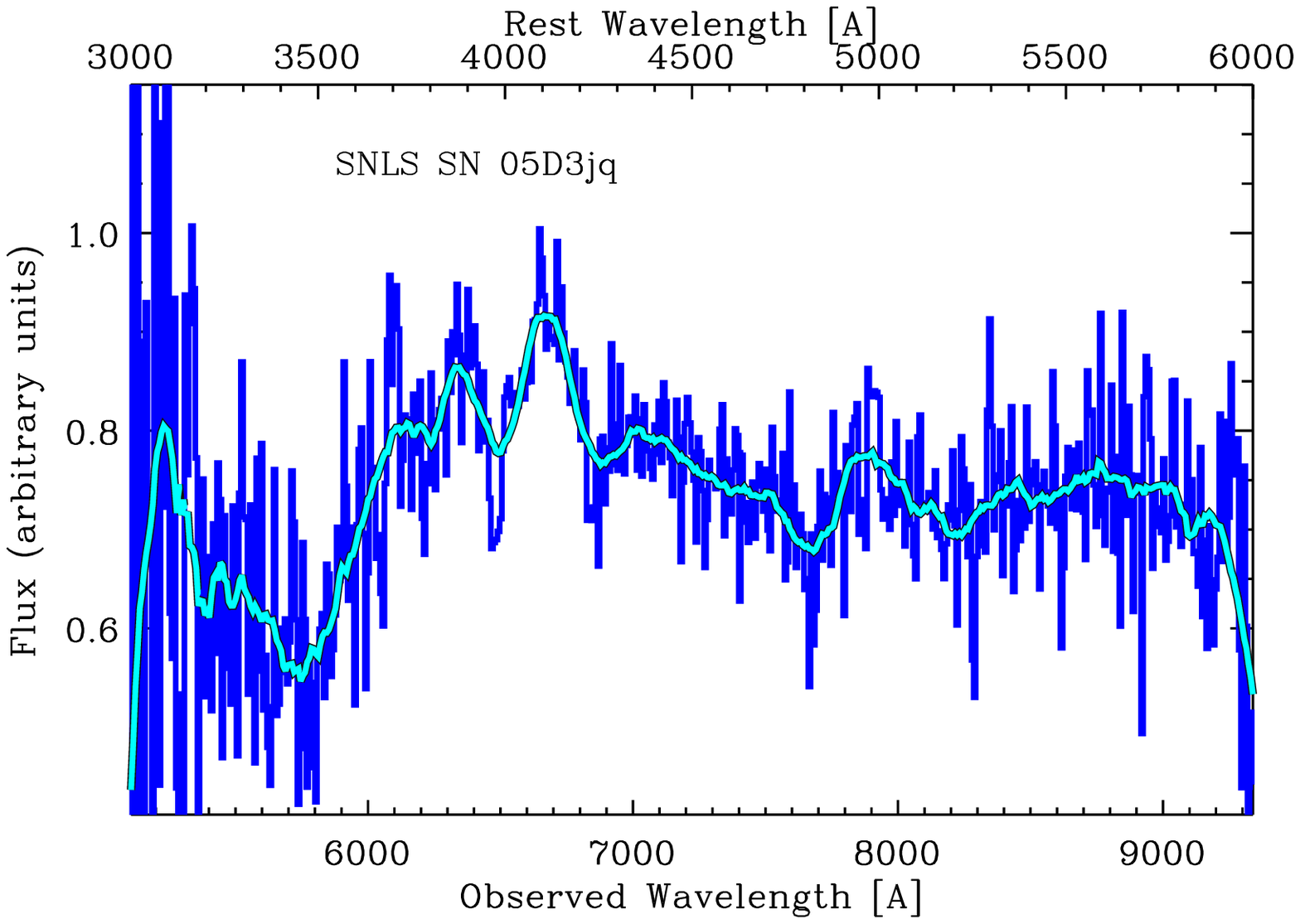} & \includegraphics[width=5.25 cm] {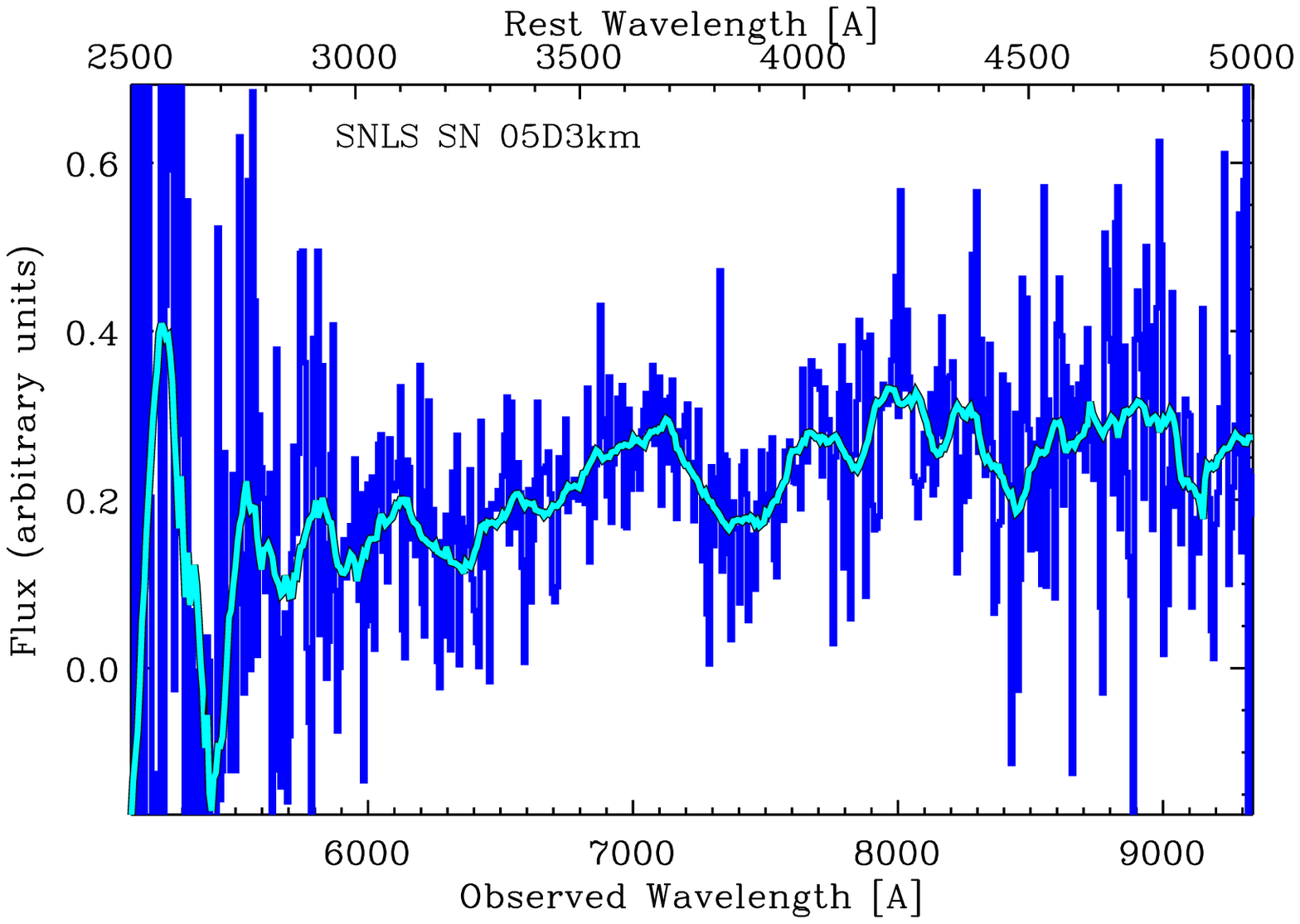} & \includegraphics[width=5.25 cm] {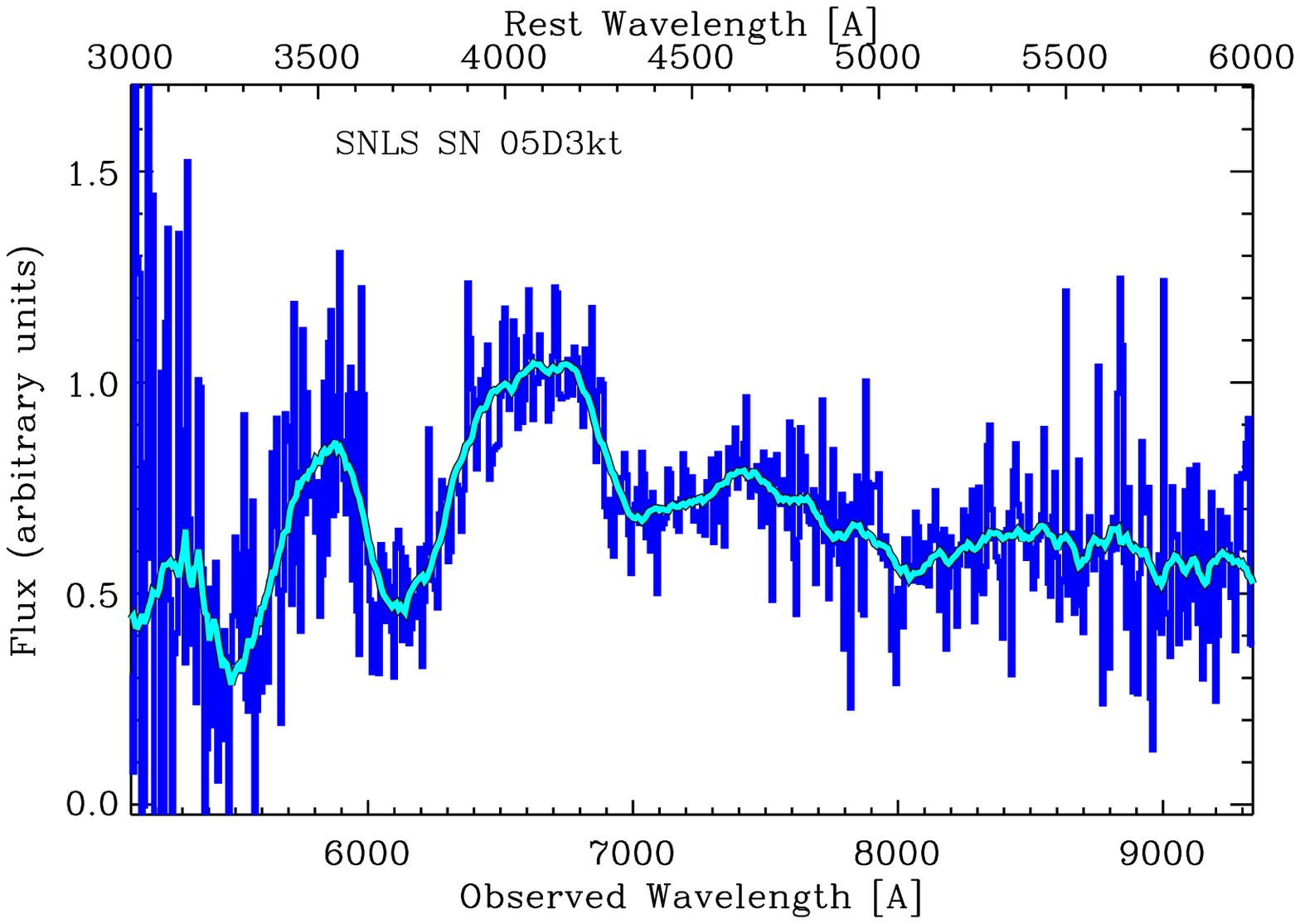} \\
\includegraphics[width=5.25 cm] {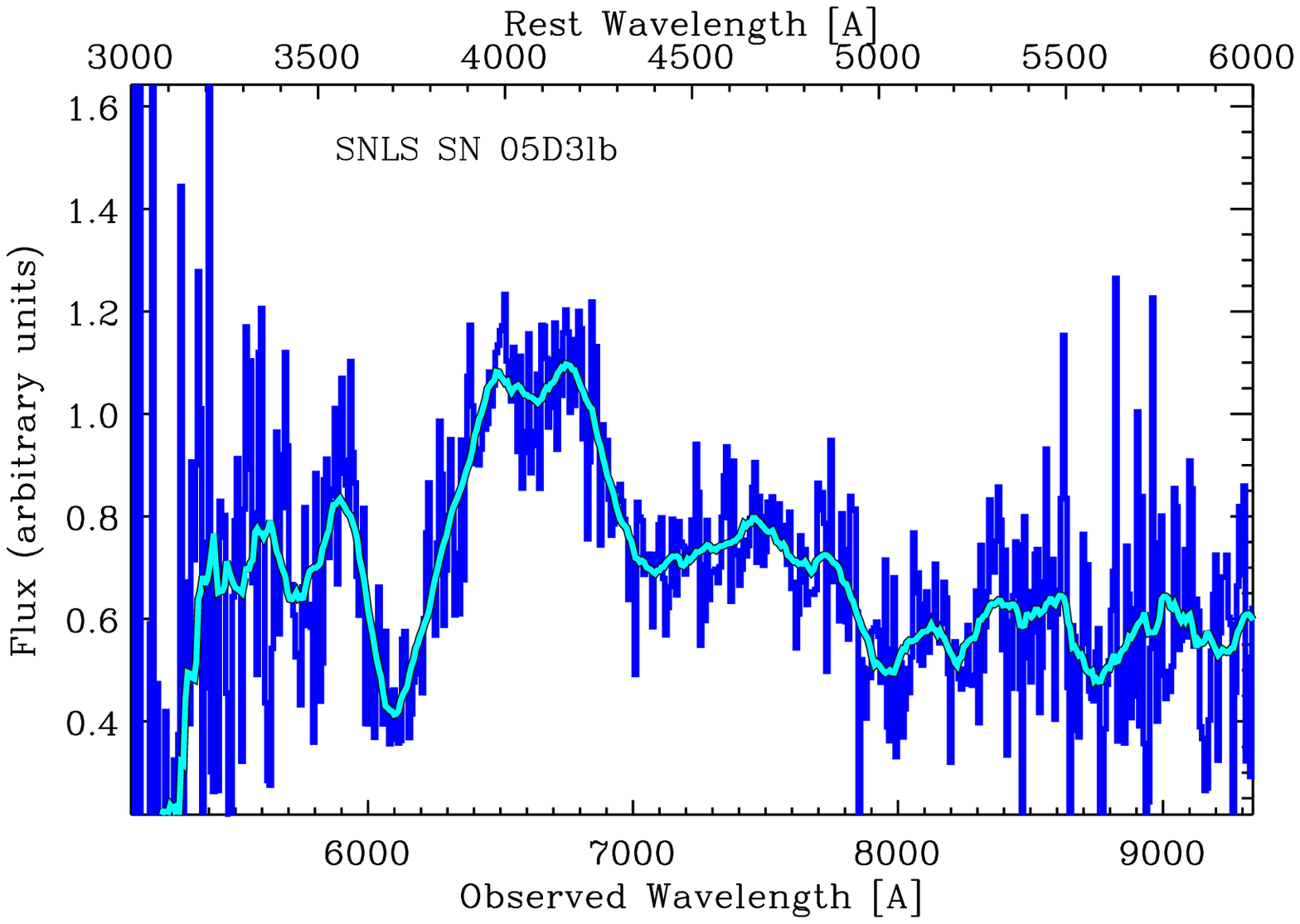} & \includegraphics[width=5.25 cm] {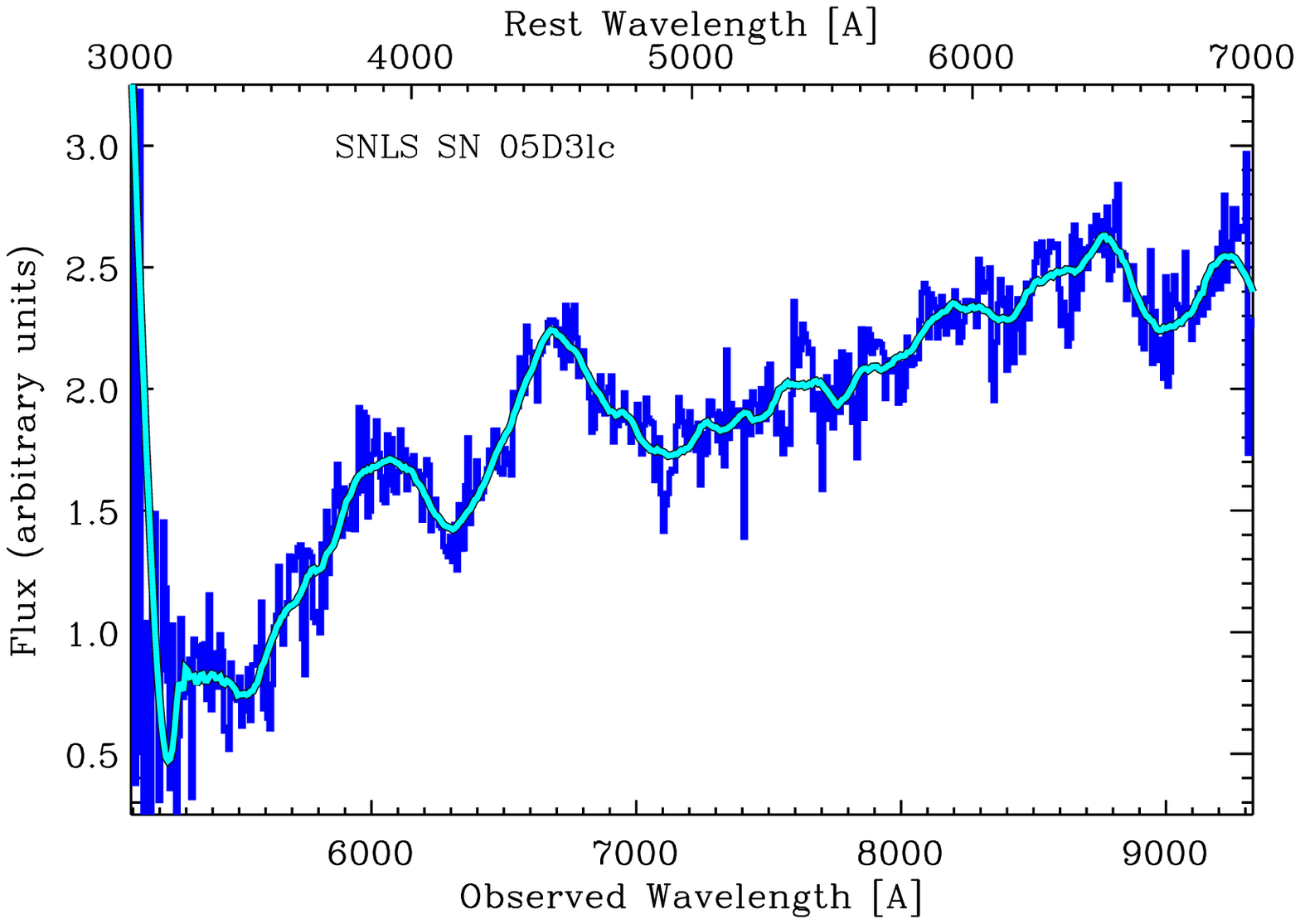} & \includegraphics[width=5.25 cm] {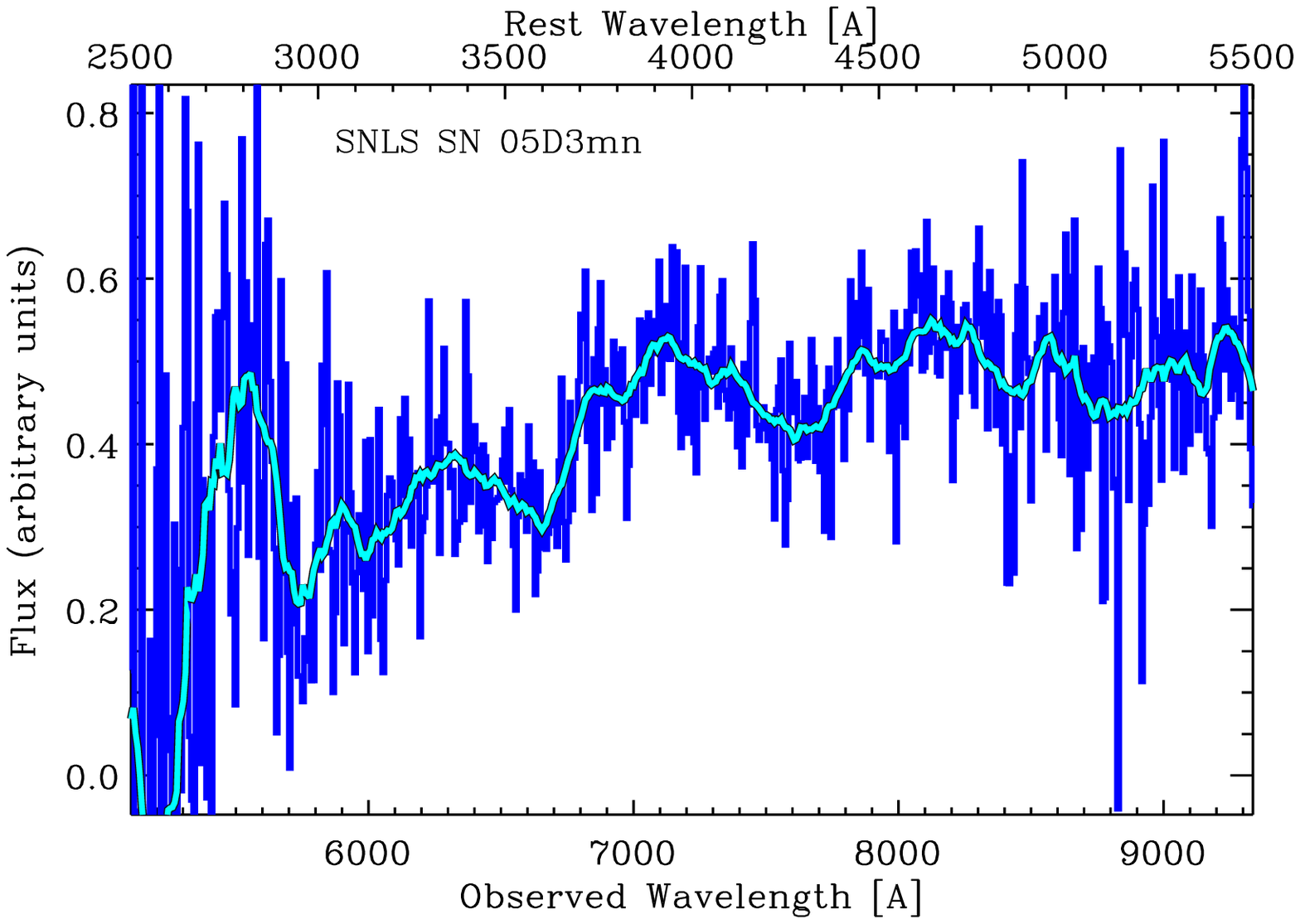} \\
\includegraphics[width=5.25 cm] {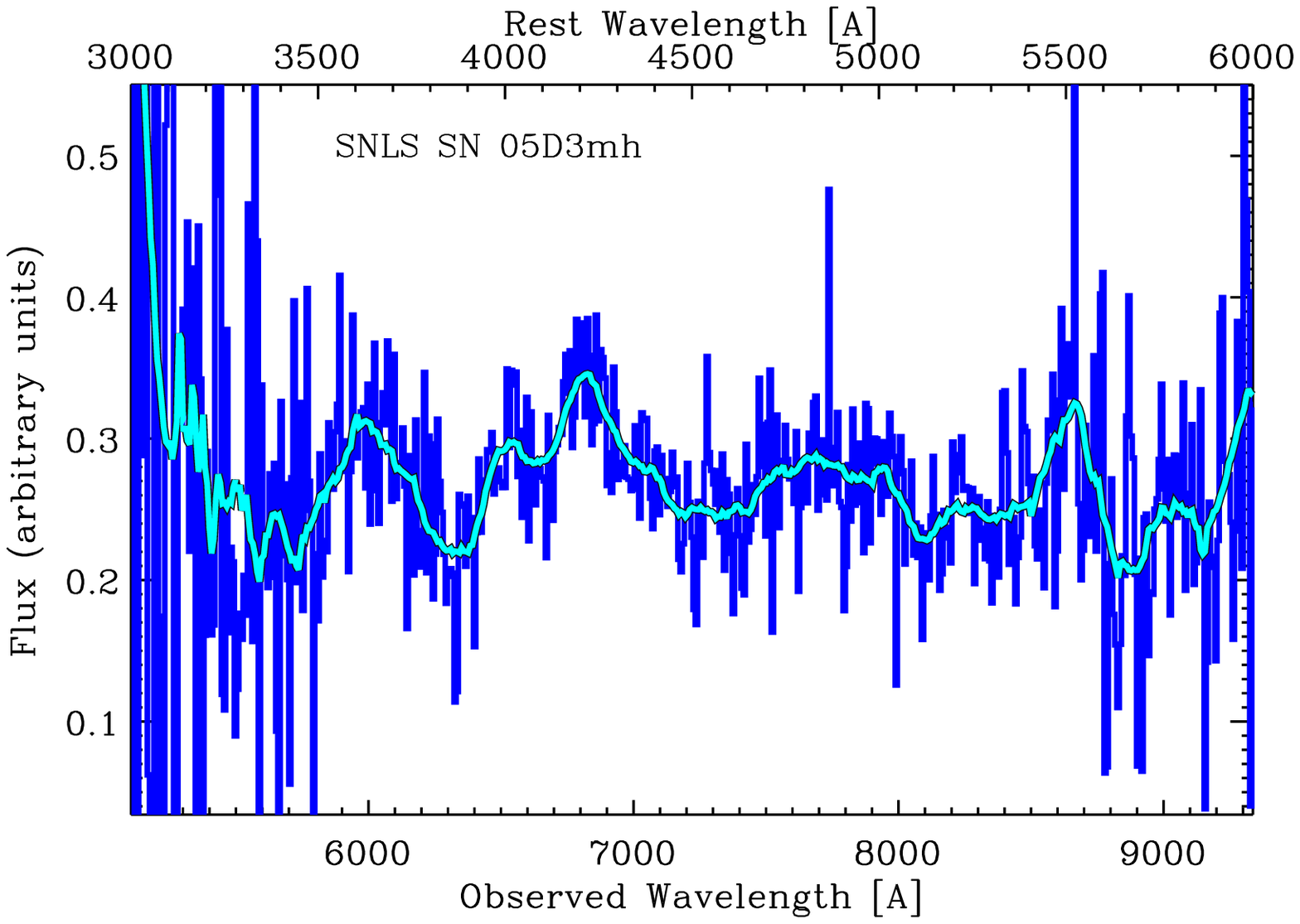} & \includegraphics[width=5.25 cm] {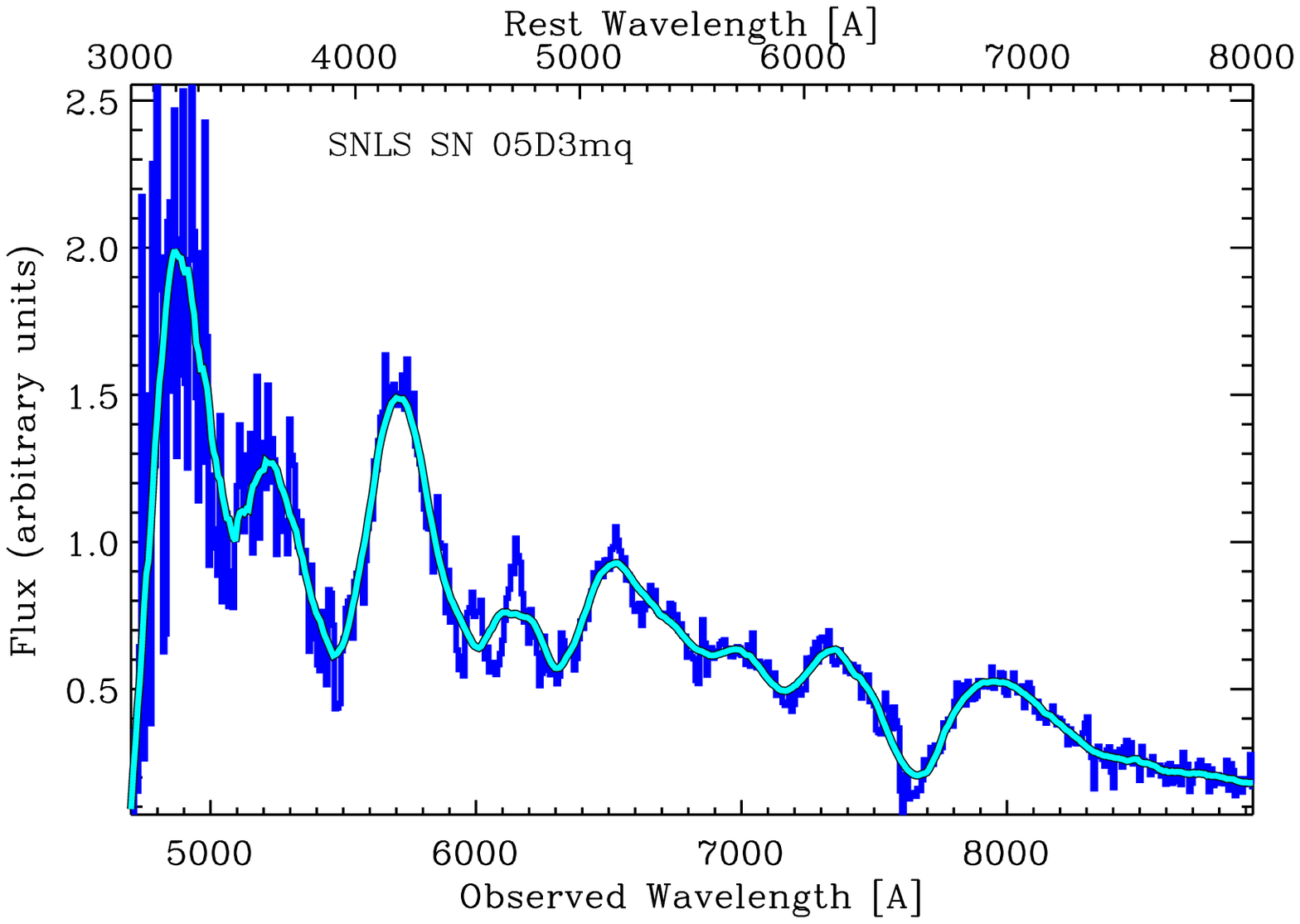} & \includegraphics[width=5.25 cm] {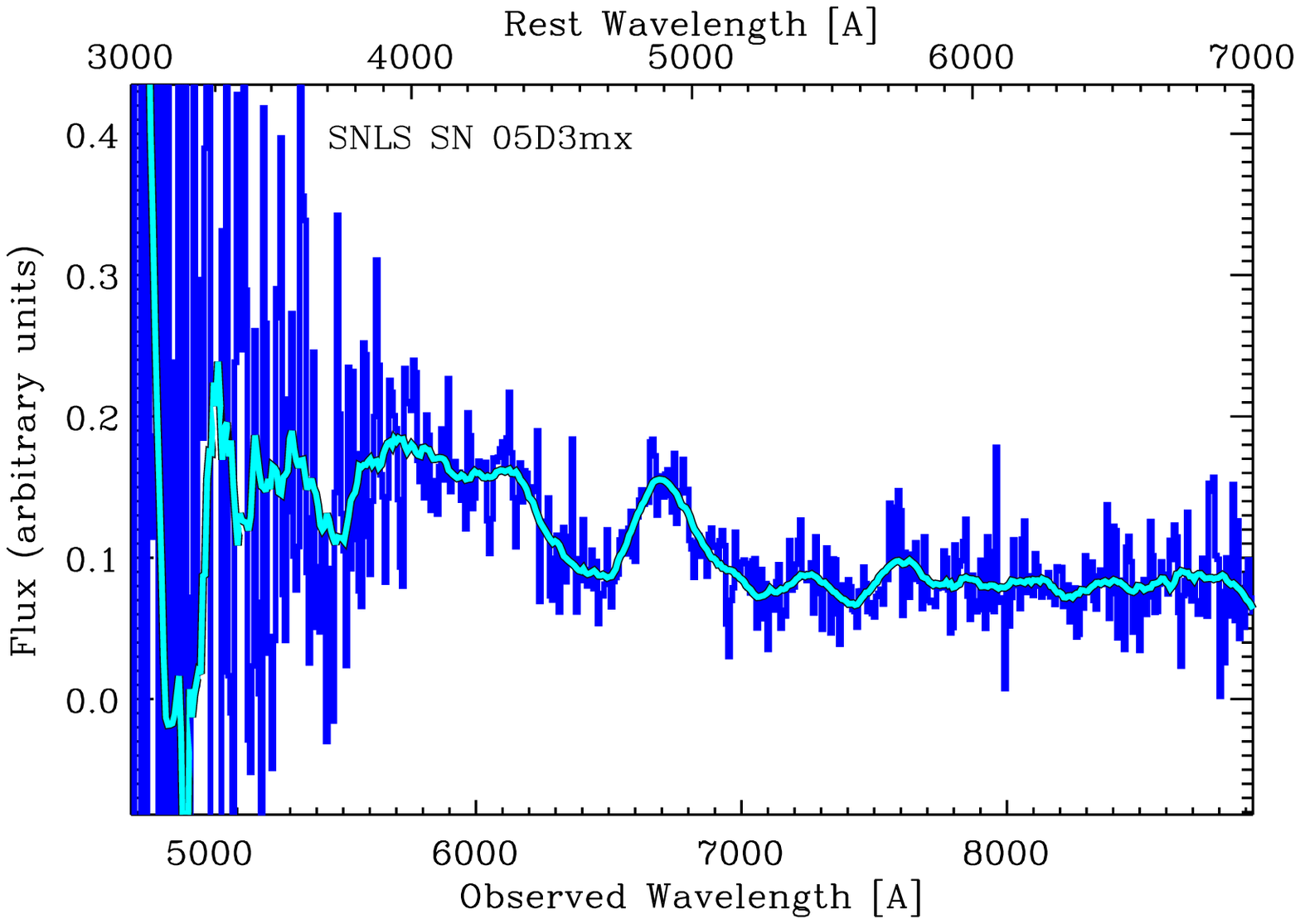} \\
\includegraphics[width=5.25 cm] {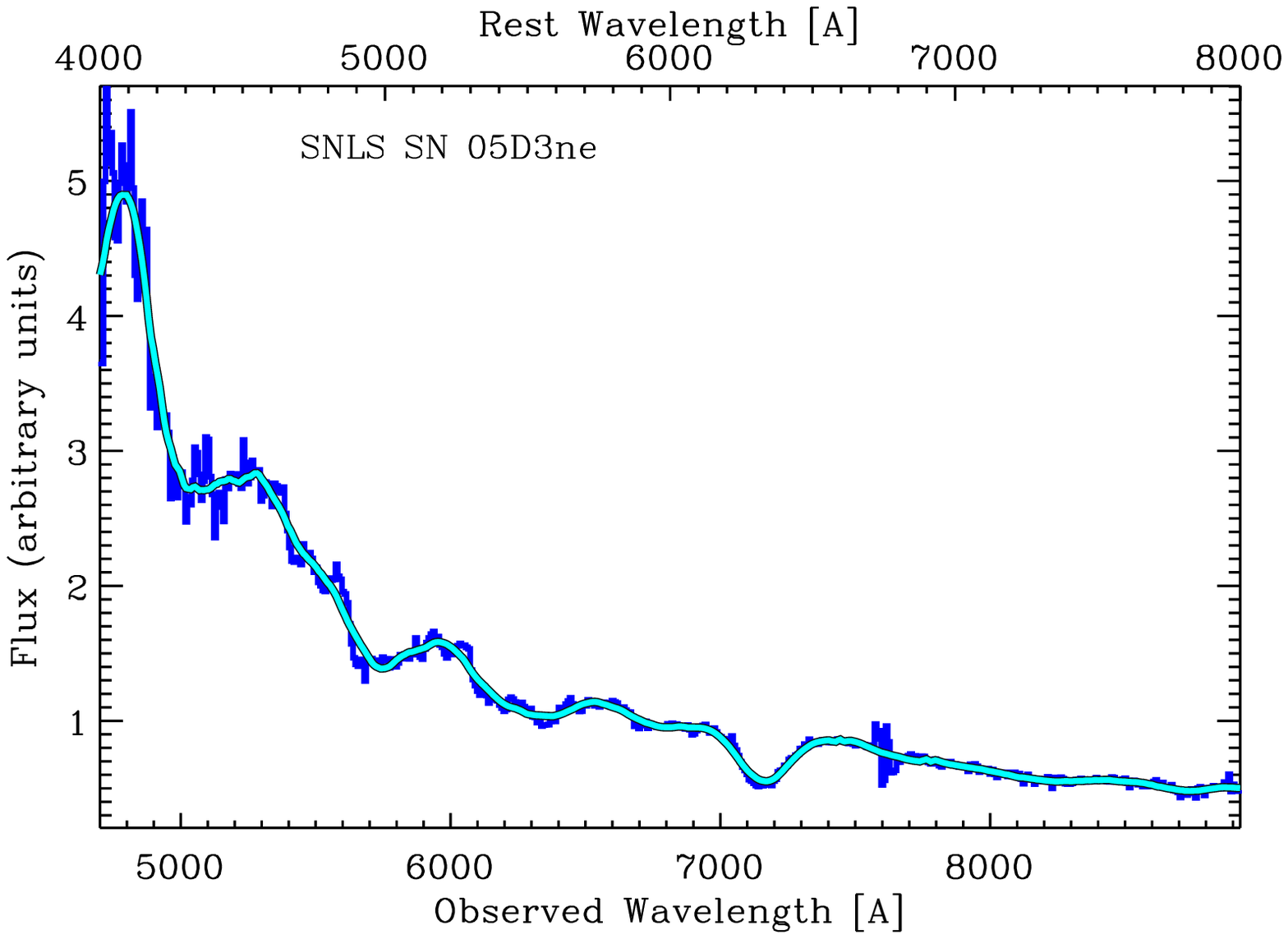} & \includegraphics[width=5.25 cm] {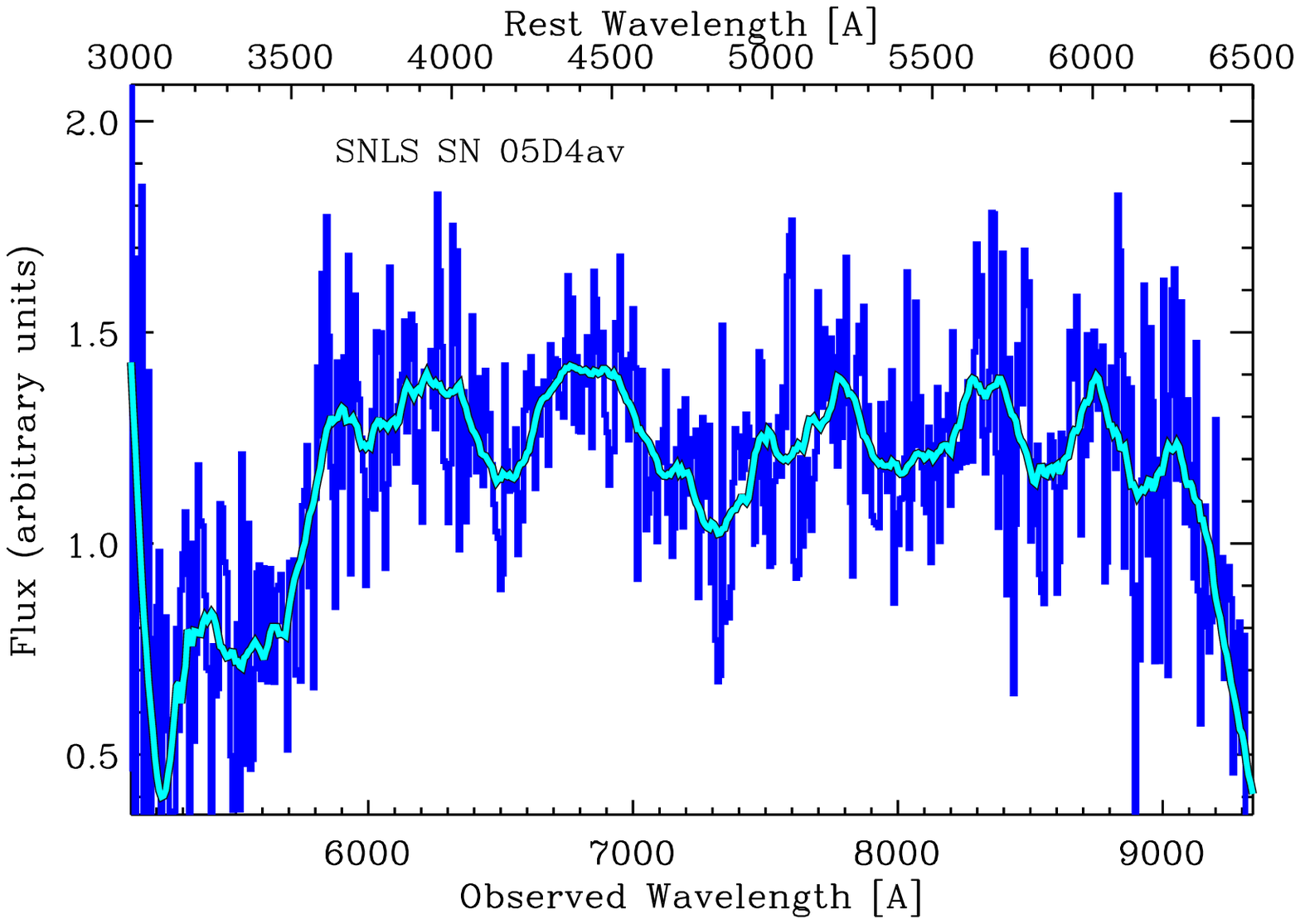} & \includegraphics[width=5.25 cm] {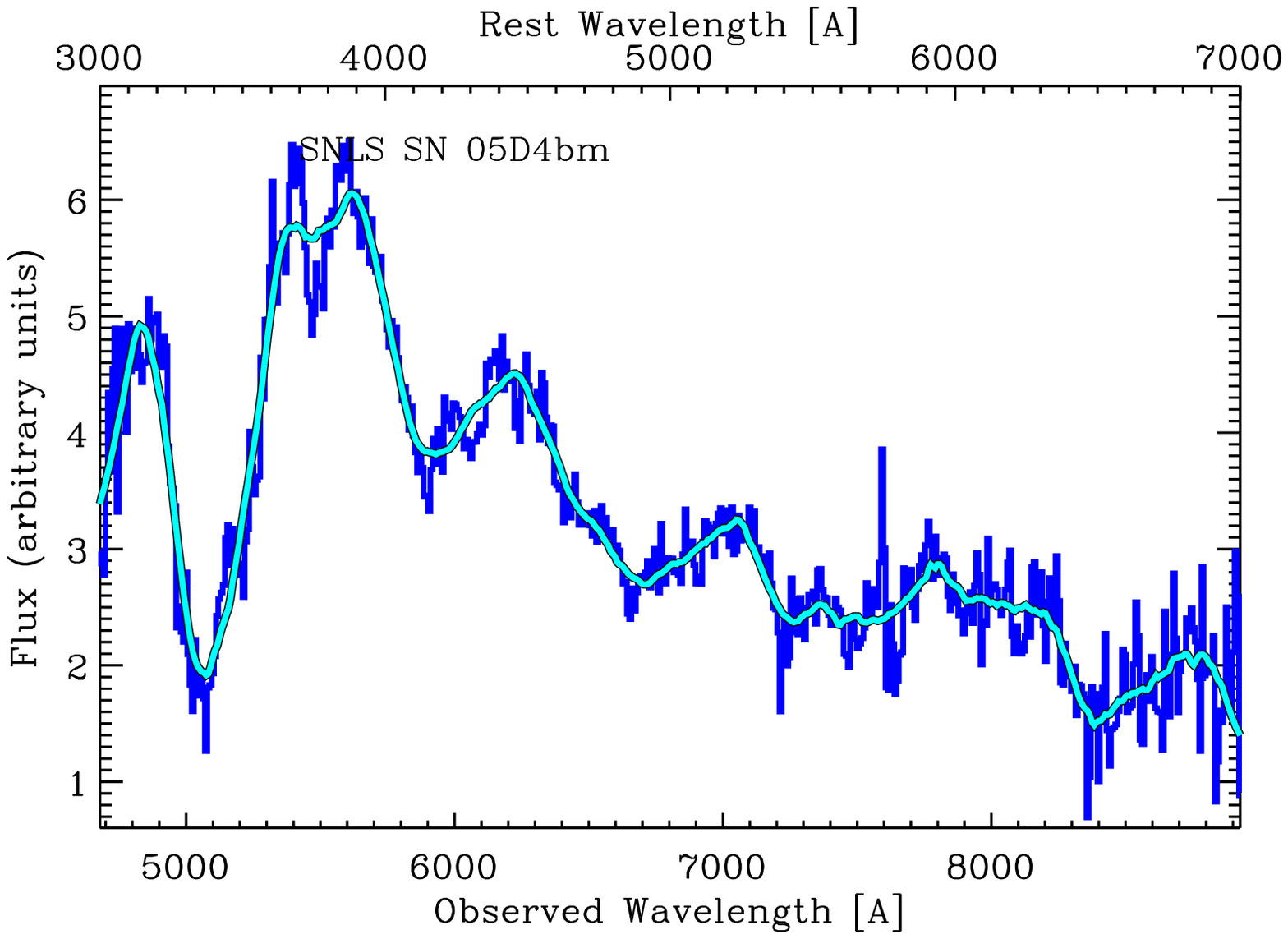} \\
\end{tabular}
\end{minipage}
\caption{Confirmed SNe Ia.  The observed, re-binned spectrum is shown in dark blue and the smoothed spectrum is overplot in light blue for illustrative purposes.}
\end{figure*}

\begin{figure*}
\begin{minipage}{2.0\textwidth}
\begin{tabular}{ccc}
\includegraphics[width=5.25 cm] {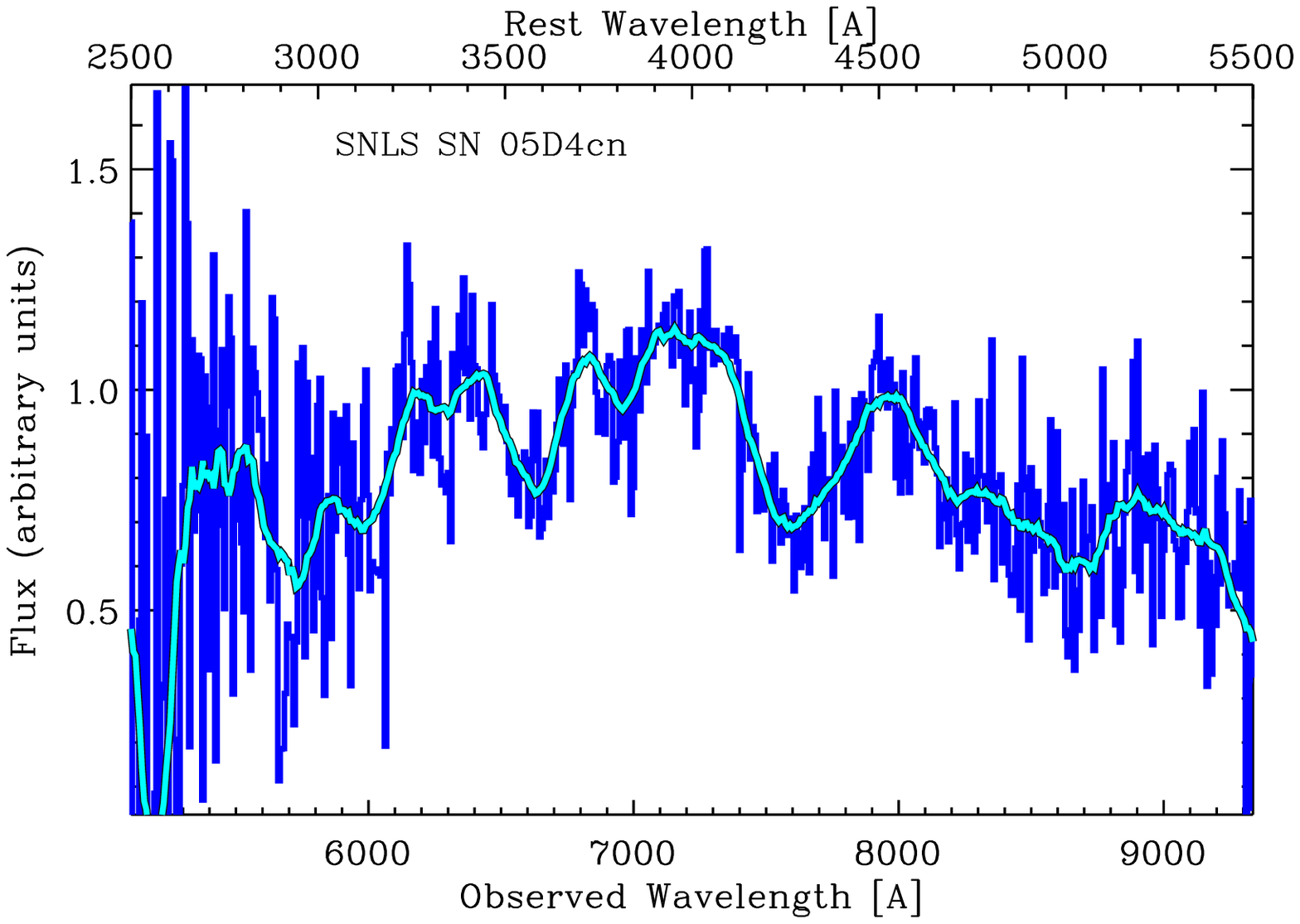} & \includegraphics[width=5.25 cm] {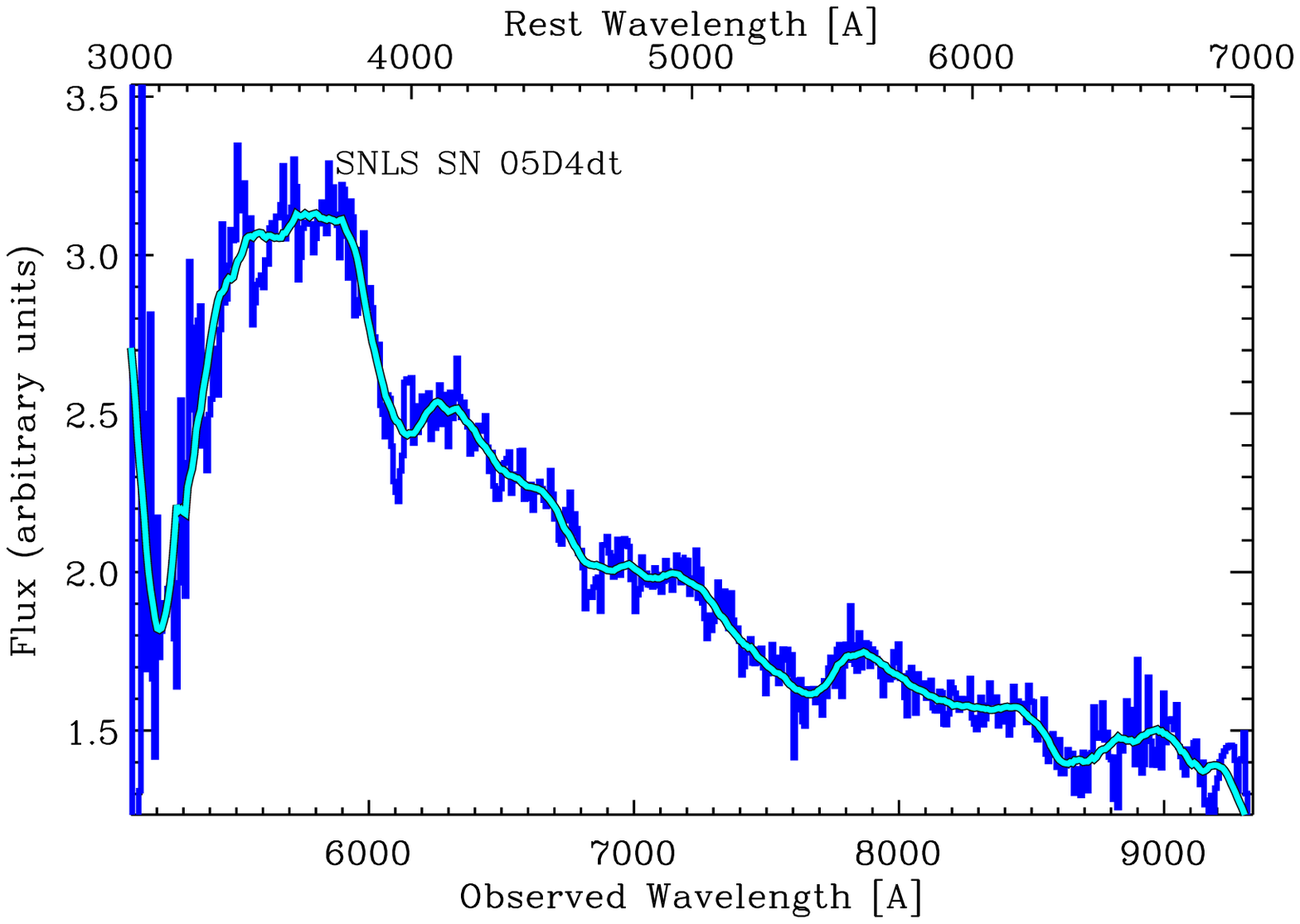} & \includegraphics[width=5.25 cm] {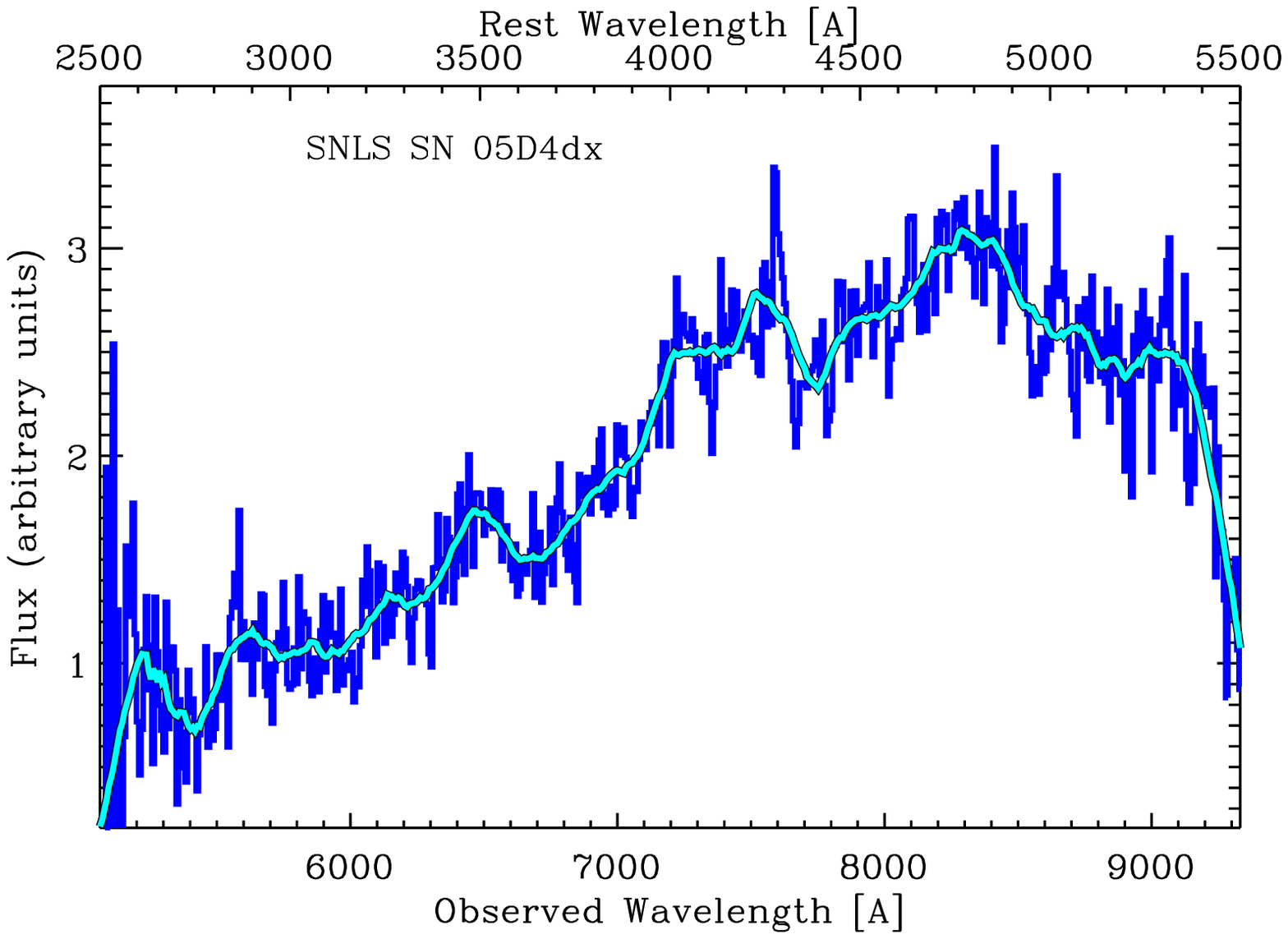} \\
\includegraphics[width=5.25 cm] {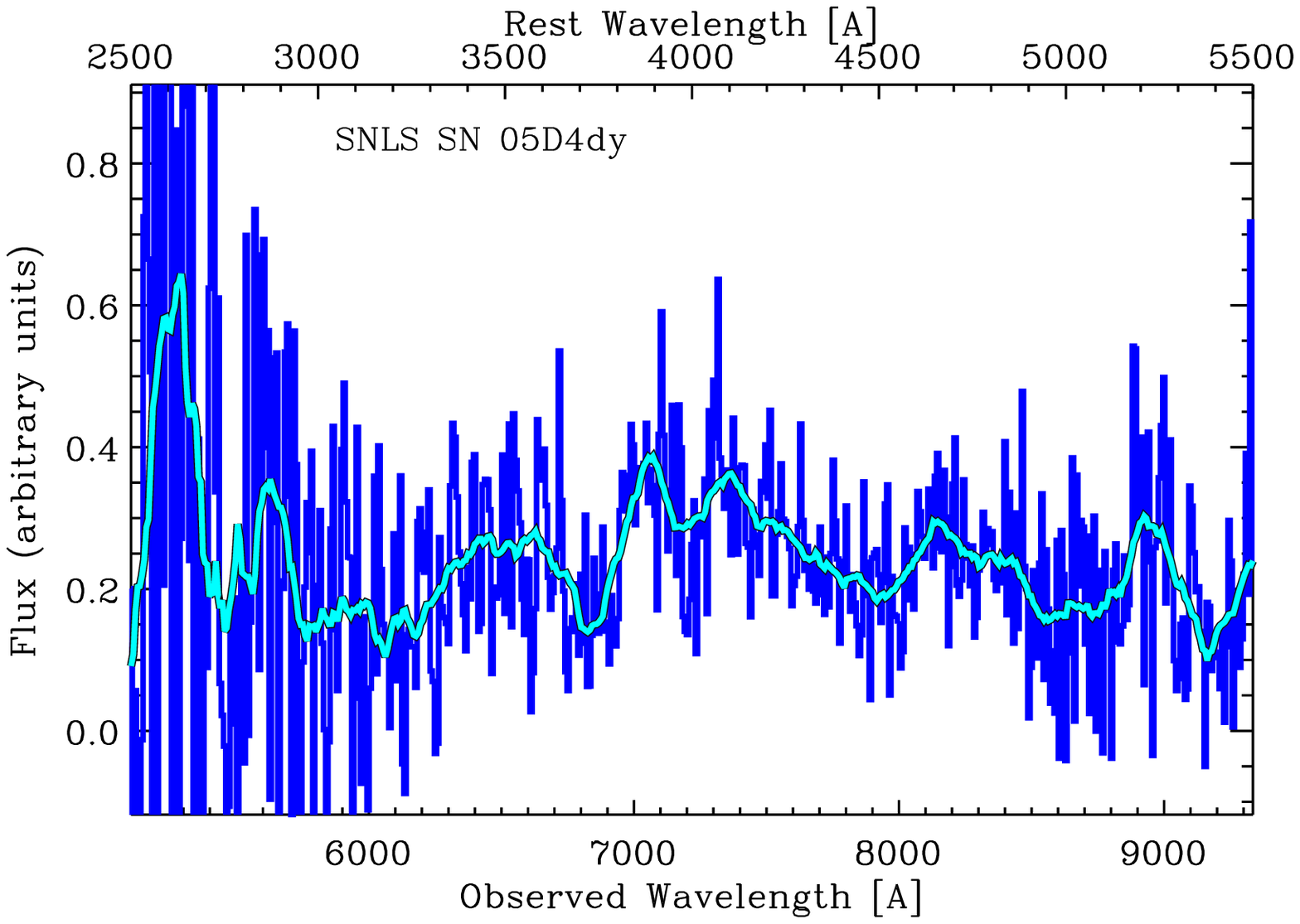} & \includegraphics[width=5.25 cm] {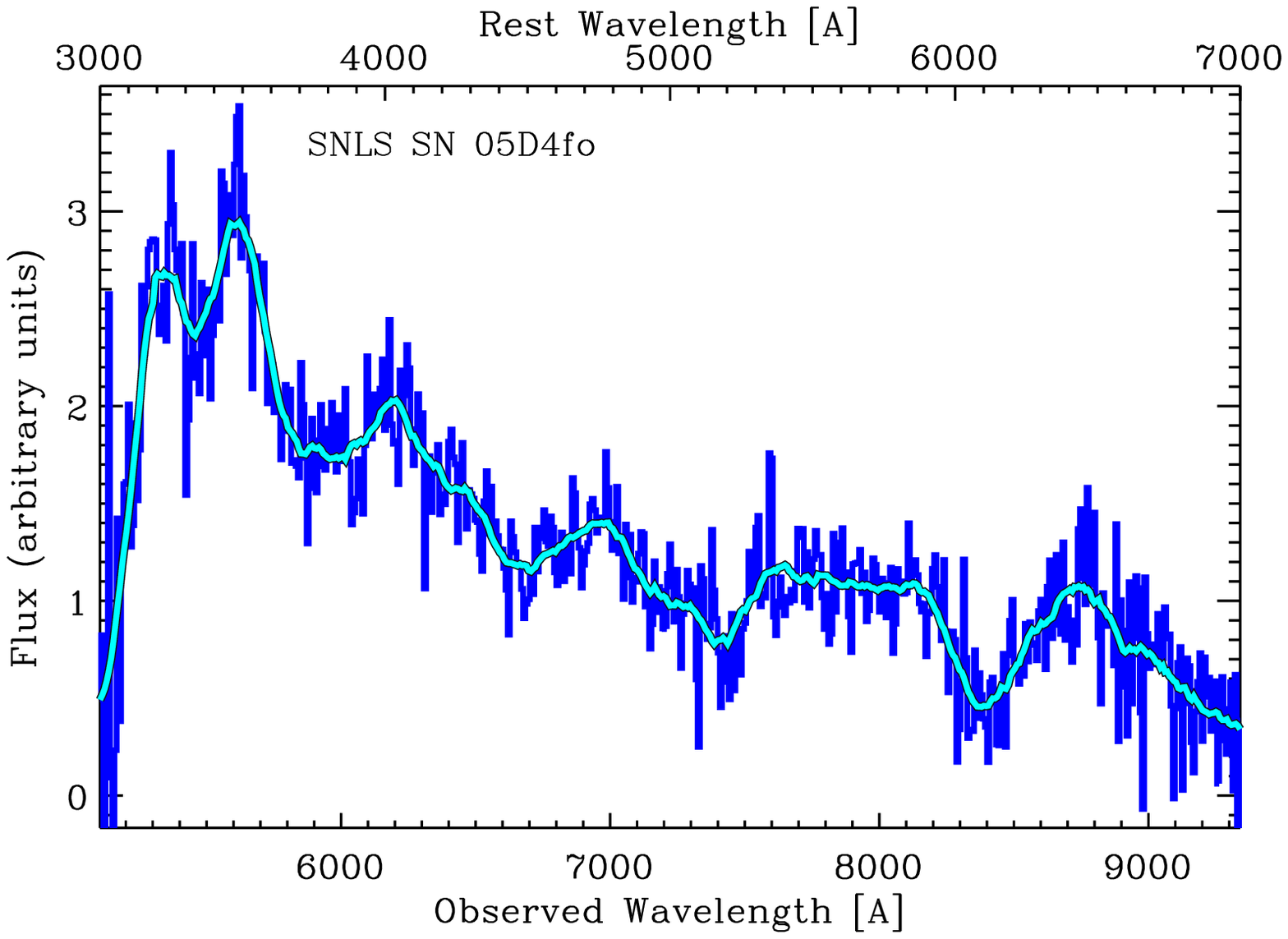} & \includegraphics[width=5.25 cm] {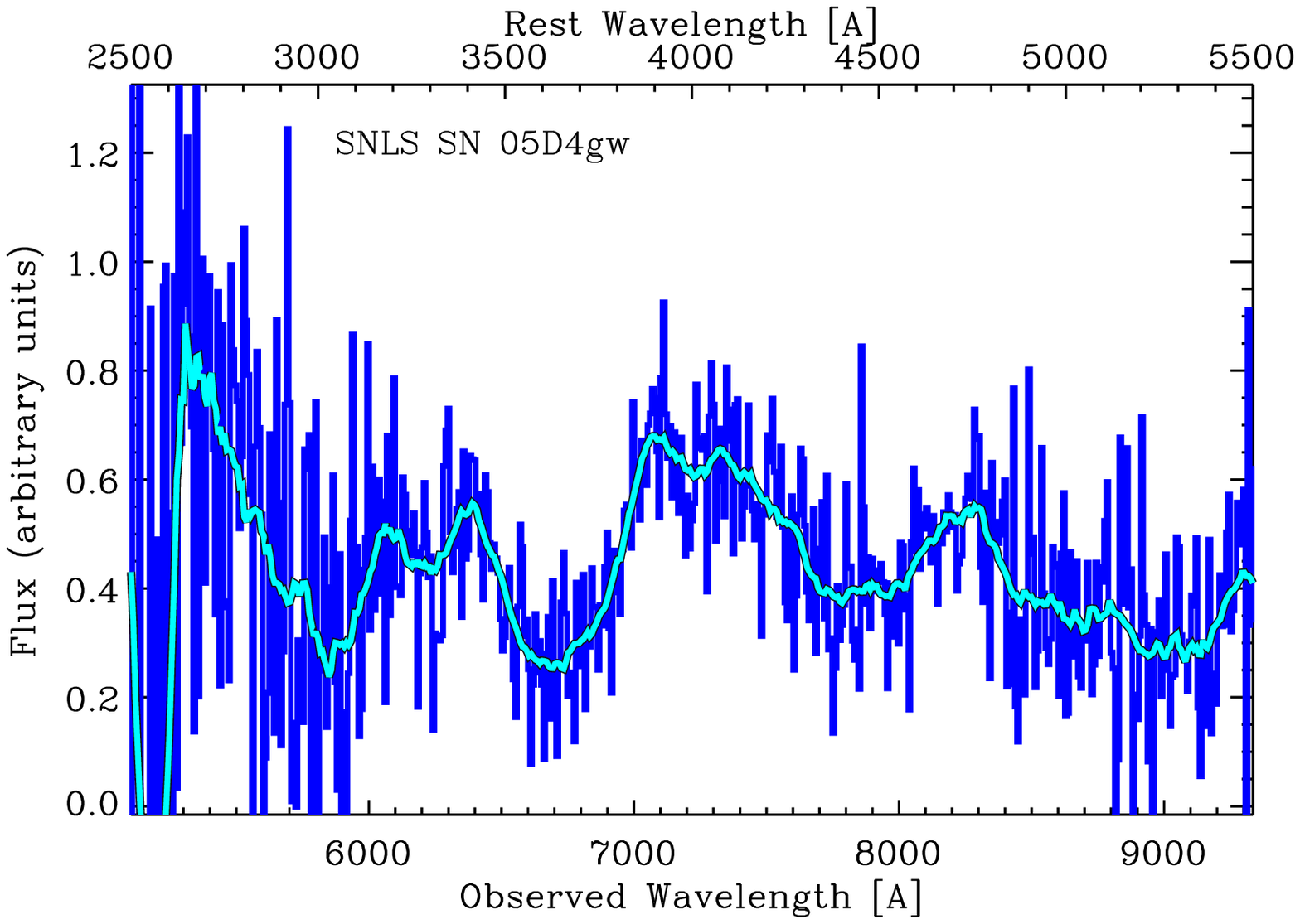} \\
\includegraphics[width=5.25 cm] {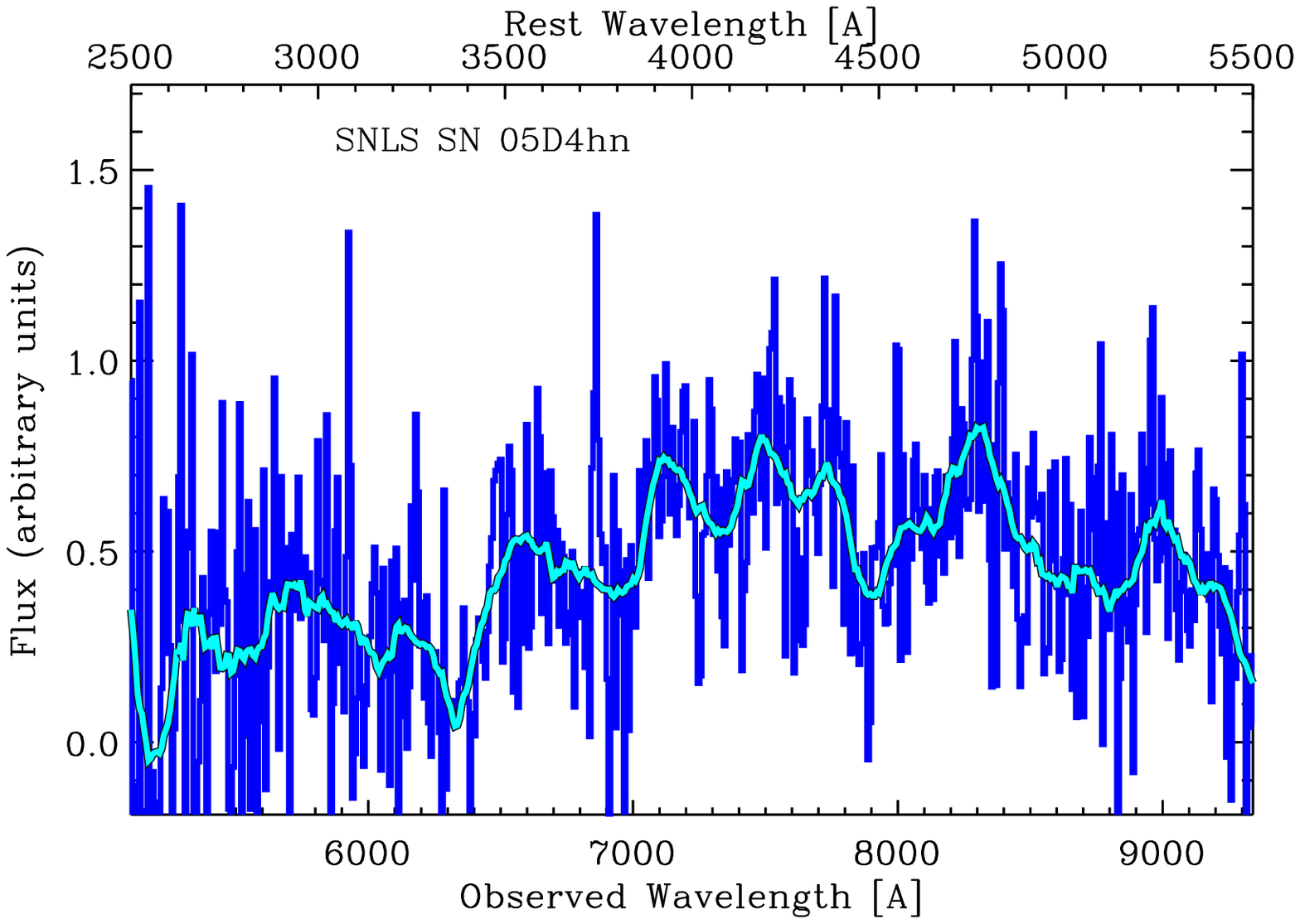} & \includegraphics[width=5.25 cm] {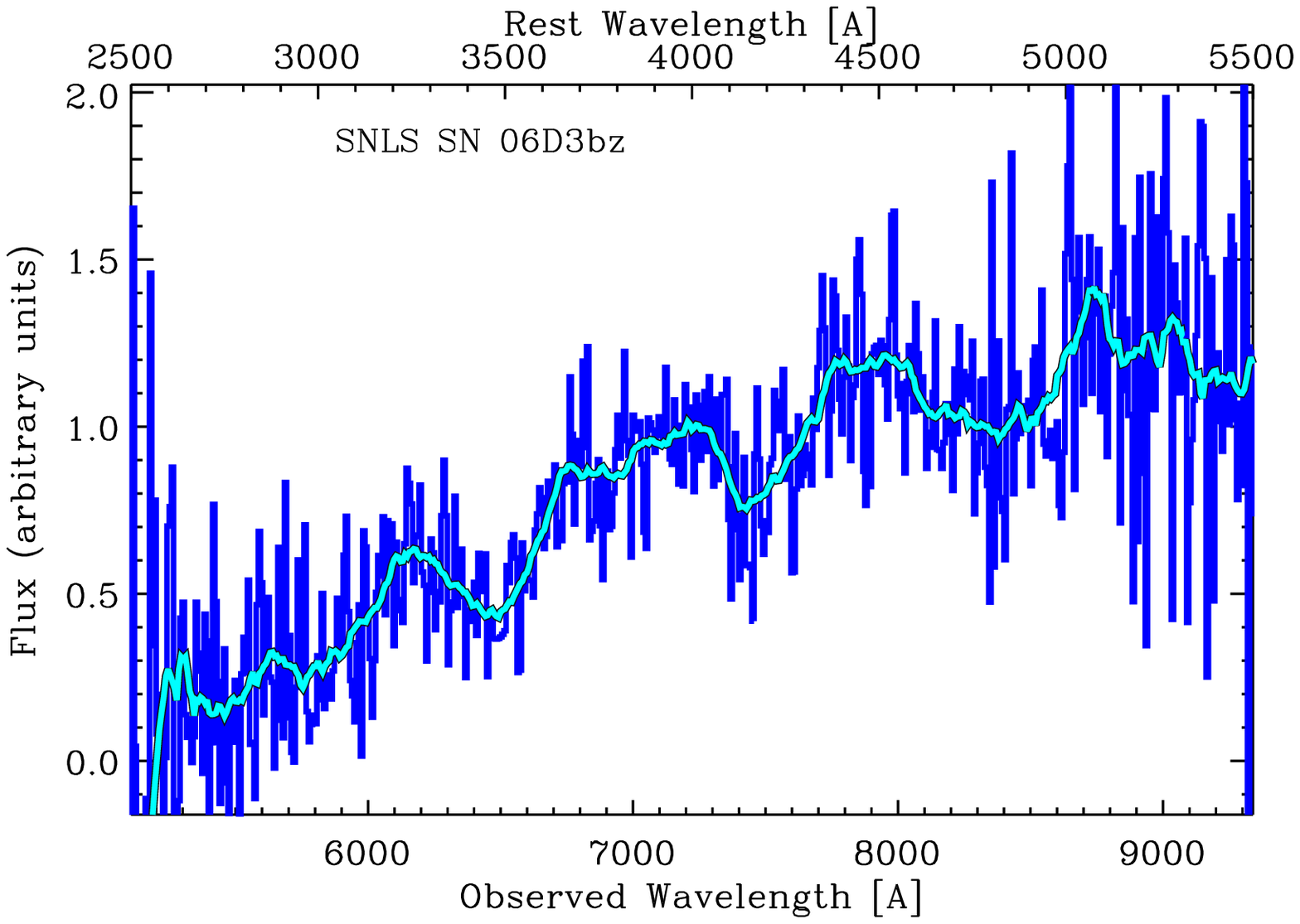} & \includegraphics[width=5.25 cm] {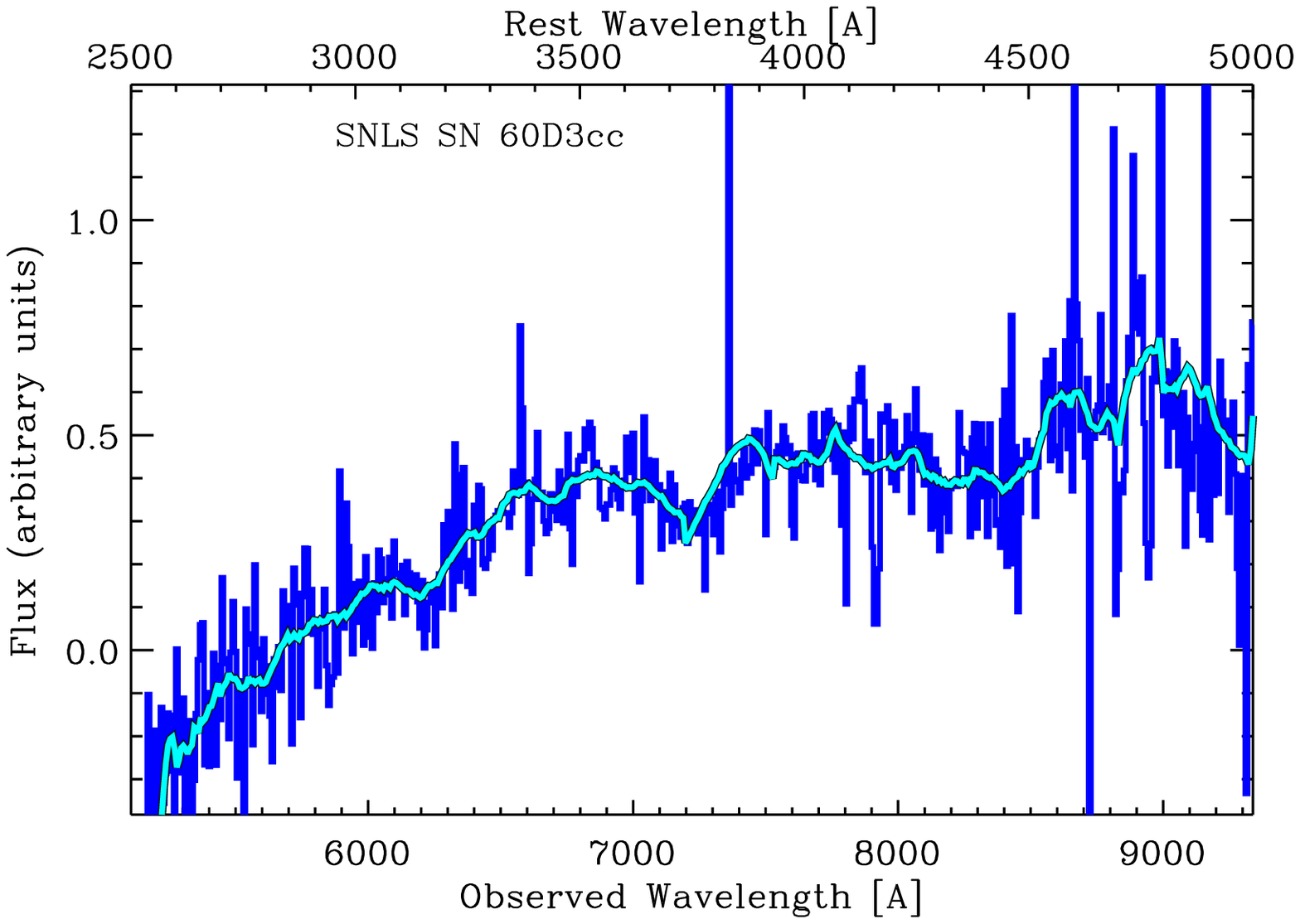} \\
\includegraphics[width=5.25 cm] {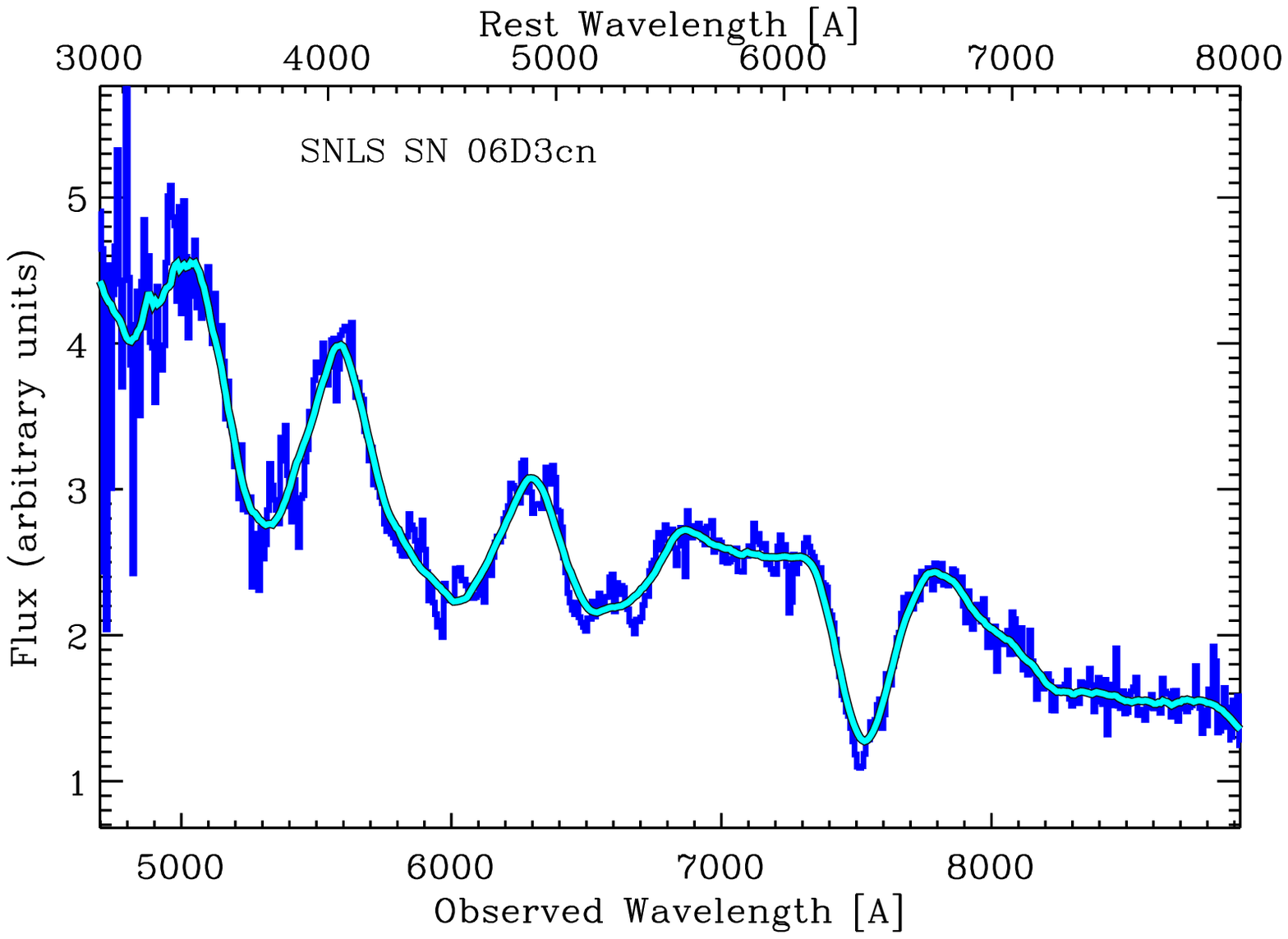} & & \\
\end{tabular}
\end{minipage}
\caption{Confirmed SNe Ia.  The observed, re-binned spectrum is shown in dark blue and the smoothed spectrum is overplot in light blue for illustrative purposes.}
\end{figure*}


\begin{figure*}
\begin{minipage}{2.0\textwidth}
\begin{tabular}{ccc}
\includegraphics[width=5.25 cm] {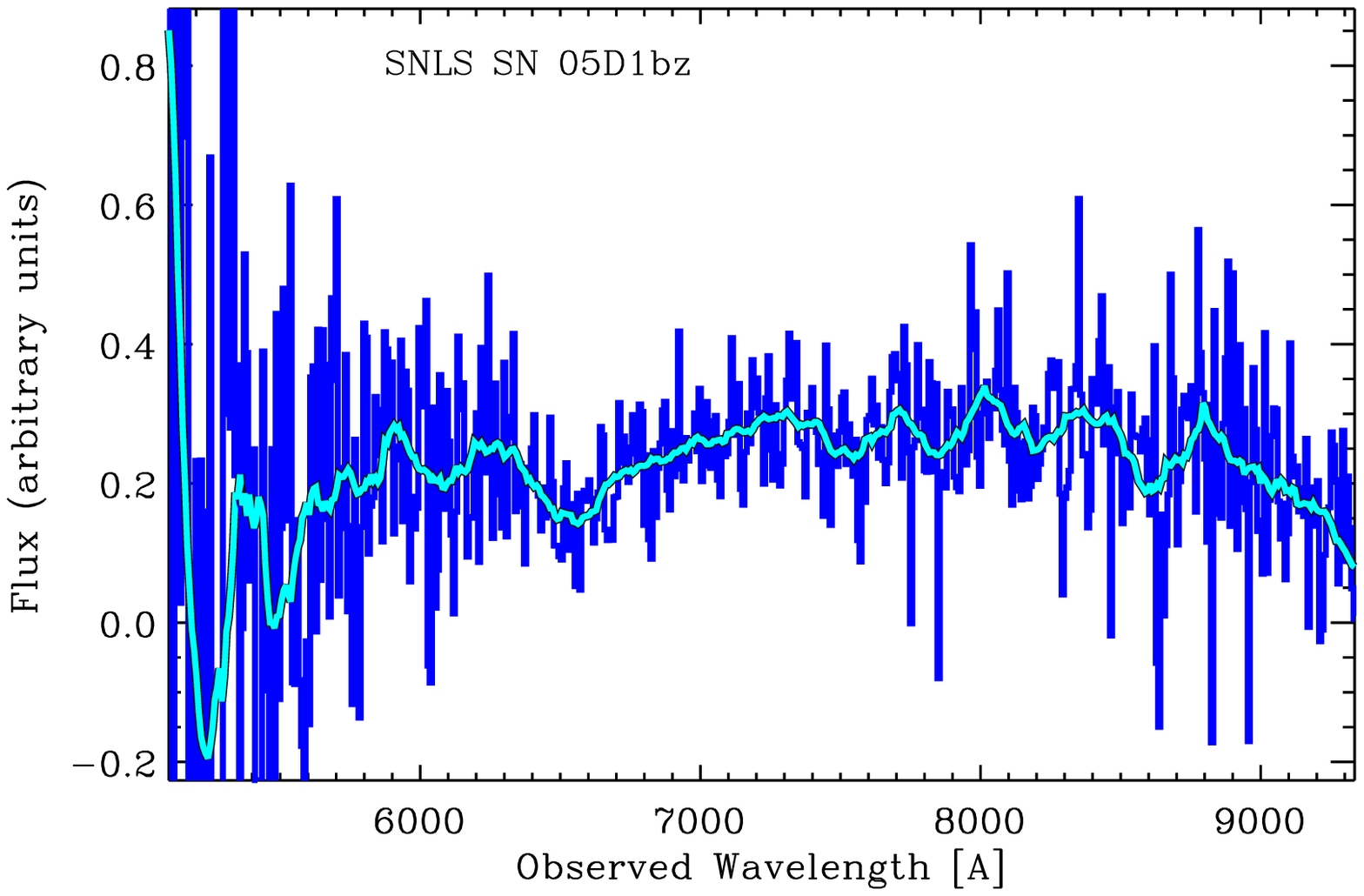} & \includegraphics[width=5.25 cm] {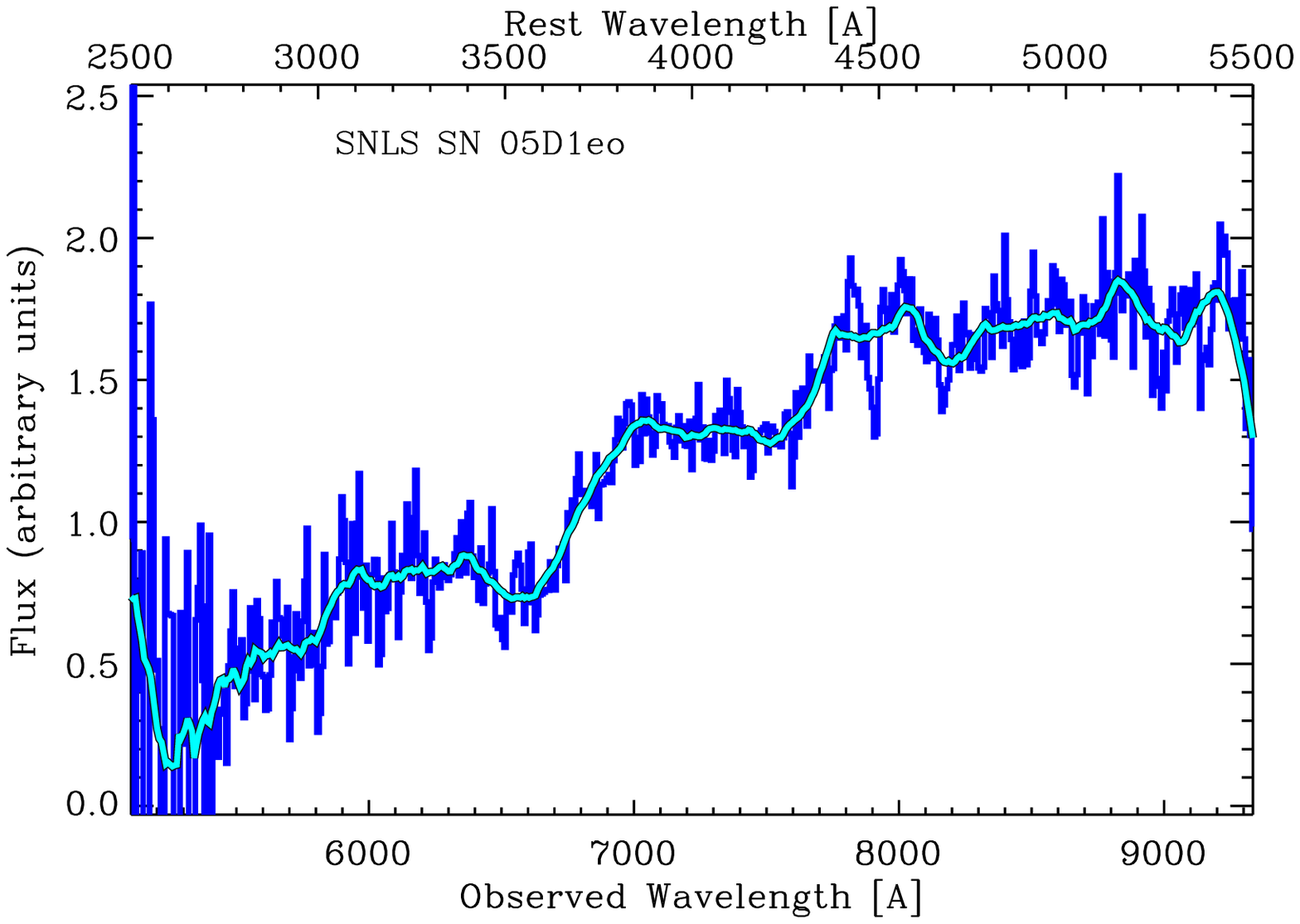} & \includegraphics[width=5.25 cm] {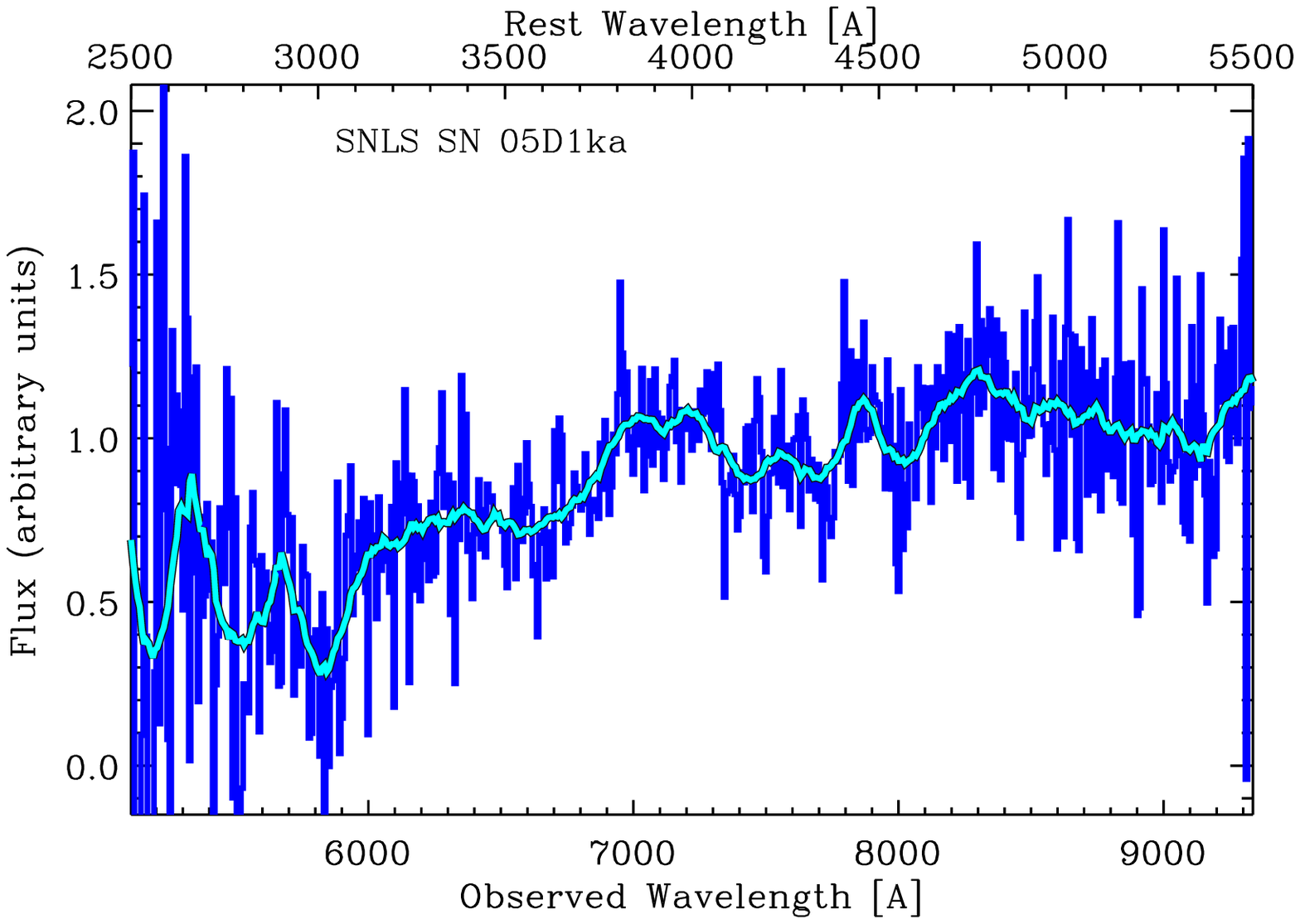} \\
\includegraphics[width=5.25 cm] {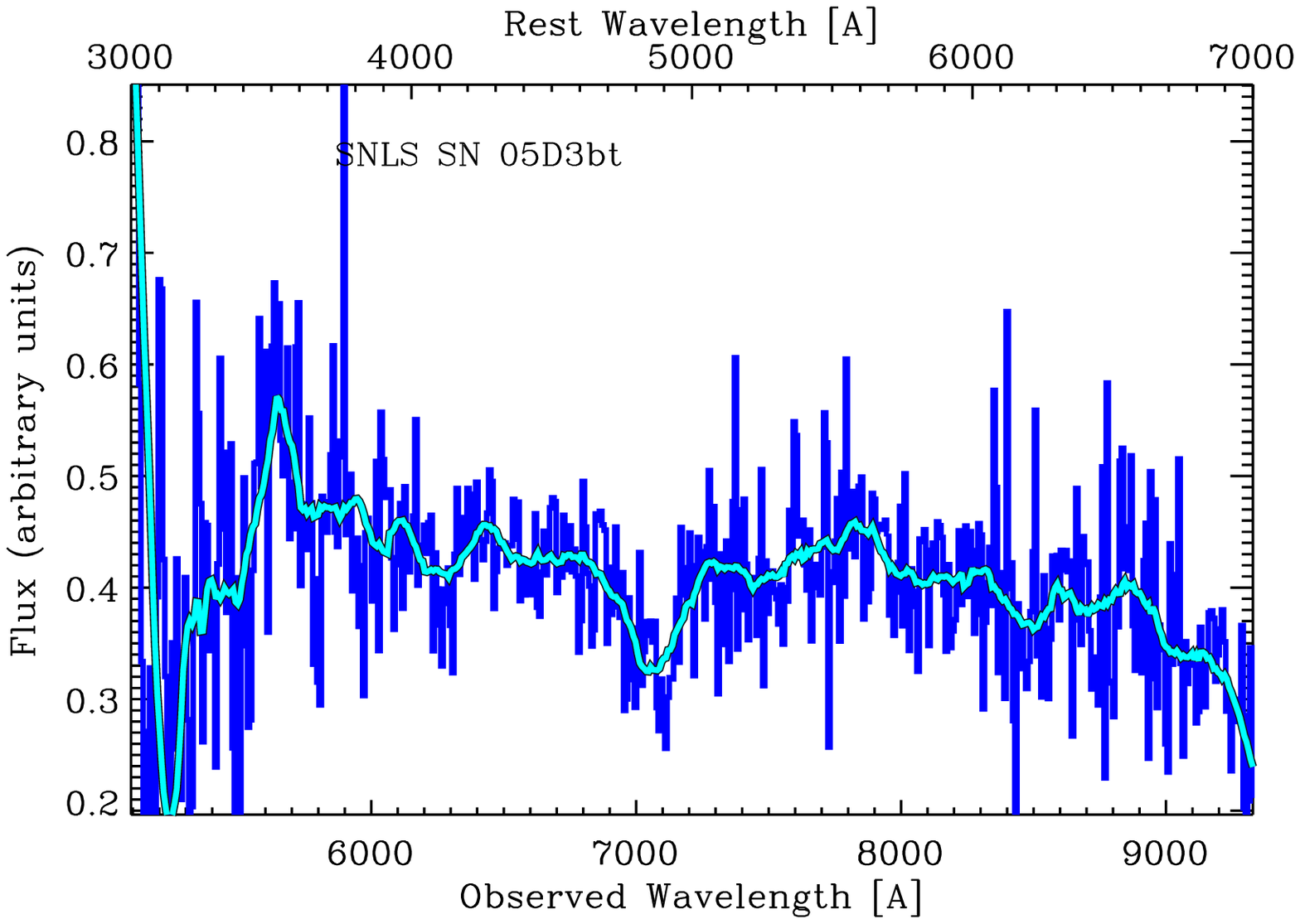} & \includegraphics[width=5.25 cm] {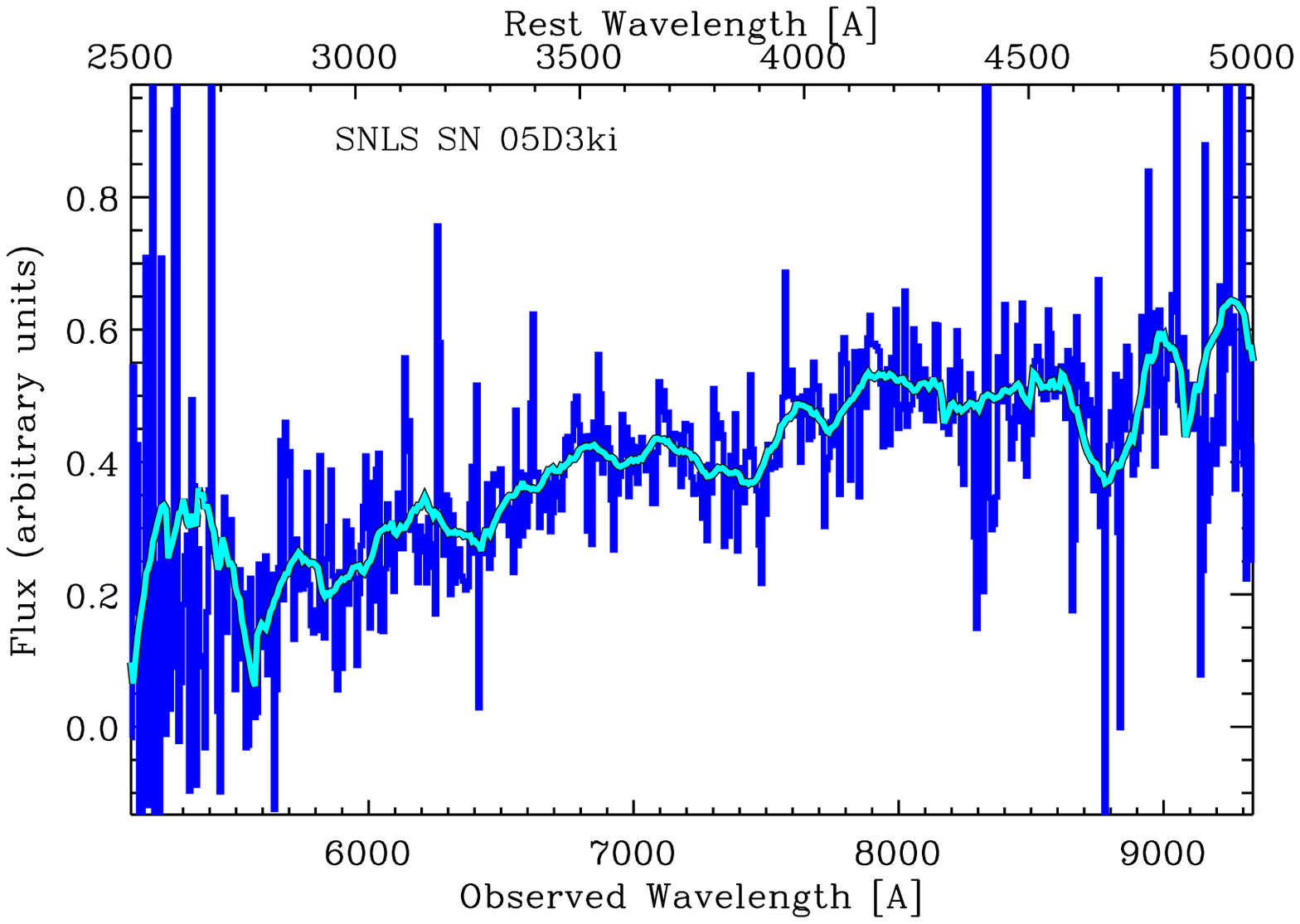} & \includegraphics[width=5.25 cm] {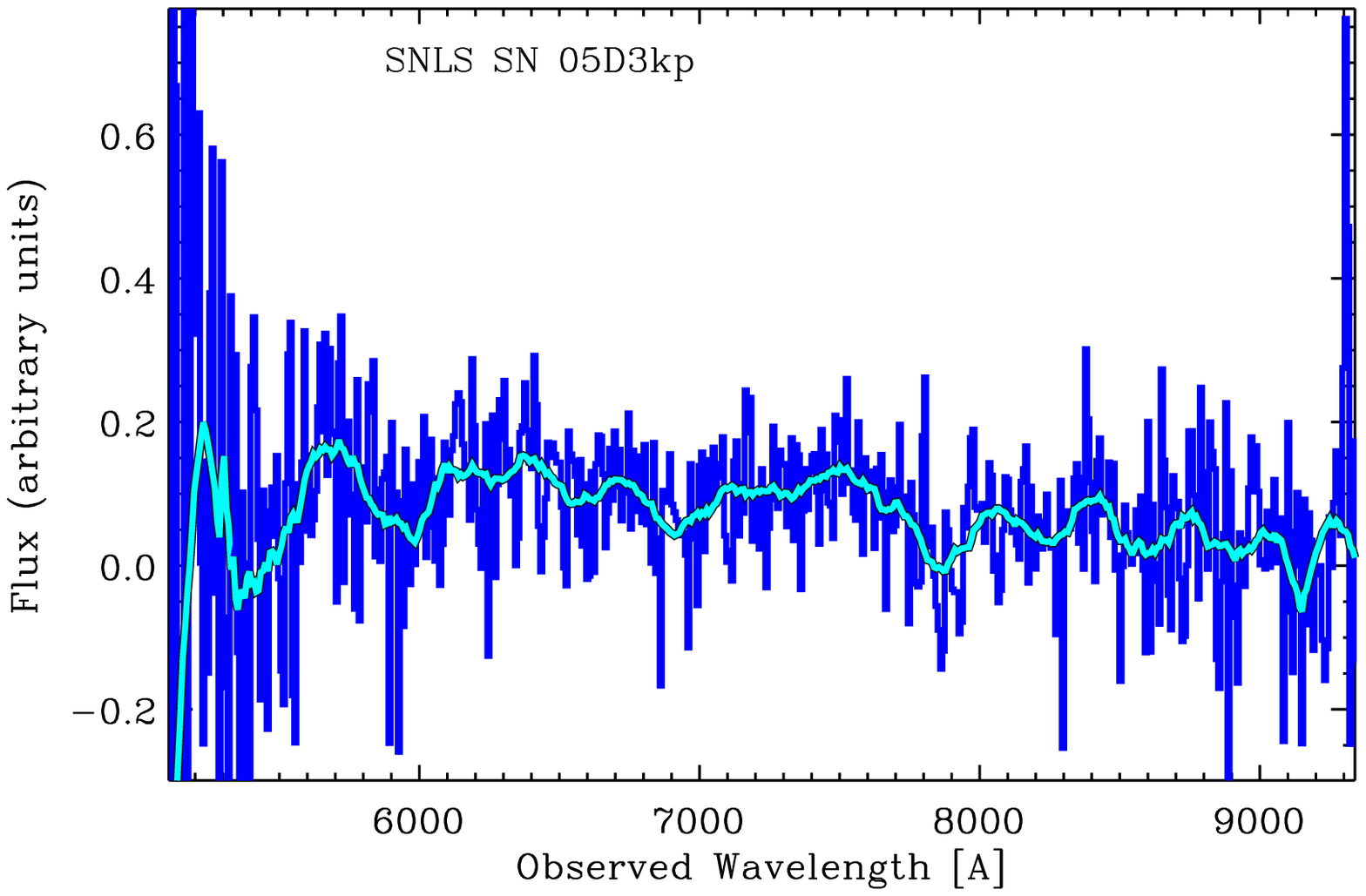} \\
\includegraphics[width=5.25 cm] {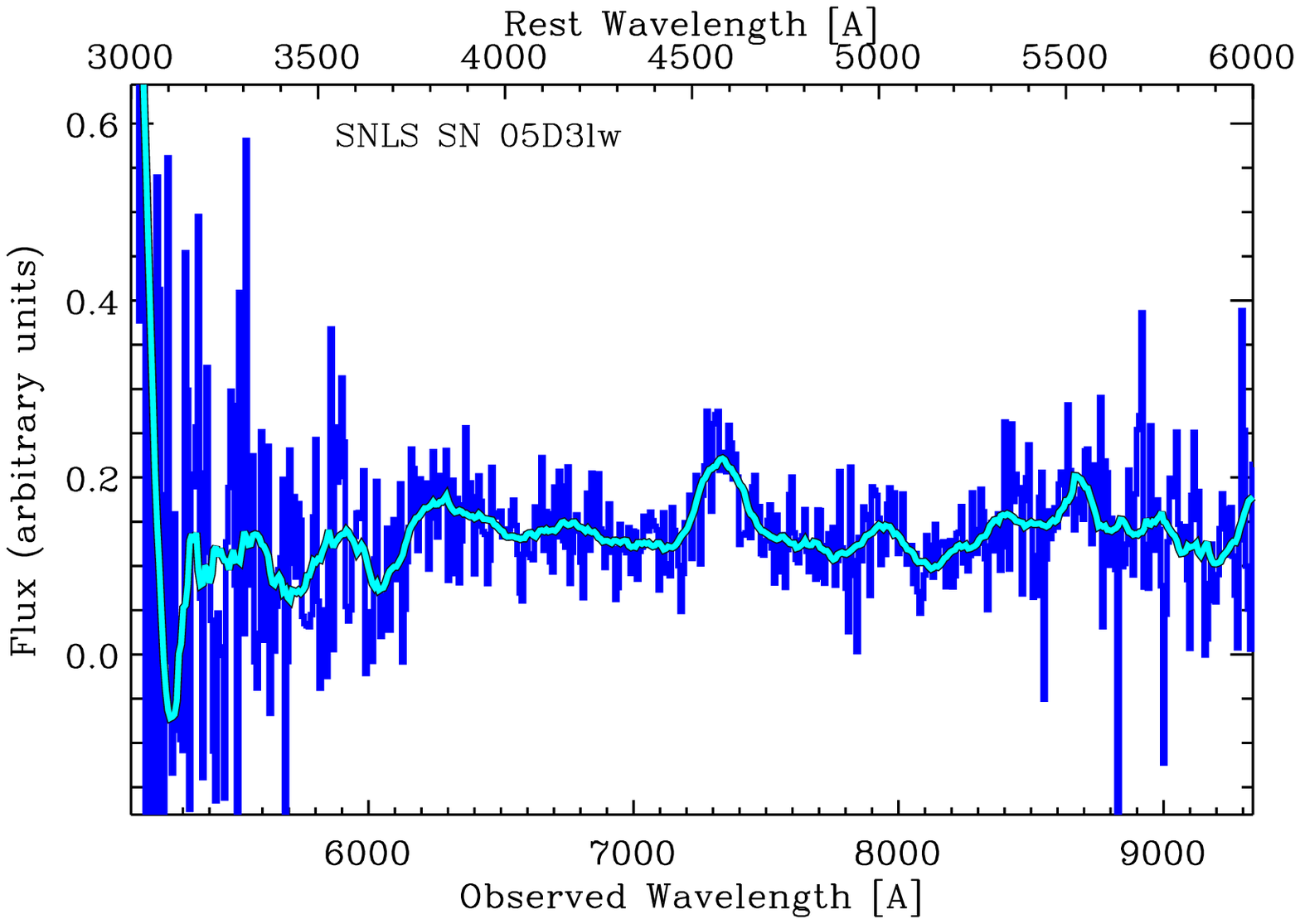} & \includegraphics[width=5.25 cm] {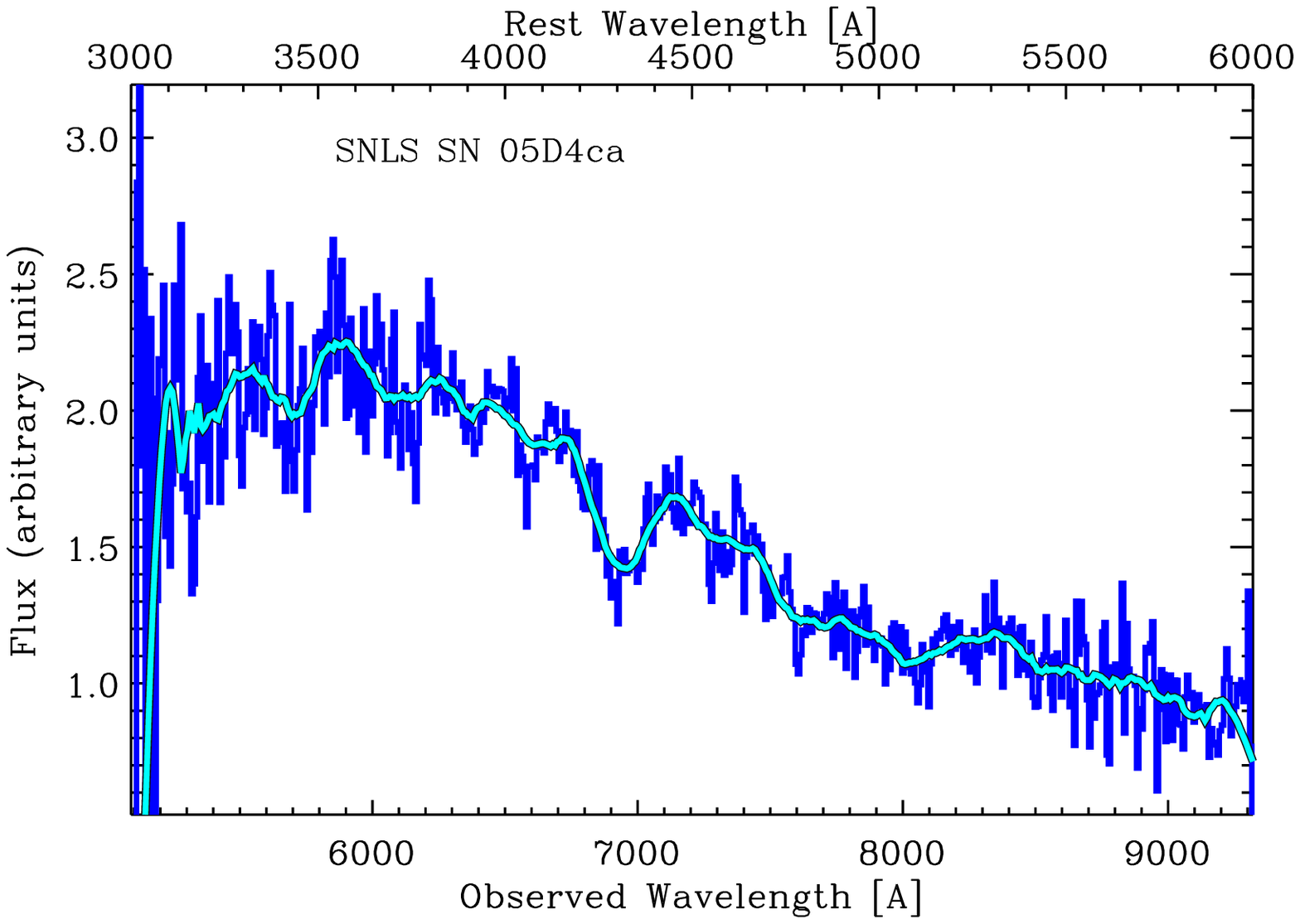} & \includegraphics[width=5.25 cm] {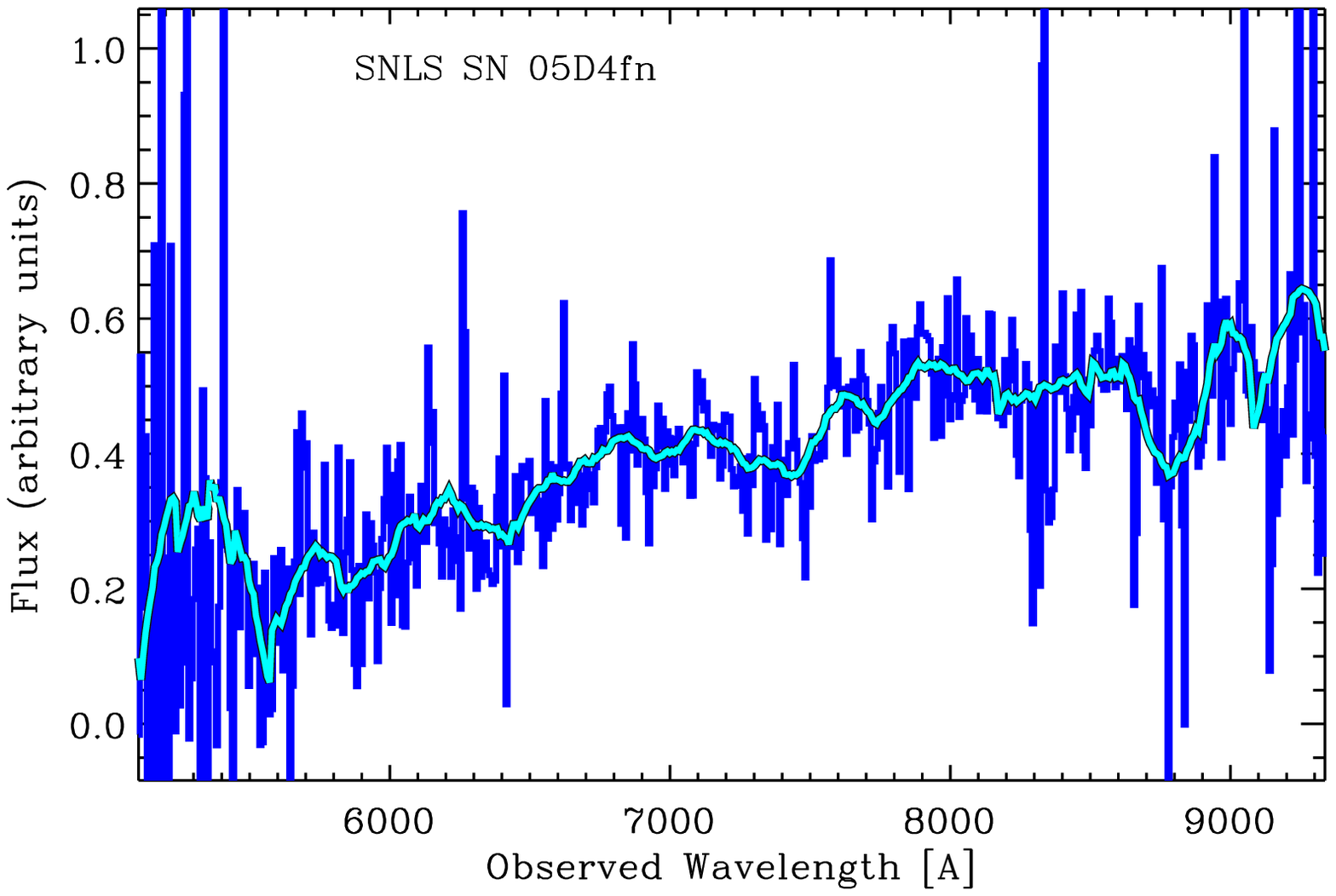} \\
\includegraphics[width=5.25 cm] {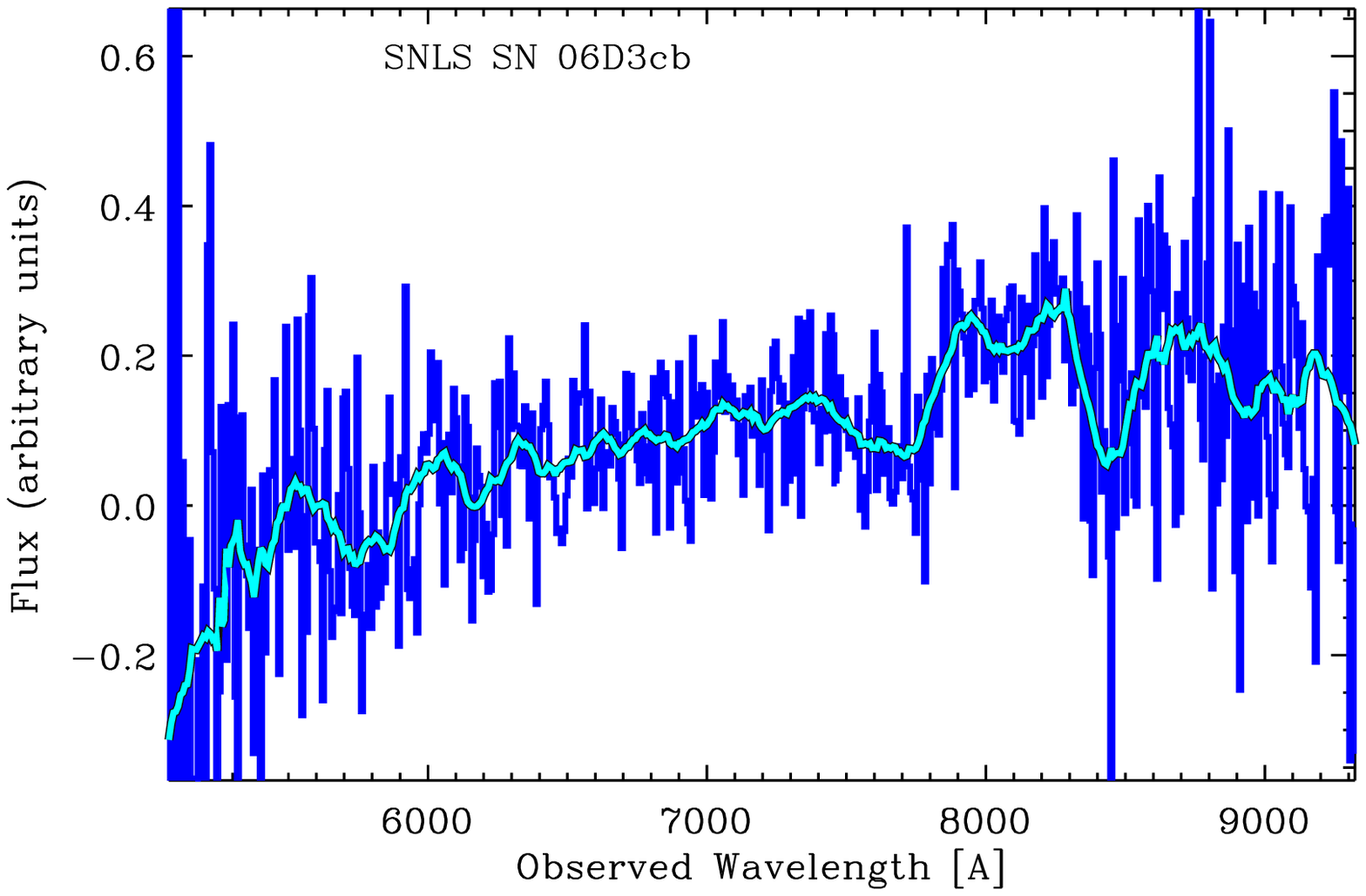} & \includegraphics[width=5.25 cm] {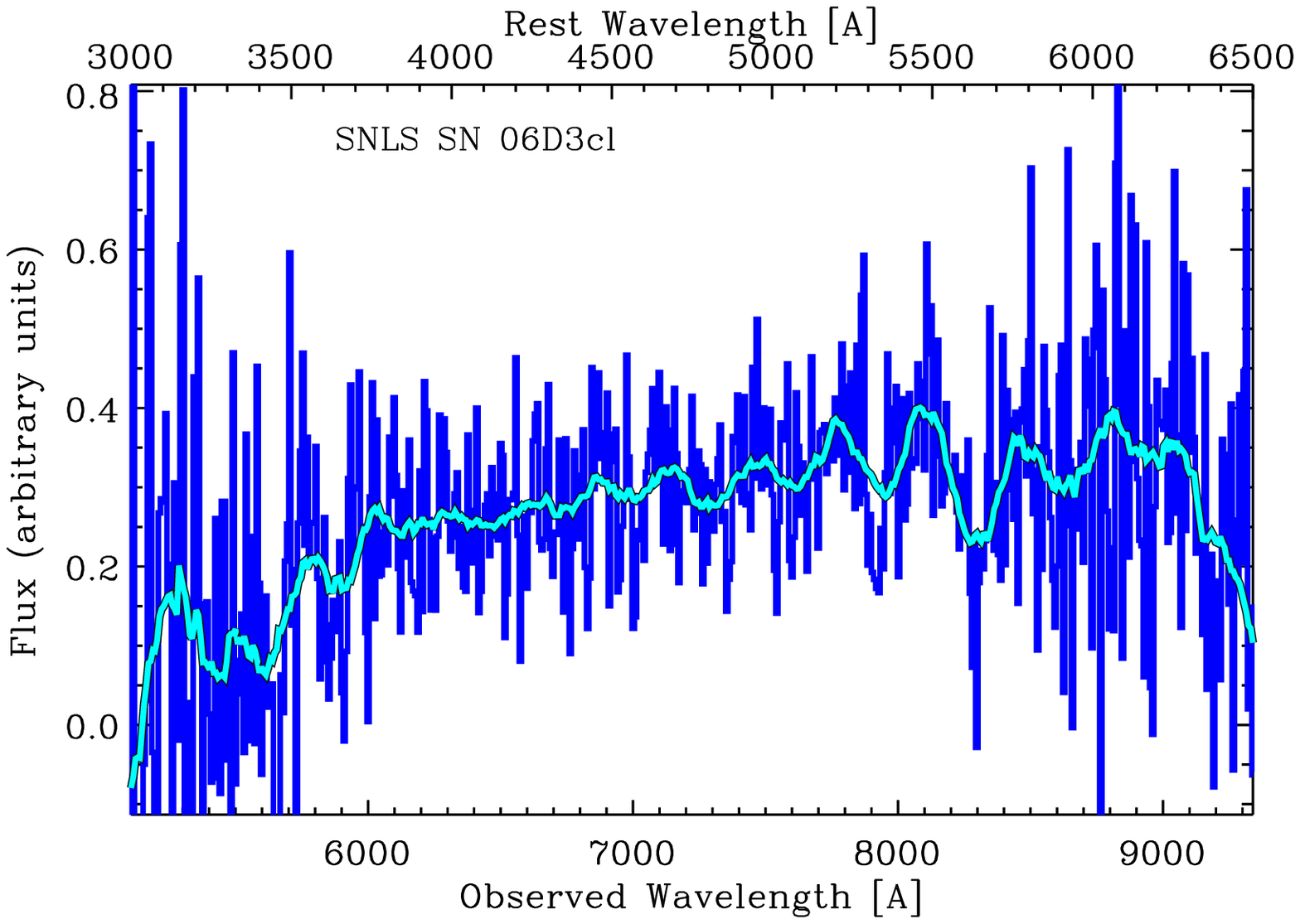} & \includegraphics[width=5.25 cm] {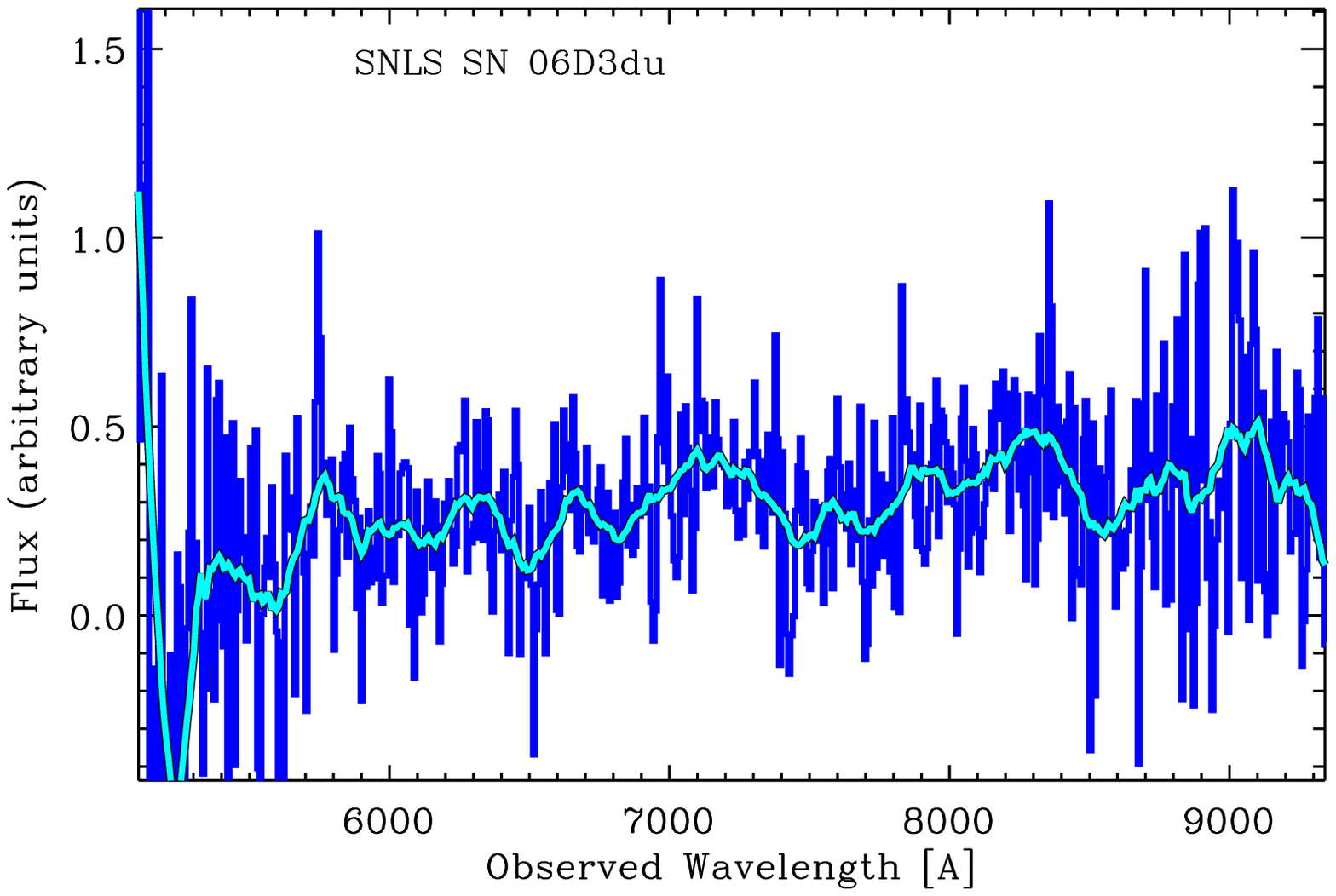} \\
\includegraphics[width=5.25 cm] {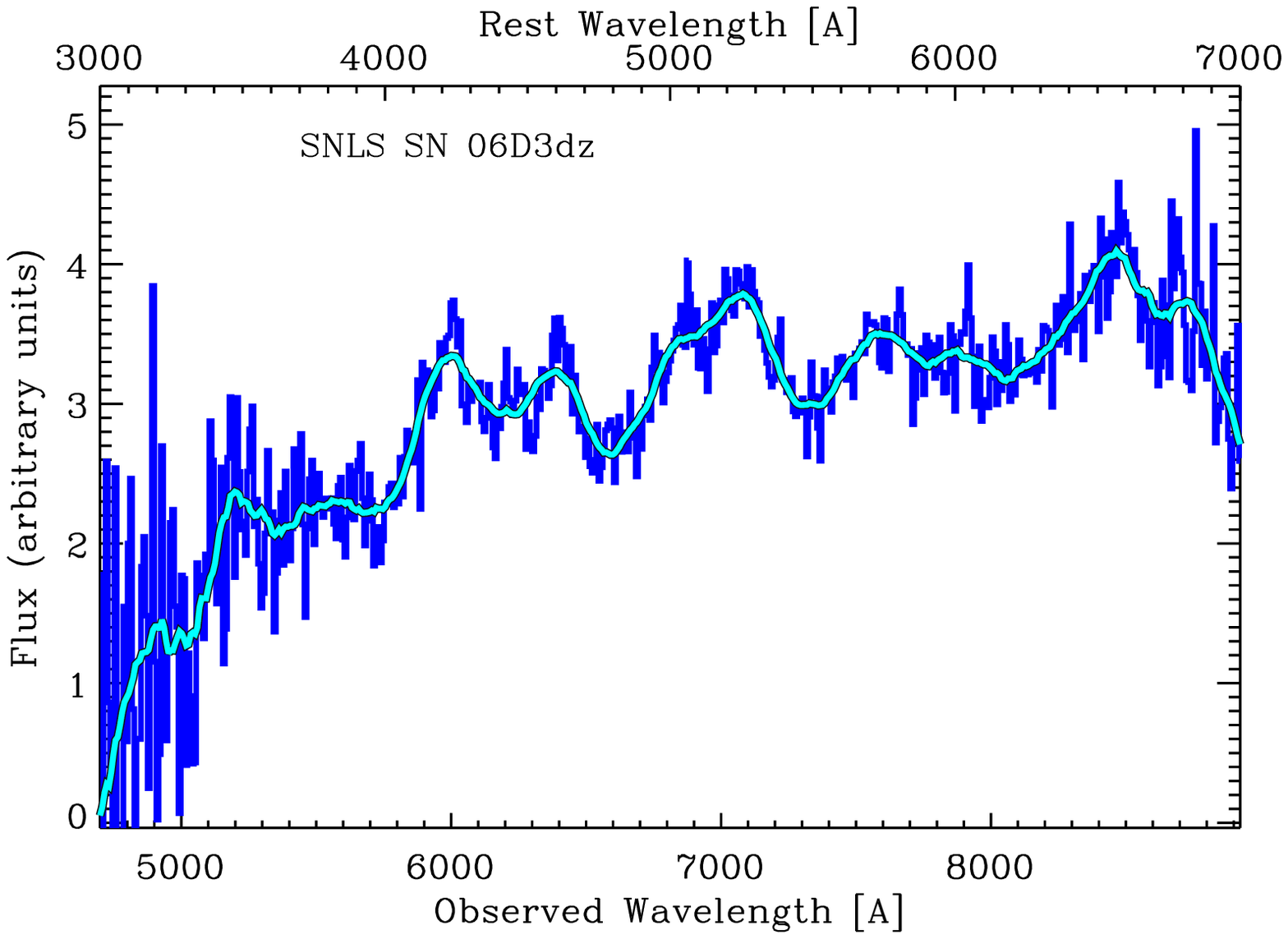} & & \\
\end{tabular}
\end{minipage}
\caption{Unidentified SN candidates.  The observed, re-binned spectrum is shown in dark blue and the smoothed spectrum is overplot in light blue for illustrative purposes.}
\end{figure*}

\end{document}